\documentclass[onecolumn]{revtex4}
\usepackage{graphicx}
\usepackage{dcolumn}
\usepackage{bm}
\usepackage{color}
\usepackage{soul}
\usepackage[english]{babel}
\usepackage{epsfig}
\usepackage{amsmath}
\usepackage{amsfonts}
\usepackage{amssymb}
\usepackage{float}
\usepackage{mathrsfs}
\usepackage{tabularx}
\usepackage{comment}

\begin{document}

\title{Vector breathers in the Manakov system}

\author{Andrey Gelash $^{1,2}$}\email{agelash@gmail.com}
\author{Anton Raskovalov $^{1,3,4}$}

\affiliation{$^{1}$Skolkovo Institute of Science and Technology, Moscow, 121205, Russia}
\affiliation{$^{2}$Institute of Automation and Electrometry SB RAS, Novosibirsk, 630090, Russia}
\affiliation{$^{4}$Mikheev Institute of Metal Physics, Ural Branch, RAS, Ekaterinburg, 620108, Russia}
\affiliation{$^{3}$ Institute of Physics and Technology, Ural Federal University, Ekaterinburg, 620002, Russia.}

\begin{abstract}
We study theoretically the nonlinear interactions of vector breathers propagating on an unstable wavefield background. As a model, we use the two-component extension of the one-dimensional focusing nonlinear Schrödinger equation -- the Manakov system. With the dressing method, we generate the multi-breather solutions to the Manakov model. As shown previously  in [D. Kraus, G. Biondini, and G. Kova{\v{c}}i{\v{c}}, Nonlinearity 28, 310, 2015] the class of vector breathers are presented by three fundamental types I, II, and III. Their interactions produce a broad family of the two-component (polarized) nonlinear wave patterns. First we demonstrate that the type I and the types II and III correspond to two different branches of the dispersion law of the Manakov system in the presence of the unstable background. Then we investigate the key interaction scenarios, including collisions of standing and moving breathers and resonance breather transformations. Analysis of the two-breather solution allows us to derive general formulas describing phase and space shifts acquired by breathers in mutual collisions. The found expressions enable us to describe the asymptotic states of the breather interactions and interpret the resonance fusion and decay of breathers as a limiting case of infinite space shift in the case of merging breather eigenvalues. Finally, we demonstrate that only type $\mathrm{I}$ breathers participate in the development of modulation instability from small-amplitude perturbations withing the superregular scenario, while the breathers of types $\mathrm{II}$ and $\mathrm{III}$, belonging to the stable branch of the dispersion law, are not involved in this process.
\end{abstract}

\maketitle

\section{Introduction}\label{Sec:1}

Breathers are coherent nonlinear pulsating wave structures living on an unstable background, which theoretical description represents a generalization of the soliton theory \cite{NovikovBook1984,AblowitzBook1981,akhmediev1997nonlinear}. The interest in studies breathers is both theoretical and practical. On one side, these nonlinear wave groups can be described by exactly solvable, i.e., integrable models, for example by the one-dimensional focusing nonlinear Schrödinger equation (NLSE) \cite{zakharov1972exact,kuznetsov1977solitons,kawata1978inverse,ma1979perturbed,NovikovBook1984}. As such the class of breather solutions describes an essential part of the integrable system dynamics. On the other side, the breathers model is applicable in a wide range of physical systems as diverse as light in optical fibers, fluids, plasma and Bose-Einstein condensates \cite{kivshar2003optical,OsborneBook2010,maimistov2013nonlinear}. Many experiments have confirmed the existence of breathers in nature, encouraging theoreticians to predict novel scenarios of breather propagation and interactions.

The scalar NLSE breathers have been the focus of the studies for the past decades, revealing such fundamental building blocs of the breather dynamics as Kuznetsov, Akhmediev, Peregrine and Tajiri-Watanabe solutions \cite{kuznetsov1977solitons,akhmediev1985generation,peregrine1983water,tajiri1998breather}; as well as superregular and ghost interaction patterns \cite{frisquet2013collision,gelash2014superregular,kibler2015superregular,xu2019breather}, and breather wave molecules \cite{xu2019breather}. All these scenarios of nonlinear wavefield evolution have been confirmed experimentally with optical, hydrodynamical, and plasma setups \cite{kibler2010peregrine,kibler2012observation,frisquet2013collision,xu2019breather,xu2020ghost,chabchoub2014hydrodynamics,chabchoub2011rogue,chabchoub2019drifting,bailung2011observation}. In addition, the breathers play an essential role in the formation of rational rogue waves \cite{akhmediev2009rogue,akhmediev2009excite}, modulation instability (MI) development \cite{akhmediev1985generation,zakharov2013nonlinear} and in the dynamics and statistics of complex nonlinear random wave states \cite{narhi2016real,soto2016integrable,osborne2019breather,roberti2021numerical}.

In this work we consider the vector two-component extension of the NLSE -- the Manakov system \cite{manakov1974theory}. In the presence of a constant background fields having amplitudes $A_{1,2}=\mathrm{const}$, $A = \sqrt{A_1^2+A_2^2}$, the Manakov system can be written as follows,
\begin{eqnarray}
\label{VNLSE}
i \psi_{1t}+\frac{1}{2}\psi_{1xx}+(|\psi_1|^{2}+|\psi_2|^{2}-A^2)\psi_1=0,
\\\nonumber
i \psi_{2t}+\frac{1}{2}\psi_{2xx}+(|\psi_1|^{2}+|\psi_2|^{2}-A^2)\psi_2=0.
\end{eqnarray}
where $t$ is time, $x$ is spatial coordinate and $\psi_{1,2}$ is a two-component complex wave field. The presence of the constant background, which often refers as condensate, means the following boundary conditions: $|\psi_{1,2}|^2 \to |A_{1,2}|^2$ at $x \to \pm \infty$.

In the case of small-amplitude condensate perturbations, the linear analysis of the system (\ref{VNLSE})  reveals two branches of the dispersion law $\omega_{\mathrm{I}}(k)$ and $\omega_{\mathrm{II}}(k)$, see Appendix section \ref{Sec:Appendix:1} for the derivation details,
\begin{equation}
\label{dispersion_laws}
\omega_{\mathrm{I}}(k) = \pm k\sqrt{k^2/4 - A^2}, \qquad \omega_{\mathrm{II}}(k) = \pm k^2/2.
\end{equation}
The first branch $\omega_{\mathrm{I}}(k)$ is the same as in the scalar NLSE with one-component condensate of amplitude $A$, and leads to the long-wave MI in the spectral region $k\in (-2A,2A)$. The second one $\omega_{\mathrm{II}}(k)$ is the same as in the scalar NLSE on zero background, and correspond to stable small-amplitude linear waves.

The system (\ref{VNLSE}) first considered by S.V. Manakov in \cite{manakov1974theory} is now widely used in nonlinear optics as a model of optical pulse propagation in a birefringent optical fiber \cite{agrawal2000nonlinear,maimistov2013nonlinear}. The two components $\psi_{1}$ and $\psi_{2}$ describe different light polarizations, which nonlinear interactions produce a broad family of complex nonlinear phenomena. Remarkably, the Manakov system, like its scalar counterpart NLSE, belongs to the class of equations integrable using the Inverse Scattering Transform (IST) technique \cite{manakov1974theory}. The IST allows finding exact multi-soliton solutions and asymptotic description of an arbitrary pulse evolution \cite{AblowitzBook1981,NovikovBook1984}. The key role in the IST construction for the Manakov system plays an auxiliary linear system of $3\times 3$ matrix wave functions (Jost functions), see e.g. \cite{maimistov2013nonlinear}. In the case of zero background each complex eigenvalue of the auxiliary system corresponds to a vector (polarized) soliton in the wavefield. S.V. Manakov presented the first study of vector soliton dynamics and demonstrated that they can change polarization as a result of mutual collisions \cite{manakov1974theory}. By now, such solitons are studied in detail, see for example \cite{ablowitz2004soliton} and the monograph \cite{maimistov2013nonlinear}.

In the presence of a condensate, vector solitons transform into vector breathers, characterized by discrete eigenvalues of the same $3\times 3$ auxiliary system. The vector breathers are in the recent trends of nonlinear studies focused on increasing the level of systems complexity. The first results on vector generalizations of Kuznetsov, Akhmediev, and Peregrine breathers have been presented within the past decade in \cite{priya2013akhmediev,chen2015vector,che2022nondegenerate}, see also the recent work \cite{che2022nondegenerate}. The vector rogue waves have been studied theoretically, numerically and experimentally in \cite{baronio2012solutions,baronio2014vector,frisquet2015polarization,manvcic2018statistics,degasperis2019rogue}. In addition, the work \cite{kraus2015focusing} provided a detailed study of the initial value problem for the system (\ref{VNLSE}) based on the IST Riemann approach and suggested a general classification of the types of vector breathers based on analytical properties of the wave field Jost functions. According to this classification, the vector breathers of the Manakov system represent three fundamental types -- type I, II, and III. Type I is a direct analog of the scalar NLSE breathers, while types II and III exhibit fundamentally different dynamics specific to the vector case. Recently we have found that the vector breathers can participate in a resonance interaction, see our Letter \cite{raskovalov2022resonant}. The resonance represents a three-breather process, i.e., a fusion of two breathers of the type I and II into one breather of the type III, which we denote it schematically as $\mathrm{I} + \mathrm{II} \rightarrow \mathrm{III}$; or the opposite, i.e. the decay $\mathrm{III} \rightarrow \mathrm{I} + \mathrm{II}$.

In this work, we study interactions of the vector breathers, including the resonance situations. With the vector variant of the dressing method scheme, see \cite{raskovalov2022resonant}, we obtain a general multi-breather solution of the Eq.~(\ref{VNLSE}). First, we analyze single breathers and show that the type I breathers correspond to the first branch $\omega_{\mathrm{I}}(k)$ of the dispersion laws (\ref{dispersion_laws}), while the types II and III are linked to the second branch $\omega_{\mathrm{I}}(k)$. This correspondence takes place for the decaying tails of the breathers. Then we study the resonanse breather interactions and two-breather solutions. We derive asymptotic expressions for the position and phase shifts acquired by breathers after mutual collisions. The found formulas allow us to describe the asymptotic states of the multi-breather ensembles and reveal the mathematical nature of the resonance interactions. More precisely, we find that similar to the three-wave system case \cite{ZakharovManakov1976theory}, the resonance interaction of vector breathers can be explained as a limiting case of infinite space shift acquired by one of the breathers. Finally, we demonstrate that only type $\mathrm{I}$ breathers participate in the development of MI from small-amplitude perturbations withing the superregular scenario, see \cite{zakharov2013nonlinear,gelash2014superregular}, while the breathers of types $\mathrm{II}$ and $\mathrm{III}$, belonging to the stable branch of the dispersion law, are not involved in this process.

The paper is organized as follows. In the next section \ref{Sec:2} we construct the general scheme of the vector dressing method and find the $N$-breather solution of the Manakov system in the presence of the condensate. In section \ref{Sec:3} we analyze the single-breather solutions of the Manakov system and their relations to the branches of the dispersion law. Then in section \ref{Sec:4} we consider the case of resonant interactions. In section \ref{Sec:5} we analyze the general two-breather solution of the Manakov system (\ref{VNLSE}), which describes elastic collisions of breathers. We end up section \ref{Sec:5} with finding the shifts of the positions and phases acquired by the breathers after their interaction. Finally in section \ref{Sec:6} we investigate important particular cases of vector two-breather solution. The last section \ref{Sec:7} presents discussions and conclusions. The appendix section \ref{Sec:Appendix} provides additional computational details, a complete table of the positions and phases shifts expressions and additional illustrations.

\section{Dressing method for vector breathers}\label{Sec:2}

In this section we build a dressing method scheme for constructing multi-breather solutions to the Manakov system. Previously in the Letter \cite{raskovalov2022resonant} we presented a shorter version of this scheme limited to single-eigenvalue solutions. Note, that the dressing method, also known as the Darboux dressing scheme, being a popular tool for constructing exact solutions to integrable nonlinear PDE, has many variations, see \cite{NovikovBook1984,zakharov1978relativistically,matveev1991darboux,akhmediev1991extremely}. We use a vector analog of the scalar dressing scheme for the NLSE developed in \cite{gelash2014superregular}.

The dressing method starts from introducing the auxiliary linear system for the $3\times 3$ matrix wave function $\mathbf{\Phi}$ depending on $x$, $t$ and the complex spectral parameter $\lambda$:
\begin{eqnarray}
&&\mathbf{\Phi}_x=\mathbf{U}\mathbf{\Phi}, \label{lax system 1}
\\
&&\mathbf{\Phi}_t = \mathbf{V}\mathbf{\Phi} = -(\lambda\mathbf{U}+i\mathbf{W}/2)\mathbf{\Phi}. \label{lax system 2}
\end{eqnarray}
Here $\mathbf{U}$ and $\mathbf{W}$ are the following $3\times 3$ matrixes:
\begin{eqnarray}
\label{U and W def}
&&\mathbf{U}=
\left(
  \begin{array}{ccc}
    -i\lambda & \psi_1 & \psi_2 \\
    -\psi^*_1 & i\lambda & 0 \\
    -\psi^*_2 & 0 & i\lambda \\
  \end{array}
\right)
,
\\\nonumber
&&\mathbf{W}=
\\\nonumber
&&\left(
  \begin{array}{ccc}
    |\psi_1|^{2}+|\psi_2|^{2}-A^{2} & \psi_{1x} & \psi_{2x} \\
    \psi^*_{1x} & -|\psi_1|^{2}+A^{2} & -\psi^*_1 \psi_2 \\
    \psi^*_{2x} & -\psi_1 \psi^*_2 & -|\psi_2|^{2}+A^{2} \\
  \end{array}
\right).
\end{eqnarray}

The Manakov system (\ref{VNLSE}) represents the compatibility condition of the equations (\ref{lax system 1}) and (\ref{lax system 2}) written as,
$$
	\mathbf{\Phi}_{xt} = \mathbf{\Phi}_{tx}.
$$
From Eqs.~(\ref{lax system 1}) and (\ref{lax system 2}) we find the following auxiliary equations for $\mathbf{\Phi}^{-1}$ and $\mathbf{\Phi}^\dag$:
\begin{eqnarray}
&&\mathbf{\Phi}^{-1}_x = - \mathbf{\Phi}^{-1} \mathbf{U}, \qquad \mathbf{\Phi}^{-1}_t = - \mathbf{\Phi}^{-1} \mathbf{V}, \label{rs1}\\
&&\mathbf{\Phi}^\dag_x = \mathbf{\Phi}^\dag \mathbf{U}^\dag, \qquad\qquad \mathbf{\Phi}^\dag_t = \mathbf{\Phi}^\dag \mathbf{V}^\dag . \label{s2} 
\end{eqnarray}
Here the sign $\dag$ means Hermitian conjugation. Comparing formulas (\ref{rs1}) and (\ref{s2}), and using the symmetry properties $\mathbf{U}^\dag (\lambda^*)=-\mathbf{U}(\lambda)$, $\mathbf{V}^\dag (\lambda^*)=-\mathbf{V}(\lambda)$, we find that $\mathbf{\Phi}$ satisfies the following reduction:
\begin{equation}
\mathbf{\Phi}^\dag (\lambda^*) = \mathbf{\Phi}^{-1}(\lambda). \label{rr}
\end{equation}

At the first step of the dressing procedure we find the solution $\mathbf{\Phi}_\mathrm{c}$ of the system (\ref{lax system 1}), (\ref{lax system 2}) for the condensate background $\psi_{1,2}=A_{1,2}$:
\begin{equation}
\mathbf{\Phi}_\mathrm{c} (x, t, \lambda)= [(1+r^2) e^{-\varphi_0}]^{-1/3}\left(\begin{array}{ccc}
0, & e^{\varphi}, & -\mathrm{i}\,r e^{-\varphi}\\
-\frac{A_2}{A}\, e^{-\varphi_{0}}, & -\frac{A_1}{A}\,\mathrm{i}\, r e^{\varphi}, & \frac{A_1}{A}\,e^{-\varphi}\\
\frac{A_1}{A}\,e^{-\varphi_{0}},& -\frac{A_2}{A}\,\mathrm{i}\, r e^{\varphi}, & \frac{A_2}{A}\,e^{-\varphi}
\end{array}\right) ,
\label{Psio}
\end{equation}
where,
\begin{equation}
\label{zeta_def}
    r = A/(\lambda+\zeta), \qquad \zeta =\sqrt{\lambda^2+A^2},
\end{equation}
and the functions $\varphi_{0}$ and $\varphi$ are,
\begin{equation}
\label{q_phases}
    \varphi_{0} = -\mathrm{i}\,\lambda\,x + \frac{\mathrm{i}}{2}\,\bigl (\lambda^2 + \zeta^2 \bigr ) \, t,
    \qquad
    \varphi = -\mathrm{i} \zeta\, x + \mathrm{i}\,\lambda\, \zeta\, t.
\end{equation}
We imply that the function $\zeta(\lambda)$ has the branchcut on the interval $[-iA,iA]$, which differs from the automatic choice $\{ [-i\infty,-iA]\cup[iA,\infty i] \}$ implied in software packets such as {\it Wolfram Mathematica}. As we see later, the choice of the branchcut is essential for constructing breather solutions. 

The solution $\mathbf{\Phi}_\mathrm{c} (x, t, \lambda)$ satisfies the auxiliary system:
\begin{equation}
\mathbf{\Phi}_{\mathrm{c}\,x}=\mathbf{U}_\mathrm{c} \mathbf{\Phi}, \qquad \mathbf{\Phi}_{\mathrm{c}\,t} = \mathbf{V}_\mathrm{c} \mathbf{\Phi},  \label{la}
\end{equation}
where $\mathbf{V}_\mathrm{c} = -(\lambda\,\mathbf{U}_\mathrm{c} + \mathrm{i} \mathbf{W}_\mathrm{c})$, and
\begin{eqnarray}
\nonumber
&&\mathbf{U}_\mathrm{c}=
\left(
  \begin{array}{ccc}
    -\mathrm{i}\lambda & A_1 & A_2 \\
    -A_1 & \mathrm{i}\lambda & 0 \\
    -A_2 & 0 & \mathrm{i}\lambda \\
  \end{array}
\right) ,
\\
\nonumber
&&\mathbf{W}_\mathrm{c}=
\frac12\left(
  \begin{array}{ccc}
    0 & 0 & 0 \\
    0 & A_2^{2} & -A_1 A_2 \\
    0 & -A_1 A_2 & A_1^{2} \\
  \end{array}
\right) .
\end{eqnarray}

In accordance with \cite{gelash2014superregular}, we introduce the dressing function $\boldsymbol{\chi}$ as,
\begin{equation}
\boldsymbol{\chi} = \mathbf{\Phi} \,\mathbf{\Phi}_0^{-1}. \label{od}
\end{equation}
The dressing function satisfies the asymptotic condition:
\begin{equation}
\boldsymbol{\chi}(\lambda) \to \mathbf{E} + \frac{\mathbf{N}}{\lambda}+O(\lambda^{-2}), \qquad \textrm{at} \quad |\lambda| \to \infty ,  \label{usl}
\end{equation}
where $\mathbf{N}$ is a constant matrix and $\mathbf{E}$ is the unit matrix.
From Eq.~(\ref{rr}) we also find, that the function $\boldsymbol{\chi} (x, t, \lambda)$ satisfies the following reduction:
\begin{equation}
\boldsymbol{\chi}^\dag (\lambda^*) = \boldsymbol{\chi}^{-1}(\lambda). \label{rrr}
\end{equation}
Then, using Eqs.~(\ref{lax system 1}), (\ref{lax system 2}), (\ref{rs1}), and (\ref{s2}) we obtain the system of linear equations for the inverse dressing function: 
\begin{eqnarray}
\boldsymbol{\chi}^{-1}_x = -\boldsymbol{\chi}^{-1} \mathbf{U} + \mathbf{U}_c \boldsymbol{\chi}^{-1}, \label{rs2}\\
\boldsymbol{\chi}^{-1}_t = -\boldsymbol{\chi}^{-1} \mathbf{V} + \mathbf{V}_c \boldsymbol{\chi}^{-1}. \nonumber
\end{eqnarray}

Now, choosing matrix $\boldsymbol{\chi}$ so, that the matrices $\mathbf{U}$ and $\mathbf{V}$ are regular in the $\lambda$-plane, we obtain a new solution of the Eqs.~(\ref{lax system 1}),~(\ref{lax system 2}). Substituting the expansion (\ref{usl}) into the equation (\ref{rs2}), we find out the final formulas to calculate the components $\psi_{1,2}$:
\begin{equation}
\psi_1 = A_1 + 2 \,\mathrm{i} \,N_{12}, \qquad \psi_2 = A_2 + 2 \,\mathrm{i}\, N_{13} . \label{res}
\end{equation}

First we propose, that the function $\boldsymbol{\chi}$ has only one pole at $\lambda = \lambda_1$, so that it can be written as,
\begin{equation}
\boldsymbol{\chi} = \mathbf{E} + \frac{\mathbf{N}}{\lambda-\lambda_1}. \label{E}
\end{equation}
The constant $\lambda_1$ represents discrete eigenvalue of the system (\ref{lax system 1},\ref{lax system 2}), which means that the corresponding wave function is bounded from both sides in space, see e.g., \cite{kraus2015focusing}. Then, from (\ref{rrr}) we obtain, that:
\begin{equation}
\boldsymbol{\chi}^{-1} = \mathbf{E} + \frac{\mathbf{N}^\dag}{\lambda-\lambda_1^*}. \label{E1}
\end{equation}
From the condition $\boldsymbol{\chi} \boldsymbol{\chi}^{-1} = \mathbf{E}$ we find:
\begin{equation}
\boldsymbol{\chi}(\lambda_1)\,\mathbf{N}^\dag (\lambda_1) = 0. \label{ze}
\end{equation}
From this it follows, that the matrices $\mathbf{N}$, $\mathbf{N}^\dag$ are degenerated and can be expressed via three-component vectors $\mathbf{p}$ and $\mathbf{q}$ as follows:
\[
\mathbf{N}_{\alpha \beta} = p_\alpha q_\beta, \qquad \mathbf{N}^\dag = q^*_\alpha p^*_\beta, \qquad \alpha, \beta = 1,2,3.
\]
To eliminate an extra pole at the point $\lambda=\lambda_1^*$ in the expression (\ref{rs2}), we impose on the vector $\mathbf{q}(x, t)$ the condition:
\[
\partial_x \mathbf{q}^* - \mathbf{U}_c (x, t, \lambda_1^*) \mathbf{q}^* = 0 .
\]
From this, we find the vector $\mathbf{q}$ in the form,
\begin{equation}
\label{q_general_form}
\mathbf{q}(x, t) = \mathbf{\Phi}_c^*(x, t, \lambda_1^*) \mathbf{C},
\end{equation}
where the vector of integration constants,
\begin{equation}
\label{C_general_form}
\mathbf{C}=(C_{0},C_{1},C_{2})^{\mathrm{T}},
\end{equation}
is an arbitrary three-component complex vector with the superscript $\mathrm{T}$ meaning transposing.
Finally, from (\ref{ze}) we obtain the vector $\mathbf{p}$ in the form,
\[
\mathbf{p} = \frac{\mathbf{q^*}}{|\mathbf{q}|^2} (\lambda_1 - \lambda_1^*) .
\]

Thereby, the one-pole function $\boldsymbol{\chi}(x, t, \lambda)$ from Eq.~(\ref{E}) is completely defined. Now, using the formula (\ref{res}), we obtain the components $\psi_{1,2}$ of the single-eigenvalue solution of the Manakov system (\ref{VNLSE}) in the presence of the condensate background:
\begin{eqnarray}
\label{solution}
\psi_{1} = A_{1} +\frac{2\,\mathrm{i}\,(\lambda_1 - \lambda_1^*)\, q_{1}^* q_{2}}{|\mathbf{q}|^2},
\\\nonumber
\psi_{2} = A_{2} +\frac{2\,\mathrm{i}\,(\lambda_1 - \lambda_1^*)\, q_{1}^* q_{3}}{|\mathbf{q}|^2} .
\end{eqnarray}

Following the analogy with \cite{gelash2014superregular}, we obtain that if the dressing matrix $\boldsymbol{\chi}(x, t, \lambda)$ have $N$ poles, $\lambda=\lambda_j$, $j=1,\ldots N$, then the corresponding $N$-eigenvalue solution of the Manakov system can be found by means of the Cramer's rule as,
\begin{eqnarray}
\psi_1 = A_1+2 \mathrm{\widetilde{M}}_{12}/\mathrm{M},
\nonumber\\
\psi_2 = A_2+2\mathrm{\widetilde{M}}_{13}/\mathrm{M}.
\label{N-solitonic solution}
\end{eqnarray}
Here $\mathrm{\widetilde{M}}_{\alpha,\beta}$ and $\mathrm{M}$ are the following determinants:
\begin{equation}
\mathrm{\widetilde{M}}_{\alpha\beta}=
\left|\begin{array}{cc}
        0 & \begin{array}{ccc}
              q_{1,\beta} & \cdots & q_{n,\beta}
            \end{array}
         \\
        \begin{array}{c}
          q^*_{1,\alpha} \\
          \vdots \\
          q^*_{N,\alpha}
        \end{array}
         &  \begin{array}{c}
              \mathbf{M}^{T}
            \end{array}
      \end{array}\right|;
\label{M1}
\quad\quad
\mathrm{M}=\mathrm{det}(\mathbf{M});
\quad\quad \mathcal{}
\mathbf{M}_{nm}=\frac{\mathrm{i}(\mathbf{q}_{n}\cdot \mathbf{q}^*_{m})}{\lambda_{n}-\lambda^*_m},
\end{equation}
with $\alpha,\beta =1,2,3$ and $n,m =1,2..N$. Note, that it is sufficient to consider the poles of the dressing function located only in the upper half of the $\lambda$-plane, i.e. 
\begin{equation}
\label{upper_plane}
    \mathrm{Im}[\lambda_j] > 0, \qquad j=1,\ldots N,
\end{equation}
since the choices $\mathrm{Im}[\lambda_j]<0$ lead to the same class of multi-breather solutions. Here and below, $\mathrm{Re}$ and $\mathrm{Im}$ means real and imaginary parts of a complex number. Recall that the eigenvalue set $\{\lambda_j\}$ represents discrete spectrum of the system (\ref{lax system 1},\ref{lax system 2}). Meanwhile the real $\lambda$-axes region belongs to the continuous part of the system (\ref{lax system 1},\ref{lax system 2}), which we do not consider here.

From Eq.~(\ref{q_general_form}) we find the the vectors $\mathbf{q}_n$ as,
\begin{eqnarray}
\mathbf{q}_n = \left(\begin{array}{ccc}
0, & e^{-\varphi_n}, & \mathrm{i}\,r_n e^{\varphi_n}\\
-\frac{A_2}{A} e^{\varphi_{0n}}, & \frac{A_1}{A}\,\mathrm{i}\, r_n e^{-\varphi_n}, & \frac{A_1}{A}\,e^{\varphi_n}\\
\frac{A_1}{A}\,e^{\varphi_{0n}},& \frac{A_2}{A}\,\mathrm{i}\, r_n e^{-\varphi_n}, & \frac{A_2}{A}\,e^{\varphi_n}
\end{array}\right) \left(\begin{array}{c}
 C_{n0} \\ C_{n1} \\ C_{n2}
\end{array}\right),
\end{eqnarray}
or, in component-wisely form,
\begin{eqnarray}
q_{n1}&=& e^{-\varphi_n} C_{n1} + \mathrm{i}\, r_n e^{\varphi_n}\, C_{n2}
\nonumber\\
q_{n2}&=&\frac{1}{A}\,\big[-A_2 e^{\varphi_{0n}} C_{n0} + A_1 \big(\mathrm{i} \,r_n e^{-\varphi_n} C_{n1} + e^{\varphi_n} C_{n2} \big)\big],
\nonumber\\
q_{n3}&=&\frac{1}{A}\big[ A_1 e^{\varphi_{0n}} C_{n0} + A_2 \big(\mathrm{i} \,r_n e^{-\varphi_n} C_{n1} + e^{\varphi_n} C_{n2} \big)\big].
\label{q vectors(lambda)}
\end{eqnarray}
Here the functions $\varphi_{0n}$ and $\varphi_n$ are the following, see Eq.~(\ref{q_phases}),
\begin{eqnarray}
\label{phi_0_and_phi}
	&&\varphi_{0n} = -\mathrm{i}\,\lambda_n\,x + \frac{\mathrm{i}}{2}\,\bigl (\lambda_n^2 + \zeta_n^2 \bigr ) \, t = u_{0n} - iv_{0n},
	\\\nonumber
	&&\varphi_n = -\mathrm{i} \zeta_n\, x + \mathrm{i}\,\lambda_n\, \zeta_n\, t = u_n - iv_n.
\end{eqnarray}
The functions $u_n$, $v_n$, $u_{0n}$, $v_{0n}$ distinguish real and imaginary parts of the functions $\varphi_{0n}$ and $\varphi_{n}$ as,
\begin{eqnarray}
\label{uvu0v0}
	&&u_{0n} = \mathrm{Im}[\lambda_n] x - \frac{1}{2}\mathrm{Im}[\lambda_n^2+\zeta_n^2]t,
	\\\nonumber
	&&v_{0n} = \mathrm{Re}[\lambda_n] x - \frac{1}{2}\mathrm{Re}[\lambda_n^2+\zeta_n^2]t,
	\\\nonumber
	&&u_n = \mathrm{Im}[\zeta_n] x - \mathrm{Im}[\lambda_n\zeta_n]t,
	\\\nonumber
	&&v_n = \mathrm{Re}[\zeta_n] x - \mathrm{Re}[\lambda_n\zeta_n]t.
\end{eqnarray}

In further calculations we will use the square of the modulus $\mathbf{q}_n$, which can be written as,
\begin{eqnarray}
|\mathbf{q}_n|^2 = |e^{-\varphi_n} C_{n1} + \mathrm{i}\, r_n e^{\varphi_n}\, C_{n2}|^2 +|\mathrm{i}\, r_n e^{-\varphi_n} C_{n1} + e^{\varphi_n} C_{n2}|^2 + |e^{\varphi_{0n}} C_{n0}|^2 .
\end{eqnarray}
Note that the transformation,
\begin{equation}
\label{Cn_transform}
    \mathbf{q}_n \rightarrow \kappa \mathbf{q}_n,
\end{equation}
where $\kappa$ is an arbitrary complex constant, does not change the solution (\ref{solution}). The latter means that an arbitrary choice of the vector $\mathbf{C}$ corresponds to four real-valued solution parameters. The nontrivial solutions of the Manakov system appear when the vector $\mathbf{C}$ has at least two nonzero components. We discuss the physical meaning of the breather parameters and different choices of the vector $\mathbf{C}$ in the following paragraphs.

\section{Vector breathers of types I, II and III}\label{Sec:3}
In this section, we describe the elementary building blocks of the vector breather dynamics -- the single breathers of the fundamental types I, II, and III. This classification based on analytical properties of the wavefield Jost functions was proposed in the work \cite{kraus2015focusing}. Type I coincides with the breather solutions to the scalar NLSE, while types II and III exhibit fundamentally different nonlinear wave dynamics specific to the vector (polarized) case. Previously in \cite{raskovalov2022resonant} we have established that on the language of the dressing method, see also Sec.~\ref{Sec:2}, the types I, II, and III correspond to the subsequent setting to zero one of the components of the vector $\mathbf{C}$, see Eq.~(\ref{C_general_form}). Here, following \cite{raskovalov2022resonant}, we present an analytical description for all three types of vector breathers and consider important particular cases which have not been touched in \cite{raskovalov2022resonant}. Also, we emphasize that type II and type III solutions can be transformed into each other by changing the Riemann surface sheets of the spectral parameter plane. Finally, we demonstrate that type I correspond to the first branch of the dispersion law $\omega_{\mathrm{I}}(k)$ while types II and III to the second branch $\omega_{\mathrm{II}}(k)$.

To avoid the sign issues of square root function $\zeta(\lambda)$ and also to simplify computations, we use the following parametrizations for the spectral parameter and associated to it functions,
\begin{eqnarray}
\label{uniformization}
\lambda &=& A\,\sinh(\xi+\mathrm{i}\,\alpha),\\\nonumber
\zeta &=& A\,\cosh(\xi+\mathrm{i}\,\alpha),\\\nonumber
r &=& e^{-\xi - \mathrm{i} \alpha}.
\end{eqnarray}
This transformation of the two-sheeted Riemann surface of $\zeta(\lambda)$ into one-sheeted plane with coordinates $(\xi,\alpha)$ is called uniformization and often used in the breathers studies, see e.g. \cite{gelash2014superregular,kraus2015focusing}. According to (\ref{upper_plane}) we consider only the regions $\xi\in (0,\infty)$ and $\alpha\in (0,\pi)$ for the breather parameters.

We use the general single-eigenvalue solution (\ref{solution}) with $N=1$, and start with the case $C_{0} = 0$ and $C_{1,2}\ne 0$ corresponding to the breather of type I. When dealing with single breathers we omit the subscripts $n$ for the formulas of the previous Sec.~\ref{Sec:2}. Substituting $C_1=0$ into (\ref{solution}) we find that the breather represents a simple vector generalization of the solution of the scalar NLSE, when the two components of the Manakov system do not interact, satisfying the relation,
\begin{equation}
\label{typeI_relation}
    \psi_{2} = (A_2/A_1)\psi_{1}.
\end{equation}

The latter means, that each wavefield component represents a well-known breather solution of the scalar NLSE, see, e.g., \cite{pelinovsky2008book}. When the scalar NLSE for one-component wavefield $\psi$ is written in the form,
\begin{eqnarray}
i \psi_{t}+\frac{1}{2}\psi_{xx}+(|\psi|^{2}+|\psi|^{2}-A_0^2)\psi=0,
\label{NLSE}
\end{eqnarray}
the vector breathers of type I can be obtained from the known breather solutions of the Eq.~(\ref{NLSE}) using the following transformation,
\begin{eqnarray}
\label{typeI_transformation}
\psi_1(x,t) = \frac{A_1}{A_0}\psi\left(\frac{A^2}{A^2_0} t,\frac{A}{A_0} x\right),
\\
\psi_2(x,t) = \frac{A_2}{A_0}\psi\left(\frac{A^2}{A^2_0} t,\frac{A}{A_0} x\right).
\end{eqnarray}

From the Eq.~(\ref{Cn_transform}) we see, that the vector $\mathbf{C}$ has only one independent complex parameter, which allows us to parametrize its components as follows:
\begin{eqnarray}
\label{C_I_param}
    C_{0} = 0, \quad C_{1} = C_{2}^{-1} = e^{\mathrm{Im}[\zeta]\delta + i\theta/2},
\end{eqnarray}
where $\delta$ and $\theta$ are real valued parameters controlling space position of the breather and its phase. From the general solution (\ref{solution}), the real and imaginary wave field components for type I breather can be written as,
\begin{eqnarray}
\label{s1}
&&\mathrm{Re} \,\psi_{1,2}^\mathrm{I} = A_{1,2} -
\\\nonumber
&&\frac{2 A_{1,2} \sin \alpha\, \cosh \xi\,[\cos(2 v_{\mathrm{I}})\cosh{\xi}+\cosh(2 u_{\mathrm{I}}) \sin \alpha]}{\cosh \xi\,\cosh(2 u_{\mathrm{I}})+\sin \alpha\, \cos(2v_{\mathrm{I}})},
\\\nonumber
&&\mathrm{Im} \,\psi_{1,2}^\mathrm{I} =
\\\nonumber
&&\frac{2 A_{1,2}\,\sin\alpha\cosh{\xi}\,[\sinh{(2u_{\mathrm{I}})}\cos\alpha + \sin{(2v_{\mathrm{I}})}\sinh{\xi}]}{\cosh \xi\,\cosh(2 u_{\mathrm{I}})+\sin \alpha\, \cos(2 v_{\mathrm{I}})},
\end{eqnarray}
where
\begin{eqnarray}
    2u_{\mathrm{I}} = l_{\mathrm{I}}^{-1}(x  -V_{\mathrm{I}} t - \delta), \qquad 2v_{\mathrm{I}}= k_{\mathrm{I}} x - \omega_{\mathrm{I}} t + \theta,
\end{eqnarray}
are expressed via the breather characteristic length $l_{\mathrm{I}}$, group velocity $V_{\mathrm{I}}$, characteristic wave vector $k_{\mathrm{I}}$, and characteristic frequency $\omega_{\mathrm{I}}$:
\begin{eqnarray}
    \nonumber
    &&l_{\mathrm{I}} = (2 \mathrm{Im}[\zeta])^{-1} = (2 A \sin{\alpha}\sinh{\xi})^{-1},
    \\\nonumber
    &&V_{\mathrm{I}} = \frac{\mathrm{Im}[\lambda\zeta]}{\mathrm{Im}[\zeta]} = \frac{A\cos{\alpha}\cosh{2\xi}}{\sinh{\xi}},
     \\\nonumber
    &&k_{\mathrm{I}} = 2\mathrm{Re}[\zeta] = 2A\cos{\alpha}\cosh{\xi},
     \\
    &&\omega_{\mathrm{I}} = 2\mathrm{Re}[\lambda\zeta] = A^2\cos{2\alpha}\sinh{2\xi}.
    \label{characteristic_values1}
\end{eqnarray}
The type I breather has the following spatial asymptotics,
\begin{eqnarray}
\label{asymptotics_I}
\psi_{1,2}^\mathrm{I} &\to& A_{1,2}\, e^{\pm 2\, \mathrm{i}\, \alpha}; \quad\quad\quad\quad x \to \pm\infty,
\end{eqnarray}
so that the total phase shift of the background field caused by the presence of the breather is $4 \alpha$.

\begin{figure} 
\centering
    \includegraphics[width=0.3\linewidth]{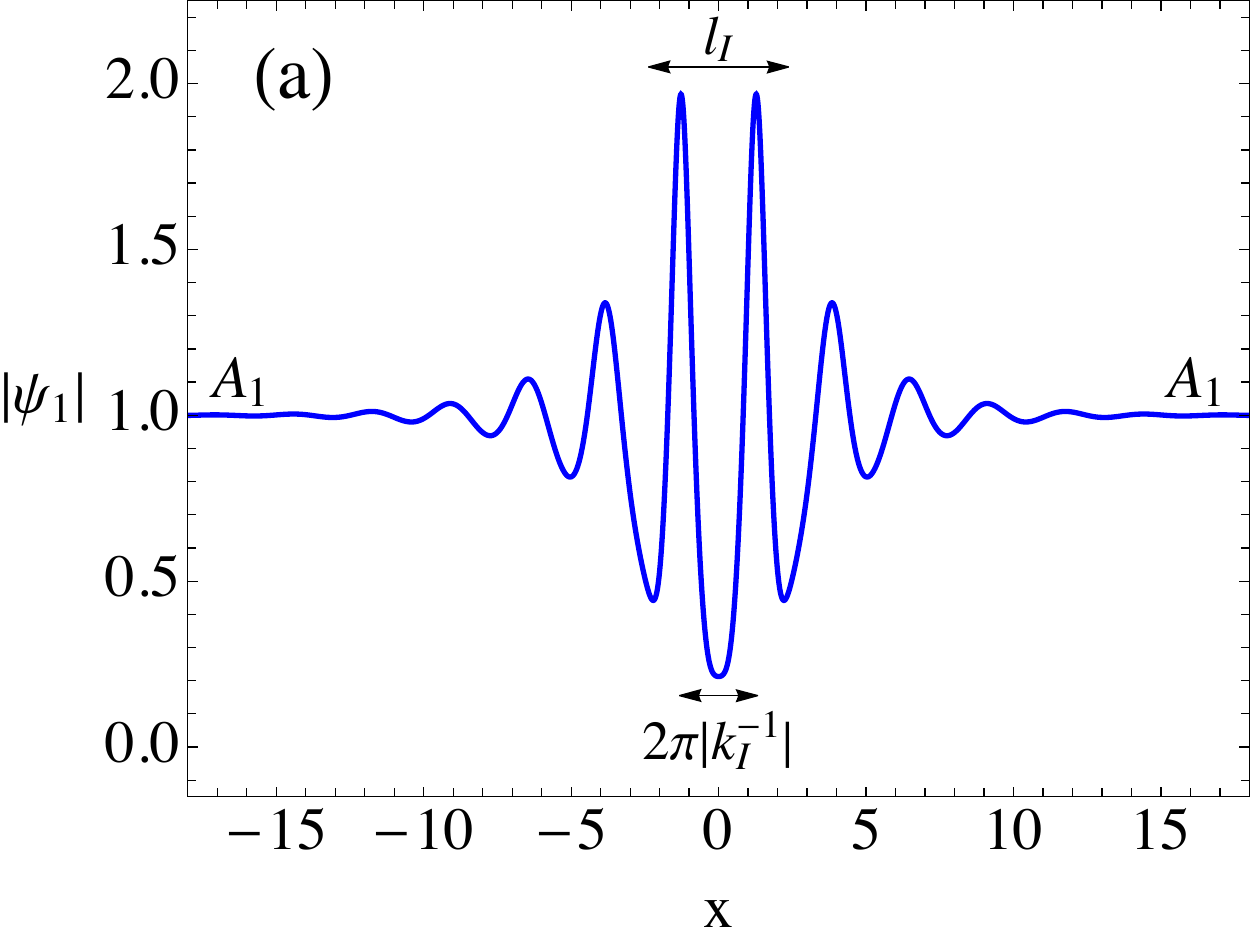}\,\,\,\,\,
    \includegraphics[width=0.3\linewidth]{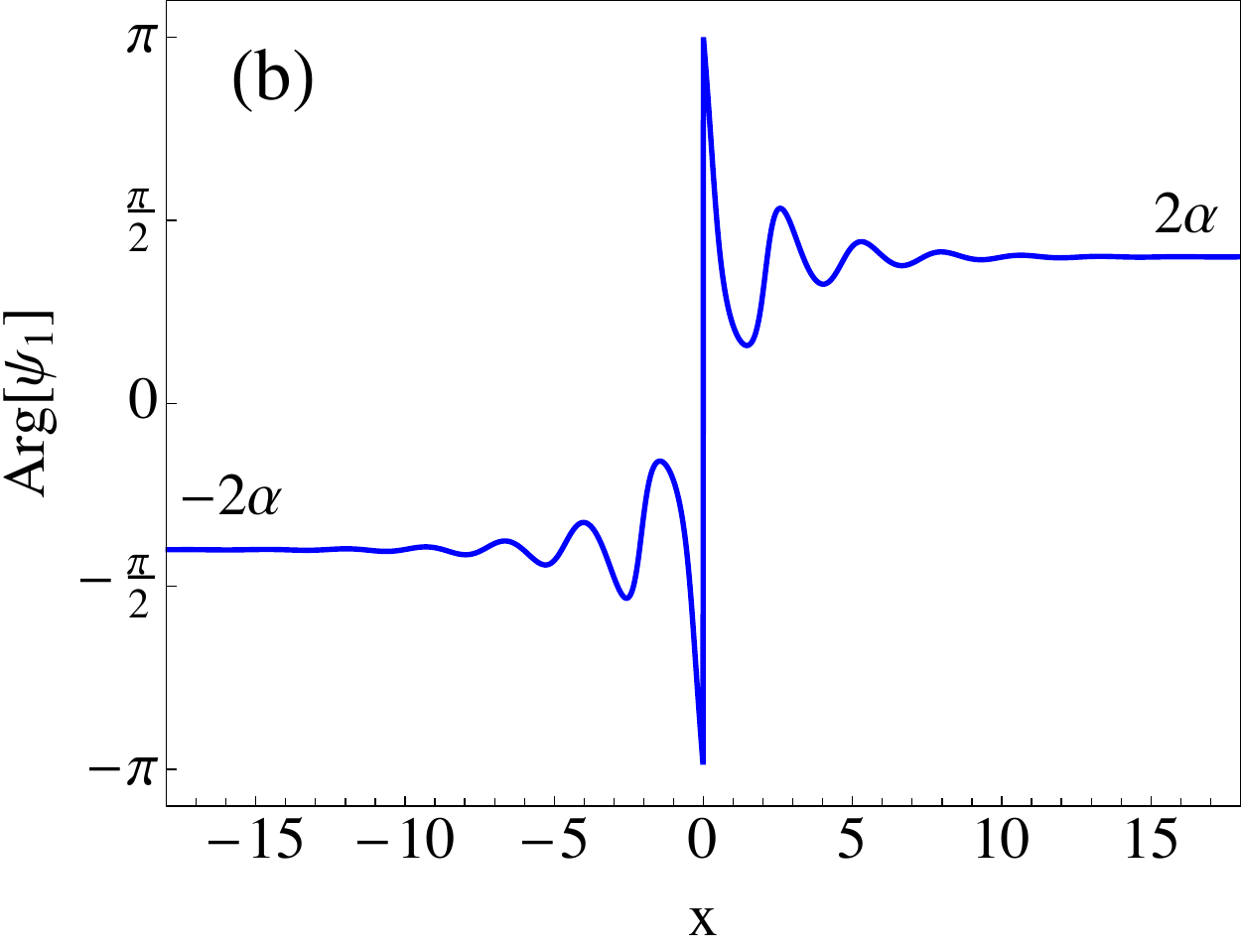}\,\,\,\,\,
    \includegraphics[width=0.31\linewidth]{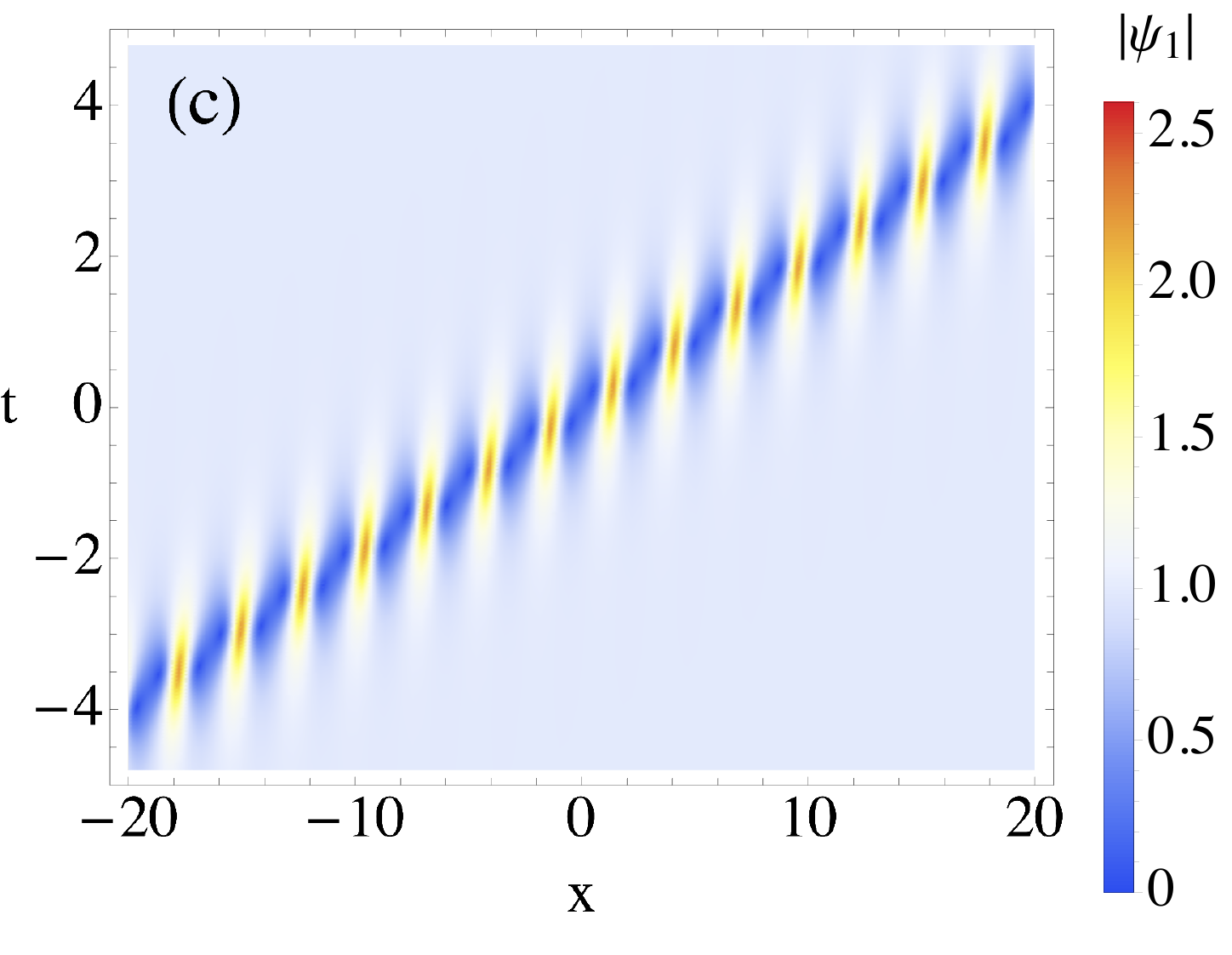}
\caption{
Vector breather of type I, see Eq.~(\ref{s1}), which can be obtained from scalar breather solution using the transformation (\ref{typeI_transformation}). The solution parameters are defined in (\ref{parameters}). (a) $|\psi_1|$ at $t=0$ with indicated asymptotic values, see Eq.~(\ref{asymptotics_I}), characteristic size and wavelength computed according to Eq.~(\ref{characteristic_values1}). (b) $\mathrm{Arg}[\psi_1]$ at $t=0$ with indicated asymptotic values, where $\mathrm{Arg}$ means complex phase. (c) spatio-temporal evolution  of $|\psi_1|$. The wave field component $\psi_2$ is not shown since it coincides with the first one after rescaling the amplitude amplitude, see Eq.~(\ref{typeI_relation}). Here and in next figures the function $\mathrm{Arg}$ is defined in the cyclic interval $[-\pi,\pi )$, so that the function value can exhibit a jump as in panel (b).
}
\label{fig_01}
\end{figure}

Fig.\ref{fig_01} demonstrates a general case, when the breather is localized in space and moves with a nonzero group velocity. In the figure we indicate the characteristic $l_{\mathrm{I}}$ and $k_{\mathrm{I}}$, and the asymptotic values (\ref{asymptotics_I}). We choose the following set of breather parameters,
\begin{eqnarray}
\label{parameters}
&&A_1 = 1, \quad A_2=1;
\\\nonumber
&&\alpha=\pi/5, \quad \xi=1/4, \quad \theta = 0, \quad \delta = 0,
\end{eqnarray}
which later we also use to show examples of type II and type III breathers. The choice of $\alpha$ and $\xi$ in (\ref{parameters}) correspond to a moving spatially localized breather, see (\ref{characteristic_values1}).

When $C_{1} =0$ we obtain another nontrivial solution of the Manakov system, which we call as type II breather again referring to the classification from \cite{kraus2015focusing}.
Writing the components of the vector $\mathbf{C}$ as follows,
\begin{eqnarray}
\label{C_II_param}
    C_{0} = e^{-\mathrm{Im}[\lambda]\delta - i\theta /2}, \quad C_{1} = 0, \quad C_{2} = e^{-\mathrm{Im}[\zeta]\delta + i\theta/2},
\end{eqnarray}
from (\ref{solution}) we obtain:
\begin{eqnarray}
\psi_{1}^{\mathrm{II}} = A_{1} +\frac{4\,\mathrm{i}\,e^{\mathrm{i} \alpha} \sin \alpha\,\cosh \xi (A_1 - A_2 e^{u_{\mathrm{II}} - iv_{\mathrm{II}}})}{e^{2u_{\mathrm{II}} + \xi}+2\, \cosh \xi}, \nonumber\\
\psi_{2}^{\mathrm{II}} = A_{2} +\frac{4\,\mathrm{i}\,e^{\mathrm{i} \alpha} \sin \alpha\,\cosh \xi (A_2 + A_1 e^{u_{\mathrm{II}} - iv_{\mathrm{II}}})}{e^{2u_{\mathrm{II}} + \xi}+2\, \cosh \xi},
\label{sol2}
\end{eqnarray}
where
\begin{eqnarray}
\label{uv_II}
    2u_{\mathrm{II}} = l_{\mathrm{II}}^{-1}(x  -V_{\mathrm{II}} t - \delta), \qquad v_{\mathrm{II}}= k_{\mathrm{II}} x - \omega_{\mathrm{II}} t + \theta,
\end{eqnarray}
are expressed via the physical characteristics of type II breather,
\begin{eqnarray}
\label{characteristic_values2}
    &&l_{\mathrm{II}} = (2 (\mathrm{Im}[\lambda] - \mathrm{Im}[\zeta]))^{-1} = (2 A e^{-\xi} \sin{\alpha})^{-1},
    \\\nonumber
    &&V_{\mathrm{II}} = \frac{\mathrm{Im}[\lambda^2+\zeta^2]/2 - \mathrm{Im}[\lambda\zeta]}{\mathrm{Im}[\lambda] - \mathrm{Im}[\zeta]} = -A\cos{\alpha}e^{-\xi},
    \\\nonumber
    &&k_{\mathrm{II}} = \mathrm{Re}[\lambda] - \mathrm{Re}[\zeta] = -A\cos\alpha\,e^{-\xi},
    \\\nonumber
    &&\omega_{\mathrm{II}} = \frac{1}{2}\mathrm{Re}[\lambda^2+\zeta^2] - \mathrm{Re}[\lambda\zeta] = \frac{A^2}{2}e^{-2\xi}\cos 2\alpha.
\end{eqnarray}
\begin{figure}[!t]
\centering
    \includegraphics[width=0.3\linewidth]{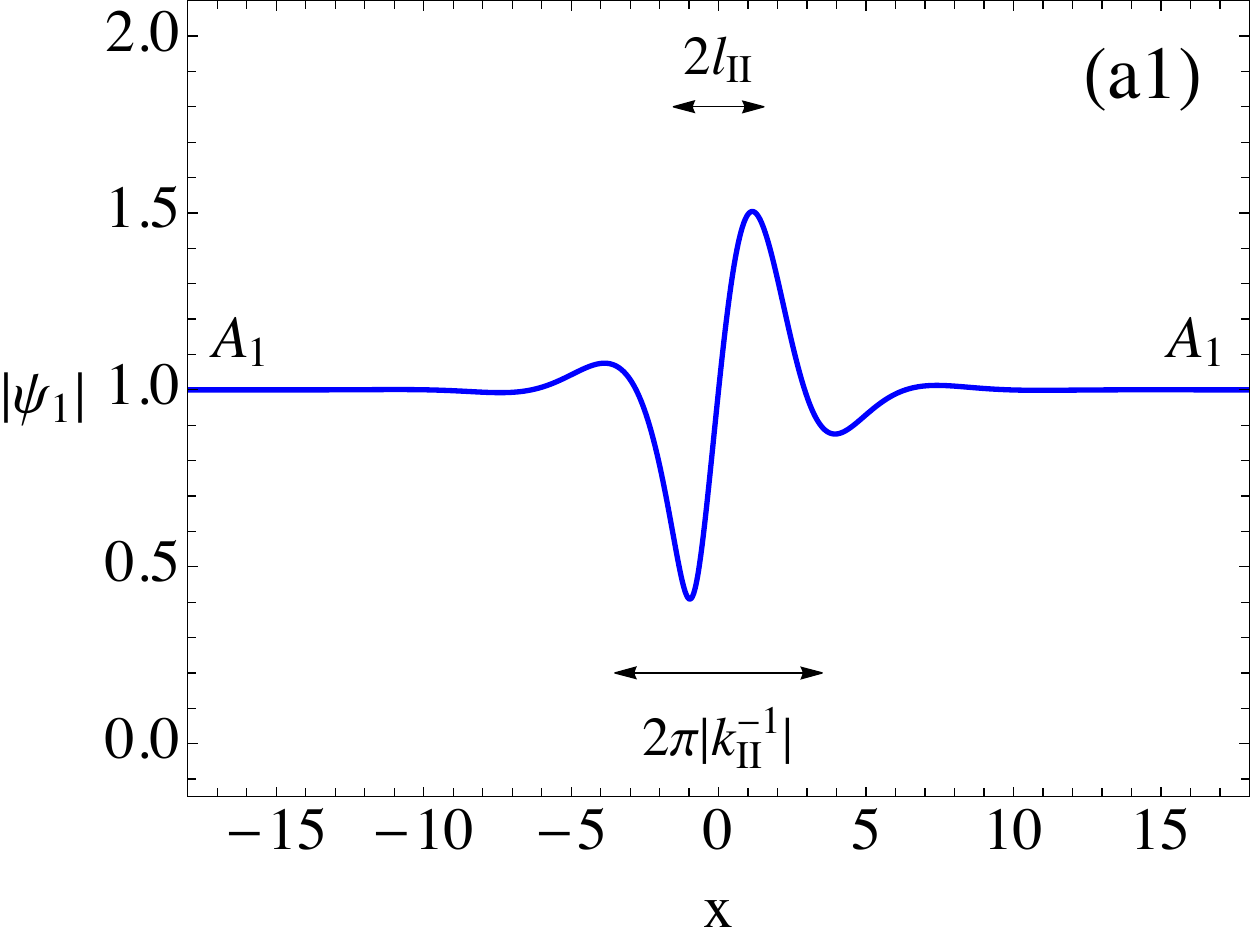}\,\,\,\,\,
    \includegraphics[width=0.3\linewidth]{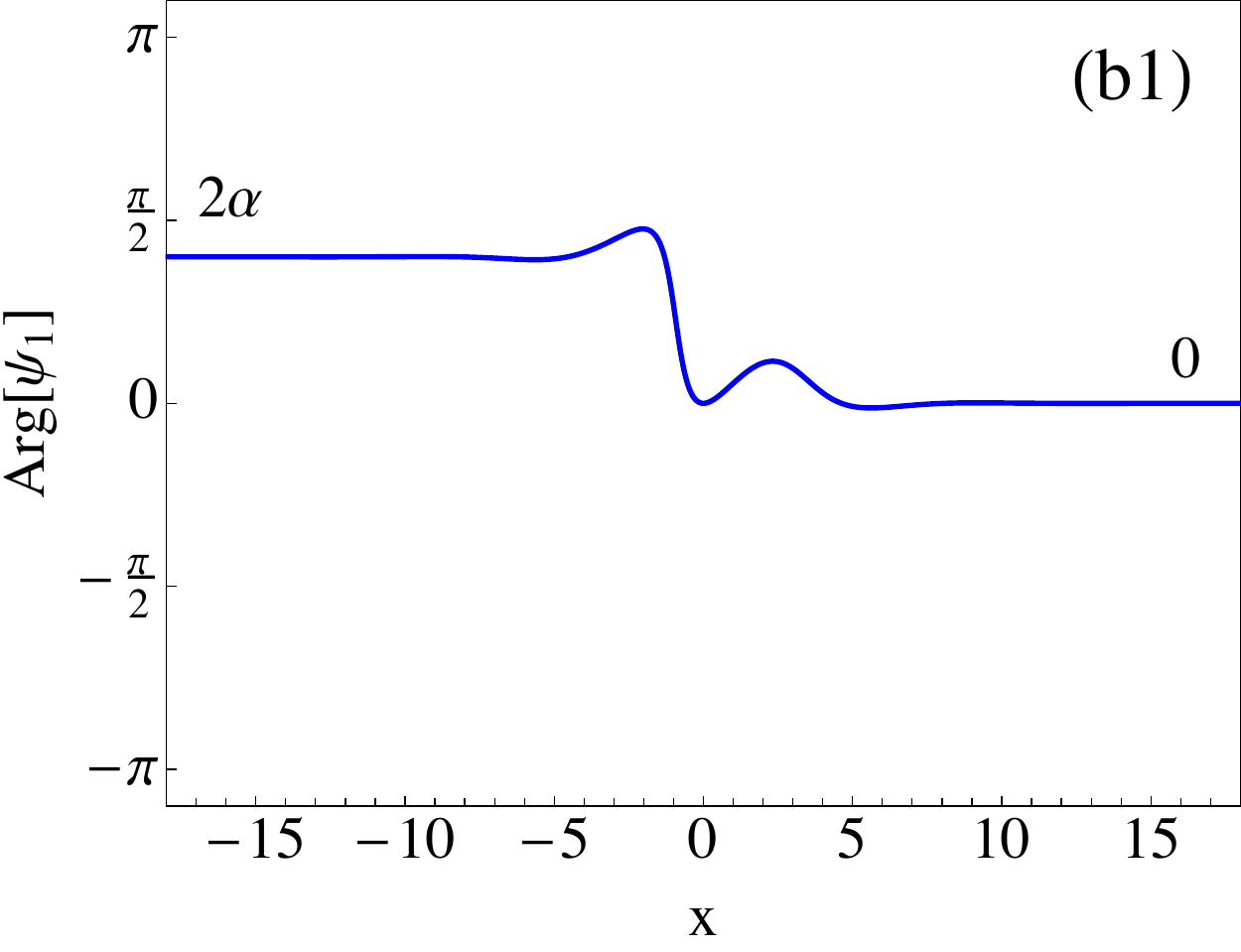}\,\,\,\,\,
    \includegraphics[width=0.31\linewidth]{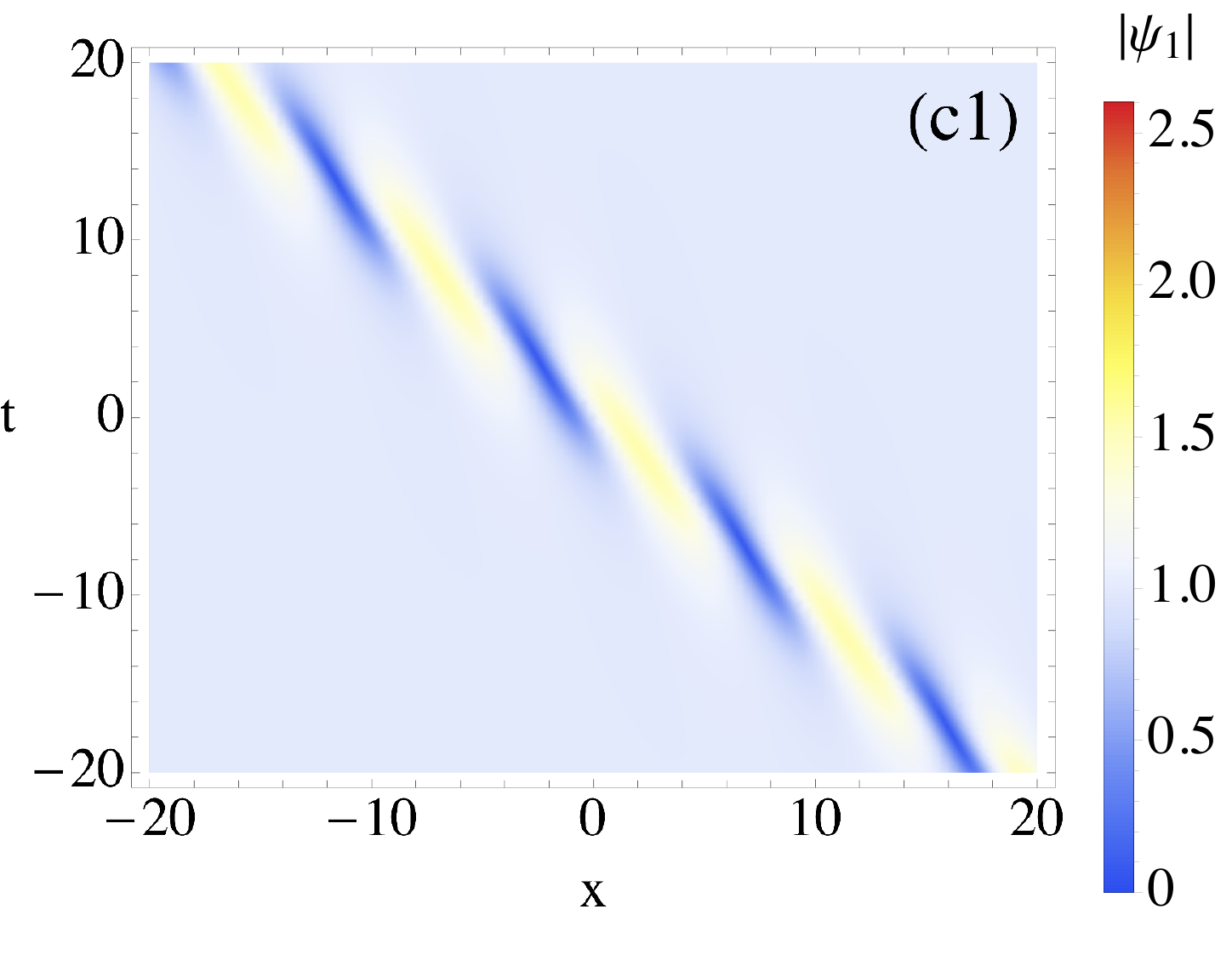}\\
    \includegraphics[width=0.3\linewidth]{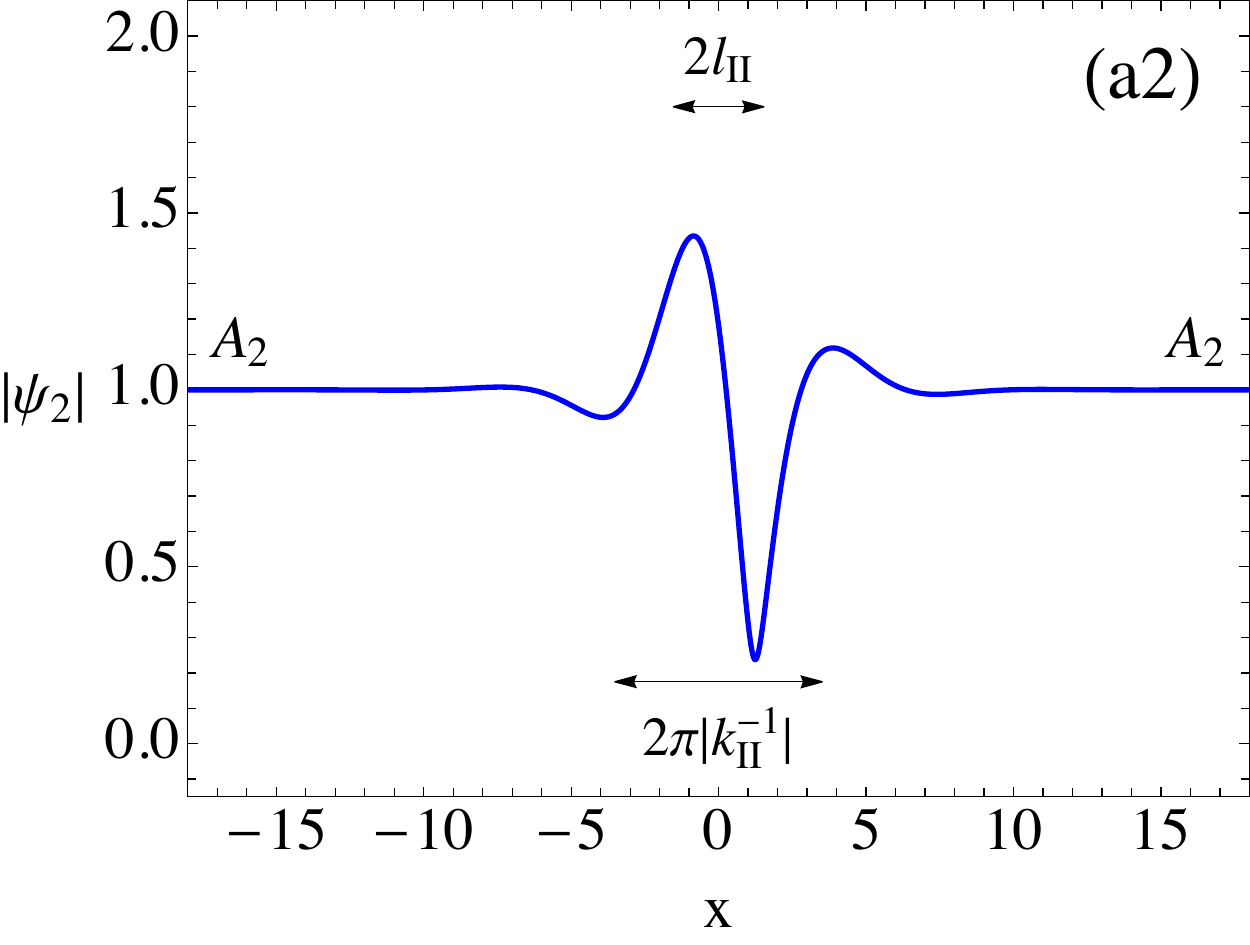}\,\,\,\,\,
    \includegraphics[width=0.3\linewidth]{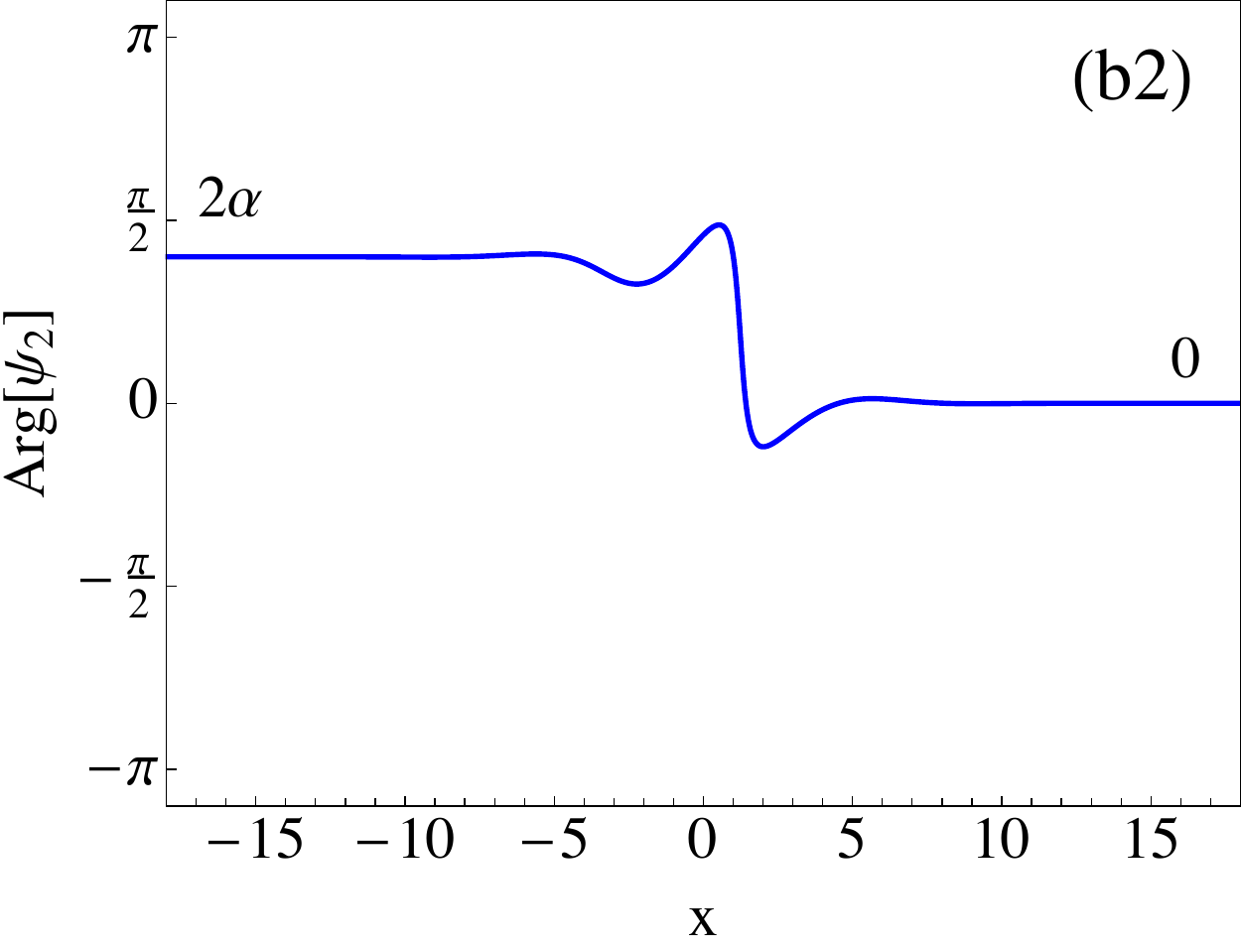}\,\,\,\,\,
    \includegraphics[width=0.31\linewidth]{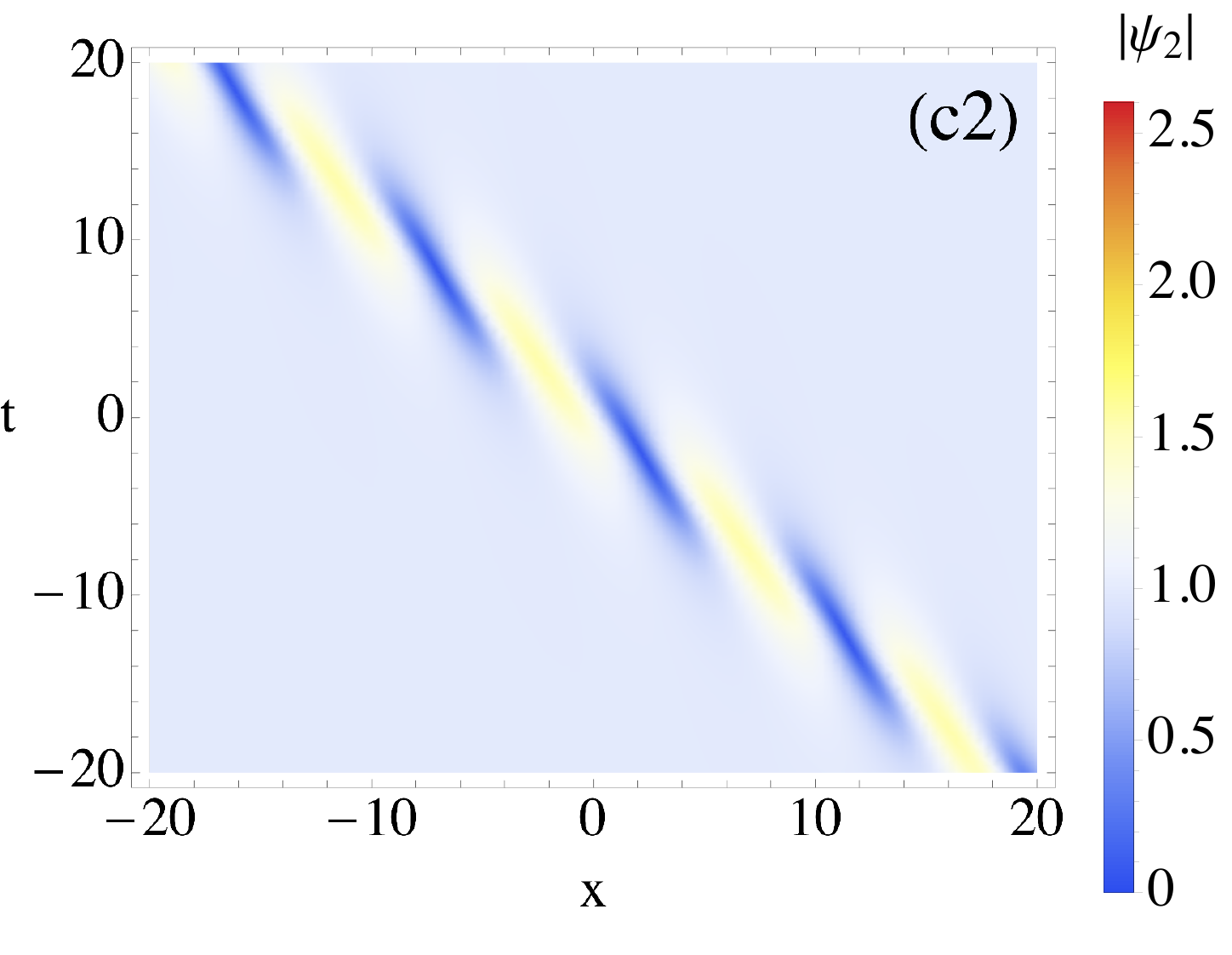}
\caption{
Vector breather of type II. The solution parameters are defined in (\ref{parameters}). (a1,a2) $|\psi_{1,2}|$ at $t=0$ with indicated asymptotic values, see Eq.~(\ref{asymptotics_II}), and characteristic size and wavelength computed according to Eq.~(\ref{characteristic_values2}). (b1,b2) $\mathrm{Arg}[\psi_{1,2}]$ at $t=0$ with indicated asymptotic values. (c1,c2) spatio-temporal evolution  of $|\psi_{1,2}|$.}
\label{fig_02}
\end{figure}
The type II breather has the following asymptotics: 
\begin{eqnarray}
\label{asymptotics_II}
&&\psi_{1,2}^{\mathrm{II}} \to A_{1,2} e^{2\,\mathrm{i}\,\alpha}; \qquad\qquad x \to -\infty
\\\nonumber
&&\psi_{1,2}^{\mathrm{II}} \to A_{1,2};  \qquad\qquad x \to +\infty.
\end{eqnarray}
Fig.~\ref{fig_02} shows an example of type II localized breather having parameters (\ref{parameters}), which moves with a nonzero group velocity.

\begin{figure}[!t]
\centering
    \includegraphics[width=0.3\linewidth]{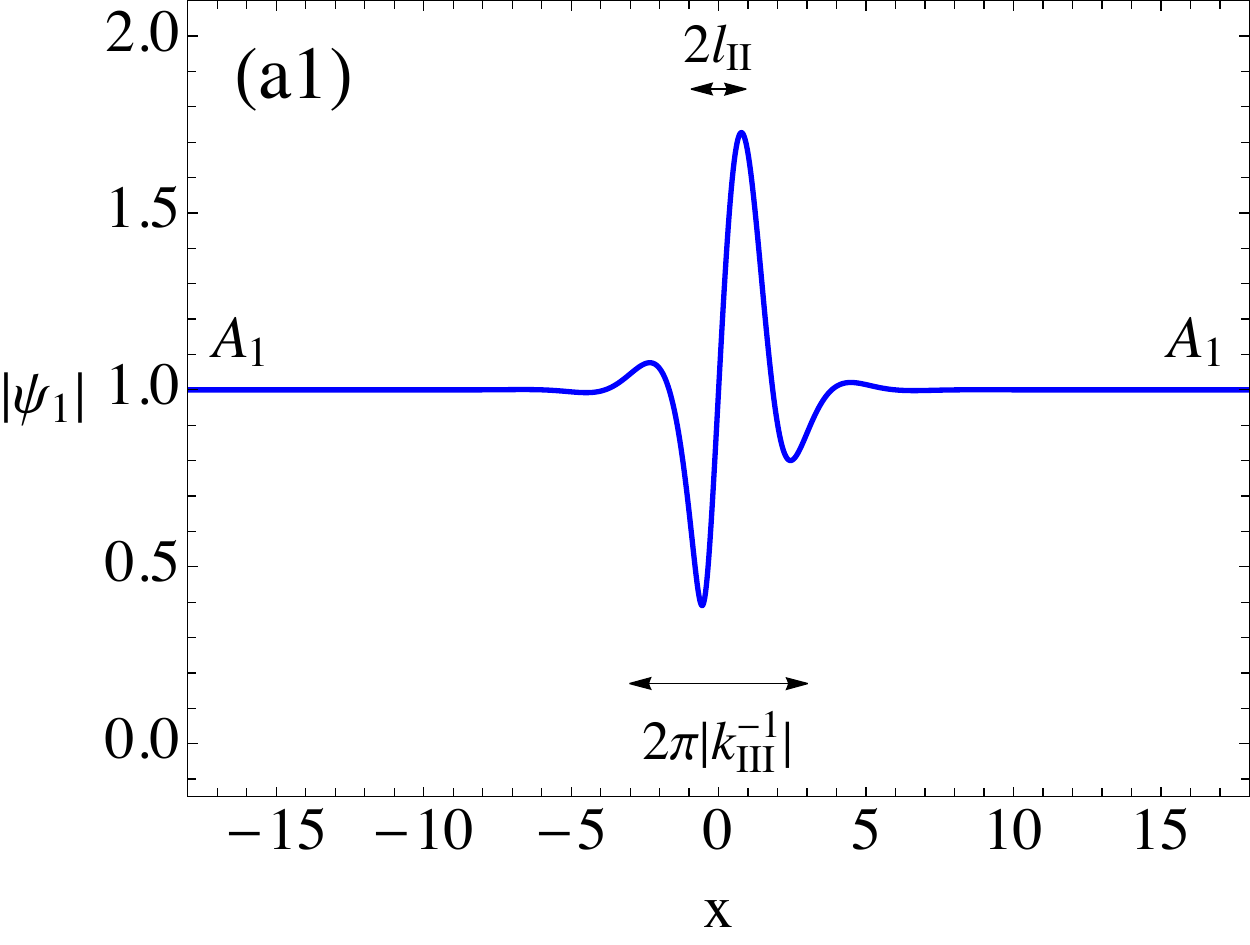}\,\,\,\,\,
    \includegraphics[width=0.3\linewidth]{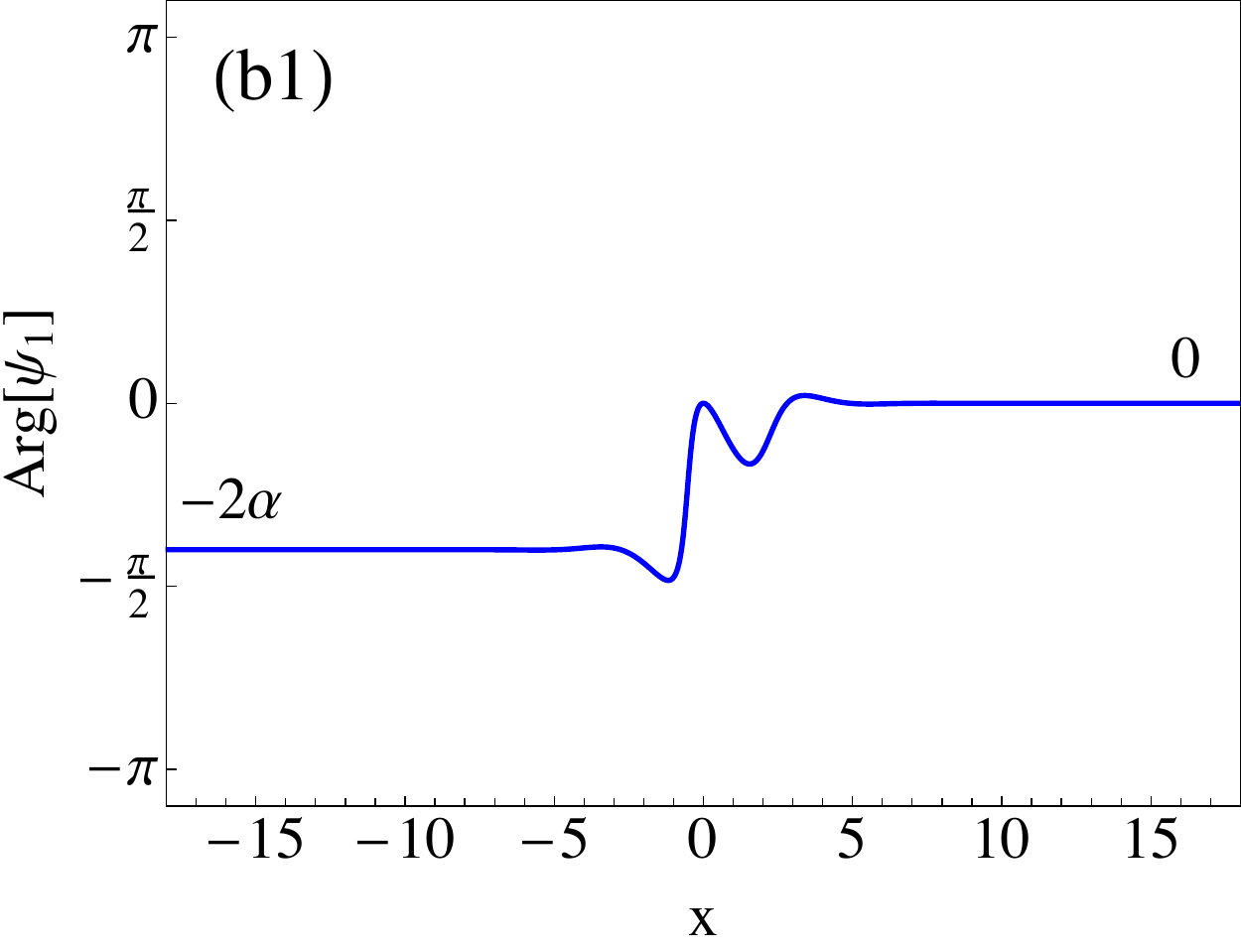}\,\,\,\,\,
    \includegraphics[width=0.31\linewidth]{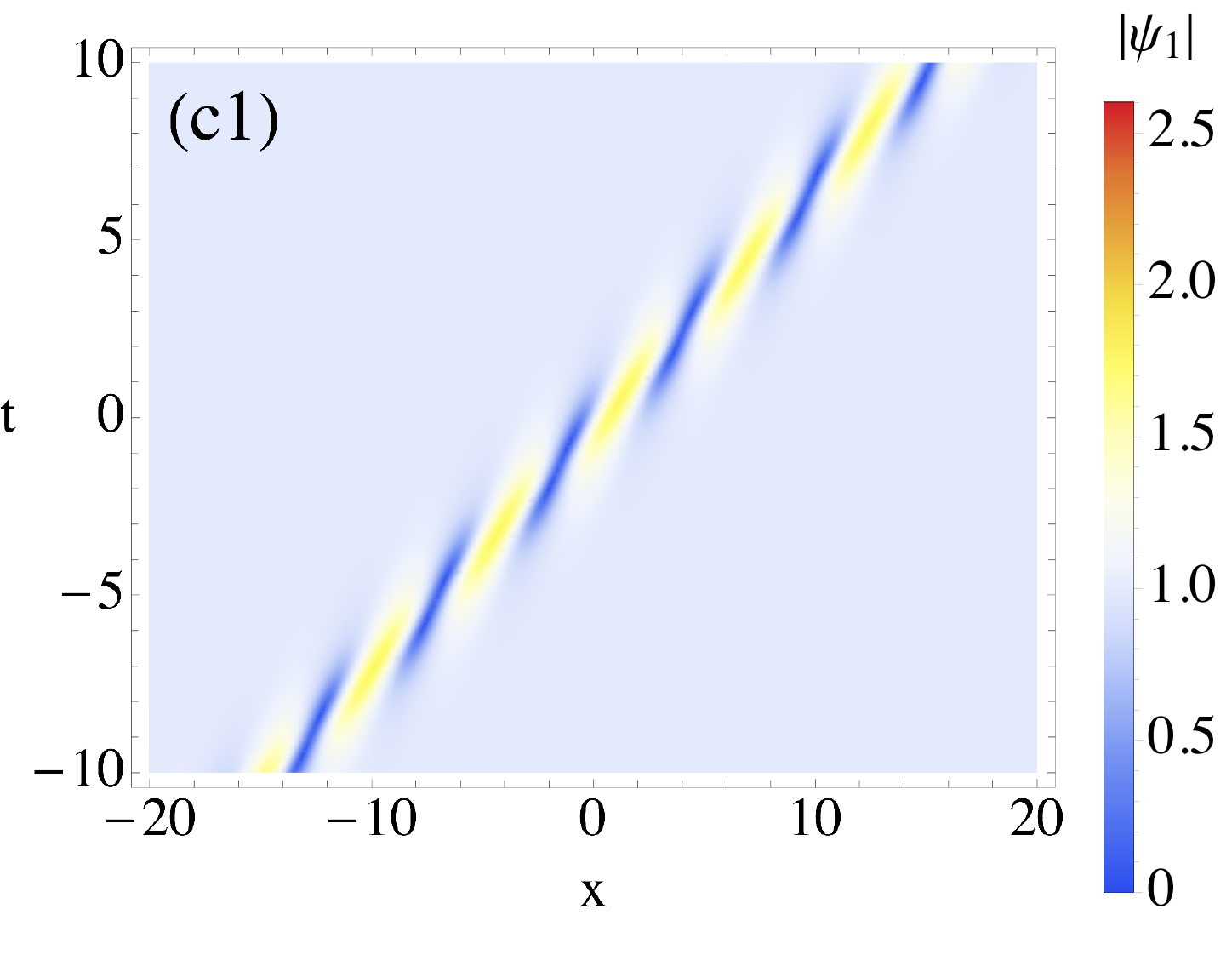}\\
    \includegraphics[width=0.3\linewidth]{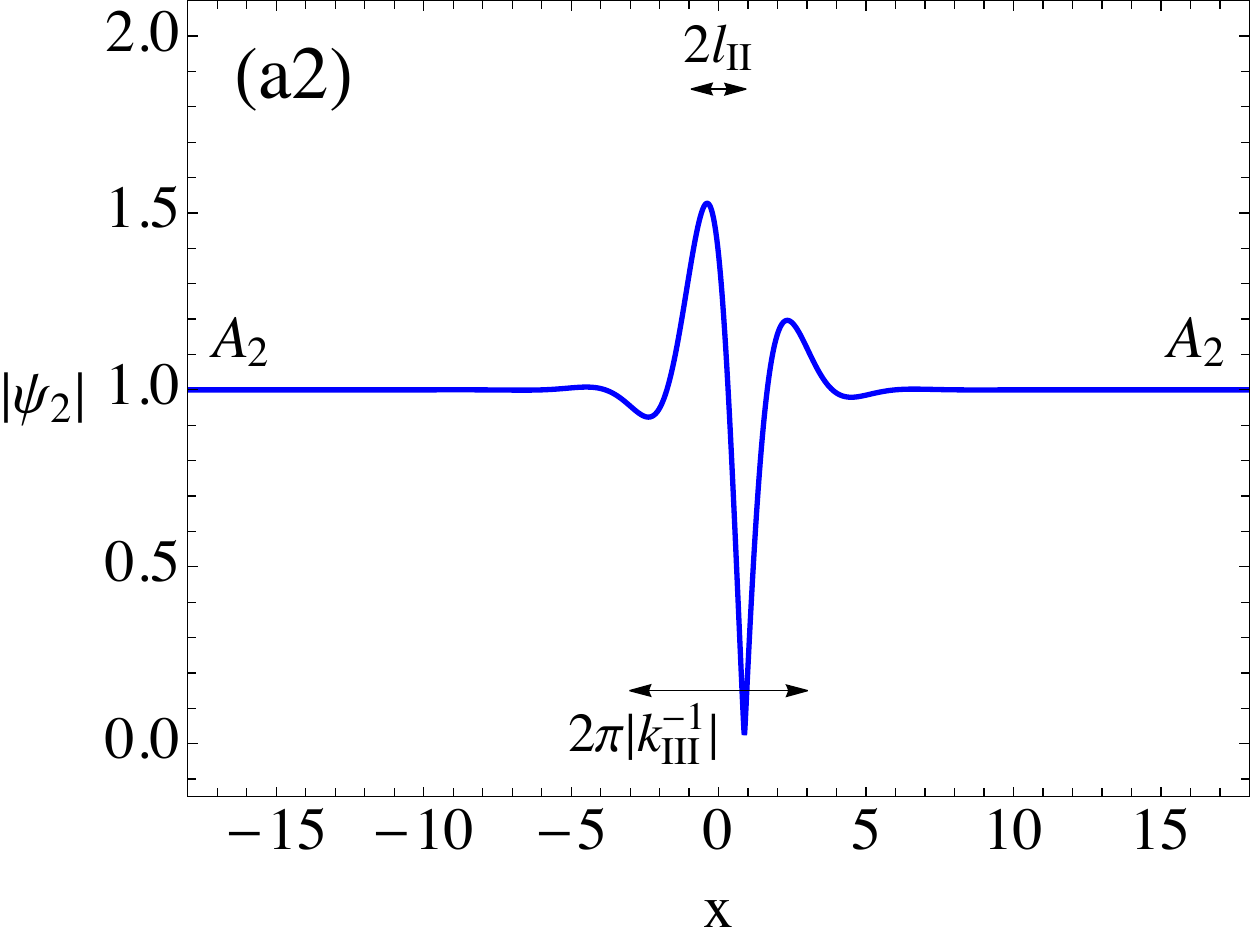}\,\,\,\,\,
    \includegraphics[width=0.3\linewidth]{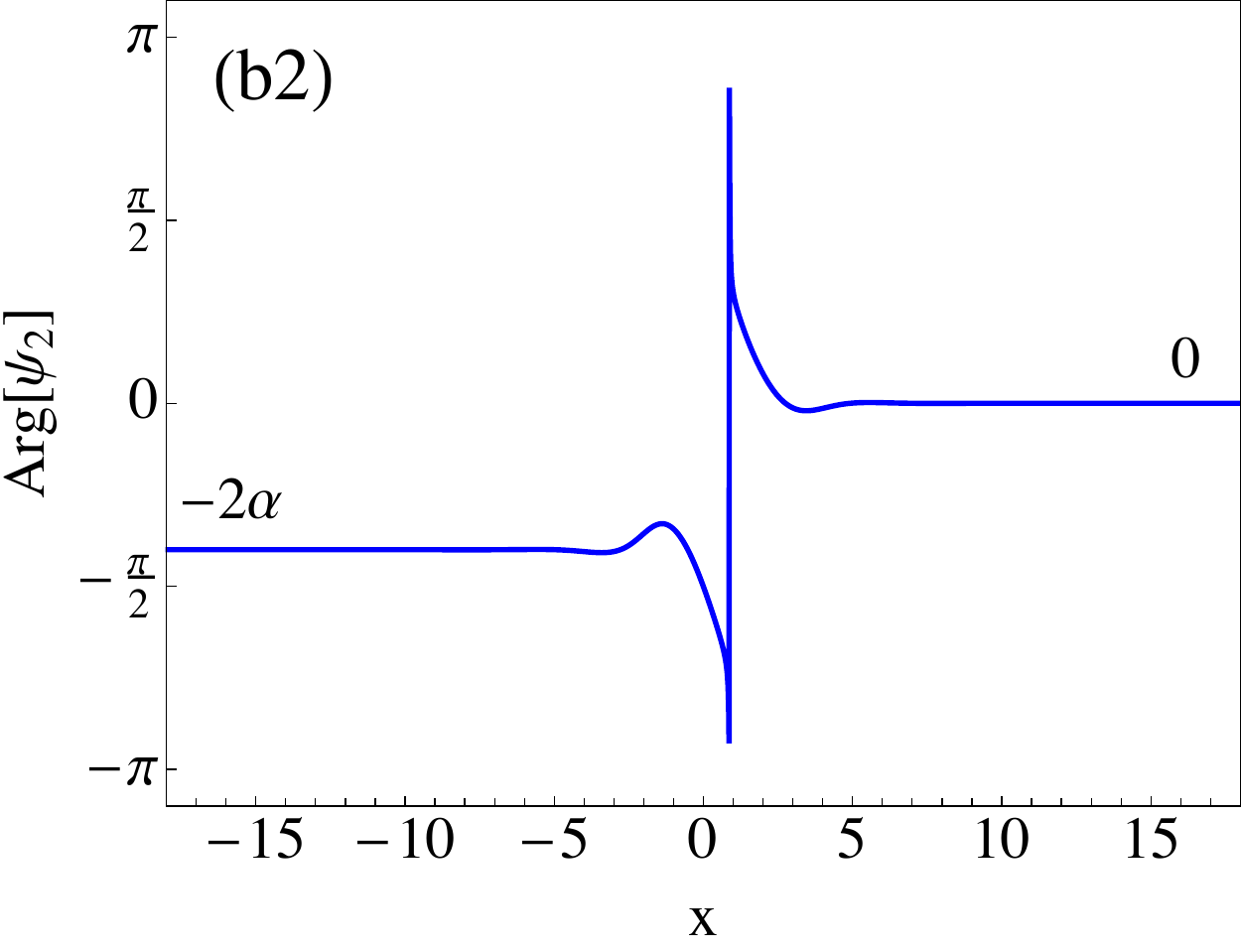}\,\,\,\,\,
    \includegraphics[width=0.31\linewidth]{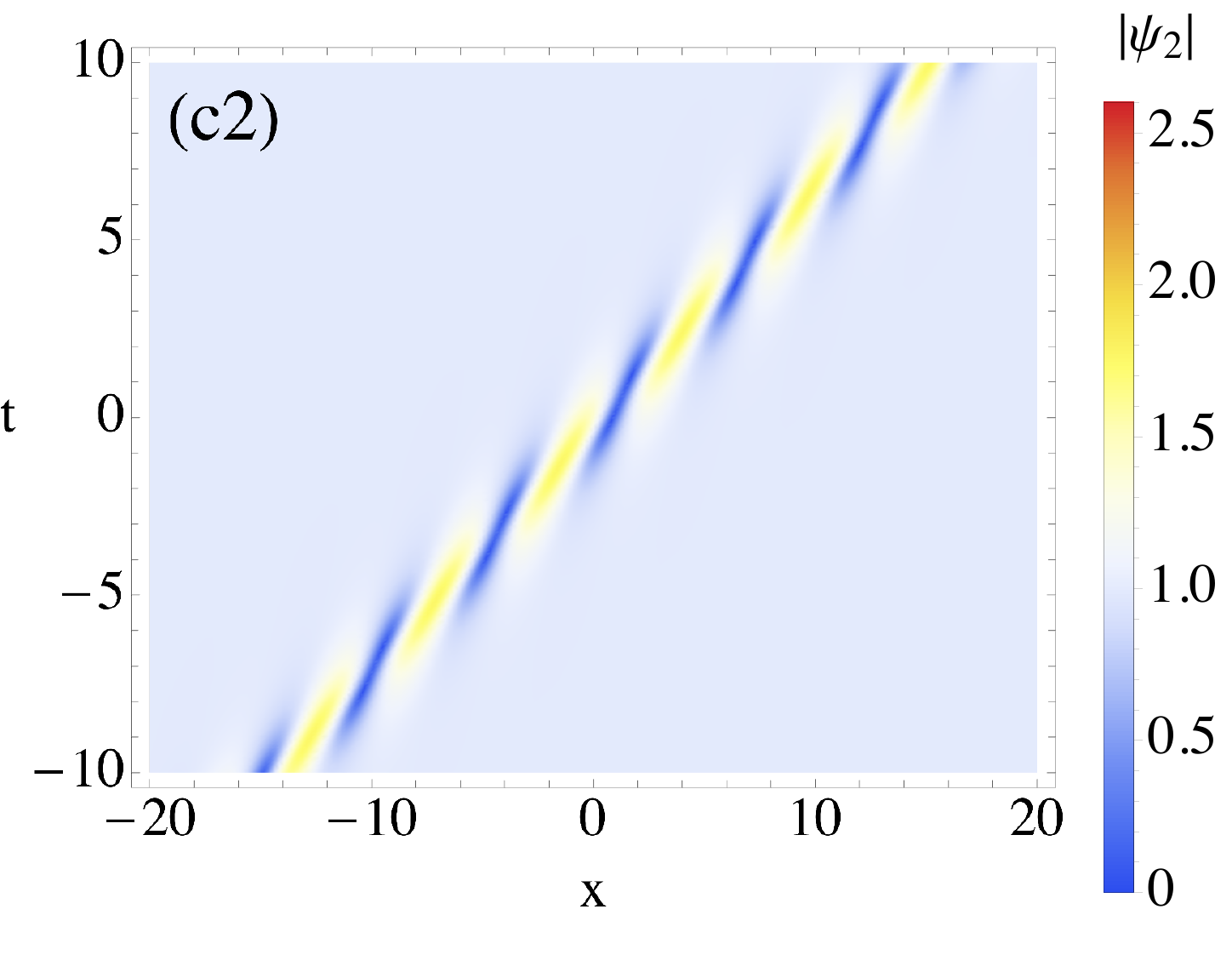}
\caption{
Vector breather of type III. The solution parameters are defined in (\ref{parameters}). (a1,a2) $|\psi_{1,2}|$ at $t=0$ with indicated asymptotic values, see Eq. (\ref{asymptotics_III}), and characteristic size and wavelength computed according to Eqs. (\ref{characteristic_values3}). (b1,b2) $\mathrm{Arg}[\psi_{1,2}]$ at $t=0$ with indicated asymptotic values. (c1,c2) spatio-temporal evolution  of $|\psi_{1,2}|$.}
\label{fig_03}
\end{figure}

Finally, for $C_{2} =0$ we obtain type III breather. Writing components $\mathbf{C}$ in the form:
\begin{eqnarray}
\label{C_III_param}
    &&C_{0} = e^{-\mathrm{Im}[\lambda]\delta - i\theta /2}, 
    \\\nonumber
    &&\mathrm{i}rC_{1} = e^{\mathrm{Im}[\zeta]\delta + i\theta/2}, \quad C_{2} = 0,
\end{eqnarray}
from (\ref{solution}) we obtain:
\begin{eqnarray}
	\psi_{1}^{\mathrm{III}} = A_1 - 4\mathrm{i}\sin\alpha e^{-\mathrm{i}\alpha} \cosh\xi
	\frac{A_1 - A_2 e^{u_{\mathrm{III}} - \mathrm{i}v_{\mathrm{III}} }}{e^{2u_{\mathrm{III}}-\xi} + 2\cosh\xi},
	\\
	\psi_{2}^{\mathrm{III}} = A_2 - 4\mathrm{i}\sin\alpha e^{-\mathrm{i}\alpha} \cosh\xi
	\frac{A_2 + A_1 e^{u_{\mathrm{III}} - \mathrm{i}v_{\mathrm{III}} }}{e^{2u_{\mathrm{III}}-\xi} + 2\cosh\xi},
	\label{sol3}
\end{eqnarray}
where
\begin{eqnarray}
    u_{\mathrm{III}} = \frac{l_{\mathrm{III}}^{-1}(x  -V_{\mathrm{III}} t - \delta)}{2}, \,\, v_{\mathrm{III}}= k_{\mathrm{III}} x - \omega_{\mathrm{III}} t + \theta .
\end{eqnarray}
Physical characteristics of type III breather are the following:
\begin{eqnarray}
\label{characteristic_values3}
    &&l_{\mathrm{III}} = (2 (\mathrm{Im}[\lambda] + \mathrm{Im}[\zeta]))^{-1} = (2 A e^{\xi} \sin{\alpha})^{-1},
    \\\nonumber
    &&V_{\mathrm{III}} = \frac{\mathrm{Im}[\lambda^2+\zeta^2]/2 + \mathrm{Im}[\lambda\zeta]}{\mathrm{Im}[\lambda] + \mathrm{Im}[\zeta]} = Ae^{\xi}\cos{\alpha},
    \\\nonumber
    &&k_{\mathrm{III}} = \mathrm{Re}[\lambda] + \mathrm{Re}[\zeta] = Ae^{\xi}\cos\alpha,
    \\\nonumber
    &&\omega_{\mathrm{III}} = \frac{1}{2}\mathrm{Re}[\lambda^2+\zeta^2] + \mathrm{Re}[\lambda\zeta] = \frac{A^2}{2}e^{2\xi}\cos 2\alpha.
\end{eqnarray}
The type III breather has the following asymptotics:
\begin{eqnarray}
\label{asymptotics_III}
\psi_{1,2}^{\mathrm{III}} \to A_{1,2} e^{-2\,\mathrm{i}\,\alpha}; \qquad x \to -\infty
\\\nonumber
\psi_{1,2}^{\mathrm{III}} \to A_{1,2}; \qquad x \to +\infty.
\end{eqnarray}
Fig.~\ref{fig_03} shows an example of the type III breather, moving with a nonzero group velocity.

The solutions of type II and III have a similar structure, however differ in asymptotic behaviour and characteristic parameters. For the same eigenvalue these breathers always propagate in opposite directions and the breather II has lager size and characteristic wavelength according to the inequalities $|k_{\mathrm{II}}|<|k_{\mathrm{III}}|$ and $|l_{\mathrm{II}}|>|l_{\mathrm{III}}|$, see Eqs.~(\ref{characteristic_values2}) and (\ref{characteristic_values3}). The following change of the spectral parameter,
\begin{equation}
\label{II_to_III_transform}
    \xi \rightarrow - \xi, \qquad \alpha \rightarrow \pi-\alpha,
\end{equation}
transforms type II solution (\ref{sol2}) into type III solution (\ref{sol3}). In terms of $\lambda$ spectral variable, the transformation (\ref{II_to_III_transform}) means that we change the Riemann sheets of the function $\zeta(\lambda)$. This situation is not typical in the IST theory, where usually different Riemann sheets correspond to the same class of solutions. For example, in the scalar NLSE model, the jump to another Riemann sheet only changes the breather phase, leaving the solution the same. One can check that type I solution (\ref{s1}) is invariant to the transformation (\ref{II_to_III_transform}), when the additional replacement $\theta \rightarrow - \theta$ is applied.

Moving and spatially localized breathers, as shown in Fig.~\ref{fig_01} are often called as Tajiri–Watanabe, see \cite{tajiri1998breather}, or general breather. The eigenvalues of the general breather belongs to the broad region,
\begin{equation}
\label{General_par_set}
\{\mathrm{Re}[\lambda] \ne 0,\quad\mathrm{Im}[\lambda] > 0\},
\quad \{\alpha\ne\pi/2, \quad \xi>0\},
\end{equation}
so that $l_{\mathrm{I}}$ and $V_{\mathrm{I}}$ are finite and nonzero. Similar to type I, the general breathers of type II and III are localized and move on the condensate background. At the same time, the structure of type II and III solutions fundamentally differs from type I and cannot be retrieved by a solution transformation, similar to Eq.~(\ref{typeI_transformation}). One can say that the general breathers of type II and III represent a nontrivial vector counterpart of the scalar NLSE breather.

\begin{figure}[!t]
\centering
    \includegraphics[width=0.31\linewidth]{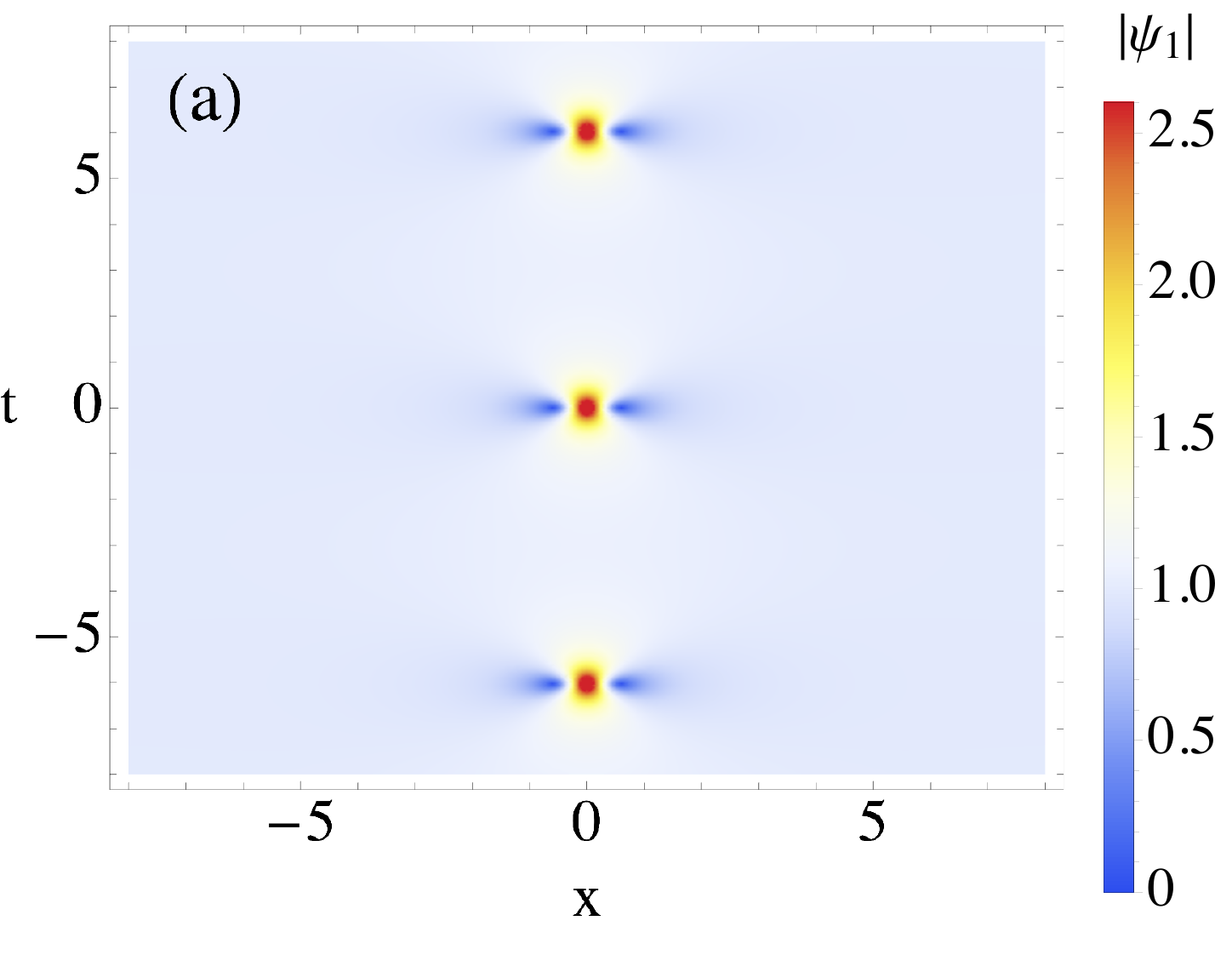}\,\,\,\,\,
    \includegraphics[width=0.31\linewidth]{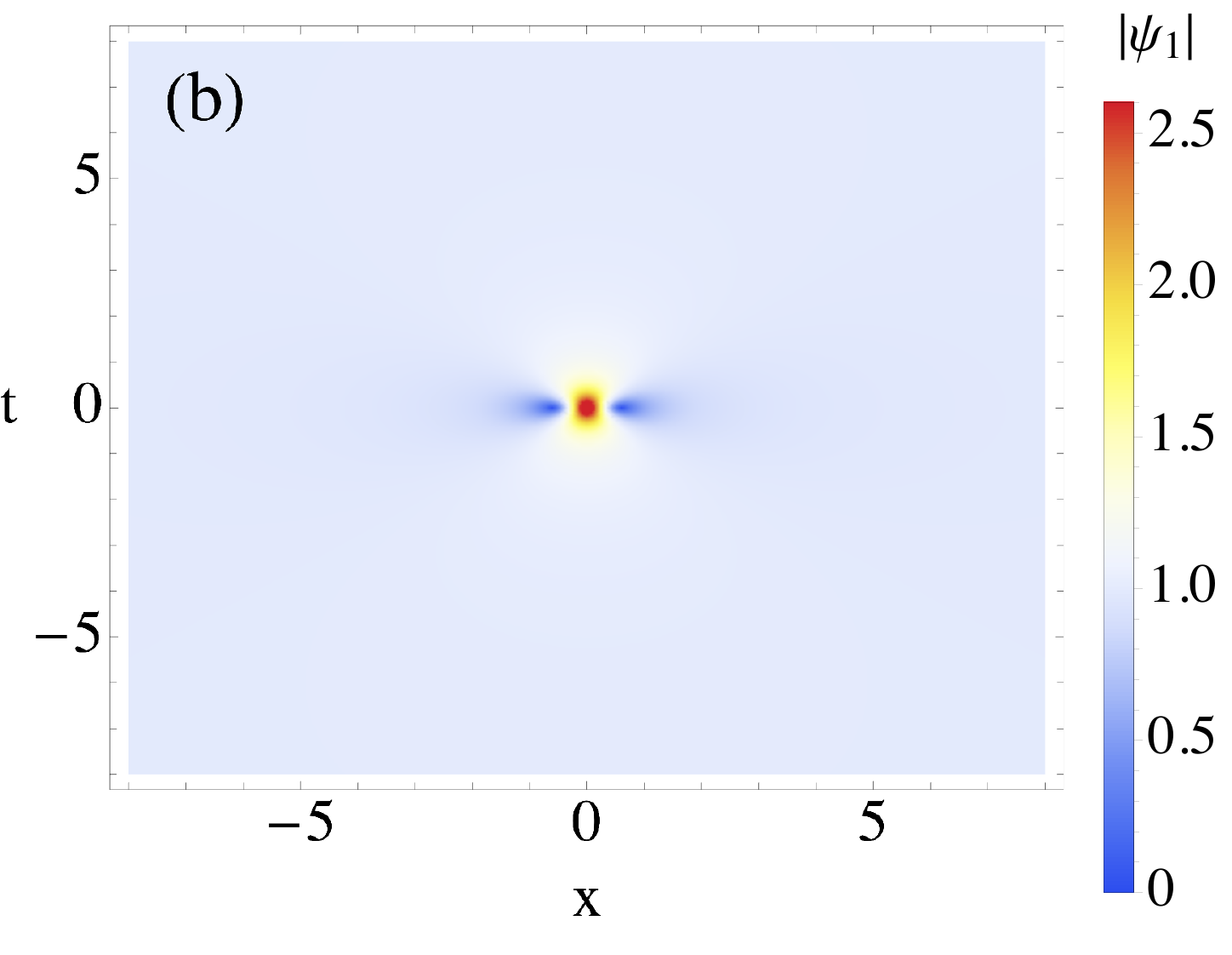}\,\,\,\,\,
    \includegraphics[width=0.31\linewidth]{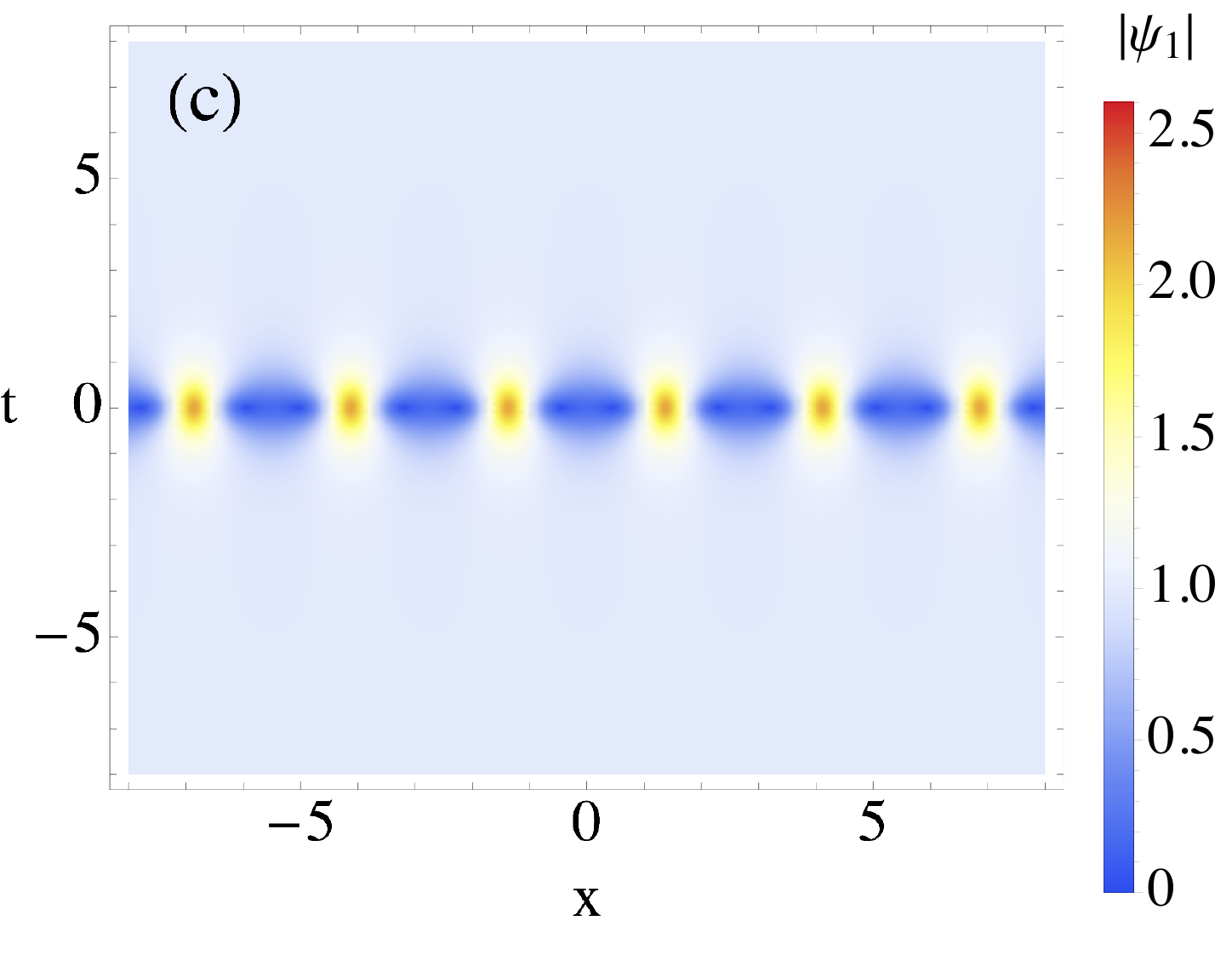}
\caption{
Vector breather of the type I, see Eq.~(\ref{s1}), which can be obtained from solutions of scalar NLSE using the transformation (\ref{typeI_transformation}). (a) Kuznetsov solution (b) Peregrine solution, (c) Akhmediev solution.}
\label{fig_04}
\end{figure}

In addition, the theory of scalar NLSE distinguishes three important particular cases: (a) Kuznetsov breather, (b) Peregrine breather, and (c) Akhmediev breather; all previously studied in detail, see, e.g., the monographs \cite{pelinovsky2008book,akhmediev1997nonlinear}. They correspond to the following choices of the spectral parameter,
\begin{eqnarray}
\label{Kuznetsov_par_set}
&&\text{(a)}\quad \{\mathrm{Re}[\lambda] = 0,\quad\mathrm{Im}[\lambda] > A\},
\quad \{\alpha=\pi/2, \quad \xi>0\},
\\
\label{Peregrine_par_set}
&&\text{(b)}\quad \{\mathrm{Re}[\lambda] = 0,\quad\mathrm{Im}[\lambda] = A\},
\quad \{\alpha=\pi/2, \quad \xi=0\},
\\
\label{Akhmediev_par_set}
&&\text{(c)}\quad \{\mathrm{Re}[\lambda] = 0,\quad\mathrm{Im}[\lambda] < A\},
\quad \{\alpha\ne\pi/2, \quad \xi=0\}.
\end{eqnarray}

Fig.~\ref{fig_04} briefly reminds the key properties of these nonlinear structures. The Kuznetsov breather is a standing one-humped wave group oscillating on the condensate background with a finite time period $T=4\pi/(A^2\sinh 2\xi)$, see Fig.~\ref{fig_04}(a). The Peregrine breather is a degenerate limit of the Kuznetsov solution appearing in the Eq.~(\ref{s1}) at the spectral parameter (\ref{Peregrine_par_set}). It can be found by resolving an uncertainty of the type $0/0$ in solution (\ref{s1}), that leads to the following rational solution,
\begin{equation}
\label{Peregrine}
    \psi_{1,2} = -A_{1,2} + 4A_{1,2} \frac{1-2iA^2 t}{1+4A^2x^2+4A^4 t^2}.
\end{equation}

The Peregrine breather (\ref{Peregrine}) emerges from a small amplitude spatially localized condensate perturbation and then disappears, see Fig.~\ref{fig_04}(b), what makes this solution a popular elementary model of rogue waves formation \cite{pelinovsky2008book,akhmediev2009extreme,shrira2010makes,OsborneBook2010}. Finally, the Akhmediev breather is a periodic solution with spatial period $L=2\pi/(A\sin\alpha)$, which, similar to the Peregrine breather, emerges only once in time, see Fig.~\ref{fig_04}(c). The Akhmediev breather describes an important scenario of the MI development of a periodically perturbed condensate \cite{pelinovsky2008book,akhmediev2009extreme,OsborneBook2010}.

Similar to as it is typically done in the linear theory of polarized light, see, e.g., \cite{gordon2000pmd}, the wavefield components can be considered as vector $(\psi_1, \psi_2)^\mathrm{T}$ which can be rotated by a rotation matrix $\mathbf{T}$ providing the same solution of the Manakov system written in a new basis. In particular, one can switch between solutions of the Manakov system $(\psi_1, \psi_2)^\mathrm{T}$ and $(\widetilde{\psi}_1, \widetilde{\psi}_2)^\mathrm{T}$ having the asymptotics,
\begin{eqnarray}
\label{rotation_basis}
\psi_{1,2}\to
\left(
  \begin{array}{cc}
    A_1 e^{i\phi^{\pm}} \\
    A_2 e^{i\phi^{\pm}} \\
  \end{array}
\right),
\qquad
\widetilde{\psi}_{1,2}\to
\left(
  \begin{array}{cc}
    e^{i\phi^{\pm}} \\
    0 \\
  \end{array}
\right);
\qquad 
x\to\pm \infty,
\end{eqnarray}
using the rotation matrix $\mathbf{T}$ and its inverse counterpart $\mathbf{T}^{-1}$ as follows:
\begin{eqnarray}
\label{rotation_T}
\left(
  \begin{array}{cc}
    \psi_1 \\
    \psi_2 \\
  \end{array}
\right)
= \mathbf{T}
\left(
  \begin{array}{cc}
    \widetilde{\psi}_1 \\
    \widetilde{\psi}_2 \\
  \end{array}
\right),
\quad
\left(
  \begin{array}{cc}
    \widetilde{\psi}_1 \\
    \widetilde{\psi}_2 \\
  \end{array}
\right)
= \mathbf{T}^{-1}
\left(
  \begin{array}{cc}
    \psi_1 \\
    \psi_2 \\
  \end{array}
\right);
\qquad
\mathbf{T}=
\frac{1}{A^2}
\left(
  \begin{array}{cc}
    A_1 & -A_2 \\
    A_2 & A_1 \\
  \end{array}
\right),
\quad
\mathbf{T}^{-1}=
\frac{1}{A^2}
\left(
  \begin{array}{cc}
    A_1 & A_2 \\
    -A_2 & A_1 \\
  \end{array}
\right) .
\end{eqnarray}

Our solutions for the breathers of types $\mathrm{I}$, $\mathrm{II}$ and $\mathrm{III}$ with asymptotics $A_{1,2} e^{i\phi^{\pm}}$ defined by Eqs.~(\ref{asymptotics_I}), (\ref{asymptotics_II}) and (\ref{asymptotics_III}) can be transformed using the inverse matrix $\mathbf{T}^{-1}$ into solutions having zero condensate level in the second component, see Eqs.~(\ref{rotation_basis}) and (\ref{rotation_T}). Most interestingly, for type $\mathrm{I}$ breathers the second component in the new bases is exactly cancelled, i.e. $\widetilde{\psi}^{\mathrm{I}}_2(x,t)\equiv 0$, due to the symmetry (\ref{typeI_relation}). In other words, type $\mathrm{I}$ solutions are flat in the sense of polarization, meanwhile type $\mathrm{II}$ and type $\mathrm{III}$ breathers always have both wavefield components different from zero. In particular, the work \cite{kraus2015focusing} uses the polarization basis corresponding to the case $(\widetilde{\psi}_1, \widetilde{\psi}_2)^\mathrm{T}$, see the illustrations for the second wavefield component in \cite{kraus2015focusing}. One can check that the solutions (\ref{s1}), (\ref{sol2}) and (\ref{sol3}) boil down to those presented in \cite{kraus2015focusing} after the transformation with the matrix $\mathbf{T}^{-1}$.

\begin{figure}[!t]
\centering
    \includegraphics[width=0.3\linewidth]{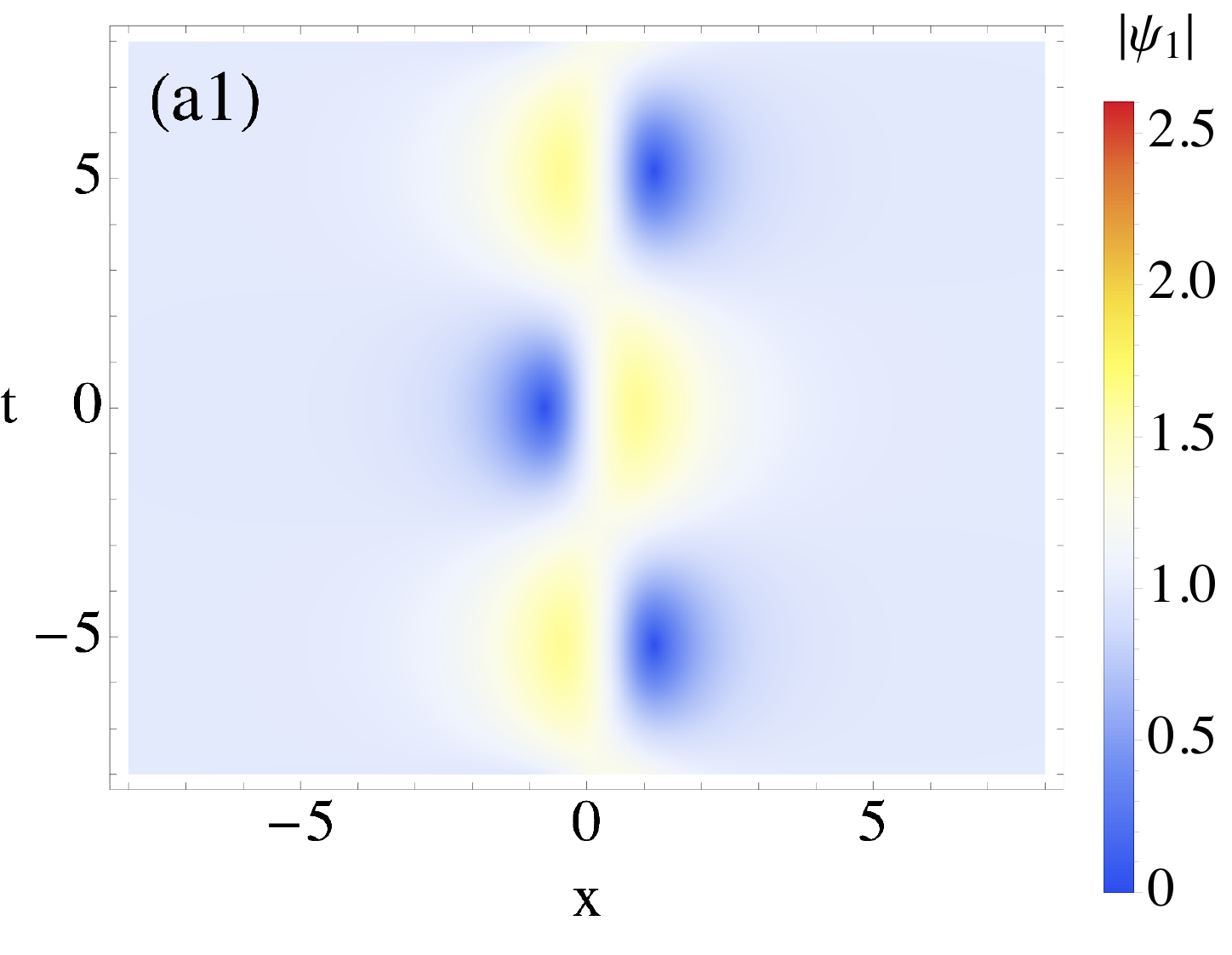}\,\,\,\,\,
    \includegraphics[width=0.3\linewidth]{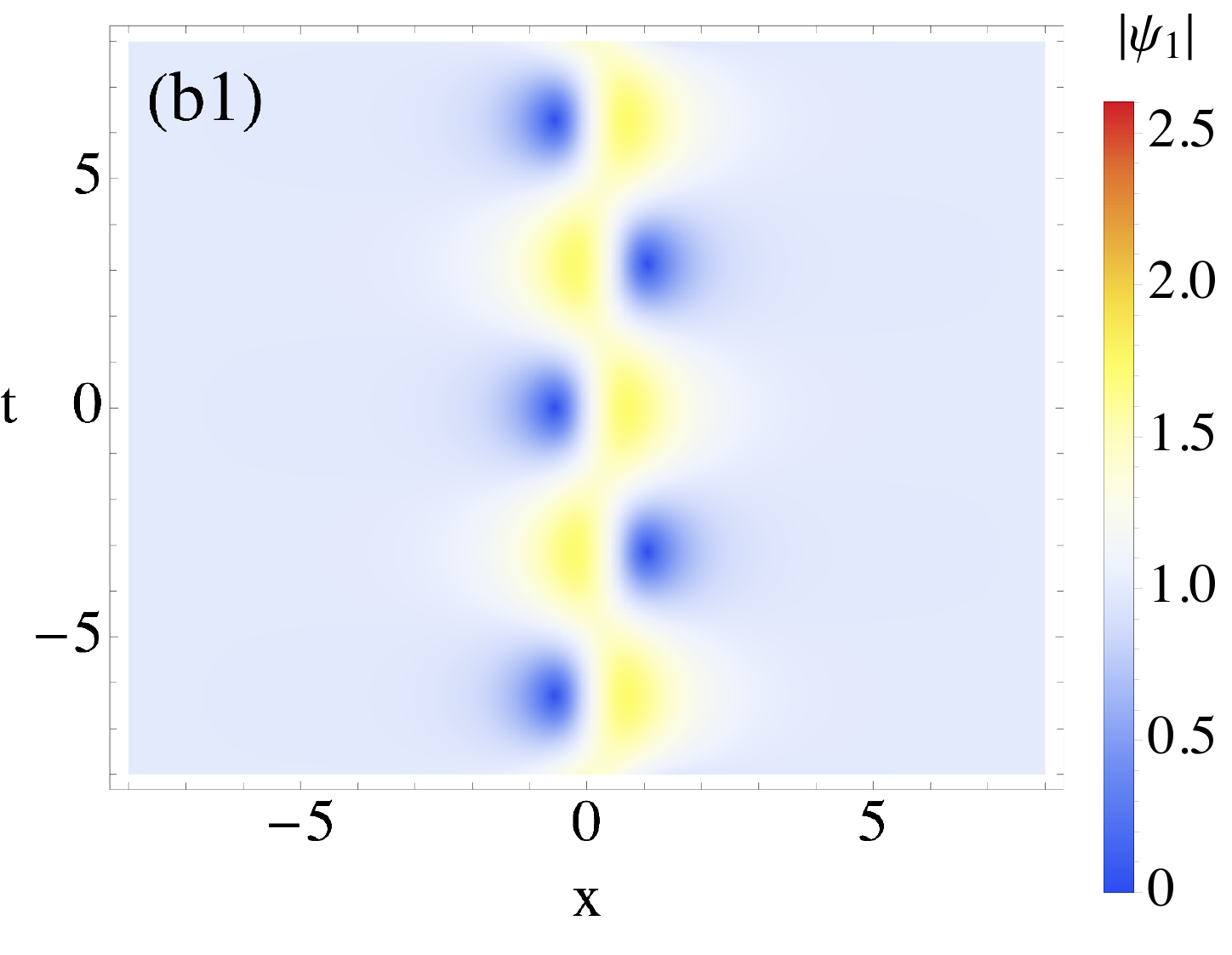}\,\,\,\,\,
    \includegraphics[width=0.31\linewidth]{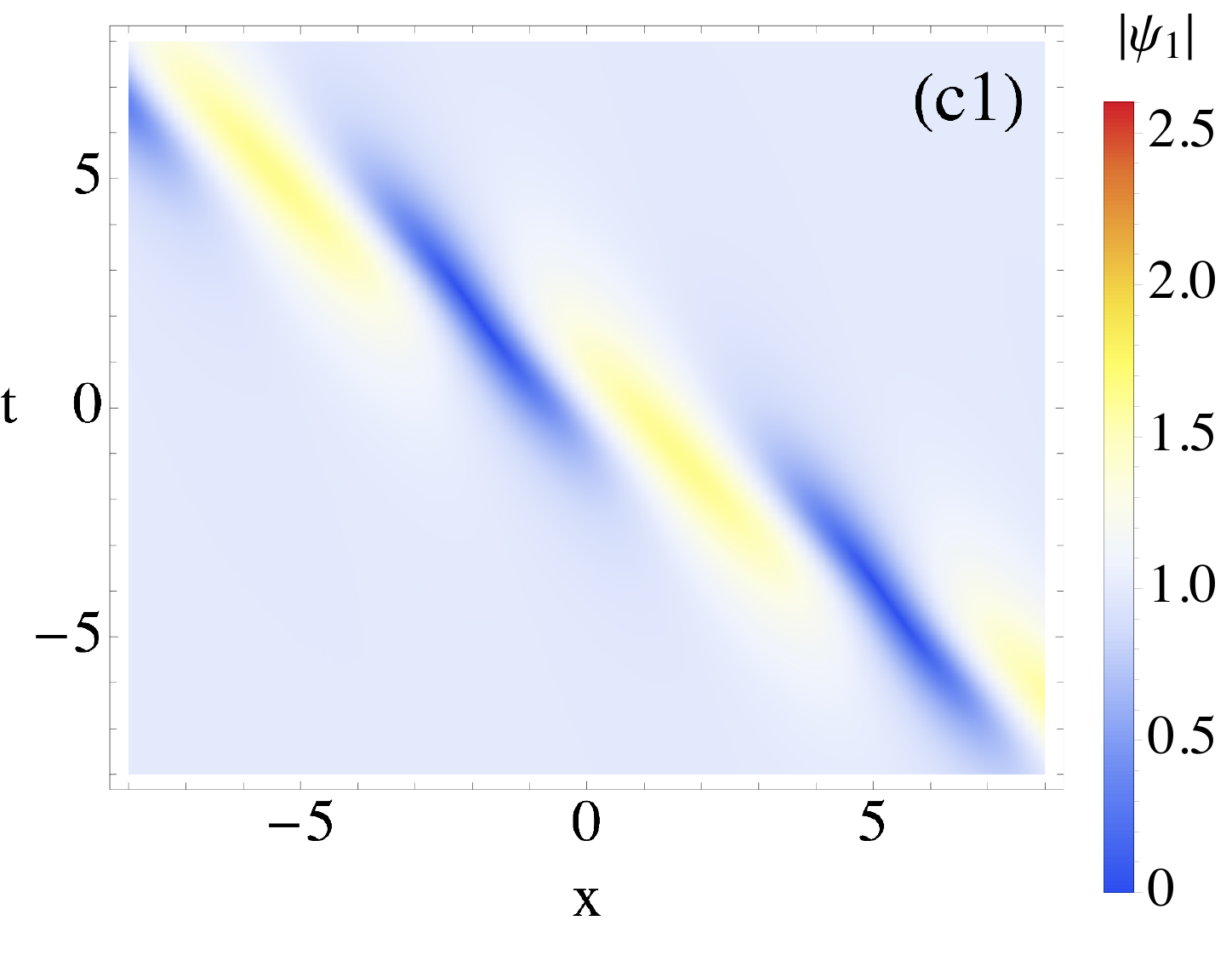}\\
    \includegraphics[width=0.3\linewidth]{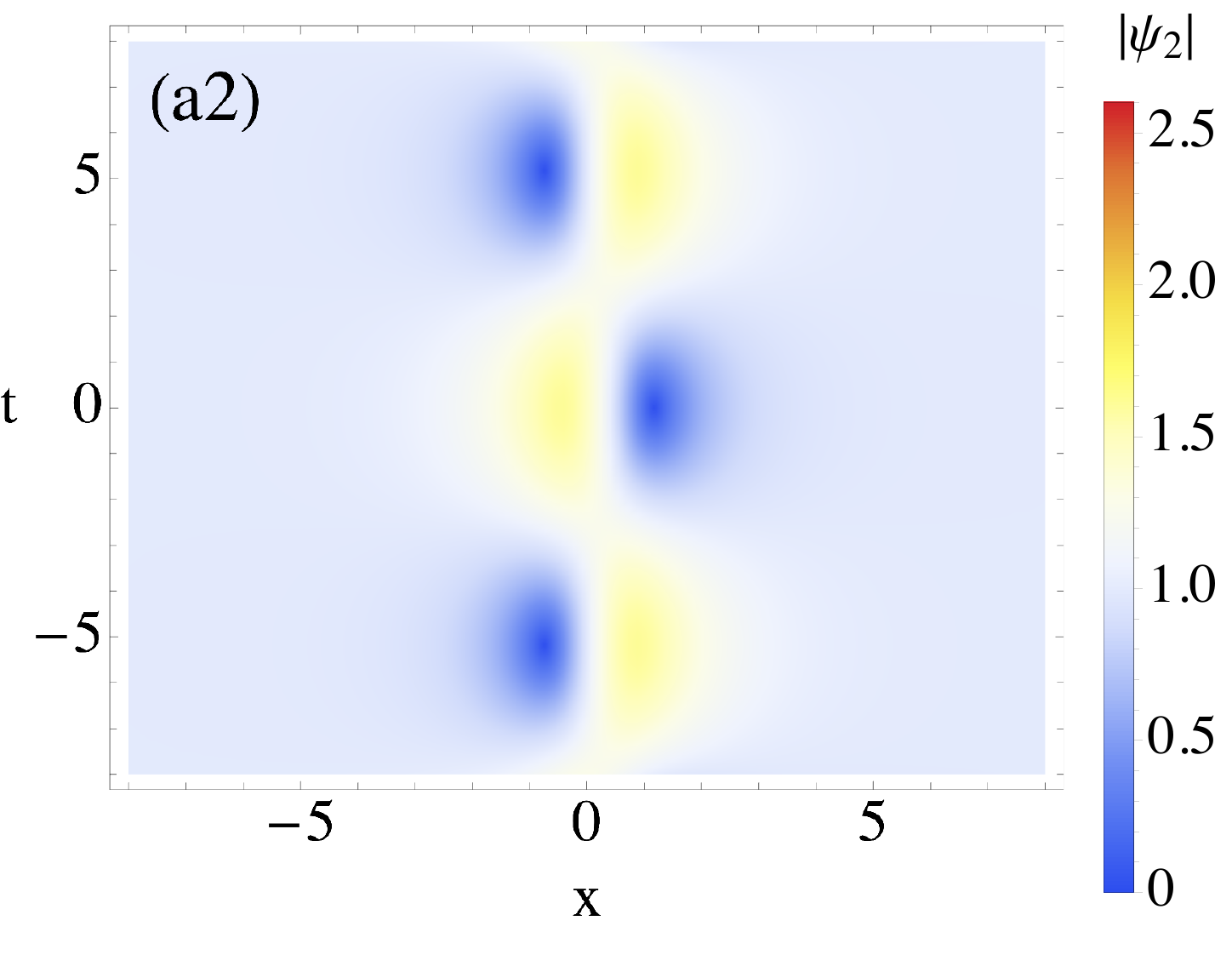}\,\,\,\,\,
    \includegraphics[width=0.3\linewidth]{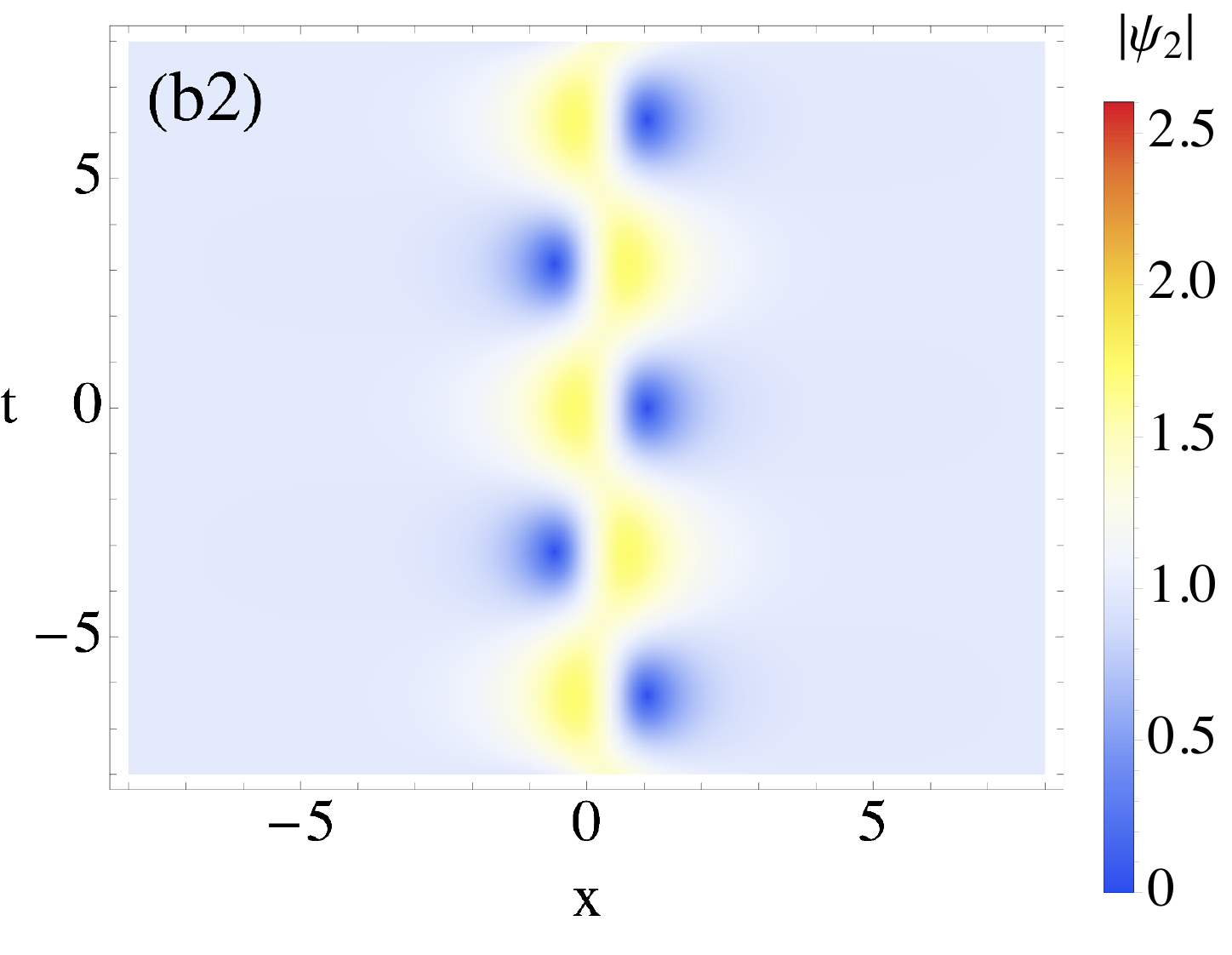}\,\,\,\,\,
    \includegraphics[width=0.31\linewidth]{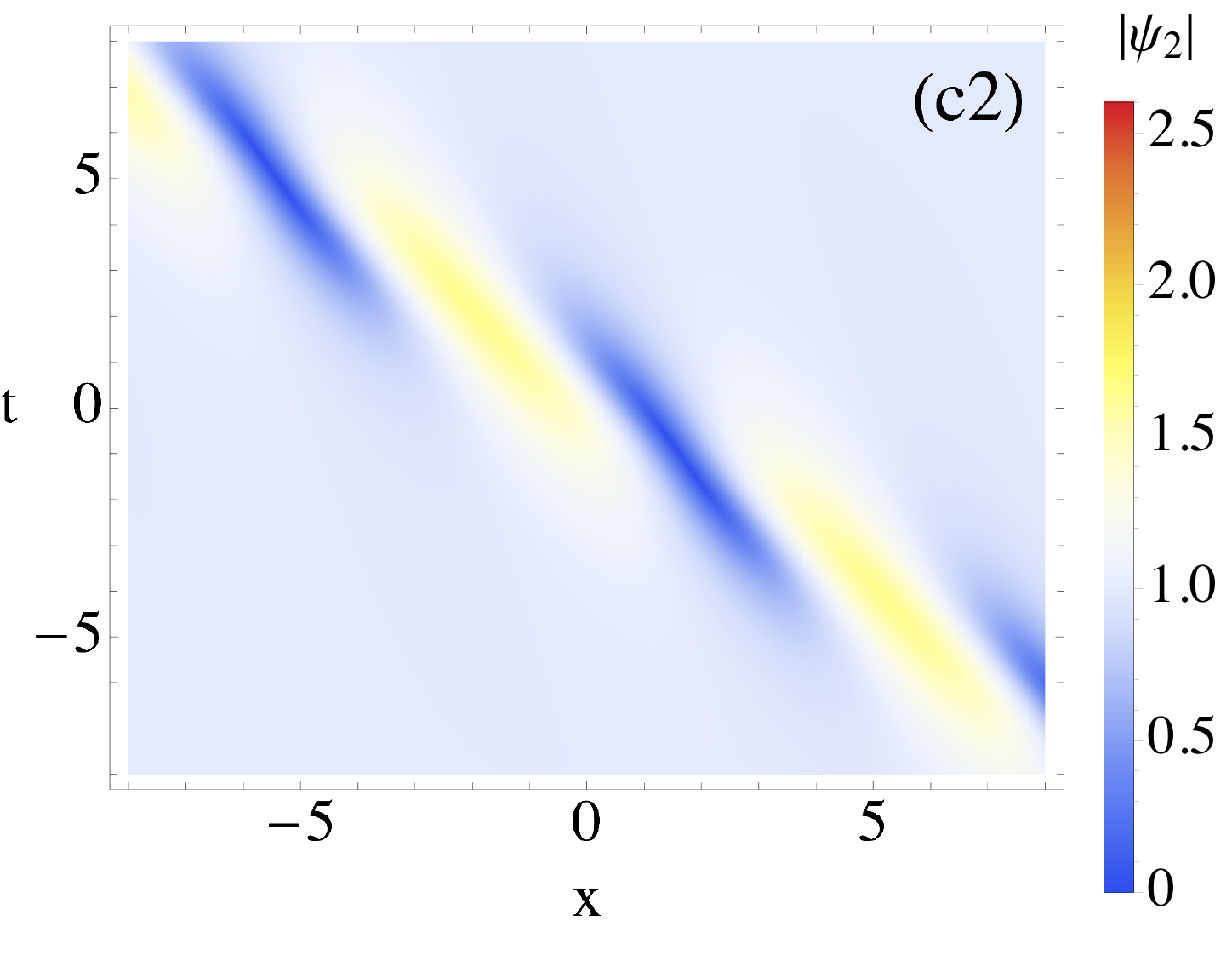}
\caption{
Vector breathers of the type II with spectral parameters defined by Eqs.~(\ref{Kuznetsov_par_set}-\ref{Akhmediev_par_set}). (a) Spectral parameter belongs to the Kuznetsov region, see Eq.~(\ref{Kuznetsov_par_set}). The corresponding solution is a standing dark-bright oscillating breather. (b) Spectral parameter belongs to the Peregrine region, see Eq.~(\ref{Kuznetsov_par_set}). The corresponding solution exhibits a particular case of type II solution (no degeneration of the solution occurs). (c) Spectral parameter belongs to the Akhmediev region, see Eq.~(\ref{Kuznetsov_par_set}). The corresponding solution exhibits a particular case of type II general dynamics (no spatially periodic dynamic occurs).}
\label{fig_05}
\end{figure}

\begin{figure}[!t]
\centering
    \includegraphics[width=0.3\linewidth]{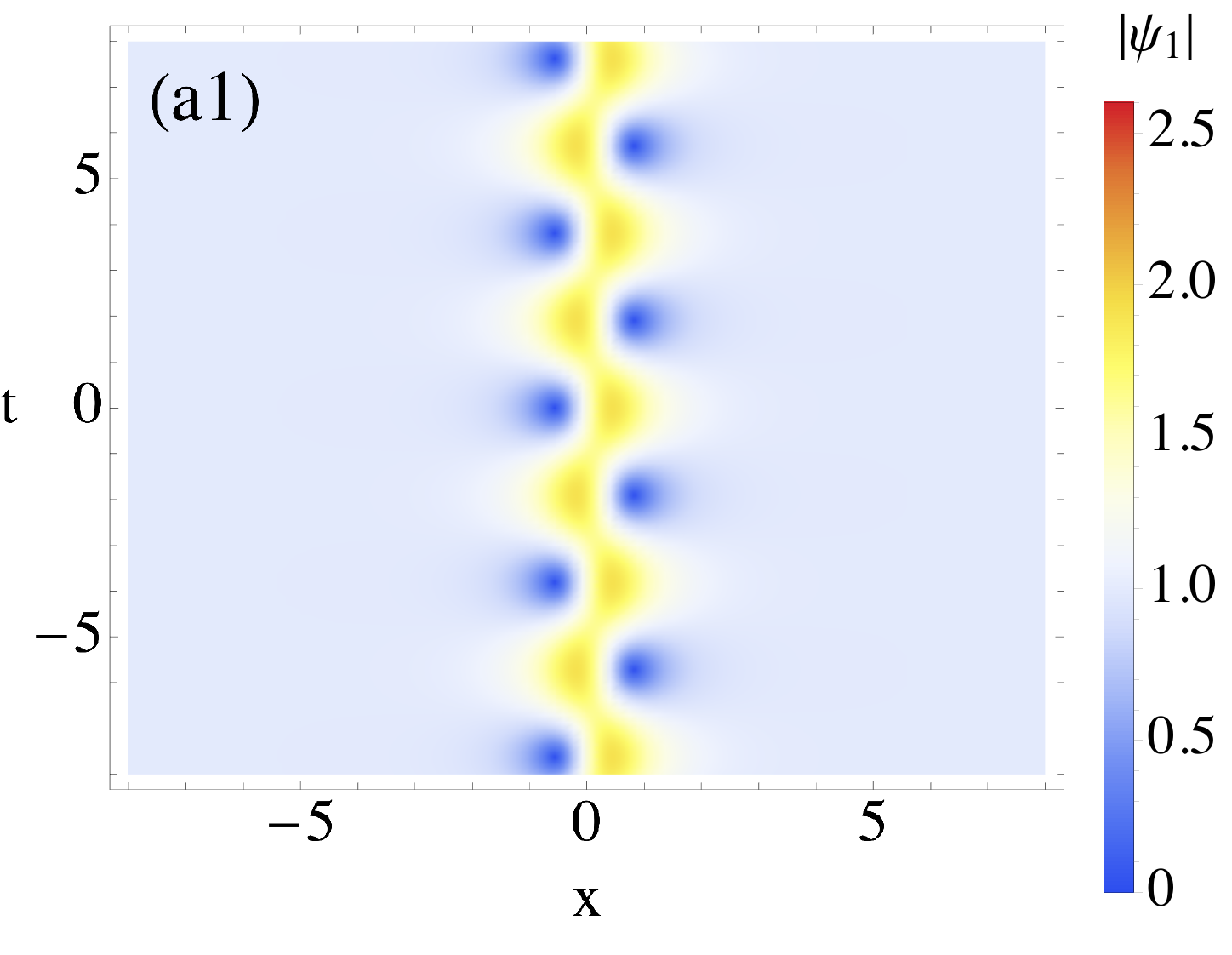}\,\,\,\,\,
    \includegraphics[width=0.3\linewidth]{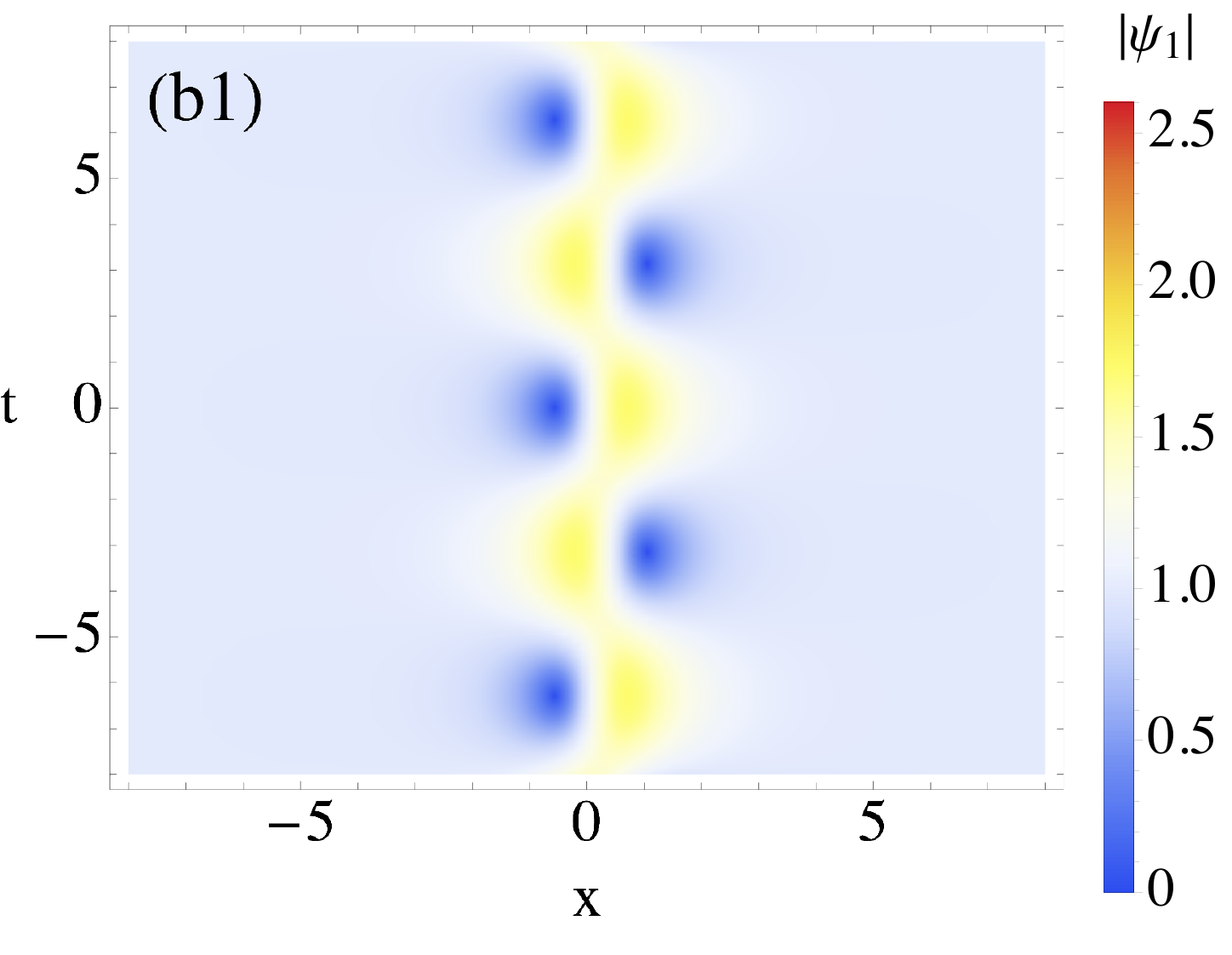}\,\,\,\,\,
    \includegraphics[width=0.31\linewidth]{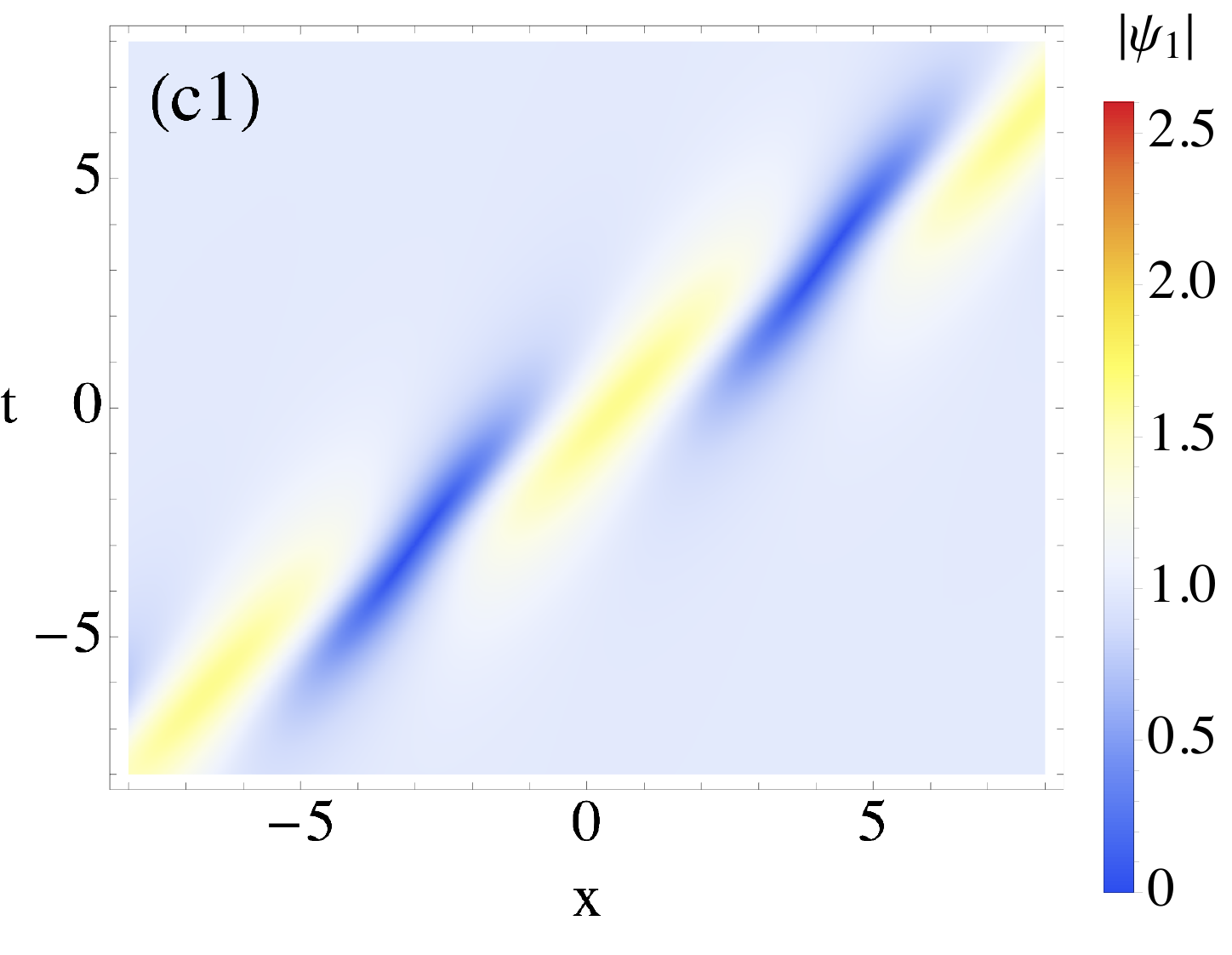}\\
    \includegraphics[width=0.3\linewidth]{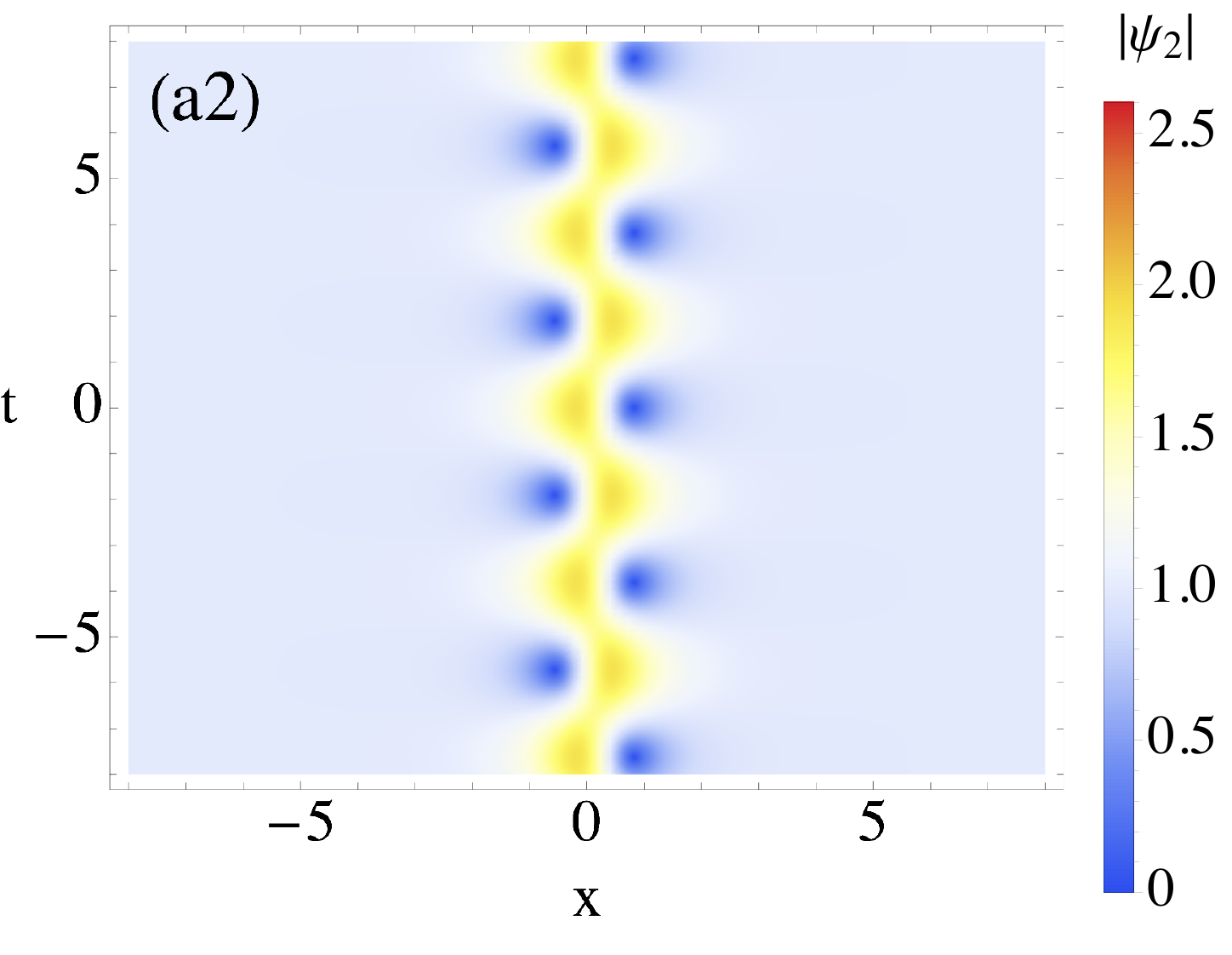}\,\,\,\,\,
    \includegraphics[width=0.3\linewidth]{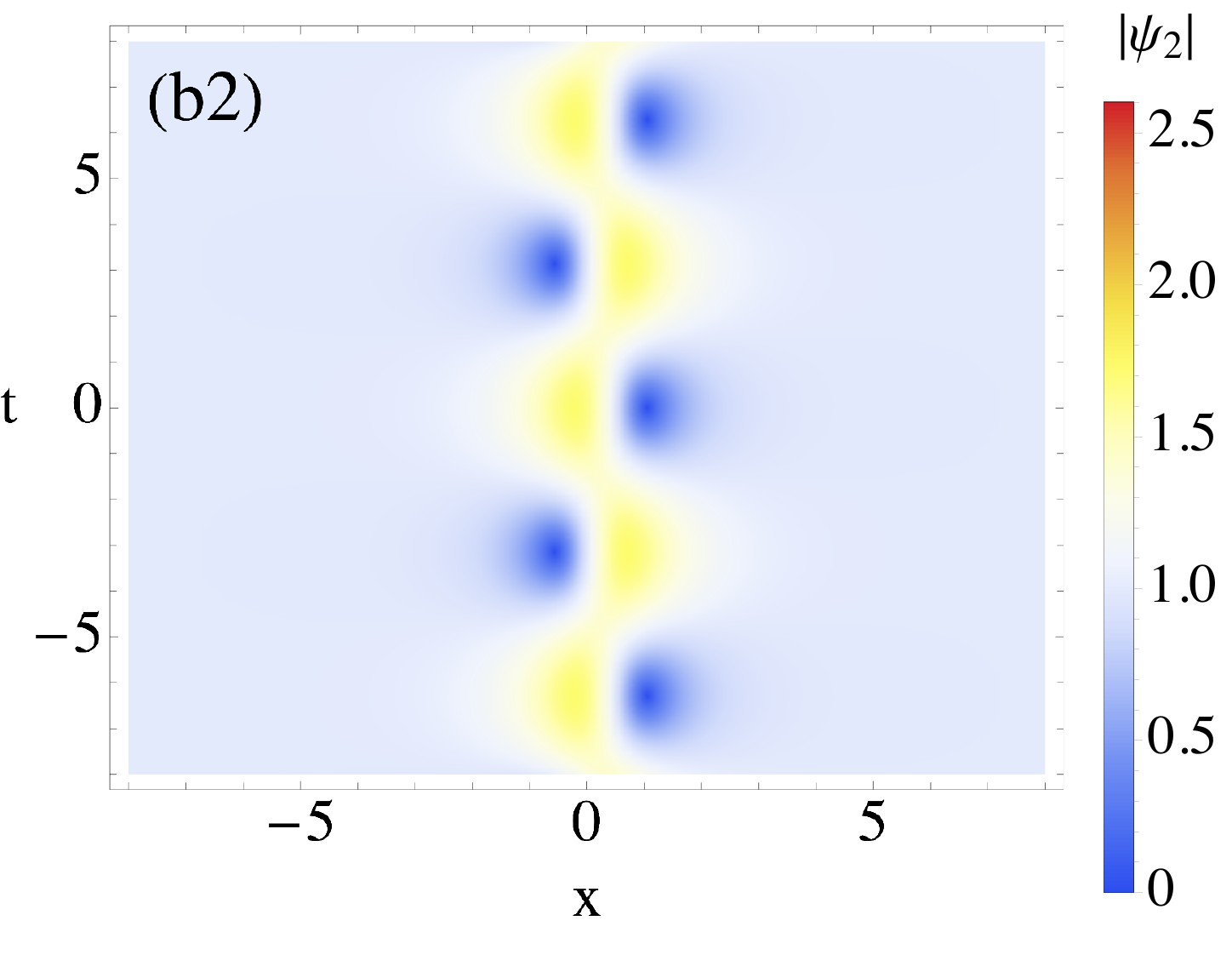}\,\,\,\,\,
    \includegraphics[width=0.31\linewidth]{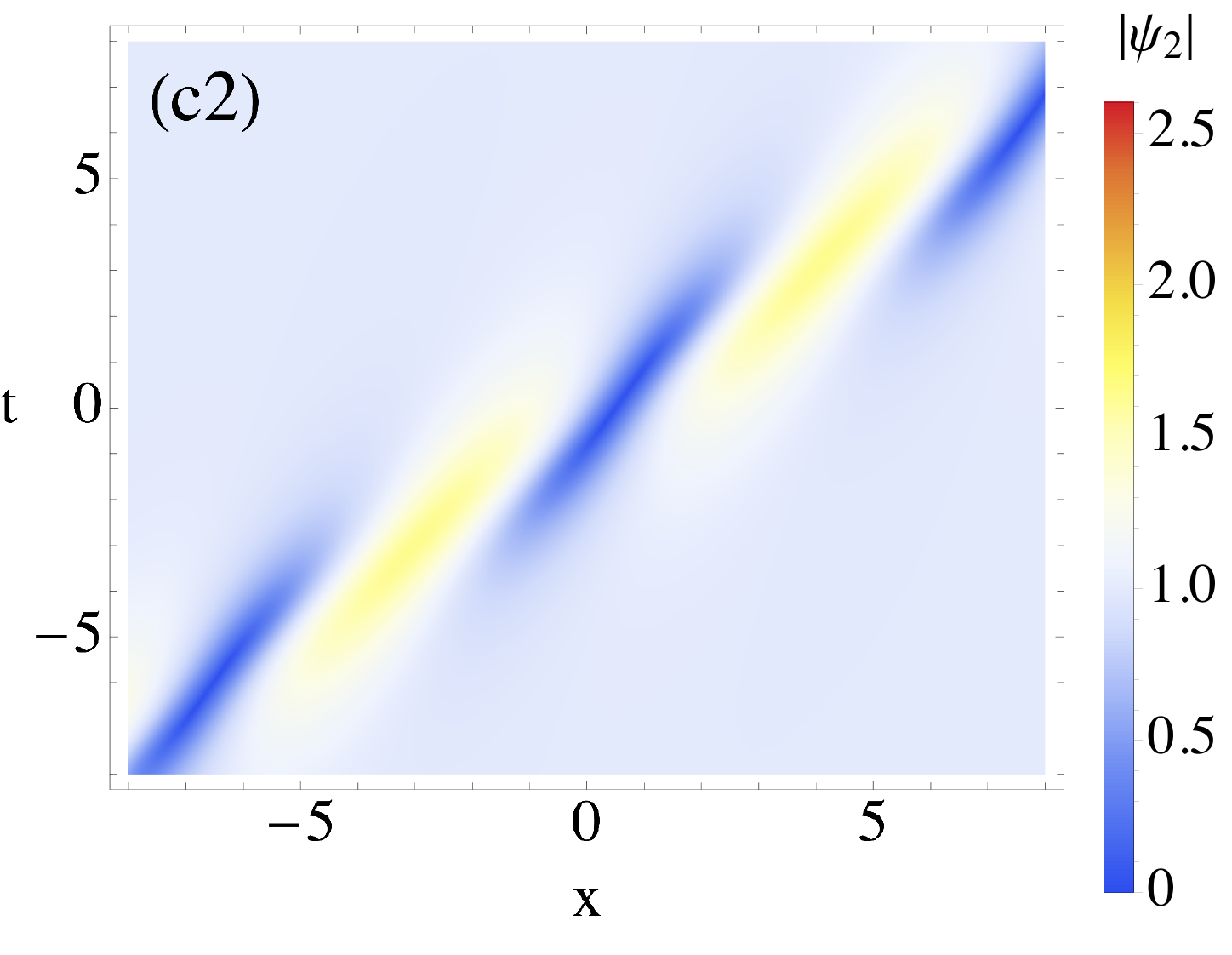}
\caption{
Vector breathers of type III with spectral parameters defined by Eqs.~(\ref{Kuznetsov_par_set}-\ref{Akhmediev_par_set}). (a) Spectral parameter belongs to the Kuznetsov region, see Eq.~(\ref{Kuznetsov_par_set}). The corresponding solution is a standing dark-bright oscillating breather. (b) Spectral parameter belongs to the Peregrine region, see Eq.~(\ref{Kuznetsov_par_set}). The corresponding solution exhibits a particular case of type III solution (no degeneration of the solution occurs). (c) Spectral parameter belongs to the Akhmediev region, see Eq.~(\ref{Kuznetsov_par_set}). The corresponding solution exhibits a particular case of type III Tajiri–Watanabe dynamics (no spatially periodic dynamic occurs).}
\label{fig_06}
\end{figure}

Mathematically, the diversity of scalar breathers emerges due to the nontrivial spectral parameter plane geometry produced by the function $\zeta(\lambda)$ branchcut. Indeed, the Kuznetsov, Akhmediev and Peregrine breathers correspond to the eigenvalue location at the imaginary $\lambda$-axis respectively outside/inside the branchcut and precisely at the branch point of the function $\zeta(\lambda)$, see Eqs.~(\ref{Kuznetsov_par_set}-\ref{Akhmediev_par_set}). Meanwhile, the general Tajiri–Watanabe breather emerges when the eigenvalue is located outside the imaginary $\lambda$-axis. As we already discussed, the general type II and type III breathers exhibit in principal a similar to the scalar case behaviour (in the sense that they are localized moving pulsating breathers). The choices of the eigenvalue locations (\ref{Kuznetsov_par_set}-\ref{Akhmediev_par_set}) lead to fundamentally different wavefield dynamics, which was recently studied in \cite{che2022nondegenerate}. We illustrate the all three cases in Fig.~\ref{fig_05} (type II) and Fig.~\ref{fig_06} (type III). For the spectral parameter choice (\ref{Kuznetsov_par_set}), the type II and III breathers represent a dark-bright standing wave group oscillating on the condensate background, see Fig.~\ref{fig_05}(a) and Fig.~\ref{fig_06}(a). Unlike their scalar counterpart, these breathers change the condensate phase to $\pi$, according to the asymptotics (\ref{asymptotics_II}) and (\ref{asymptotics_III}). Meanwhile, for the set of parameters (\ref{Peregrine_par_set}), no degeneration in the solutions (\ref{sol2}) and (\ref{sol3}) occurs. The wavefield dynamic is similar to the type II and III cases, see Fig.~\ref{fig_05}(b) and Fig.~\ref{fig_06}(b), meaning that there are no nontrivial vector analogs of the rational rogue waves. Finally, when the spectral parameter belongs to the set (\ref{Akhmediev_par_set}), the solutions of the types II and III are moving localized breathers, see Fig.~\ref{fig_05}(c) and Fig.~\ref{fig_06}(c). Accordingly, these solutions are a particular case of the general type II and III breather, and there are no nontrivial vector analogs of the periodic Akhmediev breather dynamic. In addition, we note that in the case of Akhmediev type eigenvalues (\ref{Akhmediev_par_set}), the transformation (\ref{II_to_III_transform}) boils down to a change of parameter $\alpha$ only, leaving the eigenvalue on the branchcut in the upper half of the $\lambda$-plane. In other words, for the eigenvalues (\ref{Akhmediev_par_set}), the type II and type III solutions merges into one class.

Each breather type corresponds to a certain branch of the dispersion law (\ref{dispersion_laws}). To establish this connection, we consider the breathers' tails as condensate perturbations and study them asymptotically. The role of small parameter plays the value $e^{-L}$, where $L$ is the characteristic distance from the breather center to the point where we study the breather tail. The latter can be done in the case of finite characteristic size, while for periodic Akhmediev breathers, one can similarly consider asymptotics at large times. We choose $L\gg 1$, so that we are far away from breather center and perform asymptotic expansion of the solutions (\ref{s1}), (\ref{sol2}) and (\ref{sol3}), see Appendix section \ref{Sec:Appendix:1} for the computational details. For type I solution (\ref{s1}) the first-order terms represent the main (zero-order) asymptotic (\ref{asymptotics_I}) plus a linear combinations of the first-order terms having structure $p e^{2\varphi}$ and $\tilde{p} e^{2\varphi^*}$, where $p$ and $\tilde{p}$ are constants, while the functions $\varphi$ are defined by (\ref{phi_0_and_phi}). We write these exponents in the form $e^{ikx + i\omega t}$. Considering, for example $e^{2\varphi}$, we obtain,
\begin{equation}
\label{tails1}
    e^{2\varphi} = e^{ikx+i\omega t}, \qquad k = -2\zeta, \qquad \omega=2\lambda\zeta.
\end{equation}
Now using that $\zeta(\lambda) = \sqrt{\lambda^2+A^2}$, see Eq.~(\ref{zeta_def}), we find $\omega(k) = \pm ik\sqrt{A^2-k^2/4}$. Thereby the breather tails obey the first branch of the dispersion law $\omega_{\mathrm{I}}(k)$ with complex $k$ and $\omega$. The complexity of $k$ in (\ref{tails1}) means exponential decay of the breather tail. The same result can be obtained for the terms $\tilde{p} e^{2\varphi^*}$.

For breather types II and III a similar analysis gives as zero order terms the asymptotic values (\ref{asymptotics_II}) and (\ref{asymptotics_III}) plus the first-order terms having structure $p e^{\varphi_0-\varphi}$ and $\tilde{p} e^{\varphi^*_0-\varphi^*}$ (for type II) and the terms of the structure $p e^{-\varphi_0-\varphi}$ and $\tilde{p} e^{-\varphi^*_0-\varphi^*}$ (for type III). For all the listed exponents one gets the second branch of the dispersion law $\omega_{\mathrm{II}}(k)$. For instance, in the case $e^{\varphi_0-\varphi}$, we obtain,
\begin{equation}
\label{tails2}
    e^{\varphi_0-\varphi} = e^{ikx+i\omega t}, \qquad k = \zeta - \lambda, \qquad \omega=-\frac12(\zeta - \lambda)^2,
\end{equation}
and finally easily retrieve the second branch $\omega(k) = -k^2/2$. Again, as in the type I case, the complexity of $k$  means exponential decay of the breather tails.

\section{Resonance interactions of breathers}\label{Sec:4}

In this section, we study resonant interactions of the vector breathers, i.e., a fusion of two breathers into one or decay of one breather into two, such that the characteristic wave vectors and frequencies of the breathers satisfy resonance conditions. The phenomena of inelastic mutual coherent structures transformations have been known for integrable systems since the work \cite{ZakharovManakov1976theory} devoted to solitons in the three-wave model, see also \cite{kaup1976three}. Another examples represent two-dimensional field theory \cite{zakharov1978example}, chiral fields models \cite{orlov1984Nsoliton}, and relativistic $O(1,1)$ sine-Gordon model \cite{barashenkov1988integrable,barashenkov1988exactly,barashenkov1993unified}. In the case of zero background, such nontrivial interactions are possible for solitons in the three (or more than three) component systems, such as the mentioned three-wave model. However, in the presence of a nontrivial background, the constrain on the number of dimensions can be relaxed and nontrivial interactions can be observed already in two-component systems \cite{barashenkov1988integrable,barashenkov1988exactly,barashenkov1993unified}. Mathematically speaking, the nontrivial interactions are possible when the IST auxiliary problem admits different types of the eigenvalues, which can be merged into one point without solution degeneration, see \cite{ZakharovManakov1976theory}. Recently we have observed resonance interactions for vector breathers in the Manakov system (\ref{VNLSE}), see our Letter \cite{raskovalov2022resonant} and the recent paper \cite{raskovalov2022resonanse}. The resonance represents a three-breather process of a fusion or decay, where each of the three participating breathers has a different type either I, II, or III. Here we present the theory of these nontrivial interactions in more detail than \cite{raskovalov2022resonant,raskovalov2022resonanse}.

\begin{figure}[!t]
\centering
    \includegraphics[width=0.3\linewidth]{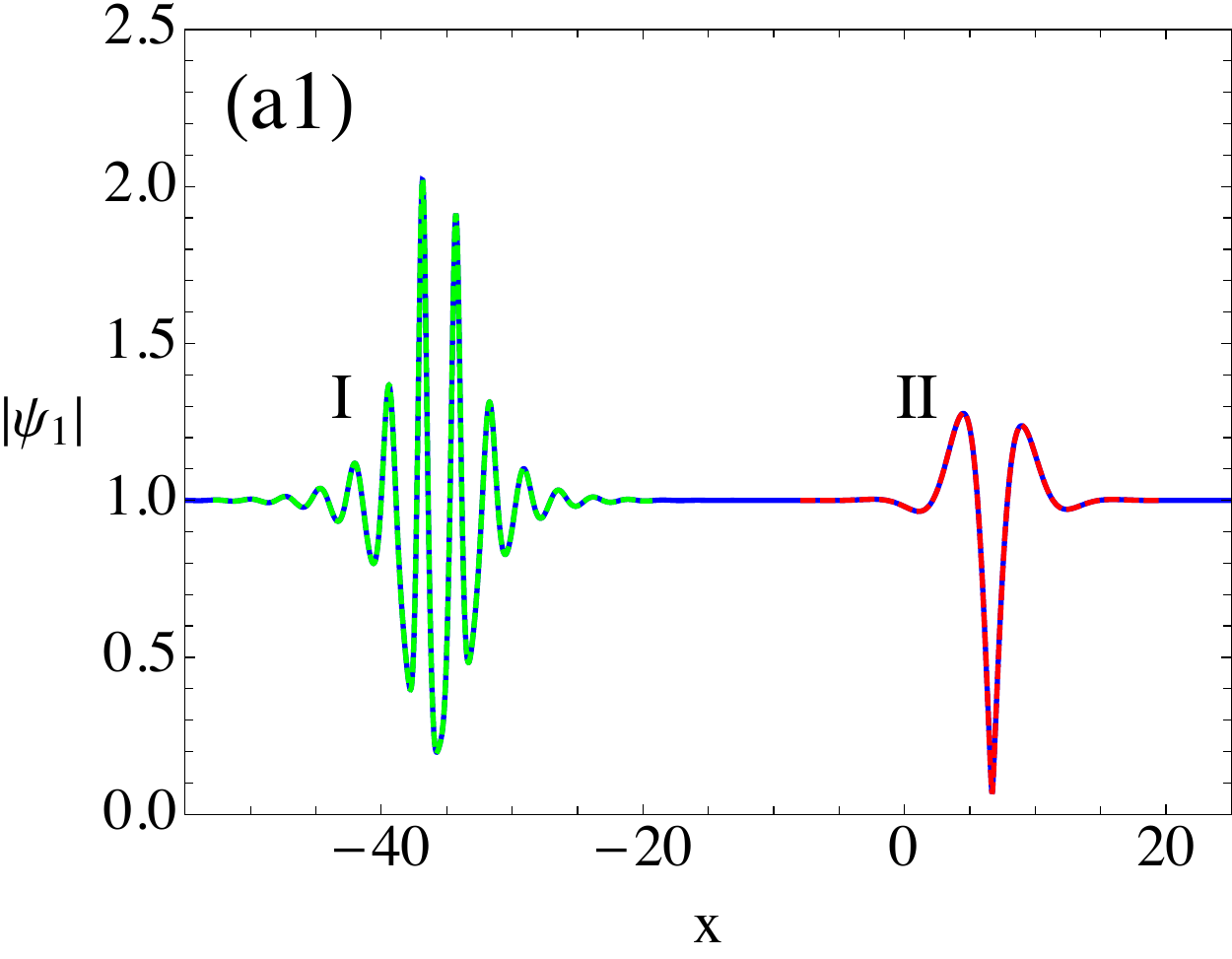}\,\,\,\,\,
    \includegraphics[width=0.3\linewidth]{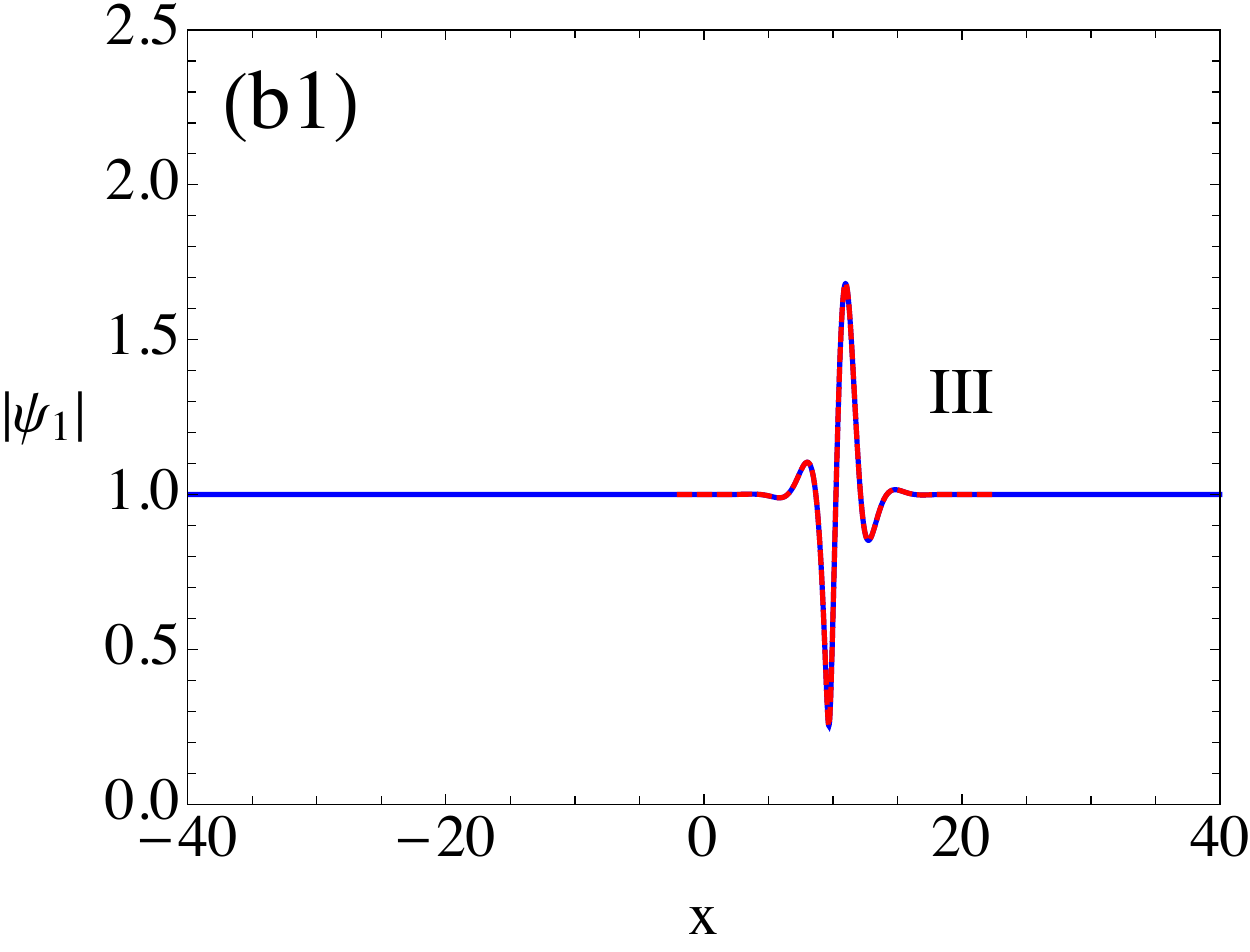}\,\,\,\,\,
    \includegraphics[width=0.31\linewidth]{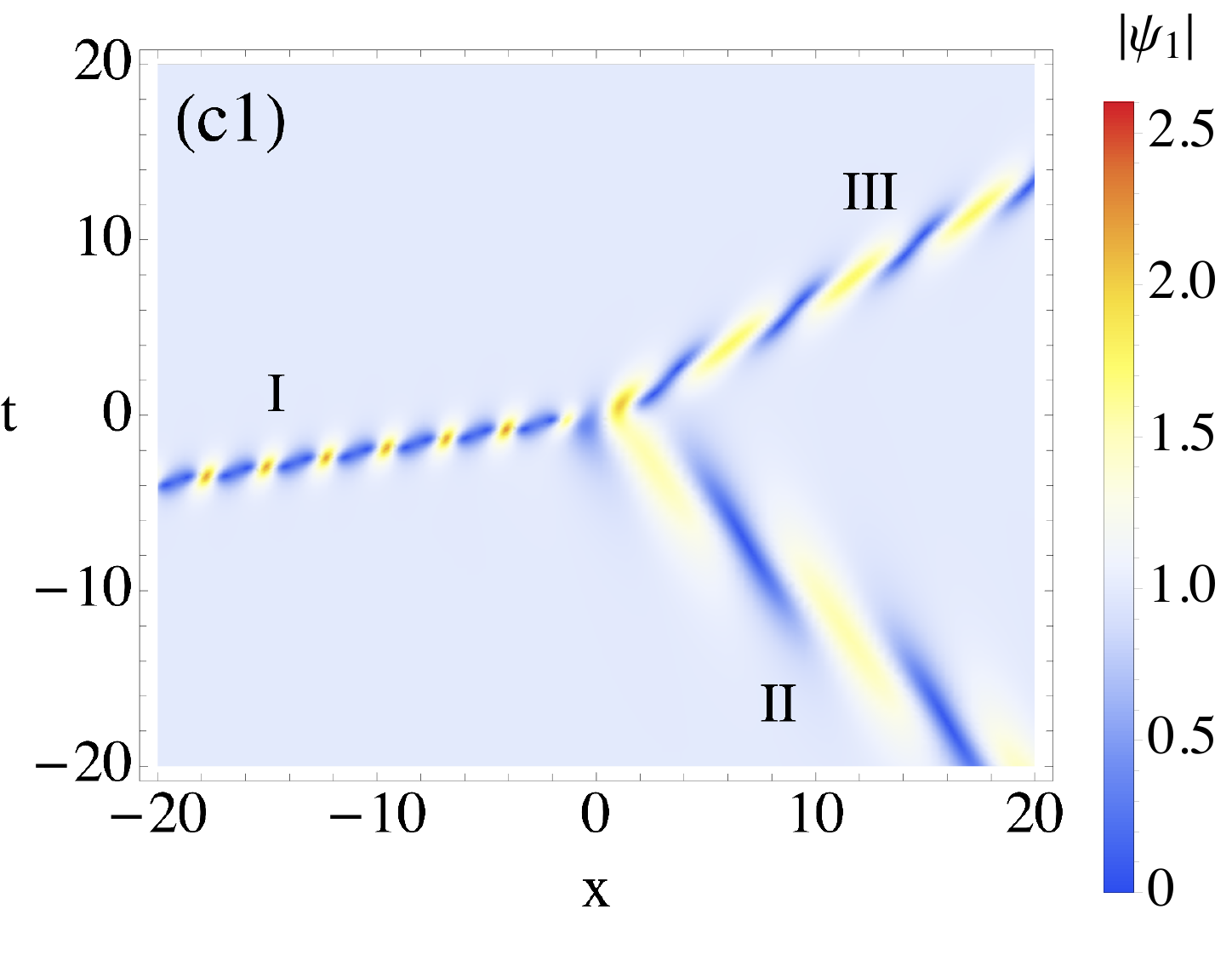}\\
    \includegraphics[width=0.3\linewidth]{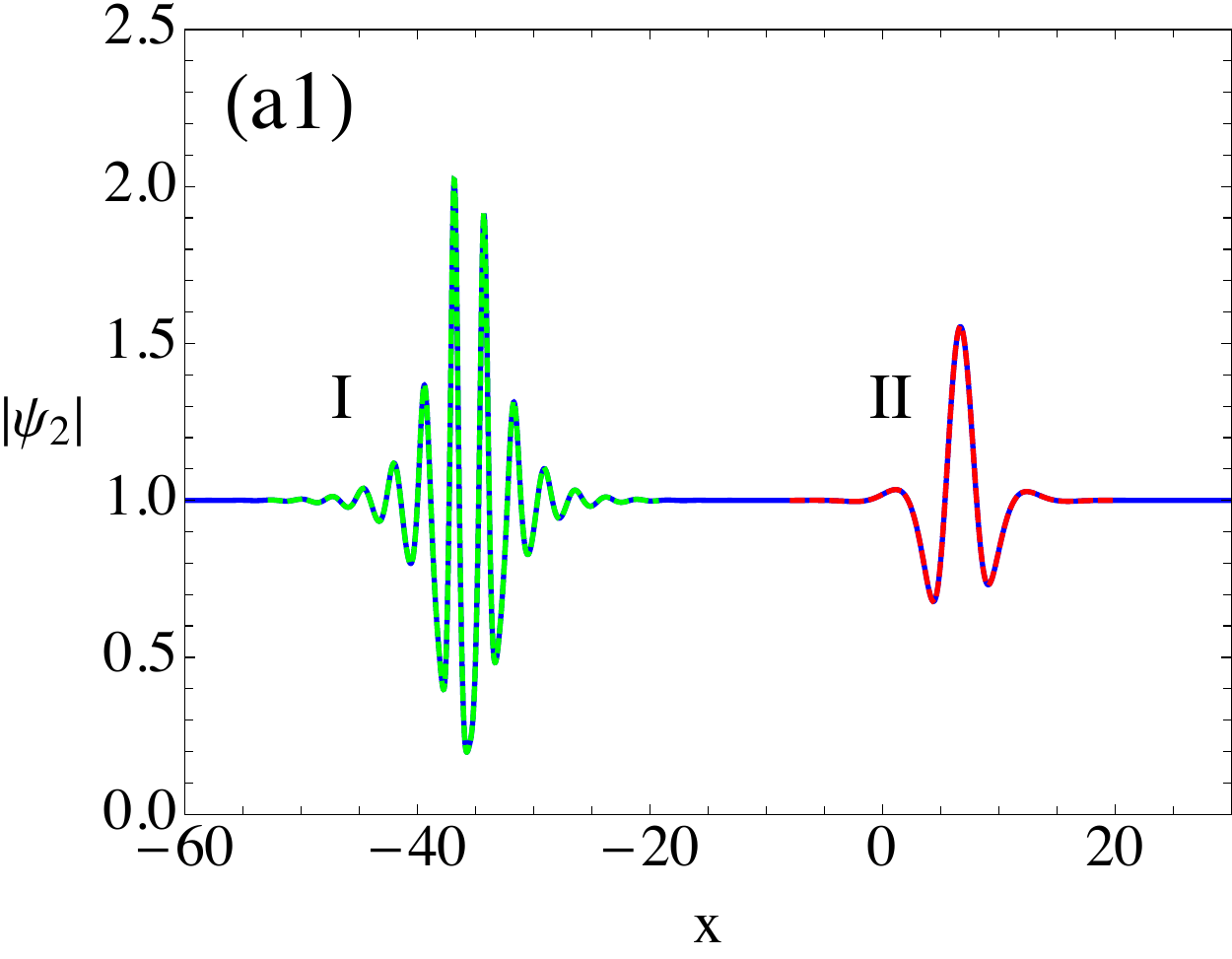}\,\,\,\,\,
    \includegraphics[width=0.3\linewidth]{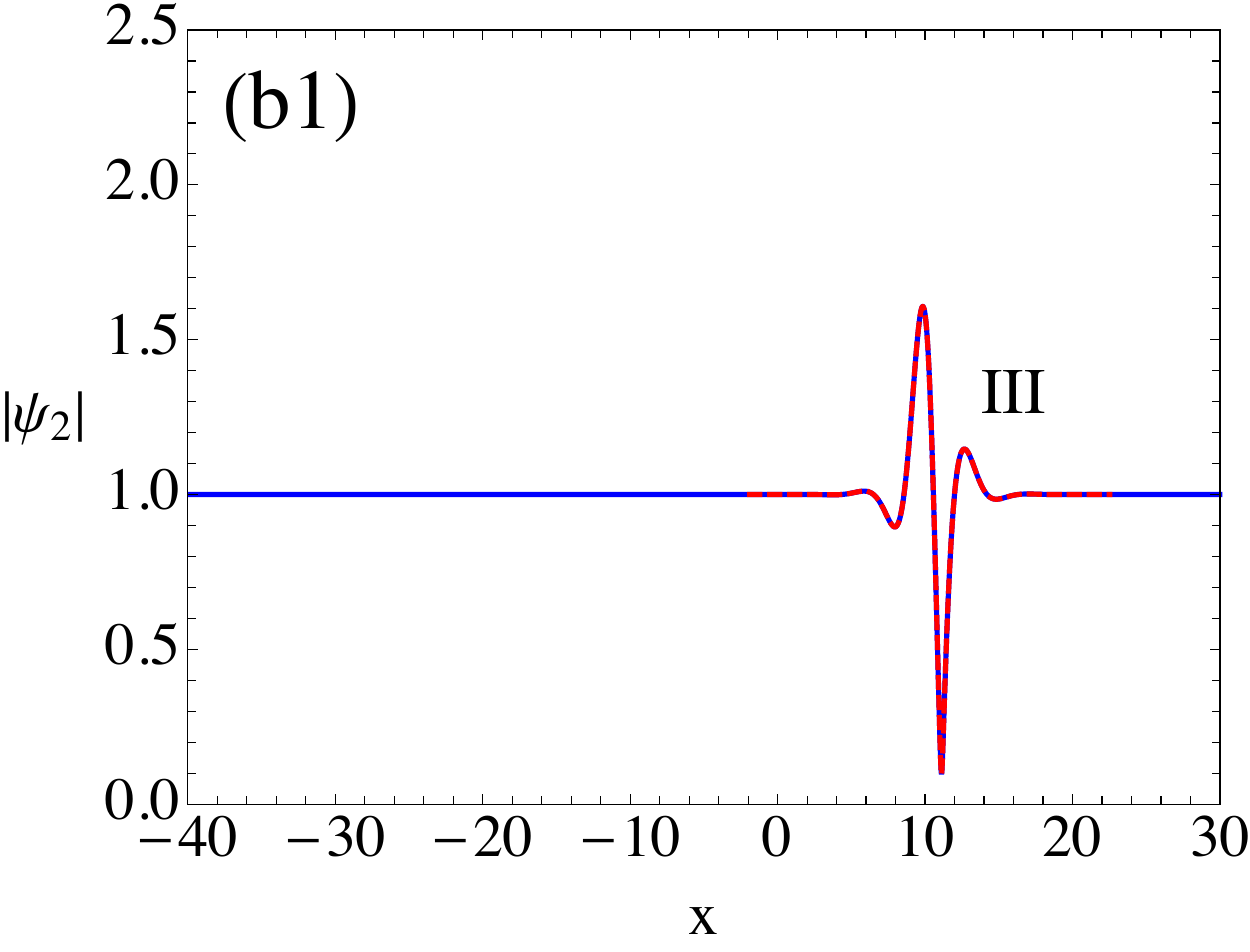}\,\,\,\,\,
    \includegraphics[width=0.31\linewidth]{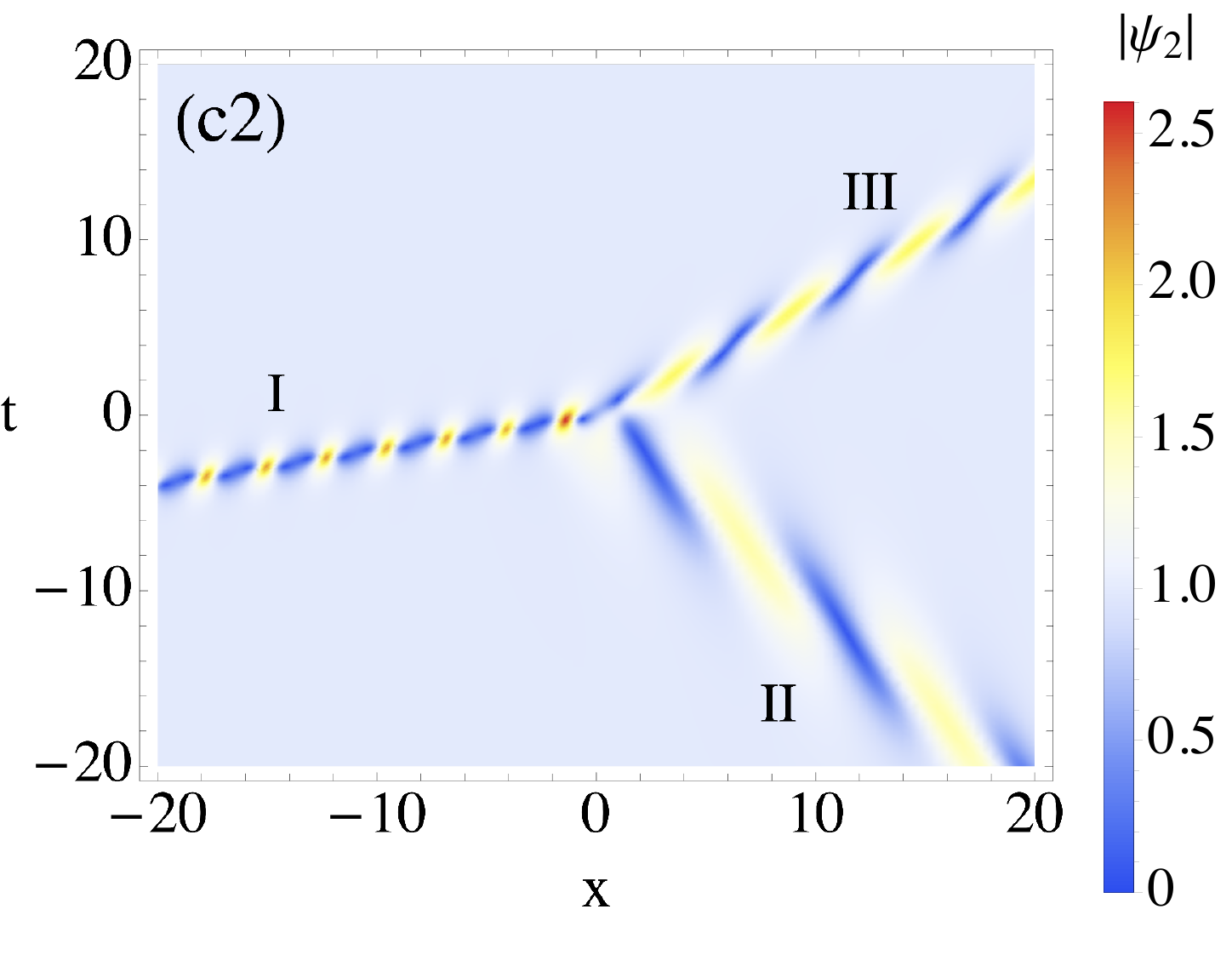}
\caption{
Resonance interaction $\mathrm{I}+\mathrm{II} \rightarrow \mathrm{III}$ of vector breathers described by single-pole solution (\ref{solution}) with $\mathbf{C} = \{1,1,1\}$. Blue lines in (a1,a2) and (b1,b2) shows $|\psi_{1,2}|$ before ($t=-7.0$) and after ($t=7.0$) the resonance interaction. The dotted green and red lines shows local approximation of the breathers of the types I, II and III (see the corresponding notations in the figures) by solution (\ref{solution}) with: I)  $\mathbf{C} = \{0,1,1\}$, green line in (a1,a2) II)  $\mathbf{C} = \{1,0,1\}$, red line in (a1,a2) III) $\mathbf{C} = \{1,1,0\}$, red line in (b1,b2). Panels (c1,c2) show spatio-temporal evolution of $|\psi_{1,2}|$ for the whole resonance interaction and its asymptotic state.}
\label{fig_07}
\end{figure}

The resonant interaction of three vector breathers is described by the one-pole solution (\ref{solution}), when all the integration constants $C_0$, $C_1$ and $C_2$ are nonzero. First we consider the situation when the eigenvalue is of general type, see Eq.~(\ref{General_par_set}), and after that we switch to the particular choices (\ref{Kuznetsov_par_set}-\ref{Akhmediev_par_set}). In the general case, the solution has the following asymptotics:
\begin{eqnarray}
\label{asymptotics_resonance}
\psi_{1,2} &\to& A_{1,2} e^{-2\,\mathrm{i}\,\alpha}; \quad\quad\quad\quad x \to -\infty,
\\\nonumber
\psi_{1,2} &\to& A_{1,2}; \quad\quad\quad\quad\quad\quad\,\,\, x \to +\infty,
\end{eqnarray}
that on one side coincides with the asymptotic III (\ref{asymptotics_III}), and on the other side can be obtained by linear superposition of the asymptotics I and II, see Eqs.~(\ref{asymptotics_II}) and (\ref{asymptotics_III}). For definiteness we consider the case $\pi/2>\alpha>0$. In order to investigate the asymptotic states of the solution with non zero integration constants, we move at $t\rightarrow -\infty$  to the reference frame,
\begin{eqnarray}
\label{ref1}
	u = \mathrm{const},
	\\
\label{ref2}
	u_0 - u = \mathrm{const},
\end{eqnarray}
while at $t\rightarrow \infty$ we move to the reference frame,
\begin{eqnarray}
\label{ref3}
	u_0 + u = \mathrm{const}.
\end{eqnarray}
Recall that $u$ and $u_0$ are defined in (\ref{uvu0v0}). Then the conditions (\ref{ref1}), (\ref{ref2}) and (\ref{ref3}) lead to that in the expressions (\ref{q vectors(lambda)}), $e^{\varphi_0}\rightarrow 0$,  $e^{-\varphi}\rightarrow 0$ and $e^{\varphi}\rightarrow 0$ correspondingly, and for each of the reference frames one can obtain exactly the single breather solutions (\ref{s1}), (\ref{sol2}) or (\ref{sol3}). The latter means that the asymptotic state of the resonance process represent single breathers of the types I, II and III with the following integration constants $\mathbf{C}$: I) $\{0,C_1,C_2\}$, II) $\{C_0,0,C_2\}$, and III) $\{C_0,C_1,0\}$. The interaction itself represent fusion of the breather I and II into the breather III, what we denote as $\mathrm{I}+\mathrm{II}\rightarrow\mathrm{III}$. We show the full resonance process and the single-breather approximation in Fig.~\ref{fig_07}. Note, that for $\pi>\alpha>\pi/2$ the resonance represent an opposite process -- the decay of the breather III into the breathers I and II, i.e., $\mathrm{III}\rightarrow \mathrm{I} + \mathrm{II}$.

From the expressions (\ref{characteristic_values1}), (\ref{characteristic_values2}) and (\ref{characteristic_values3}) describing breather characteristics, we find that the resonance process satisfy the standard resonant conditions,
\begin{eqnarray}
\label{resonance1}
k_{\mathrm{I}} + k_{\mathrm{II}}  &=& k_{\mathrm{III}},
\\
\omega_{\mathrm{I}} + \omega_{\mathrm{II}}  &=& \omega_{\mathrm{III}}.
\label{resonance2}
\end{eqnarray}
Note, that the resonance conditions (\ref{resonance1}) and (\ref{resonance2}) cannot be derived from the dispersion laws (\ref{dispersion_laws}). Indeed, the characteristic breather wave vectors and frequencies follow from the fully nonlinear solutions (\ref{s1}),~(\ref{sol2}),~(\ref{sol3}), meanwhile the dispersion laws describe only the breather tails.

The resonance is always represented by either the process $\mathrm{I}+\mathrm{II}\rightarrow\mathrm{III}$ or the process $\mathrm{III}\rightarrow \mathrm{I} + \mathrm{II}$. Other configurations, such as a fusion of breather I cannot exist, what can be seen from the structure of the solution asymptotics, as soon as the resonance asymptotic (\ref{asymptotics_resonance}) coincides with the asymptotic (\ref{asymptotics_III}). In addition, such process as $\mathrm{I}\rightarrow \mathrm{II} + \mathrm{III}$ is prohibited by the resonant conditions (\ref{resonance1}) and (\ref{resonance2}). Indeed, let us consider the region of spectral parameter where $\pi/4>\alpha>0$. Then, according to Eqs.~(\ref{characteristic_values1}), (\ref{characteristic_values2}), (\ref{characteristic_values3}), we find that $k_\mathrm{I}>0$, $k_\mathrm{II}<0$, and $k_\mathrm{III}>0$. In addition for the whole range of spectral parameter $k_\mathrm{I}>k_\mathrm{III}$, so that $k_\mathrm{I}$ cannot be represent as a sum of $k_\mathrm{II}$ and $k_\mathrm{III}$.

\begin{figure}[!t]
\centering
    \includegraphics[width=0.3\linewidth]{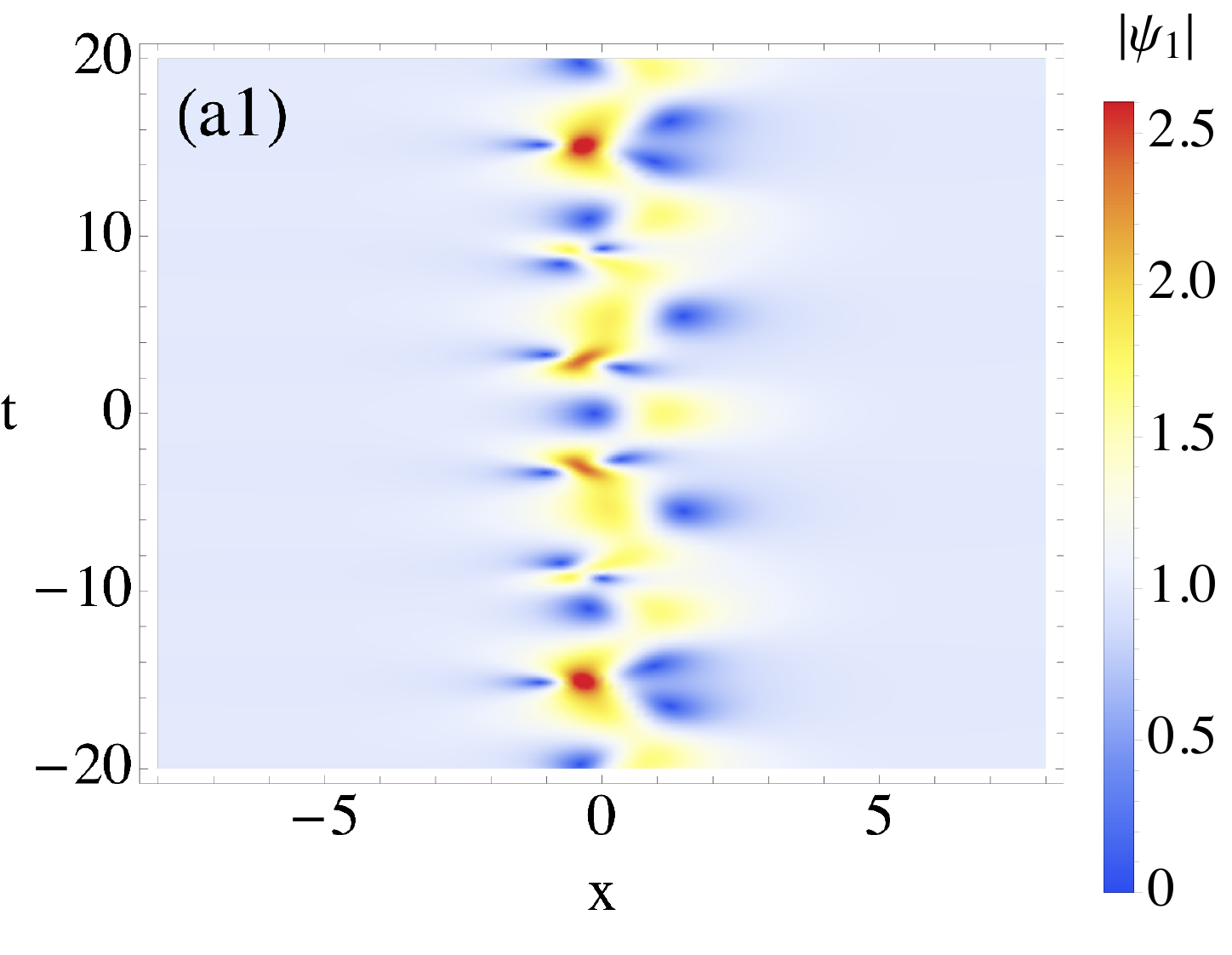}\,\,\,\,\,
    \includegraphics[width=0.3\linewidth]{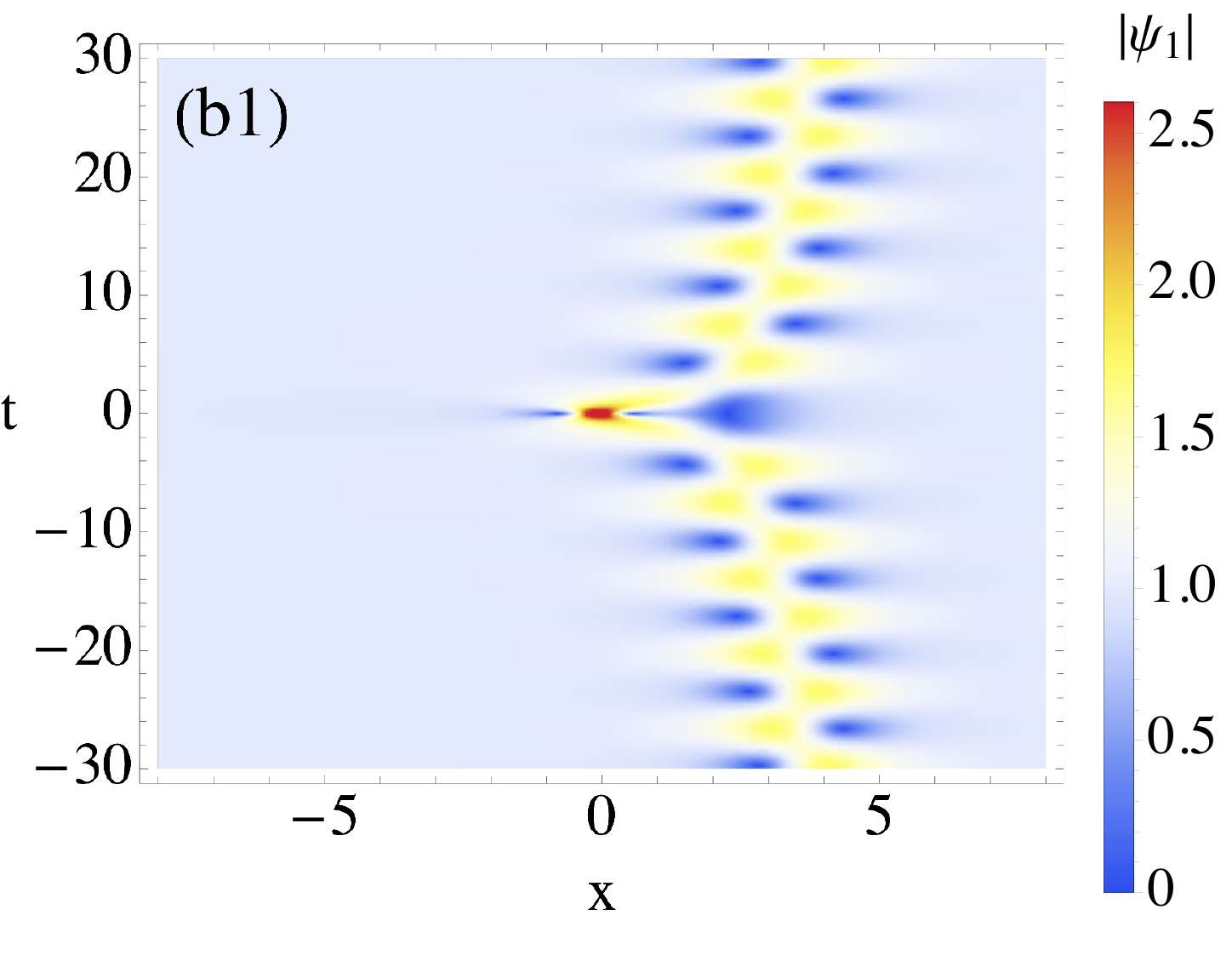}\,\,\,\,\,
    \includegraphics[width=0.31\linewidth]{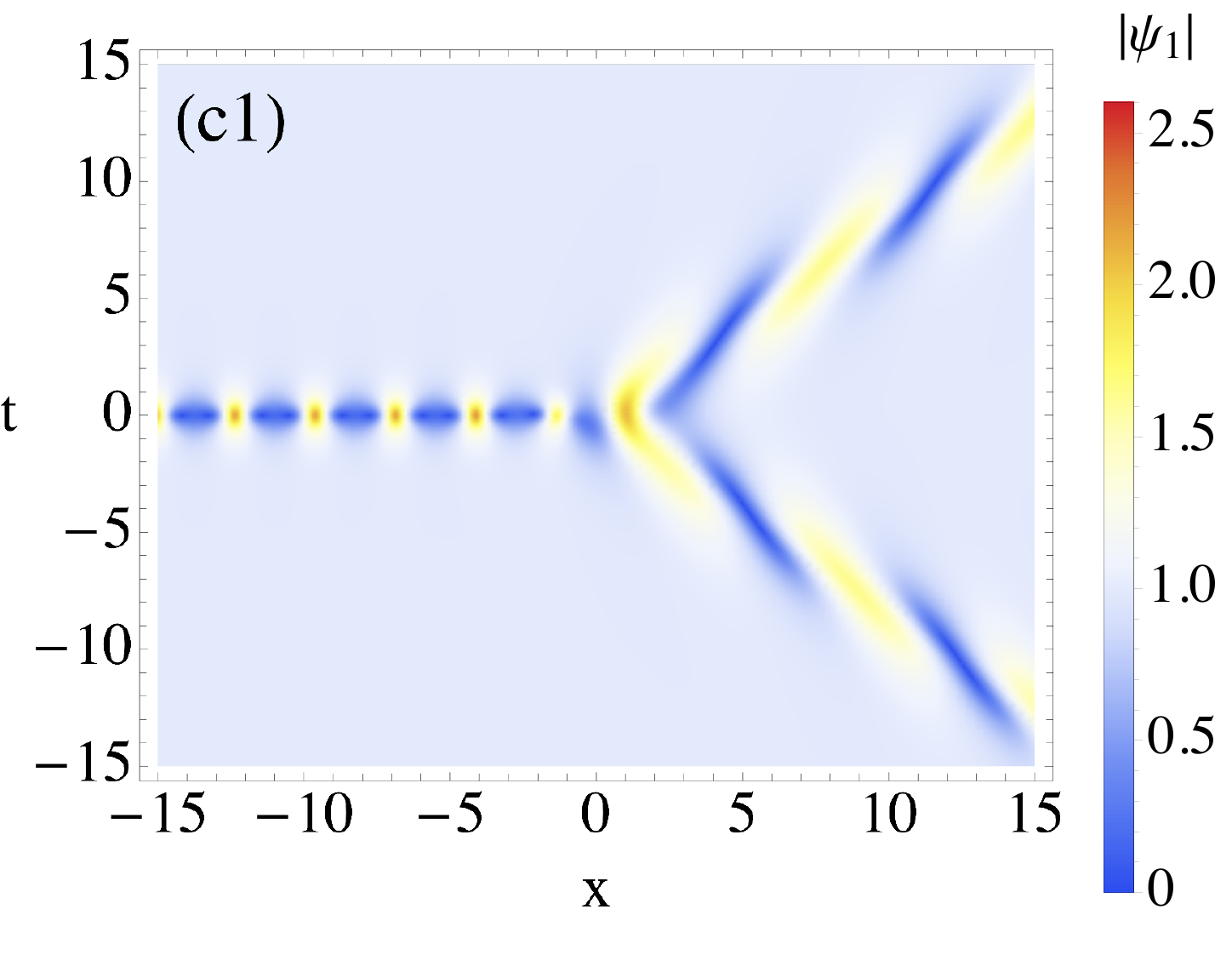}\\
    \includegraphics[width=0.3\linewidth]{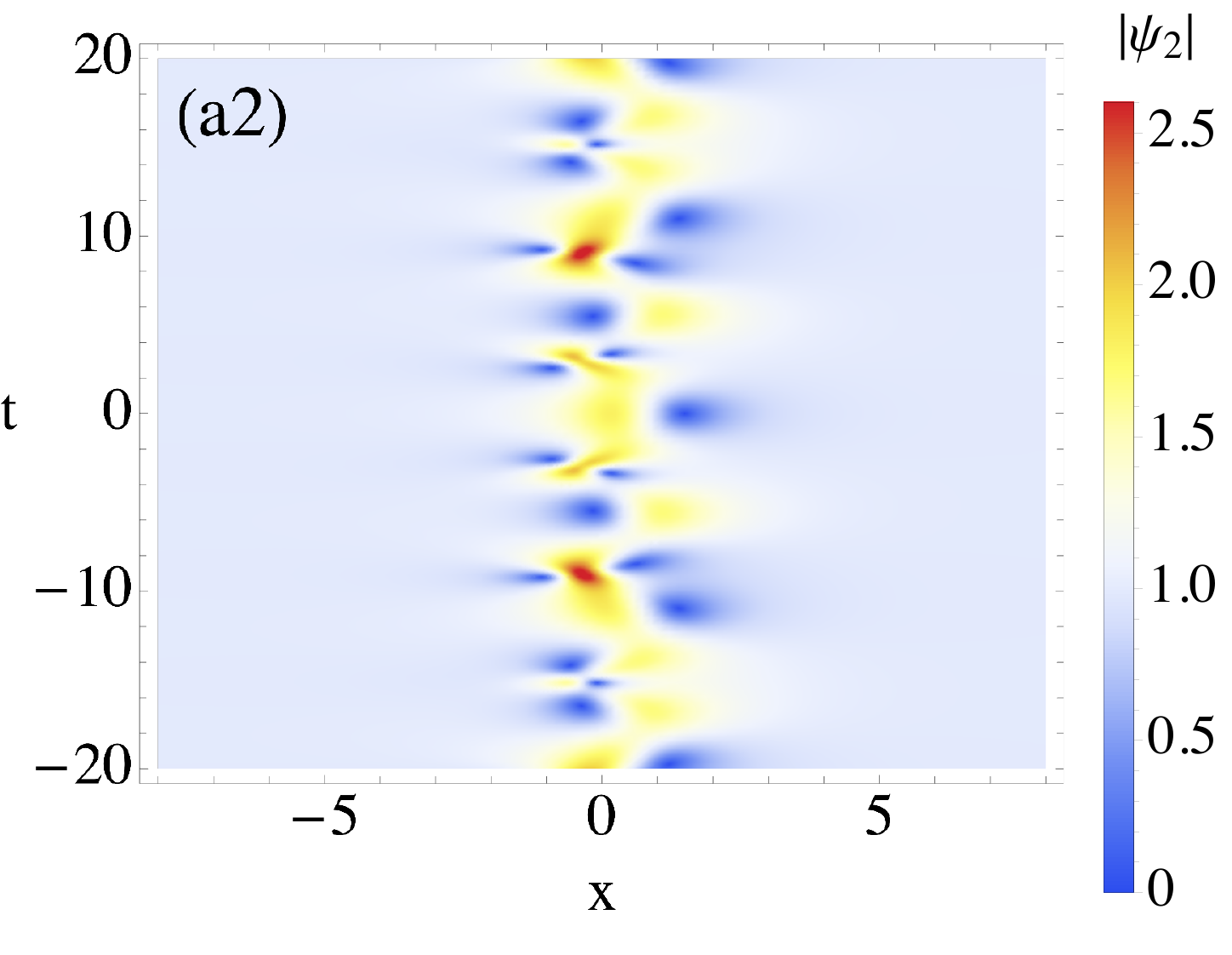}\,\,\,\,\,
    \includegraphics[width=0.3\linewidth]{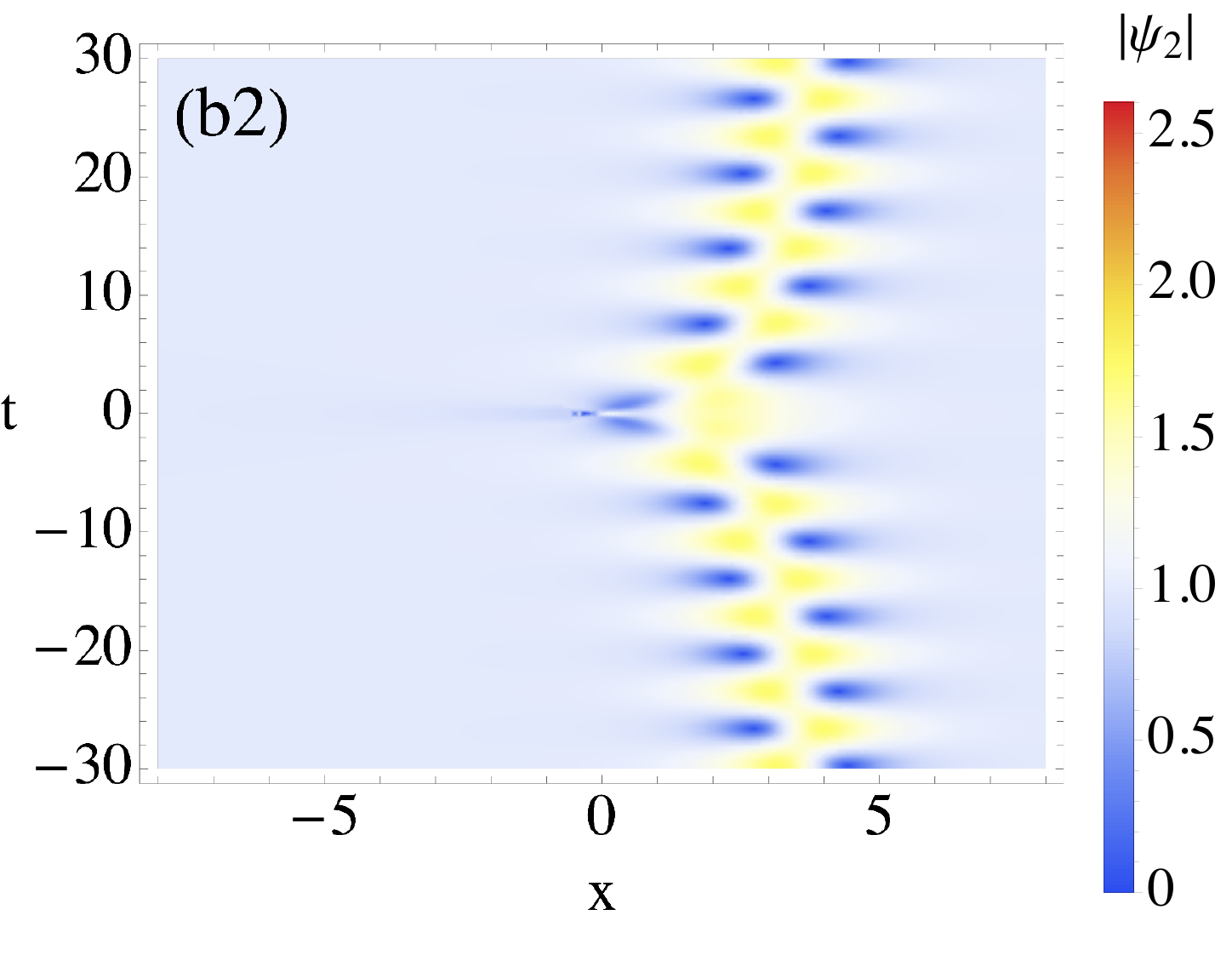}\,\,\,\,\,
    \includegraphics[width=0.31\linewidth]{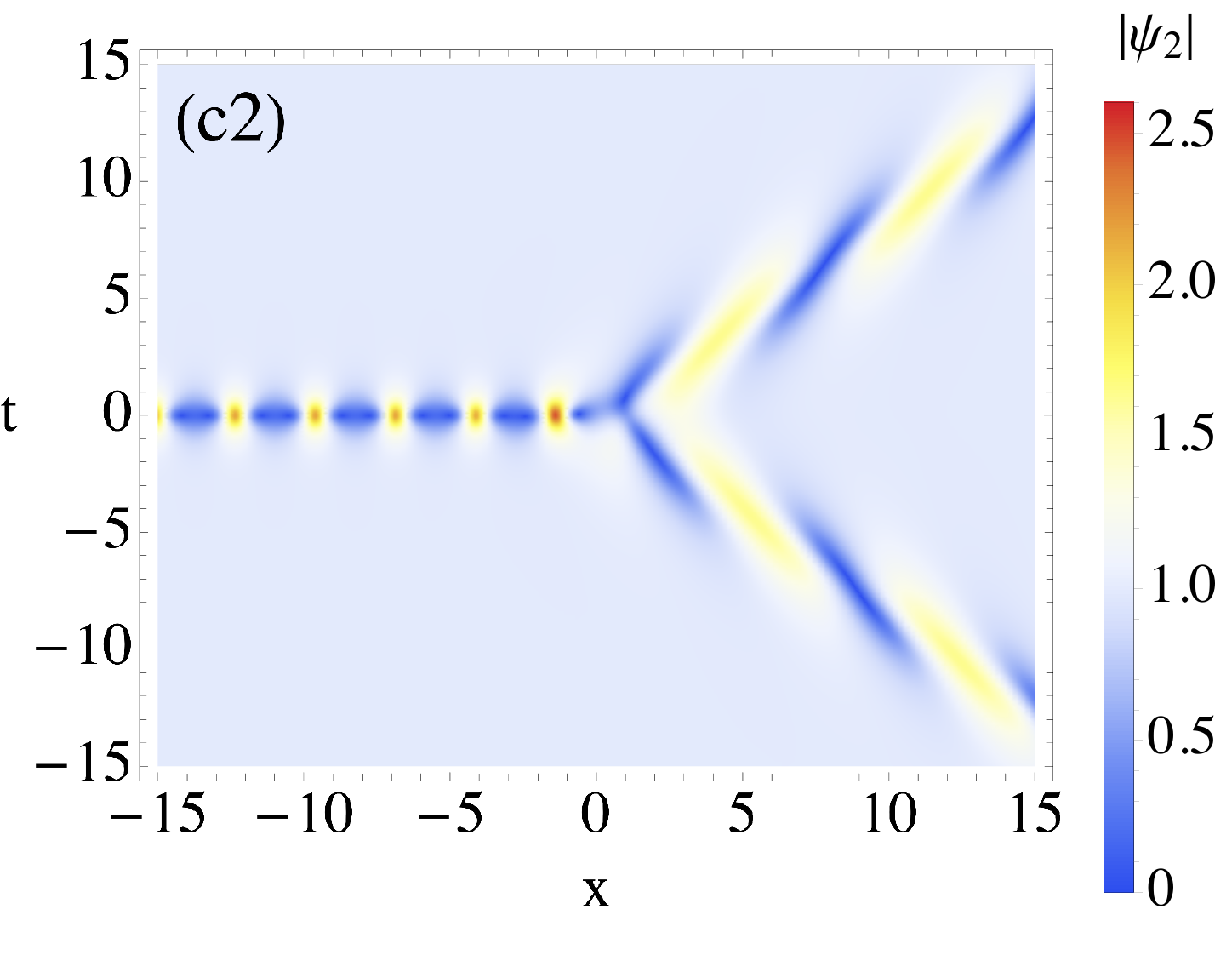}
\caption{
Resonance interaction of vector breathers with spectral parameters defined by Eqs.~(\ref{Kuznetsov_par_set}-\ref{Akhmediev_par_set}). (a) Spectral parameter belongs to the Kuznetsov region, see Eq.~(\ref{Kuznetsov_par_set}). (b) Spectral parameter belongs to the Peregrine region, see Eq.~(\ref{Kuznetsov_par_set}). (c) Spectral parameter belongs to the Akhmediev region, see Eq.~(\ref{Kuznetsov_par_set}).}
\label{fig_08}
\end{figure}

To conclude this section we consider the particular cases of the resonance interactions corresponding to the choices of the eigenvalues (\ref{Kuznetsov_par_set}-\ref{Akhmediev_par_set}). For the eigenvalue of the Kuznetsov type (\ref{Kuznetsov_par_set}), we observe a standing wave group exhibiting complex oscillations, see Fig.~\ref{fig_08}(a). In the limit (\ref{Peregrine_par_set}) solution (\ref{solution}) with nonzero $C_0$, $C_1$ and $C_2$ degenerates, leading, after resolving the uncertainty of the type $0/0$, to the following rational formula,
\begin{eqnarray}
&&\psi_1 = A_1 +\frac{4 \,[1-2 A (x -x_1)-2\, \mathrm{i}\, A^2 t]}{(e^{2 A (x-x_0)}+2+8 A^2 (x-x_1)^2+8 A^4 t^2)} \times\nonumber\\
&&\qquad\qquad\times(A_2 e^{A (x-x_0) - \mathrm{i} A^2 (t-t_0)/2}+A_1 [1+2 A (x -x_1)-2\, \mathrm{i}\, A^2 t]),\nonumber\\
&&\psi_2 = A_2 +\frac{4\, [1-2 A (x -x_1)-2\, \mathrm{i}\, A^2 t]}{(e^{2 A (x-x_0)}+2+8 A^2 (x-x_1)^2+8 A^4 t^2)}\times\nonumber \\
&&\qquad\qquad\times(-A_1 e^{A (x-x_0) - \mathrm{i} A^2 (t-t_0)/2}+A_2 [1+2 A (x -x_1)-2\, \mathrm{i}\, A^2 t]),\label{Peregrine_resonance}
\end{eqnarray}
where $x_0$, $x_1$ and $t_0$ are real valued parameters. The semirational solution (\ref{Peregrine_resonance}) represents a localized wave group, decaying with time as $t^{-2}$. At certain coordinates it exhibits a Perigrine-type bump coexisting with the rest of the solution, see Fig.~\ref{fig_08}(b), that was previously studied in \cite{baronio2012solutions} in the context of vector rogue waves formation. Finally, for the Akhmediev type eigenvalues (\ref{Akhmediev_par_set}), we observe a moving type II breather which at some point decays into a Akhmediev type I wave excitation in one-half of space, plus a type III breather moving in another half of space, see Fig.~\ref{fig_08}(c). As we noted in the previous section, for the eigenvalues (\ref{Akhmediev_par_set}), the classes of type II and III solution merges into one, which explains why the moving breathers before and after the resonance interaction in Fig.~\ref{fig_08}(c) are similar. Note that Fig.~\ref{fig_08} demonstrates a particular case of the general scenario shown in Fig.~\ref{fig_07}(c) when one uses the eigenvalues (\ref{Akhmediev_par_set}) for each of the three asymptotic breather states.

\section{Elastic collisions of breathers}\label{Sec:5}
The two-eigenvalue solution of the model (\ref{VNLSE}), see Eq.~(\ref{N-solitonic solution}) with $N=2$, has the following general form, 
\begin{eqnarray}
\psi_1 = A_1+2 \mathrm{\widetilde{M}}_{12}/\mathrm{M},
\nonumber\\
\psi_2 = A_2+2\mathrm{\widetilde{M}}_{13}/\mathrm{M},
\label{two}
\end{eqnarray}
where
\begin{eqnarray}
&&\tilde{M}_{12} =\mathrm{i} \left[-m_2 q_{12} q_{11}^* |\mathbf{q}_2|^2 +n_2 q_{12} q_{21}^* (\mathbf{q}_2, \mathbf{q}^*_1)+n_1 q_{22} q_{11}^* (\mathbf{q}_1, \mathbf{q}^*_2)-m_1 q_{22} q_{21}^* |\mathbf{q}_1|^2 \right],\nonumber\\
&&\tilde{M}_{13} =\mathrm{i} \left[-m_2 q_{13} q_{11}^* |\mathbf{q}_2|^2 +n_2 q_{13} q_{21}^* (\mathbf{q}_2, \mathbf{q}^*_1)+n_1 q_{23} q_{11}^* (\mathbf{q}_1, \mathbf{q}^*_2)-m_1 q_{23} q_{21}^* |\mathbf{q}_1|^2 \right],\nonumber\\
&& M = -[m\,|\mathbf{q}_1|^2 |\mathbf{q}_2|^2- n\, (\mathbf{q}_1, \mathbf{q}_2^*) (\mathbf{q}_2, \mathbf{q}_1^*)].
\end{eqnarray}
Here the coefficients $m_{1,2} \equiv 1/(\lambda_{1,2}-\lambda_{1,2}^*)$, $n_{1,2} \equiv 1/(\lambda_{1,2}-\lambda_{2,1}^*)$, $m=m_1 m_2$, $n = n_1 n_2 \leq 0$. The vectors $\mathbf{q}_{1,2}$ are defined in (\ref{q vectors(lambda)}). In the parametrization (\ref{uniformization}), the coefficients can be written as,
\begin{eqnarray*}
&&m_i = \frac{2 |r_i|^2}{A (r_i^*-r_i)(1+|r_i|^2)},\qquad n_i  = \frac{2 r_i r_j^*}{A (1+r_i r_j^*)(r_j^*-r_i)},\qquad i=1,2,\,\; j=3-i;\nonumber\\
&&m=\frac{4 |r_1 r_2|^2}{A^2 (1+|r_1|^2)(1+|r_2|^2)(r_1-r_1^*)(r_2-r_2^*)} =-[4 A^2 \sin \alpha_1 \sin \alpha_2\, \mathrm{cosh} \xi_1 \mathrm{cosh} \xi_2]^{-1},\nonumber\\
&&n = -\frac{4 |r_1 r_2|^2}{A^2 |1+r_1 r_2^*|^2 |r_1-r_2^*|^2} . \label{isp}
\end{eqnarray*}

\begin{figure}[!t]
\centering
    \includegraphics[width=0.31\linewidth]{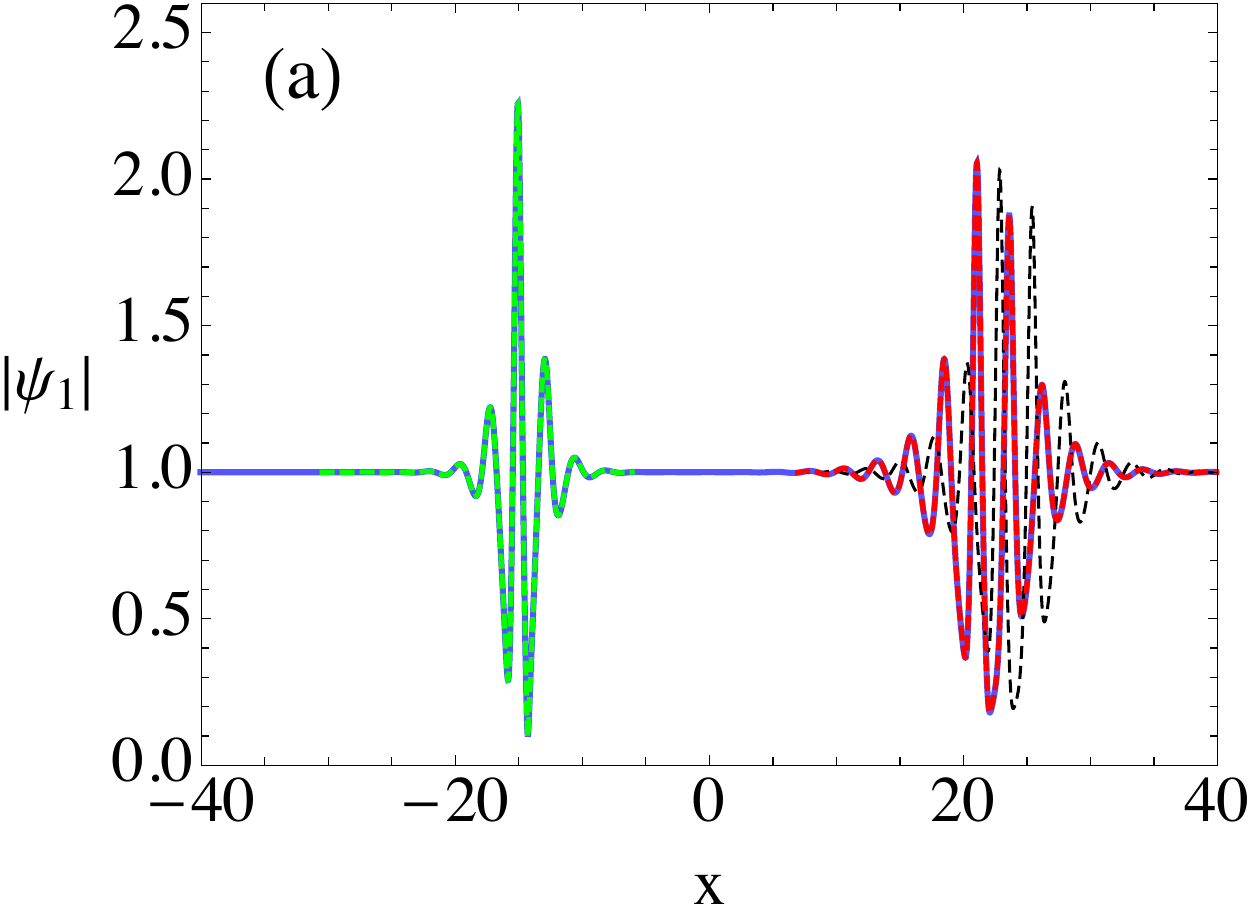}\,\,\,\,\,
    \includegraphics[width=0.31\linewidth]{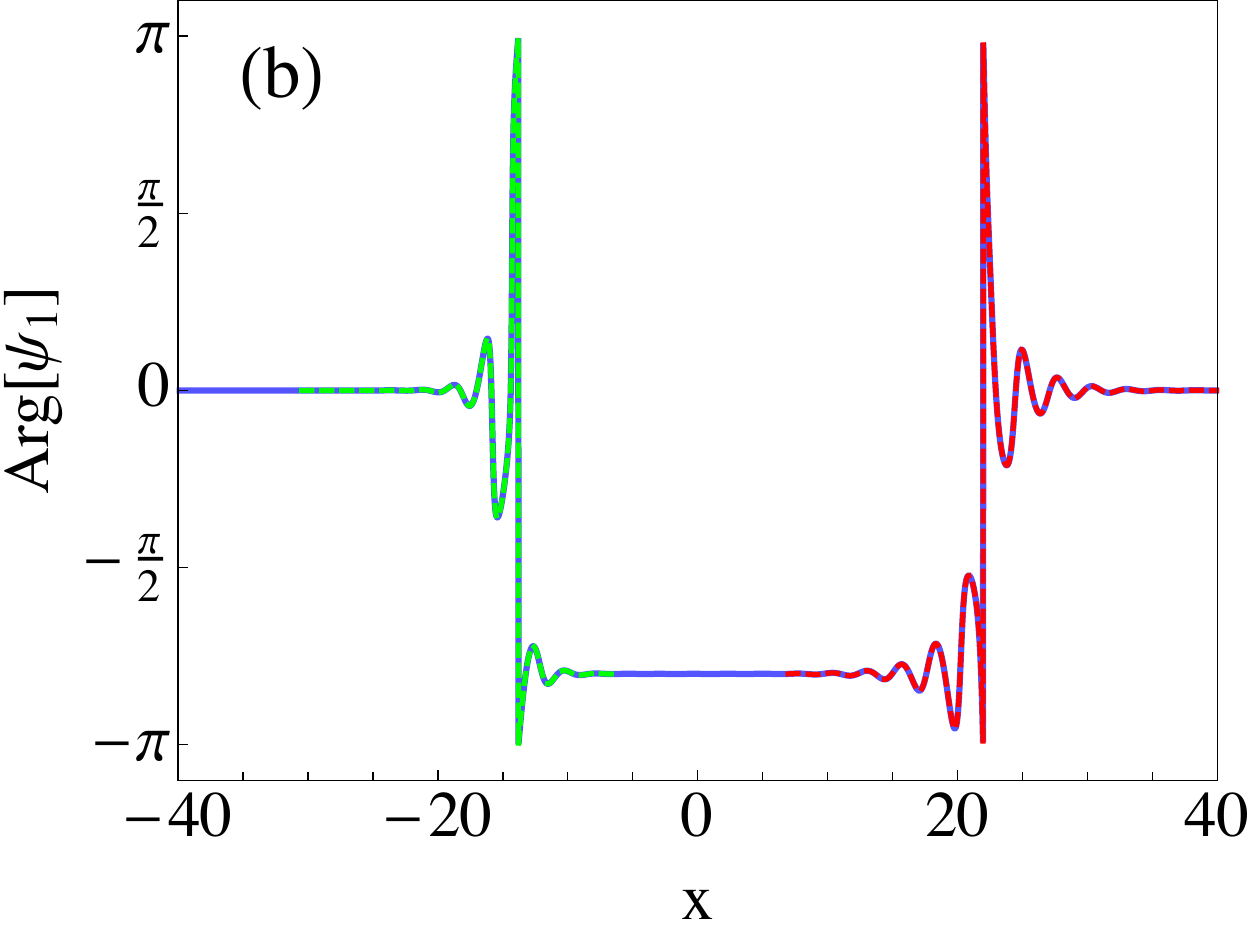}\,\,\,\,\,
    \includegraphics[width=0.32\linewidth]{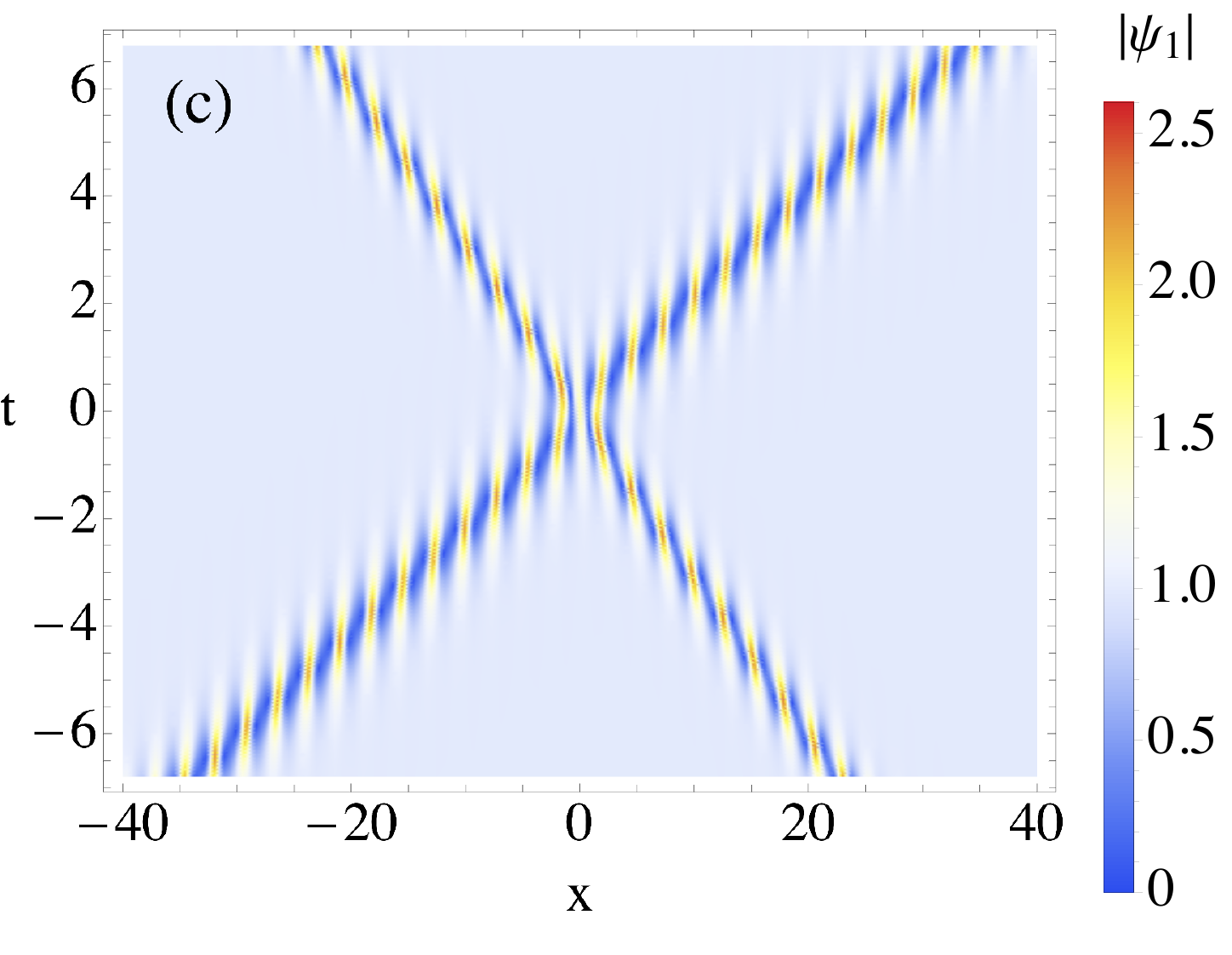}
\caption{
Elastic collision $\mathrm{I}+\mathrm{I} \rightarrow \mathrm{I} + \mathrm{I}$ of vector breathers with spectral parameters defined by Eq.~(\ref{parameters_2B}). (a,b) $|\psi_1|$ and $\mathrm{Arg}[\psi_{1}]$, where $\mathrm{Arg}$ means complex phase, after the breathers collision at $t=4.5$. (c) Spatio-temporal plot of the wave field evolution. The dotted green and red lines in (a,b) show a local approximation of the breathers after collision by single-breathers solutions from the asymptotic (\ref{2B_asymptotic_I+I}). Thin black dashed line in (a) shows how the first breather would have been if the it had been traveling alone. Both wave field components shown on the same plot since they coincide after the rescaling of the amplitude, see Eq.~(\ref{typeI_relation}).}
\label{fig_09}
\end{figure}

The solution (\ref{two}) describes a wide family of vector breathers interactions. It is characterized by two eigenvalues and 6 integration constants $C_{ij}$, $i=1,2$, $j=1,2,3$. Depending on their choice, the solution (\ref{two}) represents either two elastically colliding breathers, a breather plus fusion/decay resonance wave pattern, or a combination of two resonance wave patterns. Taking into account three fundamental types of breathers, we obtain more than ten scenarios of the breather interactions described via Eq.~(\ref{two}). Here we do not consider all of them, instead focus on fundamental aspects of the vector breathers interactions, such as collision wavefield profiles and asymptotic states of the breathers at large times. For the latter question, we compute exact formulas describing the shifts of the positions and phases acquired by the breathers after their collision.

We start with elastic collisions, i.e., we put zero one of the components in vectors $\mathbf{C}_1$ and $\mathbf{C}_2$. For each breather, we chose parametrization of its integration constant according to the breather type. We begin with the case of type I breathers collision, i.e., $\mathrm{I}+\mathrm{I} \rightarrow \mathrm{I} + \mathrm{I}$. First, we choose the parametrization for the vectors $\mathbf{C}_1$ and $\mathbf{C}_2$ according to Eq.~(\ref{C_I_param}),
\begin{equation}
    C_{i,0} = 0, \qquad C_{i,1} = C_{i,2}^{-1} = e^{\mathrm{Im}[\zeta_i]\delta_i + i\theta_i/2}, \quad i=1,2.
\end{equation}
Each of the two breathers changes the phase of the condensate according to (\ref{asymptotics_I}), so that the asymptotic of the I+I solution reads as,
\begin{eqnarray}
\label{asymptotic_I+I}
\psi_{1,2}^\mathrm{I+I} &\to& A_{1,2}\, e^{\pm 2\, \mathrm{i}\, (\alpha_1 + \alpha_2)}; \quad\quad\quad\quad x \to \pm\infty.
\end{eqnarray}

The asymptotic states of the scalar two-breather NLSE solution, which is linked to the vector case through the transformation (\ref{typeI_transformation}), have been found in \cite{gelash2018formation}, see also \cite{gelash2022breather} for additional details. Here we re-obtain this result. We consider the two-breather solution in the reference frame moving with the group velocity $V_i$ of the breather $i$ ($i=1,2$), which collides with the breather $j$ ($j=2,1$). Then we analyze solution (\ref{two}) at large times, see computational details in \cite{gelash2022breather}, and also in appendix Sec. \ref{Sec:Appendix:3}, and find the asymptotic state for each of the breathers. The full asymptotic state of the solution (\ref{two}) at $t \rightarrow \pm\infty$ represents single breathers with shifted position $\delta$ and phase $\theta$ parameters, as well as shifted general phase,
\begin{eqnarray}
\label{2B_asymptotic_I+I}
    \psi_{1,2}^{\mathrm{I+I}}(\lambda_1,\delta_1,\theta_1;\lambda_2,\delta_2,\theta_2) \rightarrow
    \begin{cases}
    e^{\mp 2\,\mathrm{i}\,s_1 \alpha_2} \psi_{1,2}^\mathrm{I}(\lambda_1,\delta_1 + \delta^{\pm}_{0,1},\theta_1 + \theta^{\pm}_{0,1}), \quad\text{at}\quad x\sim V_{\mathrm{I_1}}t, \\\\
    e^{\mp 2\,\mathrm{i}\,s_2 \alpha_1} \psi_{1,2}^\mathrm{I}(\lambda_2,\delta_2 + \delta^{\pm}_{0,2},\theta_2 + \theta^{\pm}_{0,2}), \quad\text{at}\quad x\sim V_{\mathrm{I_2}}t,
    \end{cases}
\end{eqnarray}
where the sign $s_i =\pm 1$, and the position shift $\delta^{\pm}_{0,i}$, and the phase shift $\theta^{\pm}_{0,i}$ are defined at $t \rightarrow \pm\infty$ by the following expressions,
\begin{eqnarray}
&& s_i \equiv \mathrm{sign} (V_j-V_i),
\\\nonumber
&&\delta^{\pm}_{0,i} \equiv \mp s_i \frac{l_\mathrm{I}(\lambda_i)}{2}\,\log \left|\frac{(r_i-r_j^*)(1+r_i r_j)}{(r_i-r_j)(1+r_i r_j^*)} \right|^2,
\\\nonumber
&&\theta^{\pm}_{0,i} \equiv \mp s_i\,\mathrm{Arg} \left[\frac{(r_i^*-r_j^*)(1+r_i r_j)}{(r_i^*-r_j)(1+r_i r_j^*) \sin \alpha_j}\right] .
\label{2B_asymptotic_coeff_I+I}
\end{eqnarray}

\begin{figure}[!t]
\centering
    \includegraphics[width=0.31\linewidth]{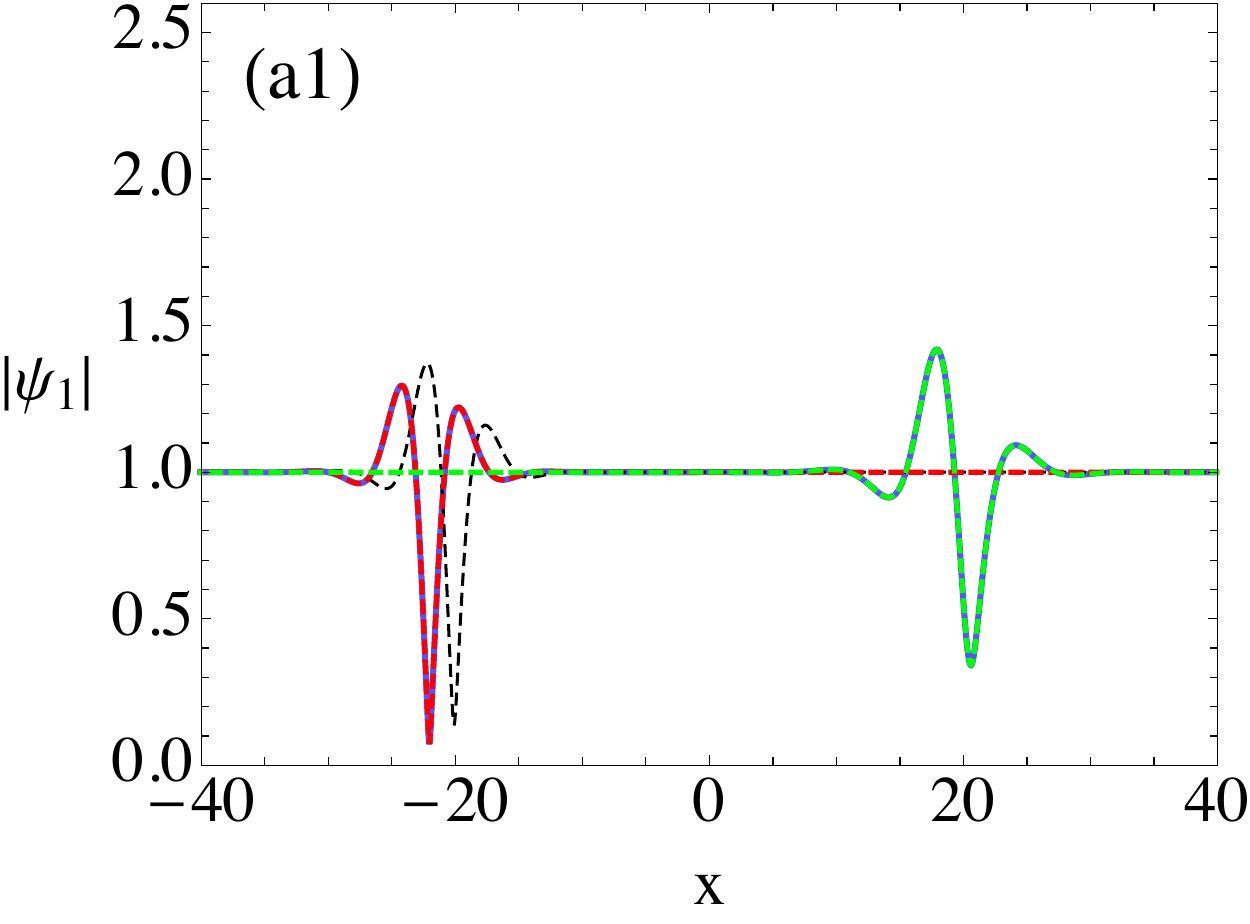}\,\,\,\,\,
    \includegraphics[width=0.31\linewidth]{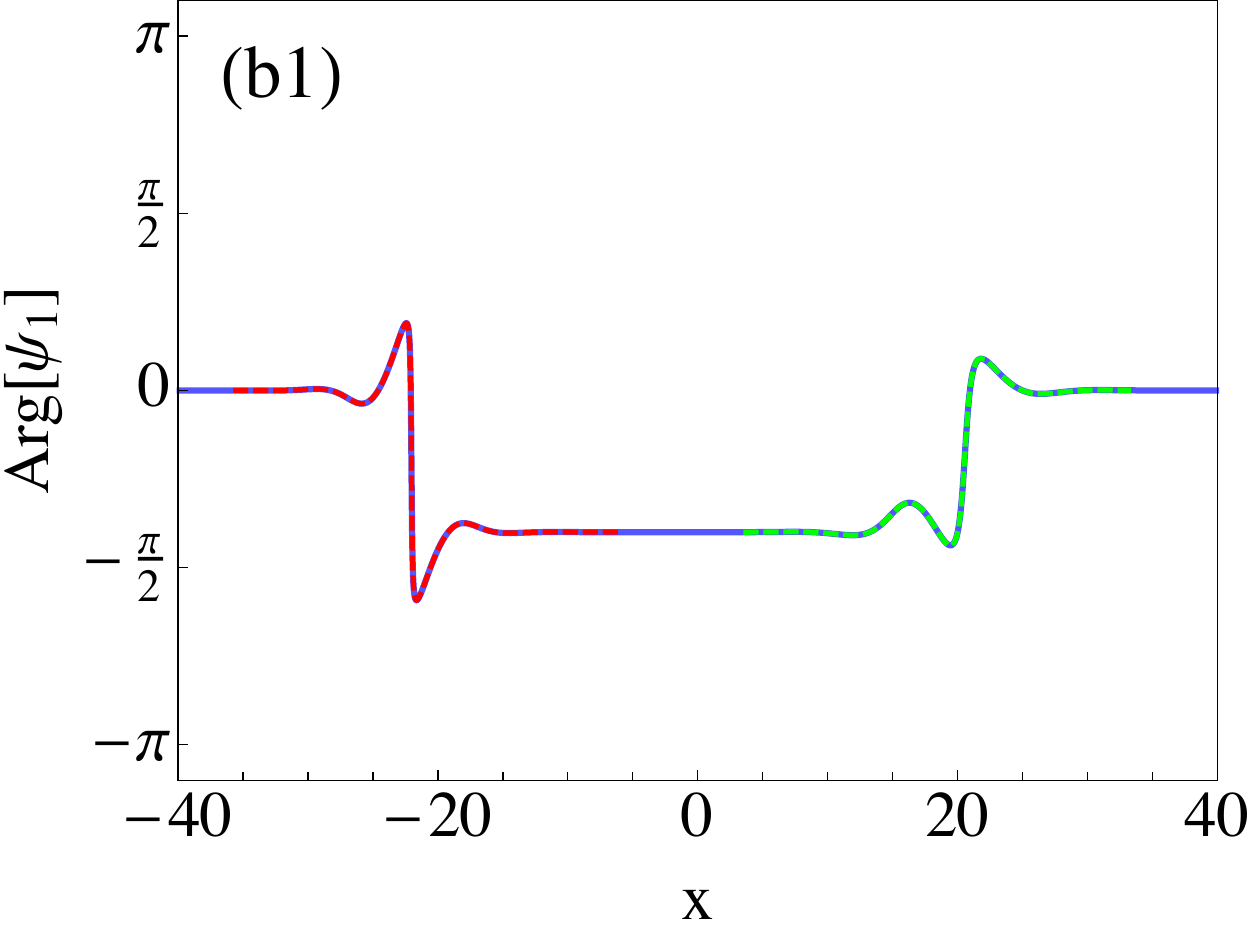}\,\,\,\,\,
    \includegraphics[width=0.32\linewidth]{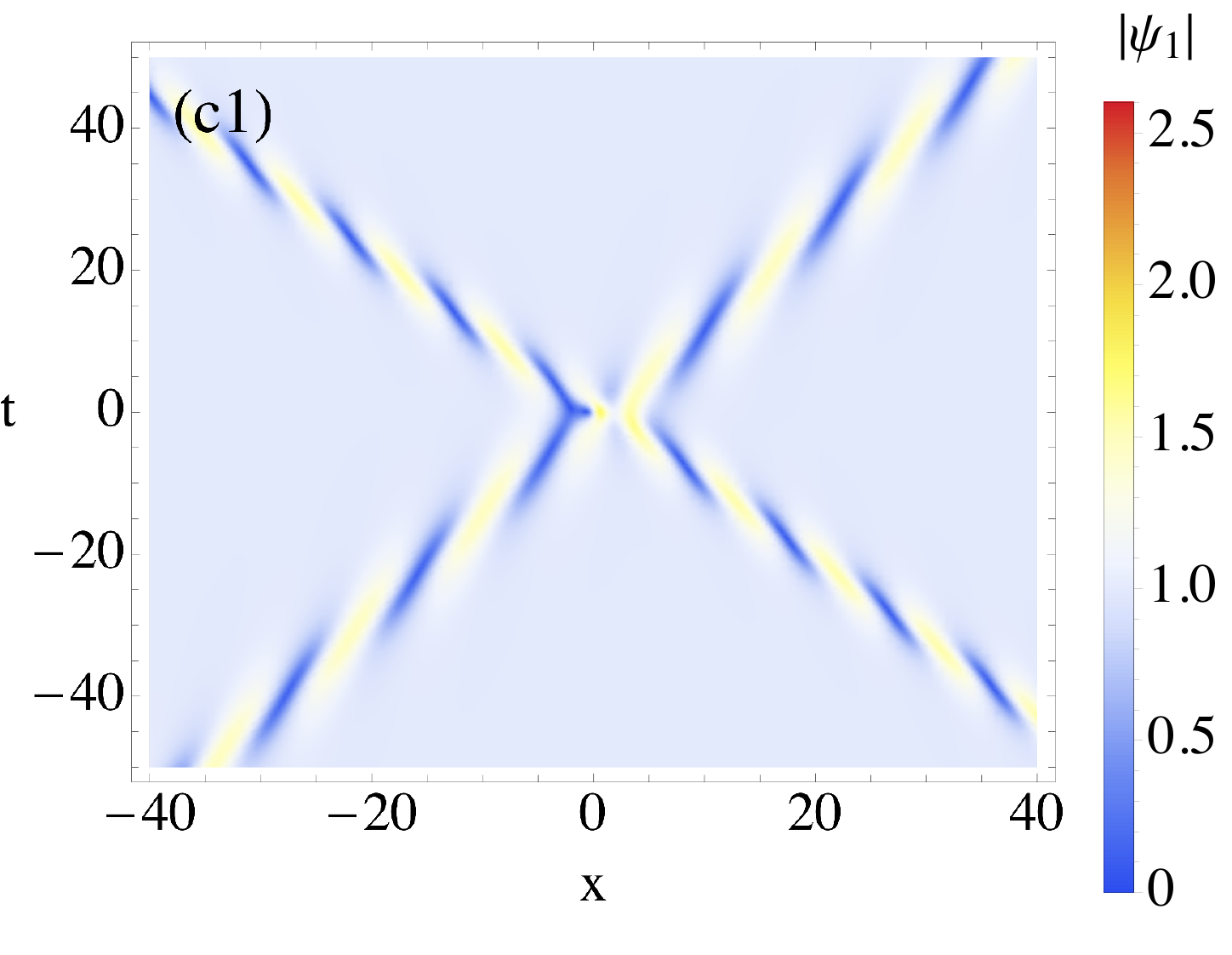}\\
    \includegraphics[width=0.31\linewidth]{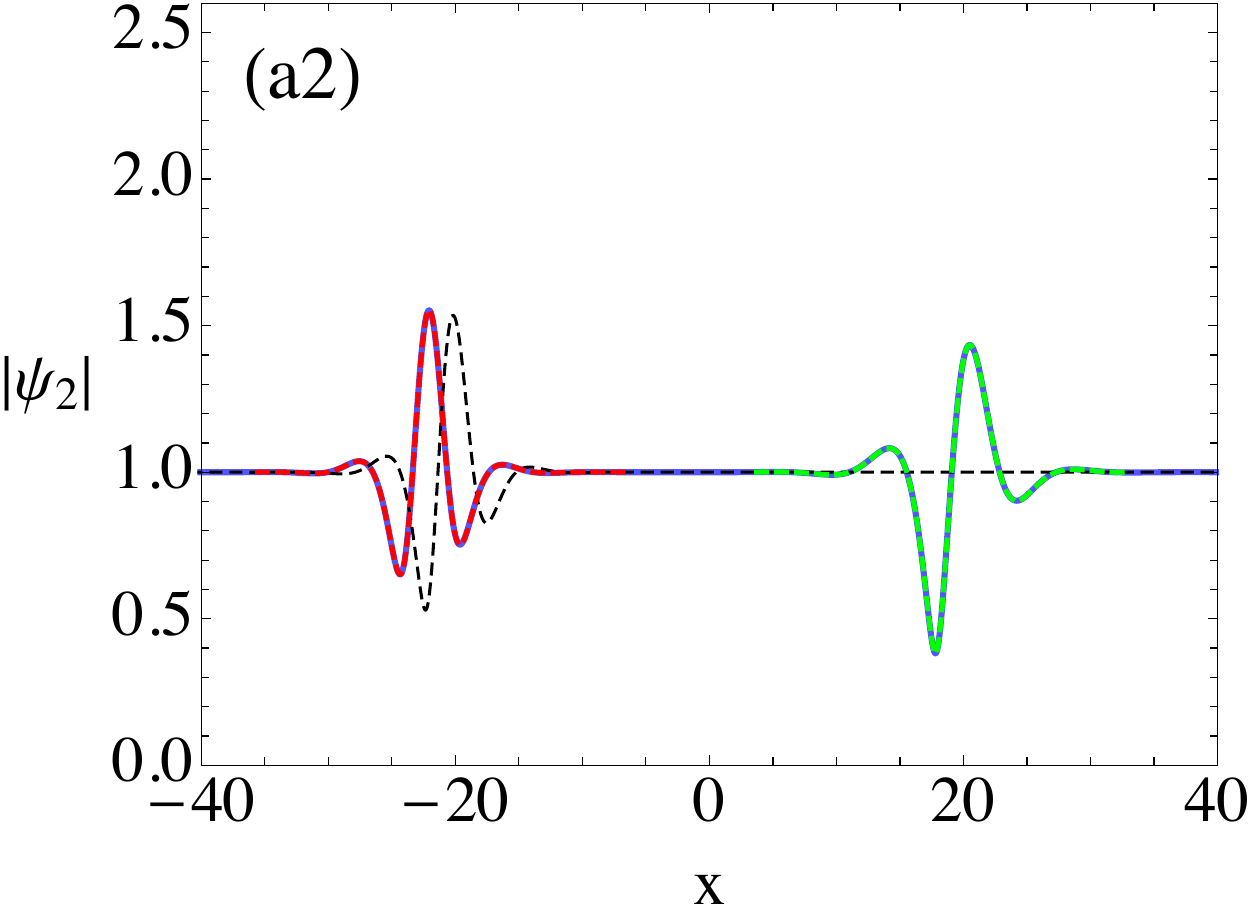}\,\,\,\,\,
    \includegraphics[width=0.31\linewidth]{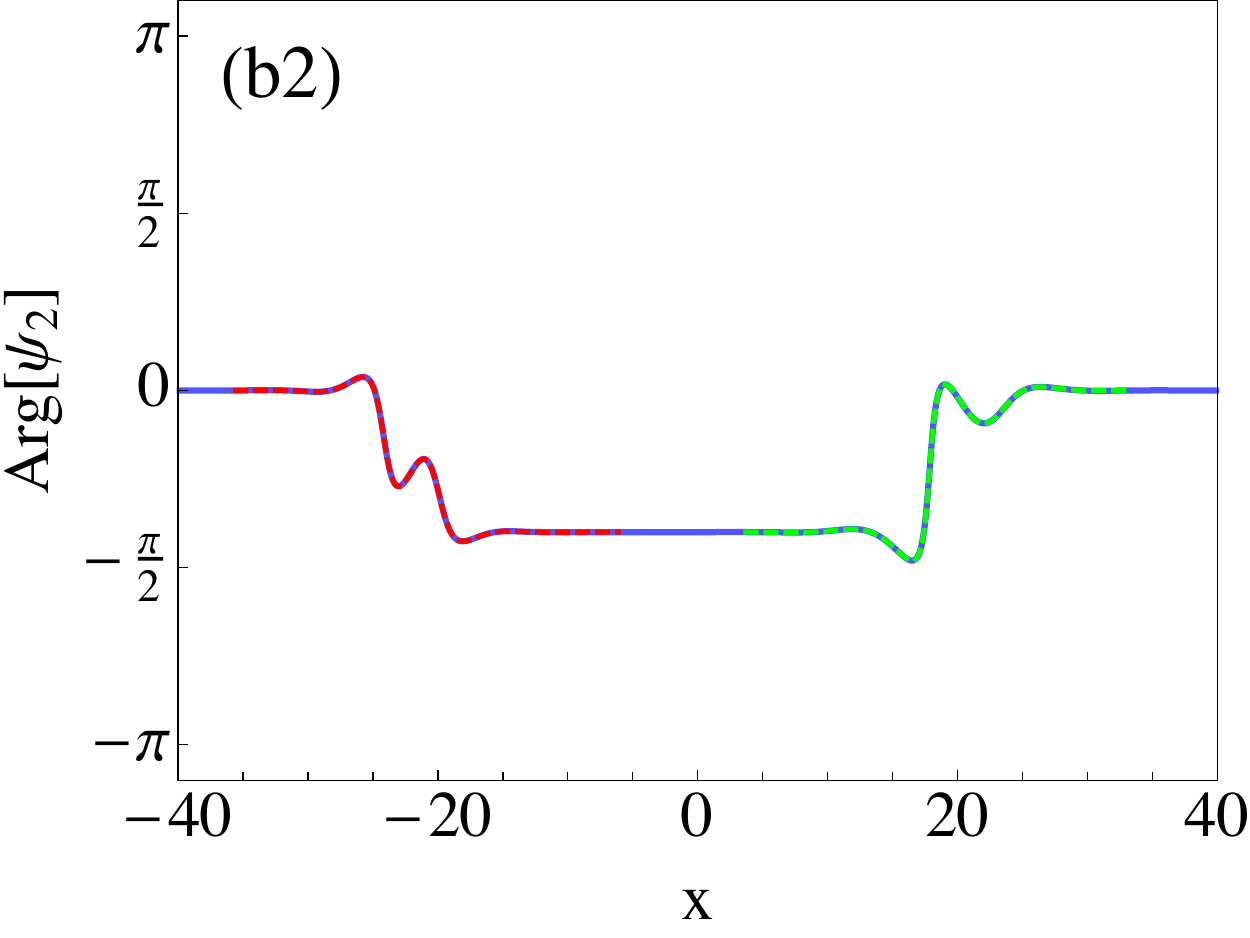}\,\,\,\,\,
    \includegraphics[width=0.32\linewidth]{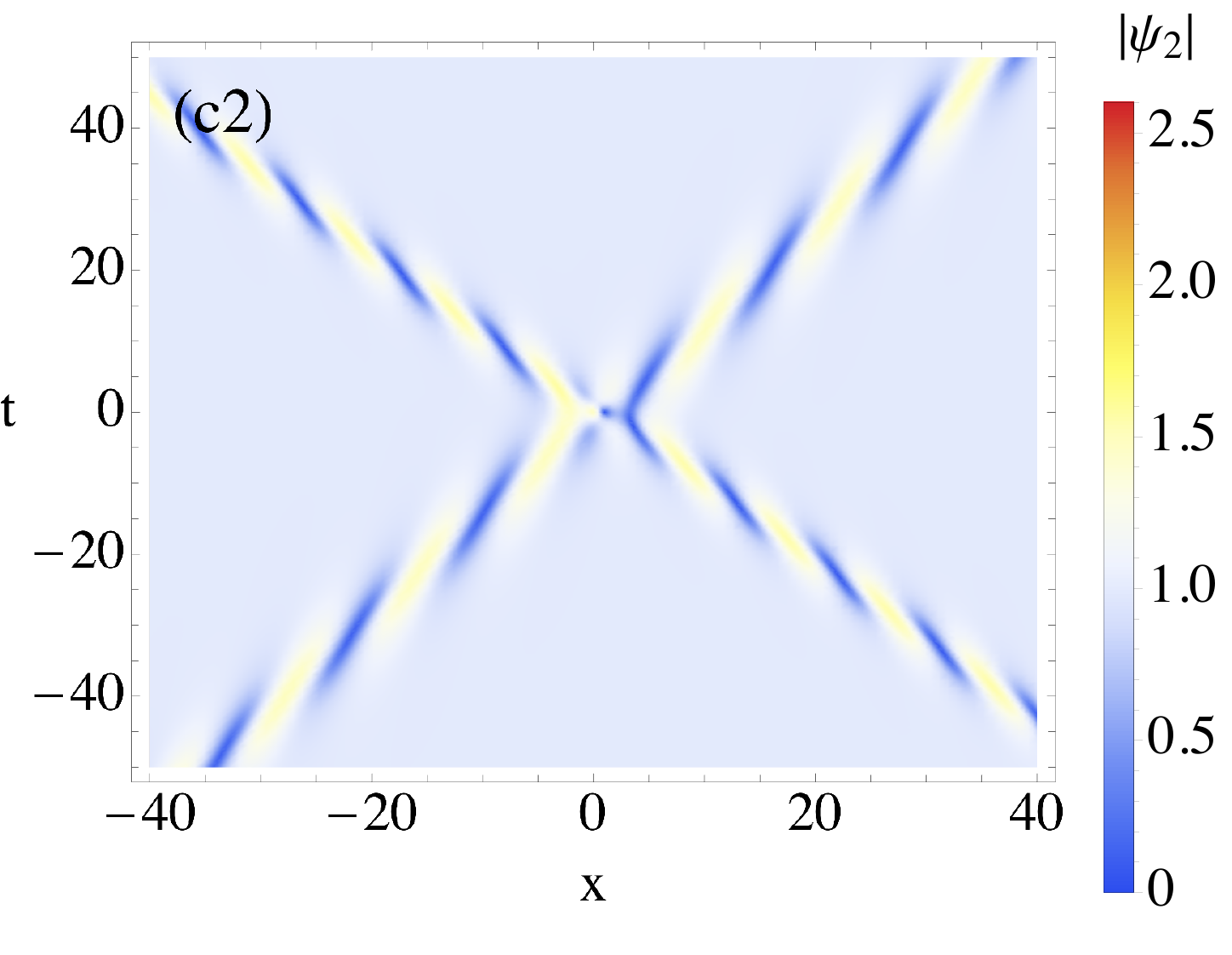}\\
\caption{
Elastic collision $\mathrm{II}+\mathrm{II} \rightarrow \mathrm{II} + \mathrm{II}$ of vector breathers with spectral parameters defined by Eq.~(\ref{parameters_2B}). (a,b) $|\psi_1|$ and $\mathrm{Arg}[\psi_{1}]$, where $\mathrm{Arg}$ means complex phase, after the breathers collision at $t=25.0$. (c) Spatio-temporal plot of the wave field evolution. The dotted green and red lines in (a,b) show a local approximation of the breathers after collision by single-breathers solutions from the asymptotic (\ref{2B_asymptotic_general}). Thin black dashed line in (a) shows how the second breather would have been if it had been traveling alone.}
\label{fig_10}
\end{figure}

Fig.~\ref{fig_09} shows an example of two breathers of types I elastic collision and also illustrates the asymptotic formula (\ref{2B_asymptotic_I+I}). One can see the change of the breathers' phases and positions by comparing the final wavefield state with the situation when one of the breathers travels along, i.e., without collision. For this and some of the subsequent examples of elastic two-breather interactions, we use the following set of parameters,
\begin{eqnarray}
\label{parameters_2B}
&&A_1 = 1, \quad A_2=1;
\\\nonumber
&&\alpha_1=\pi/5, \quad \xi_1=1/4, \quad \theta_1 = 0, \quad \delta_1 = 0,
\\\nonumber
&&\alpha_2=4\pi/5, \quad \xi_2=1/2, \quad \theta_2 = 0, \quad \delta_2 = 0.
\end{eqnarray}

In general case there are six possible combinations of the elastic two-breather interactions $\mathrm{B_i}+\mathrm{\widetilde{B}}_j \rightarrow \mathrm{B_i} + \mathrm{\widetilde{B}}_j$. Here $\mathrm{B}$ and $\mathrm{\widetilde{B}}$ stand for one of the three breather types, while the subscripts $i=1,2$ and $j=2,1$ indicate the breather index number. Note that the indexes are not related to the breather type and can be freely chosen, i.e., they indicate which breather we call the first and which the second. The asymptotic state of this interaction represents the following generalization of the Eq.~(\ref{2B_asymptotic_I+I}),
\begin{eqnarray}
\label{2B_asymptotic_general}
    \psi_{1,2}^{\mathrm{B_i+\widetilde{B}_j}}(\lambda_i,\delta_i,\theta_i;\lambda_j,\delta_j,\theta_j) \rightarrow
    \begin{cases}
    e^{2\,\mathrm{i}\beta^{\pm}_i} \psi_{1,2}^\mathrm{B}(\lambda_i,\delta_i + \delta^{\pm}_{0,i},\theta_i + \theta^{\pm}_{0,i}), \quad\text{at}\quad x\sim V_{\mathrm{B_i}}t, \\\\
    e^{2\,\mathrm{i}\beta^{\pm}_j} \psi_{1,2}^\mathrm{\widetilde{B}}(\lambda_j,\delta_j + \delta^{\pm}_{0,j},\theta_j + \theta^{\pm}_{0,j}), \quad\,\text{at}\quad x\sim V_{\mathrm{\widetilde{B}_j}}t.
    \end{cases}
\end{eqnarray}
where the shifts of positions $\delta$, phases $\theta$ and general phases are defined as follows,
\begin{eqnarray}
\{\delta_{0,i}^{-}, \delta_{0,i}^{+}, \theta_{0,i}^{-}, \theta_{0,i}^{+}, \beta^{-}_i, \beta^{+}_i\} = 
    \begin{cases}
      \{a^{\mathrm{B_i,\tilde{B}_j}}_i, b^{\mathrm{B_i,\widetilde{B}_j}}_i, c^{\mathrm{B_i,\widetilde{B}_j}}_i, d^{\mathrm{B_i,\widetilde{B}_j}}_i, e^{\mathrm{B_i,\widetilde{B}_j}}_i, f^{\mathrm{B_i,\widetilde{B}_j}}_i\}, & \text{at}\ s_i=1 \\\\
      \{b^{\mathrm{B_i,\widetilde{B}_j}}_i, a^{\mathrm{B_i,\widetilde{B}_j}}_i, d^{\mathrm{B_i,\widetilde{B}_j}}_i, c^{\mathrm{B_i,\widetilde{B}_j}}_i, f^{\mathrm{B_i,\widetilde{B}_j}}_i, e^{\mathrm{B_i,\widetilde{B}_j}}_i\}, & \text{at}\ s_i=-1.
    \end{cases}
\label{2B_asymptotic_coeff_general}
\end{eqnarray}
As before, in Eqs.~(\ref{2B_asymptotic_general}) and (\ref{2B_asymptotic_coeff_general}) the subscript indexes can be freely chosen to distinguish breathers, i.e., $i=1,2$ and $j=2,1$. Meanwhile $\mathrm{B}$ and $\widetilde{\mathrm{B}}$ indicate the breather type. The lower and the first upper indexes of the coefficients $a$, $b$, etc. from (\ref{2B_asymptotic_coeff_general}) indicate the breather for which the shift is presented, while the second upper index means the breather with which the studied one interacts. The lower index shows only what we call the studied breather, i.e., either the first or the second one. Meanwhile, the upper indexes also indicate the type of interacting breathers.
For example, $b_i^{\mathrm{B_i,\widetilde{B}_j}}$ at $s_i=1$ ($a_i^{\mathrm{B_i,\widetilde{B}_j}}$ at $s_i=-1$) represents the correction $\delta_{0,i}^{+}$ to the position of the $i$-th breather of type $\mathrm{B}$ at large time after the collision with the breather of type $\mathrm{\widetilde{B}}$.

Similar to the formulas (\ref{2B_asymptotic_coeff_I+I}), we find the rest of the coefficients in (\ref{2B_asymptotic_coeff_general}) by asymptotic analysis of the solution (\ref{two}) at large times, see details in the appendix Sec. \ref{Sec:Appendix:3}. We summarize these results in Table~\ref{table1}, which we present in appendix Sec. \ref{Sec:Appendix:B2}. In addition these tables provide asymptotic wavefield values at $x\rightarrow \pm \infty$ for each of the two-breather configuration. Consider a concrete example of how to use the notations (\ref{2B_asymptotic_coeff_general}). Suppose the process is $\mathrm{I}+\mathrm{II} \rightarrow \mathrm{I} + \mathrm{II}$ and we need to know the asymptotic shifts of positions at large positive time. We say, for instance, that $\mathrm{B_1} = \mathrm{I}$ and $\mathrm{\widetilde{B}_2} = \mathrm{II}$. For the first breather we find $\delta_{0,1}^{+}=b_1^{\mathrm{I, II}}$ if $s_1=1$ and $\delta_{0,1}^{+}=a_1^{\mathrm{I, II}}$ if $s_1=-1$, while for the second one $\delta^{+}_{0,2}=b_2^{\mathrm{II, I}}$ if $s_2=-s_1=1$ and $\delta^{+}_{0,2}=a_2^{\mathrm{II, I}}$ if $s_2=-1$. Then we go to Table \ref{table1} and find the corresponding values of the coefficients $b$ or $a$ in its forth row.

\begin{figure}[!t]
\centering
    \includegraphics[width=0.31\linewidth]{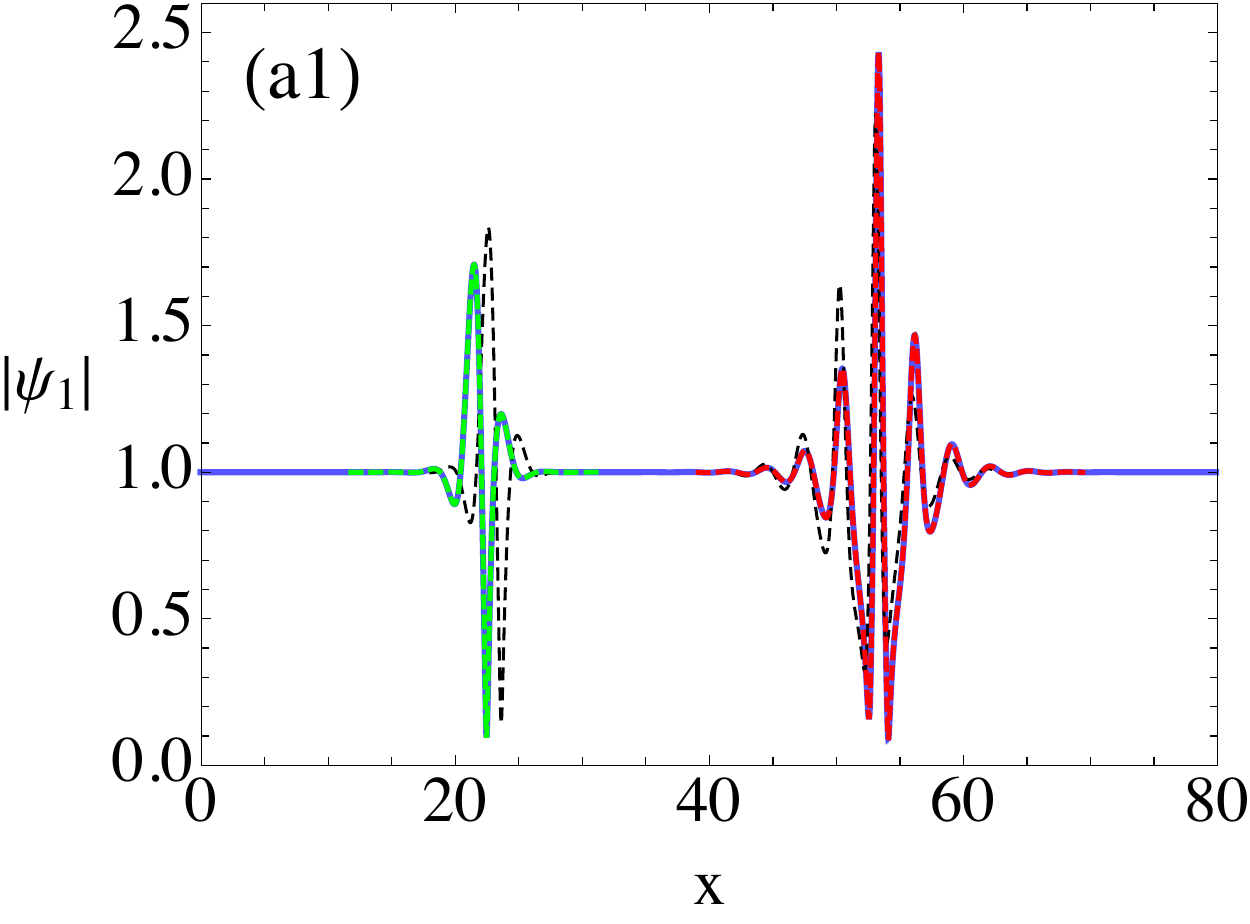}\,\,\,\,\,
    \includegraphics[width=0.31\linewidth]{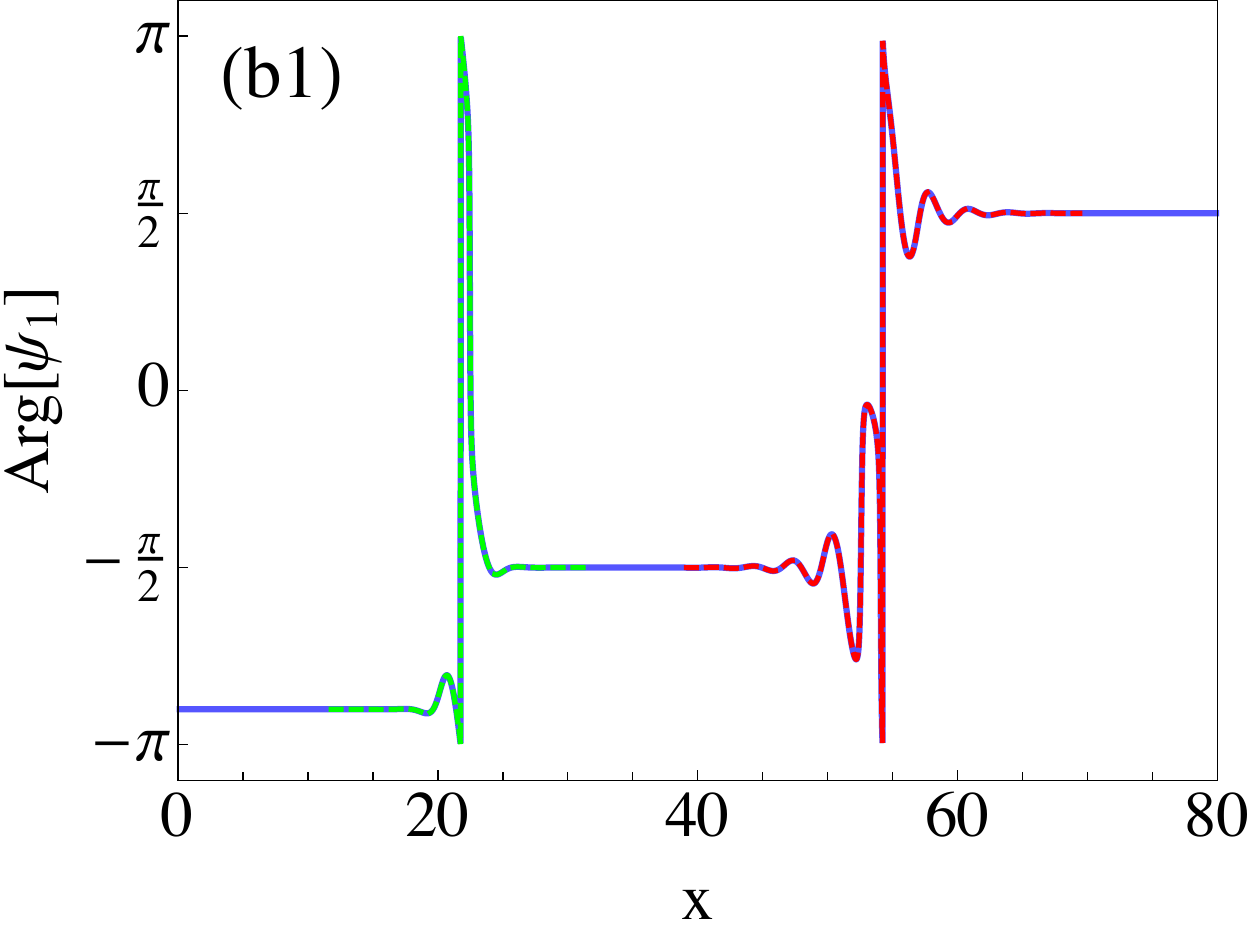}\,\,\,\,\,
    \includegraphics[width=0.32\linewidth]{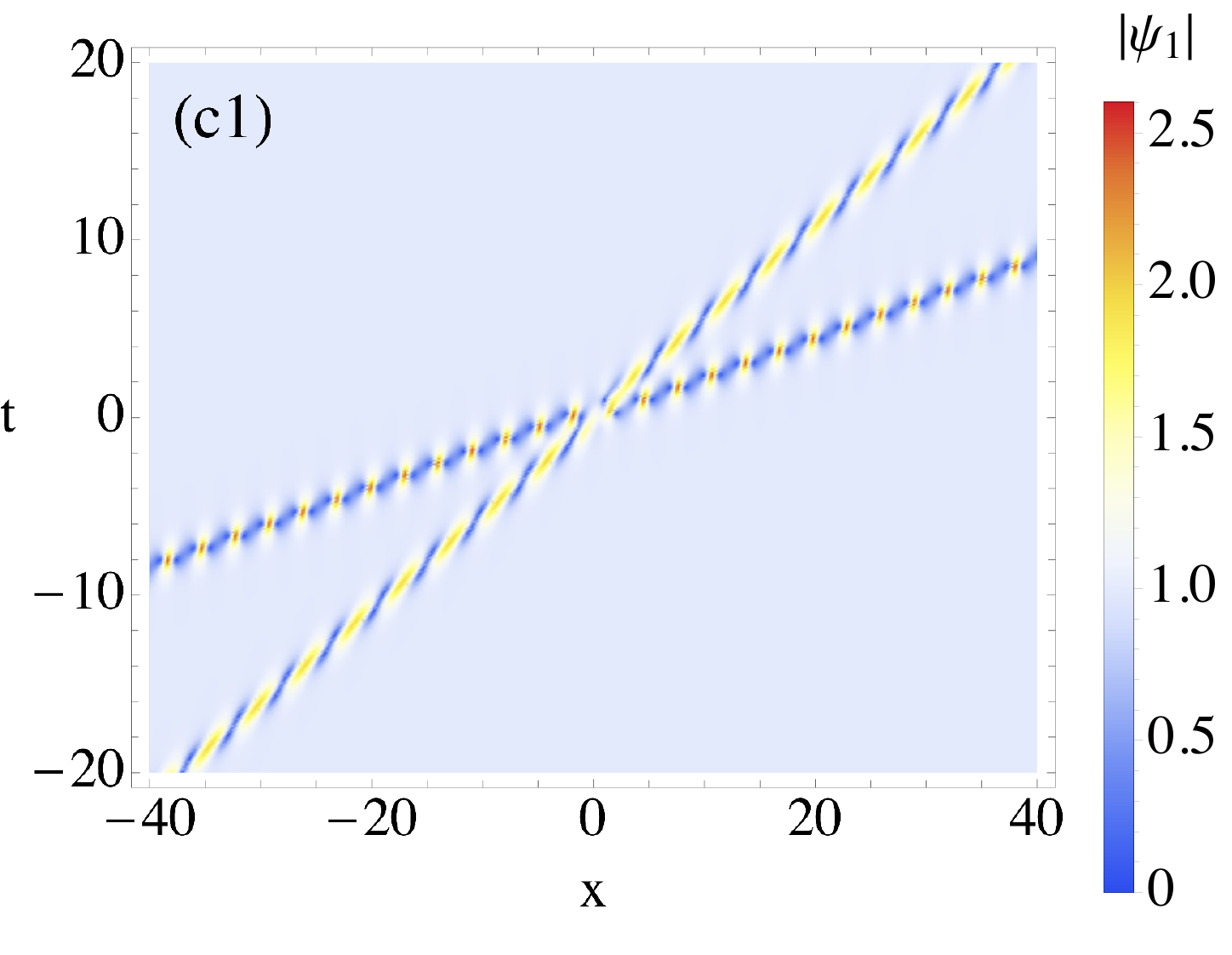}\\
    \includegraphics[width=0.31\linewidth]{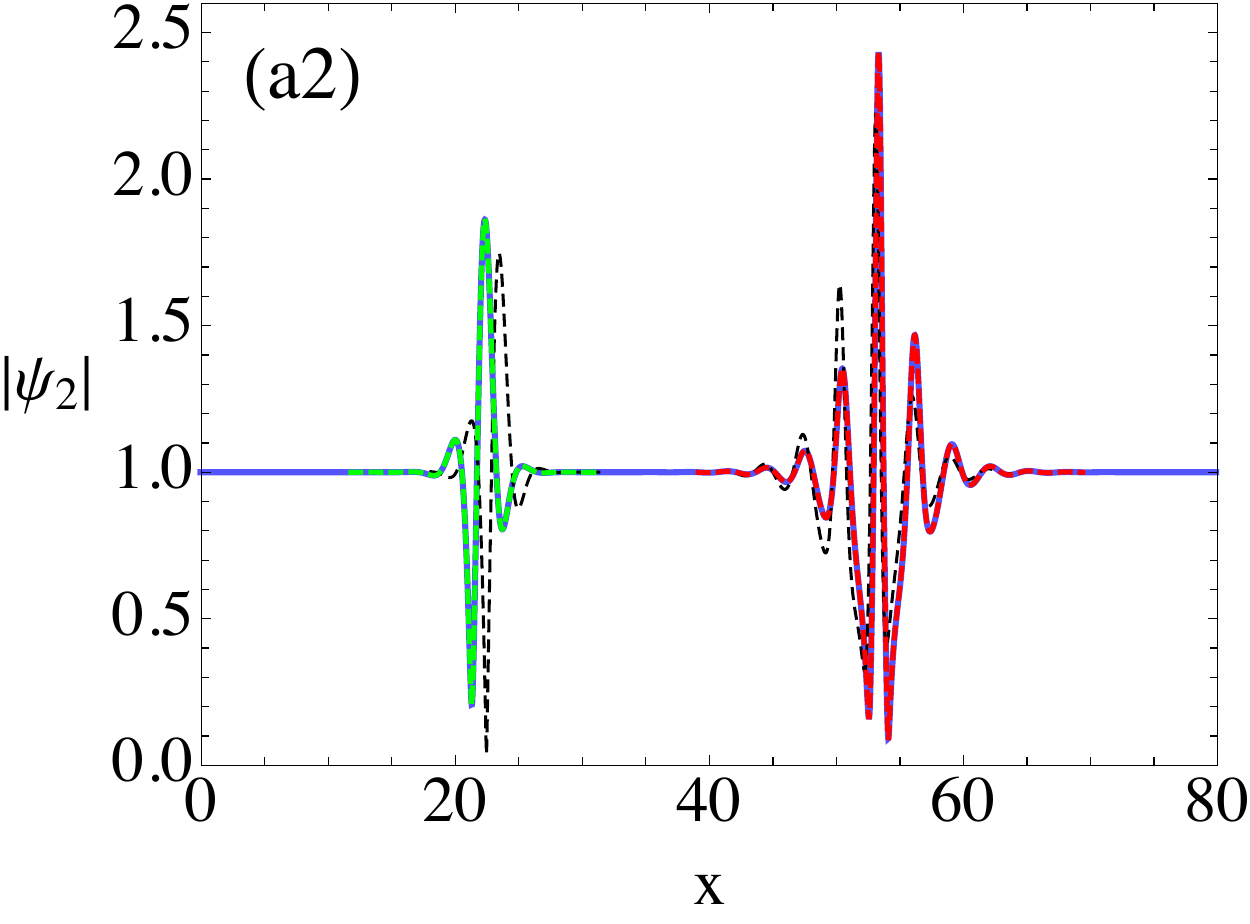}\,\,\,\,\,
    \includegraphics[width=0.31\linewidth]{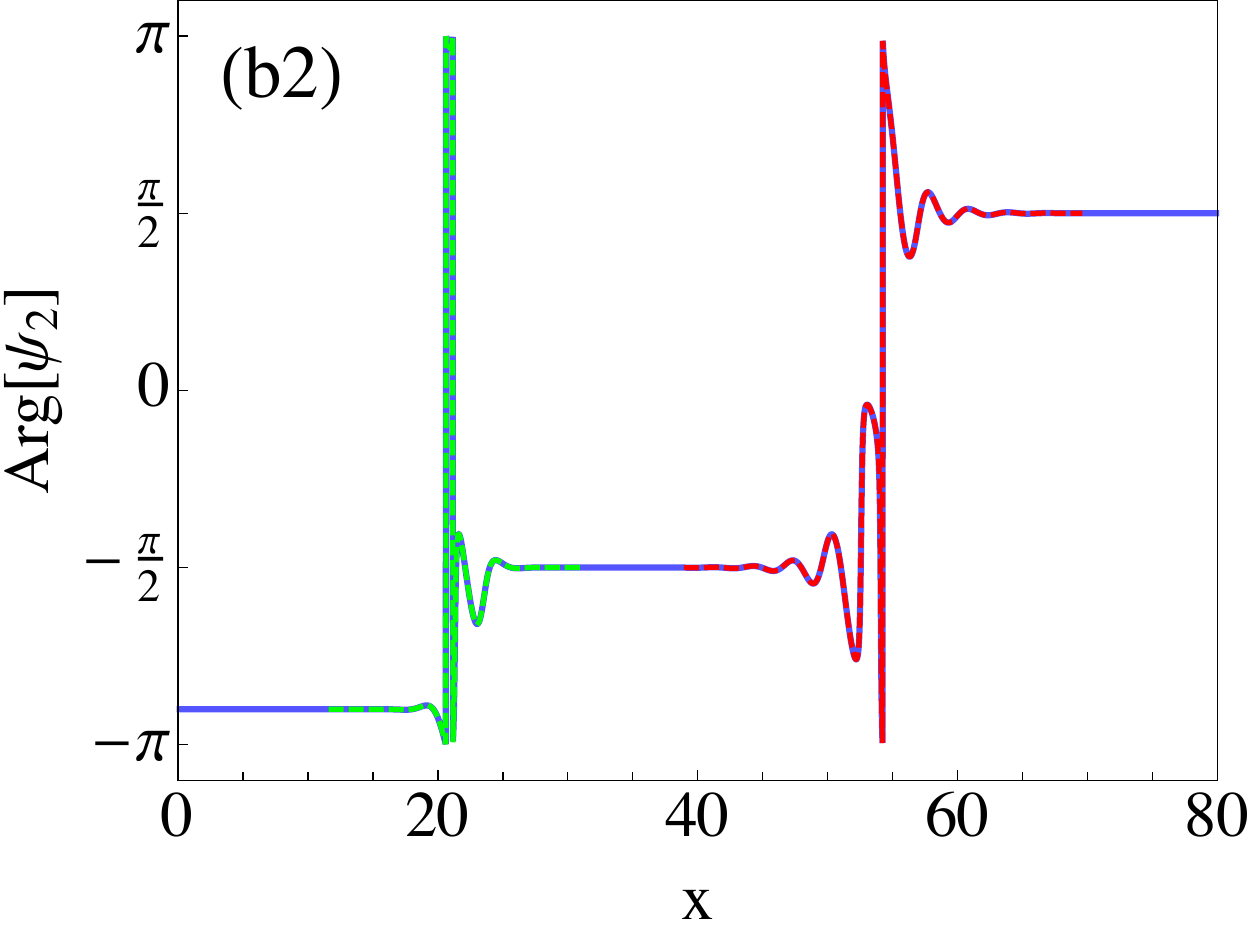}\,\,\,\,\,
    \includegraphics[width=0.32\linewidth]{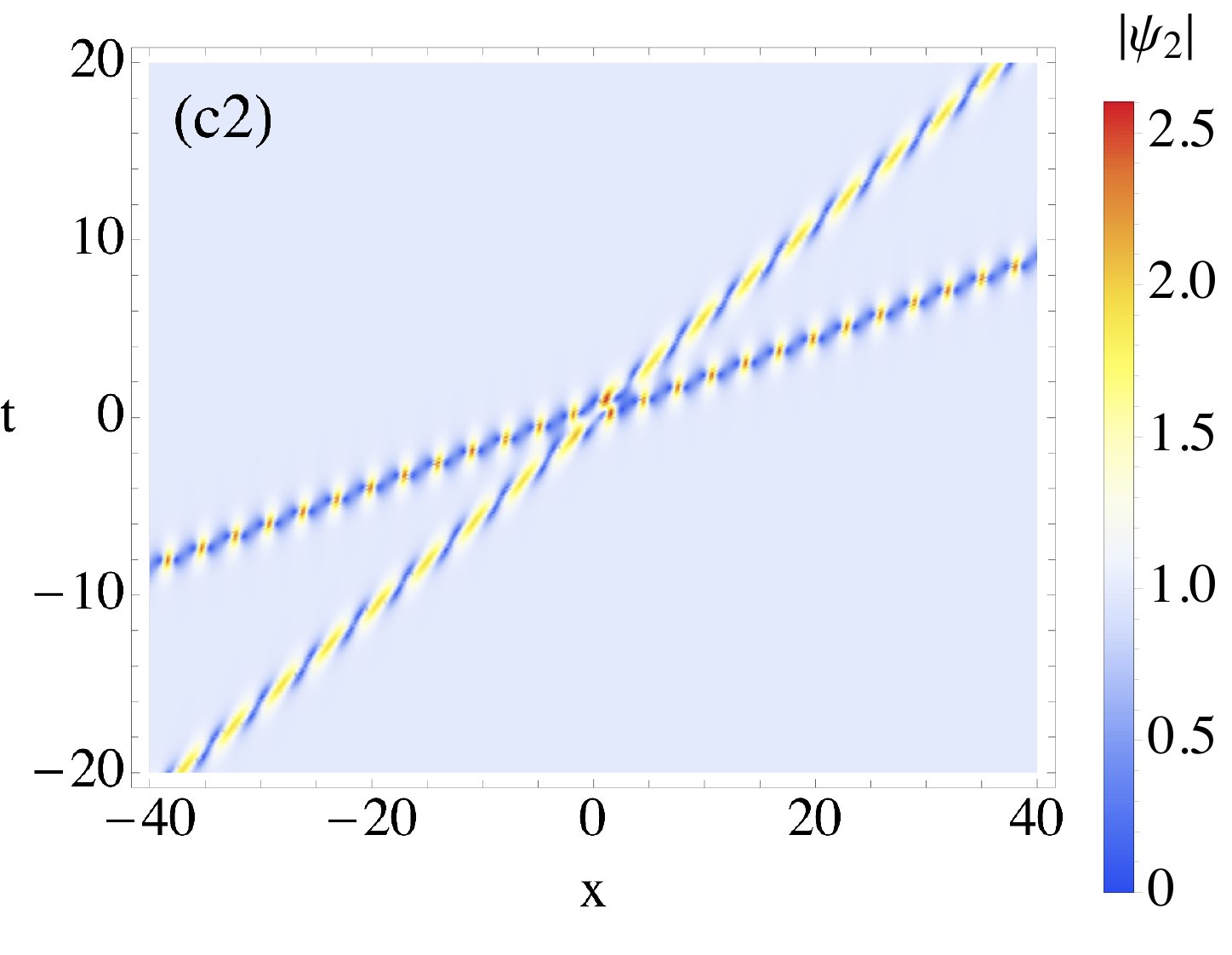}\\
\caption{
Elastic collision $\mathrm{I}+\mathrm{III} \rightarrow \mathrm{I} + \mathrm{III}$ of vector breathers with spectral parameters defined by (\ref{parameters_2B_new}). (a,b) $|\psi_1|$ and $\mathrm{Arg}[\psi_{1}]$, where $\mathrm{Arg}$ means complex phase, after the breathers collision at $t=12.0$. (c) Spatio-temporal plot of the wave field evolution. The dotted green and red lines in (a,b) show a local approximation of the breathers after collision by single-breathers solutions from the asymptotic (\ref{2B_asymptotic_general}). Thin black dashed line in (a) shows how breathers would have been if they had been traveling alone.}
\label{fig_11}
\end{figure}

By analogy with Fig.~\ref{fig_09}, Fig.~\ref{fig_10} shows one example of elastic collisions of two equal-type breathers $\mathrm{II}+\mathrm{II} \rightarrow \mathrm{II} + \mathrm{II}$, while Fig.~\ref{fig_11} shows one mixed case $\mathrm{I}+\mathrm{III} \rightarrow \mathrm{I} + \mathrm{III}$. More examples are presented in appendix Sec. \ref{Sec:Appendix:B1}. The asymptotic result (\ref{2B_asymptotic_general}) is illustrated by approximation of the two breathers after mutual collision by single breathes with appropriately shifted phases and position. Note, that in addition to (\ref{parameters_2B}) we also use for the illustrations the following set of breather parameters,
\begin{eqnarray}
\label{parameters_2B_new}
&&A_1 = 1, \quad A_2=1;
\\\nonumber
&&\alpha_1=\pi/4, \quad \xi_1=1/4, \quad \theta_1 = 0, \quad \delta_1 = 0,
\\\nonumber
&&\alpha_2=\pi/5, \quad \xi_2=1/2, \quad \theta_2 = 0, \quad \delta_2 = 0.
\end{eqnarray}

While the formulas (\ref{2B_asymptotic_general}) and (\ref{2B_asymptotic_coeff_general}) allow to describe the asymptotic state of the two-breather interaction mathematically, the physical meaning plays the total values of the position and phase shifts $\Delta\delta^{\mathrm{B_i,\widetilde{B}_j}}_i$ and $\Delta\theta^{\mathrm{B_i,\widetilde{B}_j}}_i$ acquired by the breather $\mathrm{B_i}$ as a result of collision with the breather $\mathrm{\widetilde{B}_j}$,
\begin{eqnarray}
    \label{total_shifts1}
    \Delta\delta^{\mathrm{B_i,\widetilde{B}_j}}_i = \delta_{0,i}^{+} - \delta_{0,i}^{-},
    \\\label{total_shifts2}
    \Delta\theta^{\mathrm{B_i,\widetilde{B}_j}}_i = \theta_{0,i}^{+} - \theta_{0,i}^{-}.
\end{eqnarray}
The total shift values (\ref{total_shifts1}) and (\ref{total_shifts2}) can be found using the coefficients listed in Table \ref{table1} of the appendix Sec. \ref{Sec:Appendix:B2}. They represent a broad family of expressions. Here we focus on particular features of the space shifts which are responsible for qualitative changes of the breather collision behaviour.

\begin{table}[!t]
\centering 
\renewcommand{\arraystretch}{1.3}
\begin{tabular}{|c| c| c|}
\hline 
Process & Total space shift \\[0.5ex] 
\hline 
$\mathrm{I}_i+\mathrm{I}_j \rightarrow \mathrm{I}_i + \mathrm{I}_j$ 
&
$\begin{aligned}
\Delta\delta^{\mathrm{I_i,I_j}}_i = -s_i l_\mathrm{I}(\lambda_i)\,\log \left|\frac{(r_i-r_j^*)(1+r_i r_j)}{(r_i-r_j)(1+r_i r_j^*)} \right|^2
\end{aligned}$
\\ 
\hline
$\mathrm{II}_i+\mathrm{II}_j \rightarrow \mathrm{II}_i + \mathrm{II}_j$ 
&
$\begin{aligned}
\Delta\delta^{\mathrm{II_i,II_j}}_i = s_i l_\mathrm{II}(\lambda_i) \,
\left( \log \left[1 - \frac{n |1+r_i^* r_j |^2}{m (1+|r_i|^2) (1+|r_j|^2)}\right]
+ \log \left[1 - \frac{n}{m}\right] \right)
\end{aligned}$
\\ 
\hline
$\mathrm{III}_i+\mathrm{III}_j \rightarrow \mathrm{III}_i + \mathrm{III}_j$ 
&
$\begin{aligned}
\Delta\delta^{\mathrm{III_i,III_j}}_i = s_il_\mathrm{III}(\lambda_i) \,
\left( \log \left[1 - \frac{n |1+r_i^* r_j |^2}{m (1+|r_i|^2) (1+|r_j|^2)}\right]
+ \log \left[1 - \frac{n}{m}\right] \right)
\end{aligned}$
\\ 
\hline
$\mathrm{I}_i+\mathrm{II}_j \rightarrow \mathrm{I}_i + \mathrm{II}_j$ 
&
$\begin{aligned}
\Delta\delta^{\mathrm{I_i,II_j}}_i = s_i \frac{l_\mathrm{I}(\lambda_i)}{2}\,\log \left|\frac{(r_i-r_j^*)(1+r_i r_j)}{(r_i-r_j)(1+r_i r_j^*)} \right|^2
\\
\Delta\delta^{\mathrm{II_j,I_i}}_j = s_j l_\mathrm{II}(\lambda_j)\,\log \left|\frac{(r_j-r_i^*)(1+r_j r_i)}{(r_j-r_i)(1+r_j r_i^*)} \right|^2
\end{aligned}$
\\
\hline
$\mathrm{I}_i+\mathrm{III}_j \rightarrow \mathrm{I}_i + \mathrm{III}_j$ 
&
$\begin{aligned}
\Delta\delta^{\mathrm{I_i,III_j}}_i = -s_i \frac{l_\mathrm{I}(\lambda_i)}{2}\,\log \left|\frac{(r_i-r_j^*)(1+r_i r_j)}{(r_i-r_j)(1+r_i r_j^*)} \right|^2
\\
\Delta\delta^{\mathrm{III_j,I_i}}_j = -s_j l_\mathrm{III}(\lambda_j) \,\log \left|\frac{(r_j-r_i^*)(1+r_j r_i)}{(r_j-r_i)(1+r_j r_i^*)} \right|^2
\end{aligned}$
\\
\hline
$\mathrm{II}_i+\mathrm{III}_j \rightarrow \mathrm{II}_i + \mathrm{III}_j$ 
&
$\begin{aligned}
\Delta\delta^{\mathrm{II_i,III_j}}_i = -s_i l_\mathrm{II}(\lambda_i)\,
\left( \log \left[1 - \frac{n |r_i - r^*_j |^2}{m (1+|r_i|^2) (1+|r_j|^2)}\right]
+ \log \left[1 - \frac{n}{m}\right] \right)
\\
\Delta\delta^{\mathrm{III_j,II_i}}_j = s_j l_\mathrm{III}(\lambda_j) \,
\left( \log \left[1 - \frac{n |r_j - r^*_i |^2}{m (1+|r_j|^2) (1+|r_i|^2)}\right]
+ \log \left[1 - \frac{n}{m}\right] \right)
\end{aligned}$
\\
\hline
\end{tabular}
\caption{Values of the total space shifts, see Eq.~(\ref{total_shifts1}), for elastic collisions of vector breathers. Left column indicates the collision process type in the form $\mathrm{B_i}+\mathrm{\widetilde{B}_j} \rightarrow \mathrm{B_i} + \mathrm{\widetilde{B}_j}$, where $i=1\,\text{or}\,2$ while $j = 2\,\text{or}\,1$ respectively. Right column presents the corresponding values of $\Delta\delta^{\mathrm{B_i,\widetilde{B}_j}}_i$ and $\Delta\delta^{\mathrm{\widetilde{B}_j,B_i}}_j$ for each of the two breathers participating in the interaction. When $\mathrm{B}$ has the same type as $\mathrm{\widetilde{B}}$ we leave only $\Delta\delta^{\mathrm{B_i,\widetilde{B}_j}}_i$ value. The meaning of the indexes has been explained in the main text. For example, $\Delta\delta^{\mathrm{I_i,III_j}}_i$ describes the total space shift acquired by $i$th breather of type $\mathrm{I}$ as a result of a collision with the breather of type $\mathrm{III}$.}
\label{tablemain}
\end{table}

In Table \ref{tablemain} we summarize the total spatial shifts in all possible breather interaction scenarios. As was shown in \cite{gelash2022breather}, for scalar NLSE breathers participating in the head-on collisions the value of the spatial shift in the direction of the breather propagation can be positive, negative and even zero depending on the breather parameters, which means that breather can move forward or backward relative to its initial trajectory or remain on it. In the vector case we have a similar situation for those interactions where the type $\mathrm{I}$ breathers participate. More precisely, the sign of the spatial shift depends on which breather is faster, what is controlled by the value of $s$. However, the fastest breather not necessarily move forward, and in addition in some cases the sign of $\Delta\theta^{\mathrm{B_i,B_j}}_i$ can be switched by changing breather parameters keeping $s_i$ the same. Note, that this situation is unusual taking into account that for NLSE solitons the fastest one always move forward with respect to its propagation direction, while another one move backwards \cite{NovikovBook1984}.

\begin{figure}[!h]
\centering
    \includegraphics[width=0.31\linewidth]{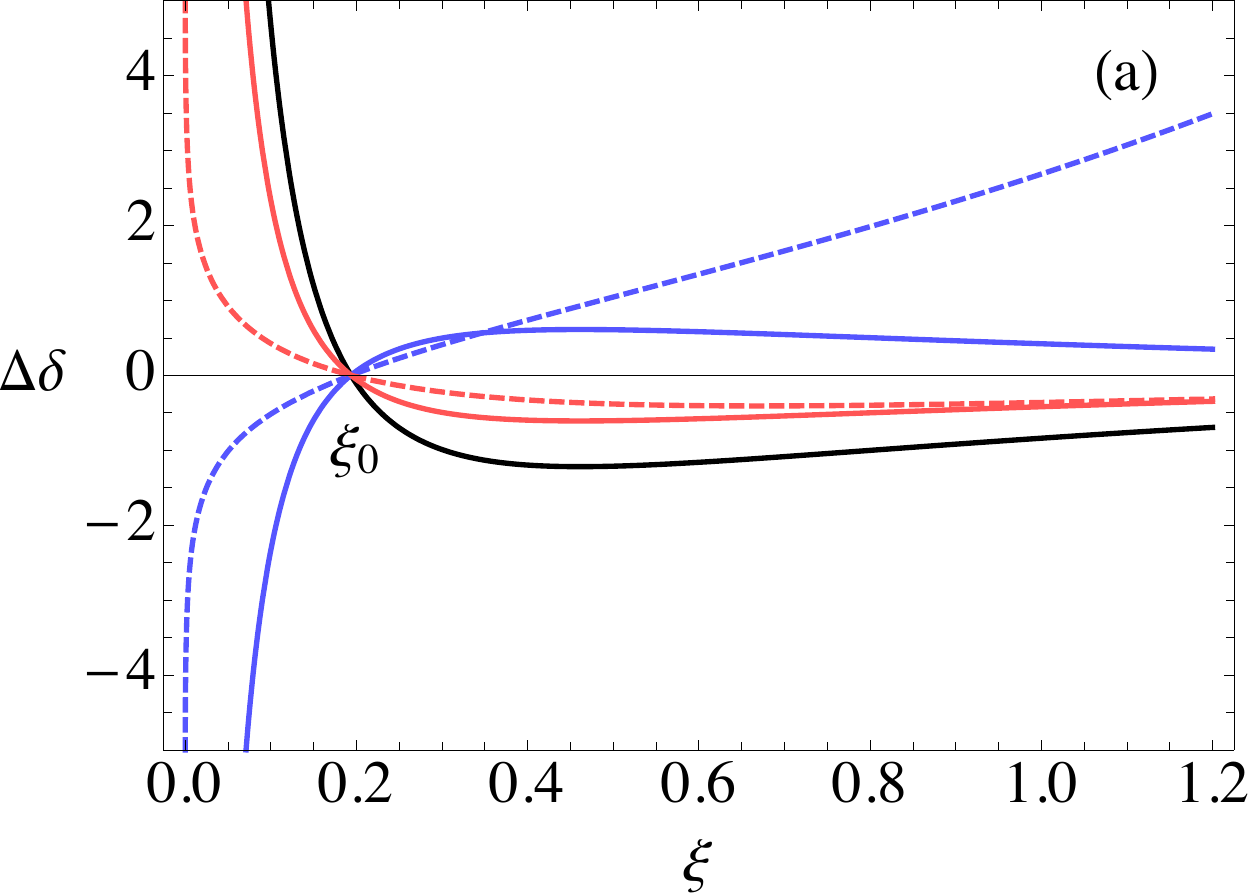}\,\,\,\,\,
    \includegraphics[width=0.31\linewidth]{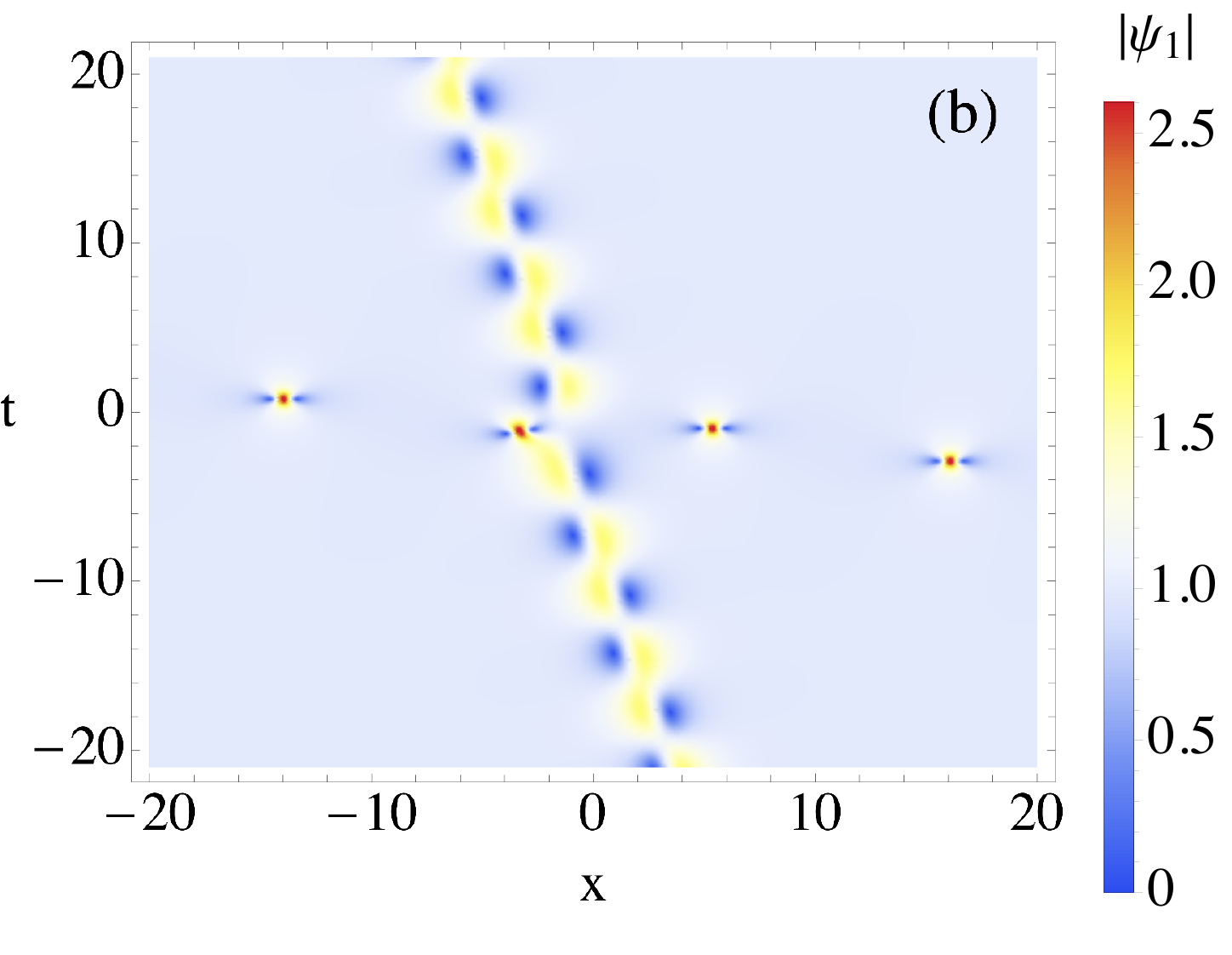}\,\,\,\,\,
    \includegraphics[width=0.32\linewidth]{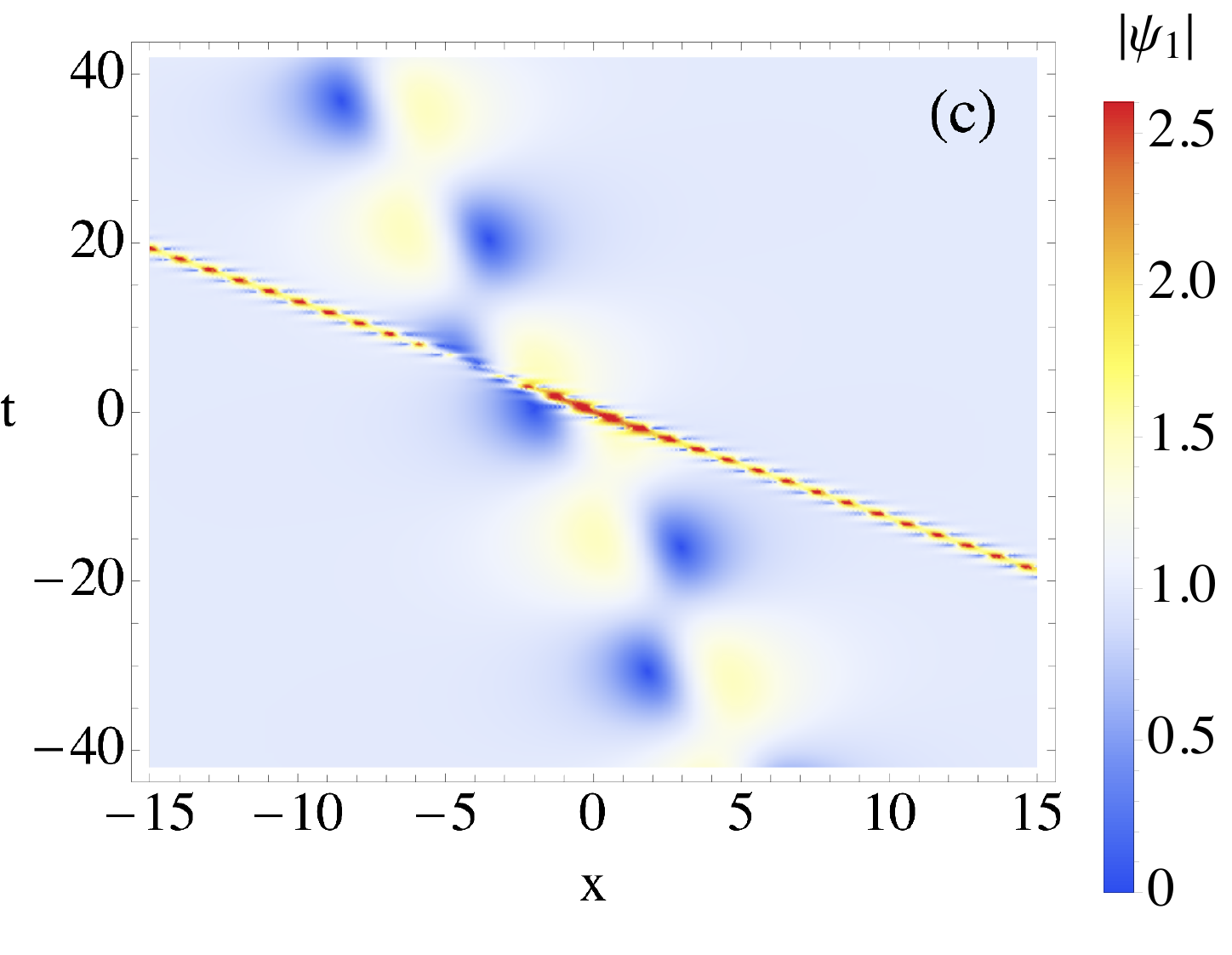}
\caption{Illustration of the total space shifts behaviour in the collisions $\mathrm{I_1}+\mathrm{B_2} \rightarrow \mathrm{I_1} + \mathrm{B_2}$, where $\mathrm{B_2}$ is either type $\mathrm{I}$, $\mathrm{II}$ or $\mathrm{III}$ breather. (a) Dependence of the total space shift for the breathers with parameters defined by (\ref{parameters_log_shifts}) as a function on $\xi$ and fixed $\alpha_1 = 6\pi/16$. Black line corresponds to the case $\mathrm{B_2} = \mathrm{I_2}$, blue lines to the process $\mathrm{B_2} = \mathrm{II_2}$ (solid blue line for the breather $\mathrm{I}_1$ and dashed one for the breather $\mathrm{II}_2$), red lines to the case $\mathrm{B_2} = \mathrm{III_2}$ (solid red line for the breather $\mathrm{I}_1$ and dashed one for the breather $\mathrm{III}_2$). The shift curves change sing at the point $\xi_0$ where the argument of the logarithm (\ref{log}) turns unity. (b,c) Spatio-temporal plots for collisions $\mathrm{I}_1+\mathrm{II}_2$ in the cases $\alpha_1=9\pi/16$, $\xi=0.05<\xi_0$ (b), and $\xi=0.8>\xi_0$ (c). 
}
\label{fig_11add}
\end{figure}

To illustrate the described above behaviour of the vector breathers we consider the following particular case of parameters,
\begin{eqnarray}
\label{parameters_log_shifts}
\alpha_1= \alpha, \quad \alpha_2=\pi - \alpha, \quad \xi_1 = \xi_2 = \xi,
\end{eqnarray}
and assume that the first breather is always of type $\mathrm{I}$, i.e. $\mathrm{B_1}=\mathrm{I}_1$. When $\mathrm{B_2}=\mathrm{I}_2$ or $\mathrm{B_2}=\mathrm{III}_2$, the breathers collide in the head-on manner, while for $\mathrm{B_2}=\mathrm{II}_2$ the collision is overtaking. In addition the absolute value of the breather $\mathrm{I}$ velocity is always bigger than for the breathers $\mathrm{II}$ and $\mathrm{III}$, see Eqs.~(\ref{characteristic_values1}), (\ref{characteristic_values2}) and (\ref{characteristic_values3}). The latter means that the sign of $s_i$ does not change when changing $\alpha$ or $\xi$. Meanwhile the sign of the total spatial shift can be changed by changing $\xi$ at a fixed value of $\alpha$. Indeed, the logarithm in the corresponding shift expressions, see Table \ref{tablemain}, can be simplified under the constrain (\ref{parameters_log_shifts}) as follows,
\begin{eqnarray}
\label{log}
    \log \left|\frac{(r_1-r_2^*)(1+r_1 r_2)}{(r_1-r_2)(1+r_1 r_2^*)} \right|^2 = 
    \log \left(\frac{\sinh^2{\xi}}
    {\cos^2{\alpha}(\cos^2{\alpha}\sinh^2{\xi} + \sin^2{\alpha}\cosh^2{\xi})} \right).
\end{eqnarray}
Now one sees that for a fixed $\alpha\ne\pi/2$ the argument of the logarithm (\ref{log}) changes from zero to $\cos^{-2}{\alpha}$, when $\xi$ changes from zero to infinity. At the point $\xi_0$ satisfying transcendental condition $\sinh^2{\xi_0} = \cos^2{\alpha}(\cos^2{\alpha}\sinh^2{\xi_0} + \sin^2{\alpha}\cosh^2{\xi_0})$, the shift change sign, what we illustrate in Fig.~\ref{fig_11add}(a). In addition Fig.~\ref{fig_11add}(b,c) show examples of the overtaking collision $\mathrm{I}+\mathrm{II} \rightarrow \mathrm{I} + \mathrm{II}$, where $\mathrm{I}$ overtakes $\mathrm{II}$, for $\xi<\xi_0$ (b) and $\xi>\xi_0$ (c). One can see that in the case (b) the first breather $\mathrm{I}$ shifts forward along its trajectory, while the second breather $\mathrm{II}$ shifts backwards. In the case (c) the geometry of collision is the same, i.e. the first breather being faster overtakes the second one, however the sings of the shift are opposite. The latter can be easily seen for the breather $\mathrm{II}$ and poorly pronounced for the breather $\mathrm{I}$ due to a relatively small absolute value of the shift, see also Fig.~\ref{fig_11add}(a).

In addition we note that the sign of the logarithmic expressions in the total space shifts formulas of the processes $\mathrm{II}+\mathrm{II} \rightarrow \mathrm{II} + \mathrm{II}$ and $\mathrm{III}+\mathrm{III} \rightarrow \mathrm{III} + \mathrm{III}$ is always negative, i.e.
\begin{eqnarray}
\label{log_signs}
\log \left[1 - \frac{n |1+r_i^* r_j |^2}{m (1+|r_i|^2) (1+|r_j|^2)}\right]
+ \log \left[1 - \frac{n}{m}\right] <0,
\\\nonumber
\log \left[1 - \frac{n |1+r_i^* r_j |^2}{m (1+|r_i|^2) (1+|r_j|^2)}\right]
+ \log \left[1 - \frac{n}{m}\right] <0.
\end{eqnarray}
The inequalities (\ref{log_signs}) mean that the sign of the total shifts $\Delta\delta^{\mathrm{II_i,II_j}}_i$ and $\Delta\delta^{\mathrm{III_i,III_j}}_i$ is determined only by the sign of $s_i$. The latter is different for the processes $\mathrm{II}+\mathrm{II} \rightarrow \mathrm{II} + \mathrm{II}$ and $\mathrm{III}+\mathrm{III} \rightarrow \mathrm{III} + \mathrm{III}$, since the signs of $V_{\mathrm{II}}$ and $V_{\mathrm{III}}$ are opposite, see Eqs.~(\ref{characteristic_values2}) and (\ref{characteristic_values3}). Thus, for the same set of eigenvalues, the signs of $\Delta\delta^{\mathrm{II_i,II_j}}_i$ and $\Delta\delta^{\mathrm{III_i,III_j}}_i$ are also opposite.

\section{Particular cases of vector two-breather solution}\label{Sec:6}
The resonant fusion and decay of breathers presented in Sec. \ref{Sec:4} is based on the single-eigenvalue solution with nonzero integration constants $C_0$, $C_1$ and $C_2$. However, the same expression can be also obtained from the two-eigenvalue solution (\ref{two}) in the case of merging eigenvalues, i.e., when $\lambda_1 = \lambda_2 = \lambda$. More specifically, one must choose the integration constants in (\ref{two}) so that each breather has a different type, either I, II, or III. In other words, each of the vectors $\mathbf{C}_1$ and $\mathbf{C}_2$ has one zero in its components, and the positions of these zeros are different. For example, consider the case I+II, i.e. $C_{0,1} = 0$ and $C_{1,2} = 0$. We substitute $\lambda_1 = \lambda_2 = \lambda$ in (\ref{two}), so that  $r_1 = r_2 = r$. In addition, we change $\alpha\rightarrow-\alpha$, and end up with the resonant solution described in Sec. \ref{Sec:4}, characterized by the following  three nonzero integration constants,
\begin{eqnarray}
\label{Resonance_param2}
    C_0 = \left [ C_{1,1}C_{2,2}(1+r^2)/C_{0,2} \right ]^*, \qquad
    C_1 = C^*_{2,1}, \qquad
    C_2 = -C^*_{1,1}.
\end{eqnarray}

\begin{figure}[!t]
\centering
    \includegraphics[width=0.31\linewidth]{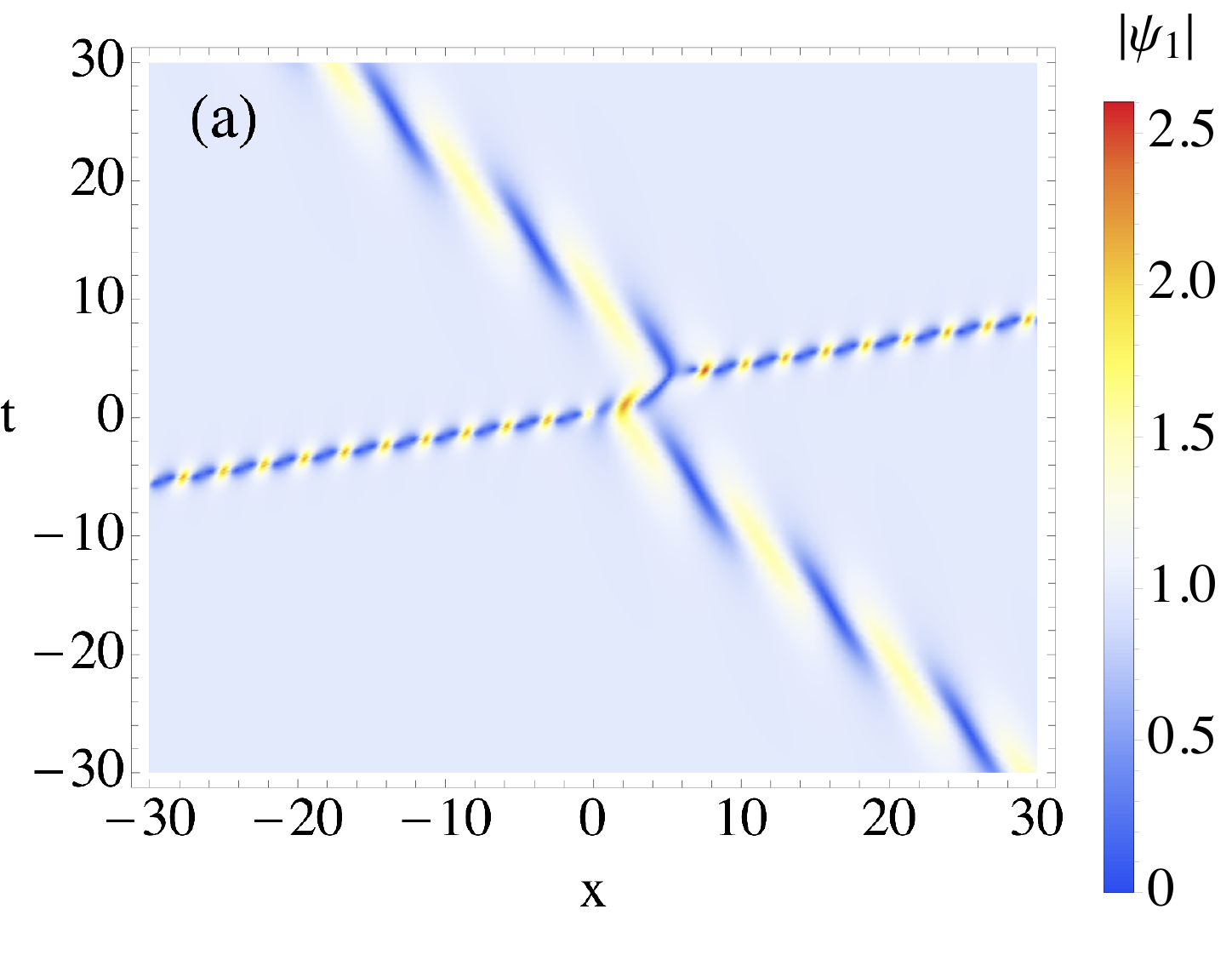}\,\,\,\,\,
    \includegraphics[width=0.31\linewidth]{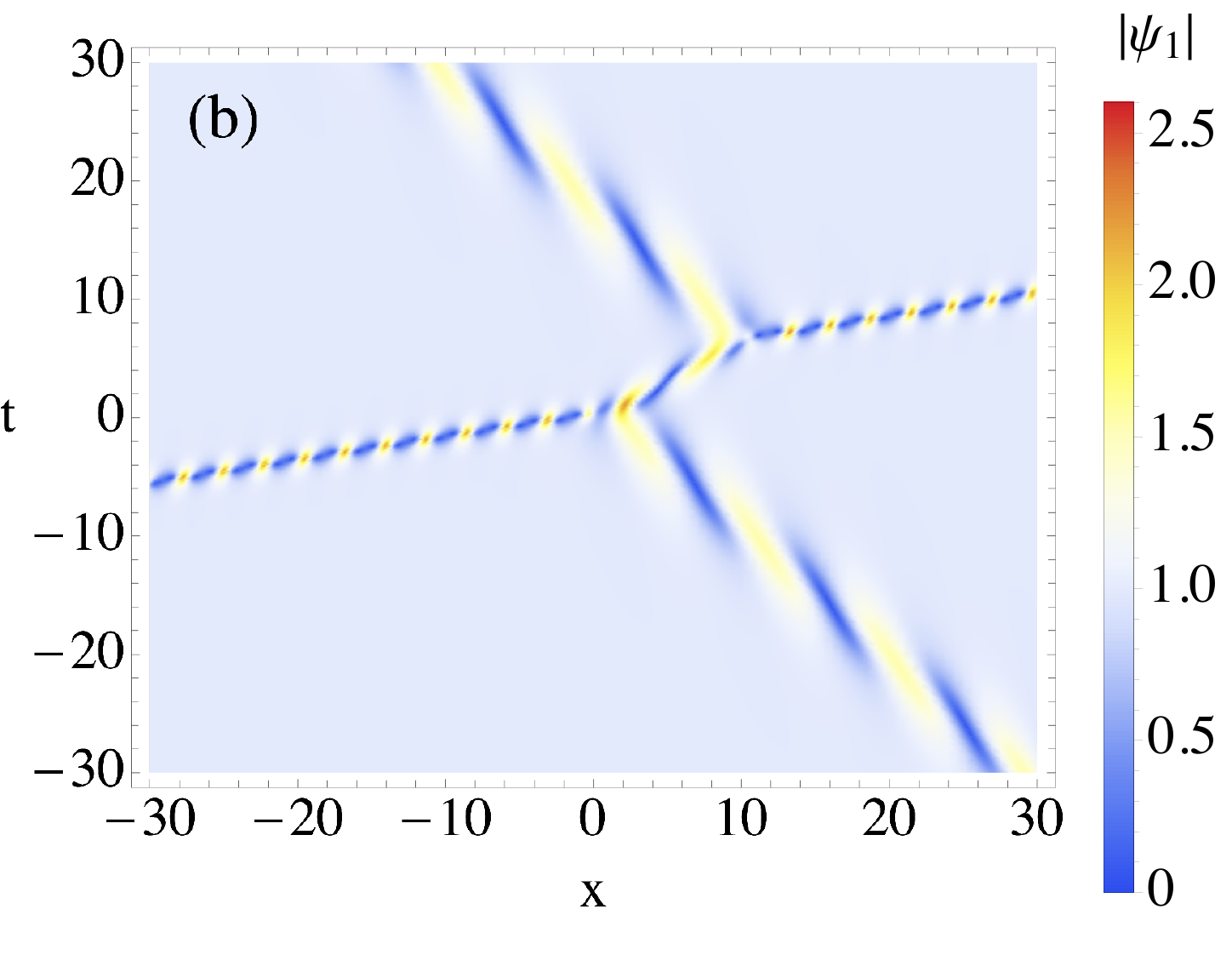}\,\,\,\,\,
    \includegraphics[width=0.32\linewidth]{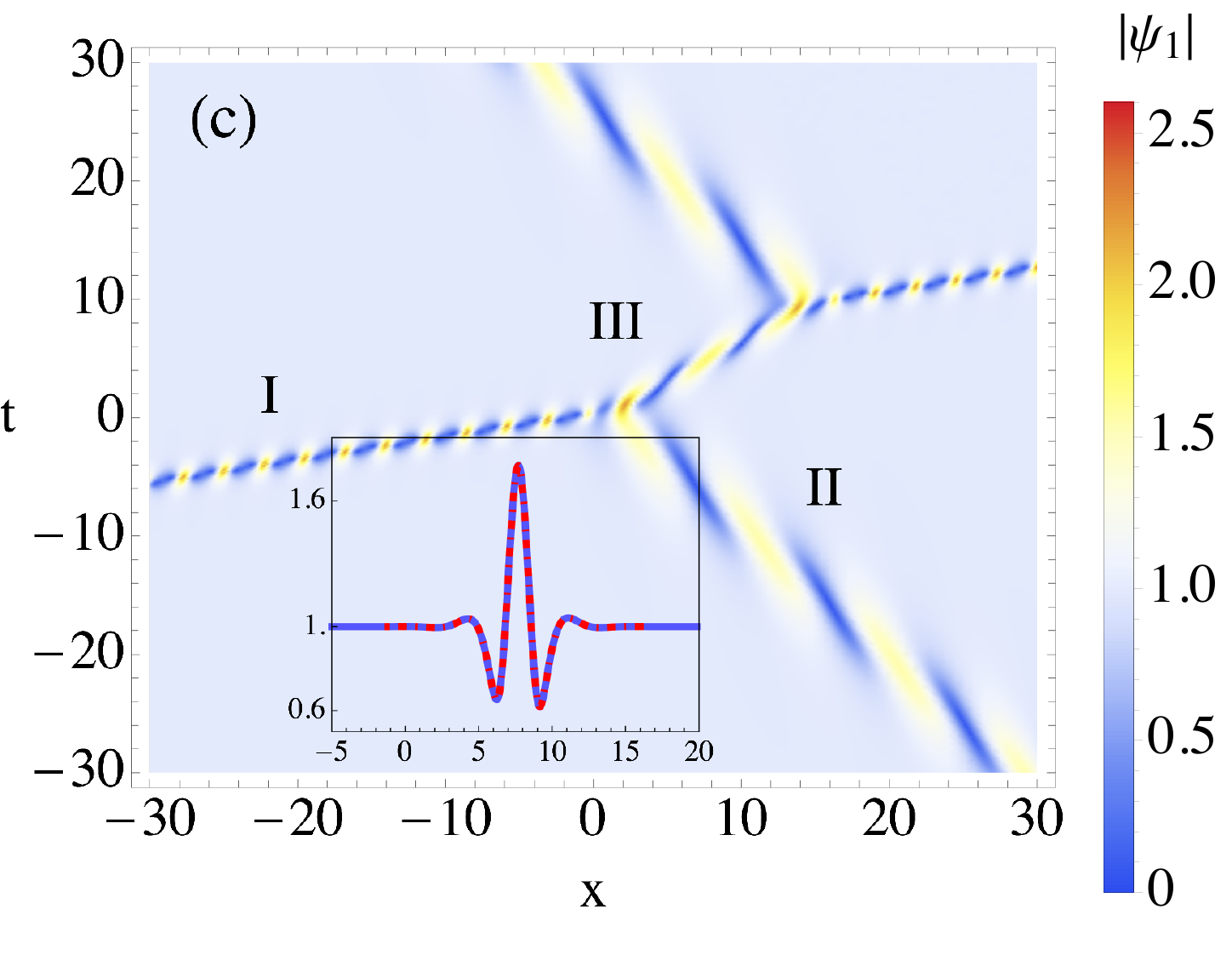}
\caption{Elastic collision $\mathrm{I}+\mathrm{II} \rightarrow \mathrm{I} + \mathrm{II}$ of vector breathers with close eigenvalues chosen according to Eq.~(\ref{epsilon_resonance}), and (a) $\varepsilon = 10^{-2}$, (b) $\varepsilon = 10^{-4}$, (c) $\varepsilon = 10^{-6}$. The spectral parameters $\xi$ and $\alpha$ taken from in Eq.~(\ref{parameters}). Characters  $\mathrm{I}$, $\mathrm{II}$, and $\mathrm{III}$ in panel (c) indicates the two colliding breathers and the collision junction. The latter transforms to the type $\mathrm{III}$ asymptotically at $\varepsilon = 0$, see also the inset in panel (c) showing comparison of the junction and type $\mathrm{III}$ breather at $t=5$.
}
\label{fig_12}
\end{figure}

Zakharov and Manakov proposed the interpretation of the resonant interaction as a result of merging eigenvalues in \cite{ZakharovManakov1976theory}. They found that when $\lambda_1 \rightarrow \lambda_2$, the three wave system soliton acquires infinite space shifts due to collision. The same situation takes place for vector breathers. Indeed, assuming $\lambda_1 \rightarrow \lambda_2$ in (\ref{total_shifts1}), see also Table \ref{tablemain}, for the processes $\mathrm{I}+\mathrm{II} \rightarrow \mathrm{I} + \mathrm{II}$ or $\mathrm{I}+\mathrm{III} \rightarrow \mathrm{I} + \mathrm{III}$ we obtain an infinite value of the space shift. To get a feeling of this limit, we plot the spatio-temporal diagrams for the breathers collisions corresponding to a set of small differences $\varepsilon$ between the breather eigenvalues, defined as,
\begin{equation}
\label{epsilon_resonance}
    \alpha_1=\alpha_2 = \alpha, \quad \xi_1 = \xi, \quad \xi_2 = \xi_1 + \varepsilon,
\end{equation}
see Fig.~\ref{fig_12}. At small $\varepsilon$, the point where the breathers collide transforms into an increasing straight junction, which begins at the point of the breathers association and later ends at the point of the breathers separation. The length of the junction is of order of the total shift $\Delta\delta^{\mathrm{I,II}}$, which dependence on $\varepsilon$ in case of eigenvalues (\ref{epsilon_resonance}) is defined by the following logarithm, see Table \ref{tablemain},
\begin{equation}
    \ln \left|\frac{(r_i-r_j^*)(1+r_i r_j)}{(r_i-r_j)(1+r_i r_j^*)} \right|^2 = \ln (1+ X),\quad X=\frac{4 \,a\, b \sin^2 \alpha\, [(a^2-1)(b^2-1)+4 \,a\, b \cos^2 \alpha]}{(a-b)^2 (1+a\, b)^2},
\end{equation}
where $a \equiv e^{-\xi_i}$, $b \equiv e^{-\xi_j}$. In case of small $\varepsilon$ one can see that $(a-b)\sim\varepsilon$ and thus $\Delta\delta^{\mathrm{I,II}}\sim \ln(1/\varepsilon)$; i.e., the junction logarithmically increases with decreasing $\varepsilon$. The logarithmic behaviour of the junction length can be also seen in Fig.~\ref{fig_12} where we plot three collision portraits (a), (b) and (c) with $\varepsilon = 10^{-2}$, $10^{-4}$, and $10^{-6}$ correspondingly. The beginning of the junction remains in the same position on the spatio-temporal diagram while its end goes to infinity. The junction itself became the breather of the type different from the types of the colliding breathers. Finally, at $\varepsilon=0$, i.e., when the eigenvalues merge precisely, the two-breather solution transforms into the resonance solution. As we noted above, a similar transformation of the two-soliton collision into the resonance pattern described in \cite{ZakharovManakov1976theory} for the three-wave system.

Now let us consider the particular case of vector breathers, which emerges when the breathers are placed close to each other, and their group velocities coincide. Such nonlinear wave complex, also called breather molecule, has been studied theoretically in the scalar case, see \cite{belanger1996bright,gelash2014superregular,li2018soliton,xu2019breather}, and recently reproduced experimentally in a nearly conservative optical fiber system \cite{xu2019breather}. Here we briefly consider its vector generalization. The group velocity condition of the two-breather is as follows,
\begin{equation}
\label{bound_state}
    V_{\mathrm{B_1}} = V_{\mathrm{B_2}}.
\end{equation}

\begin{figure}[!t]
\centering
    \includegraphics[width=0.31\linewidth]{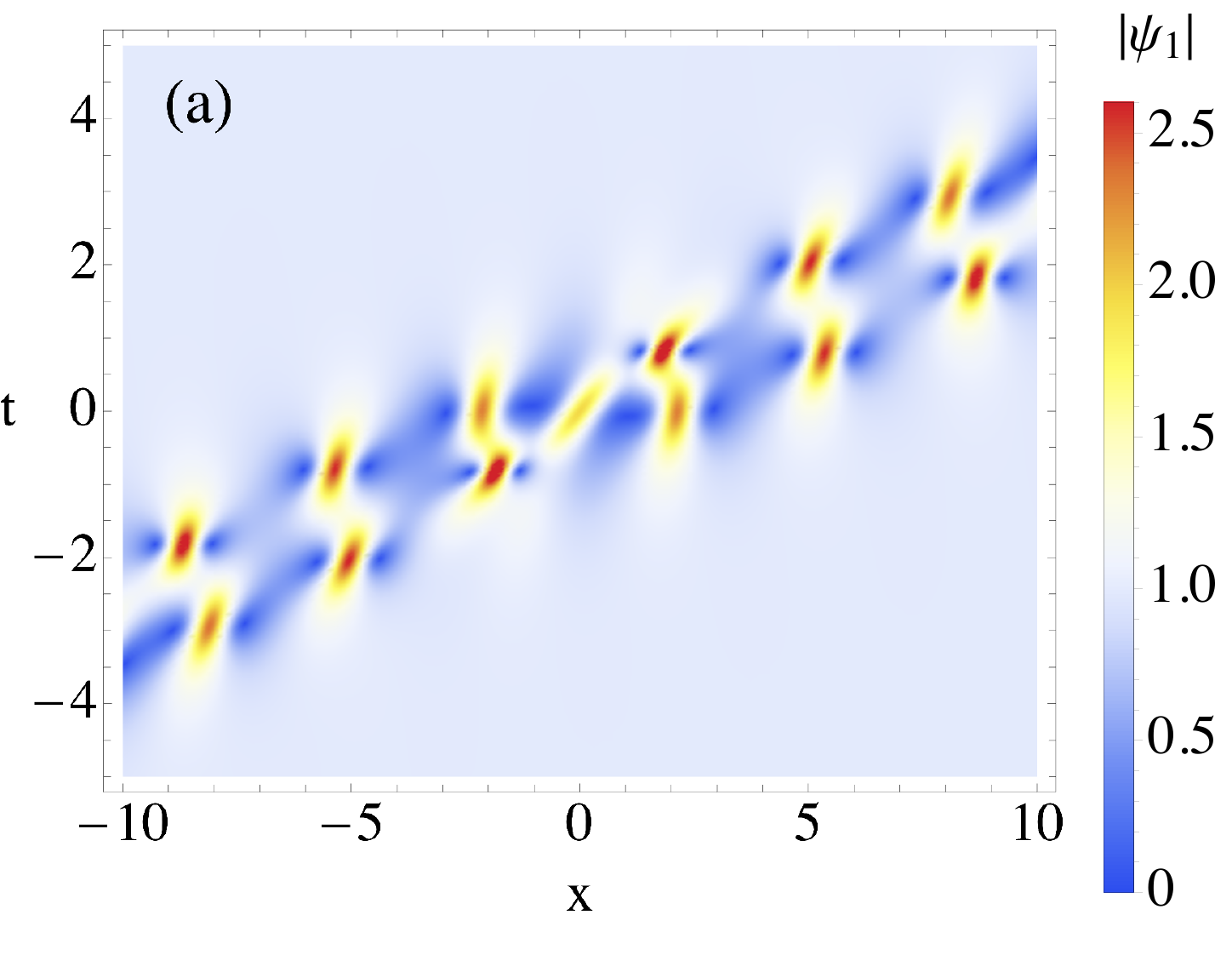}\,\,\,\,\,
    \includegraphics[width=0.31\linewidth]{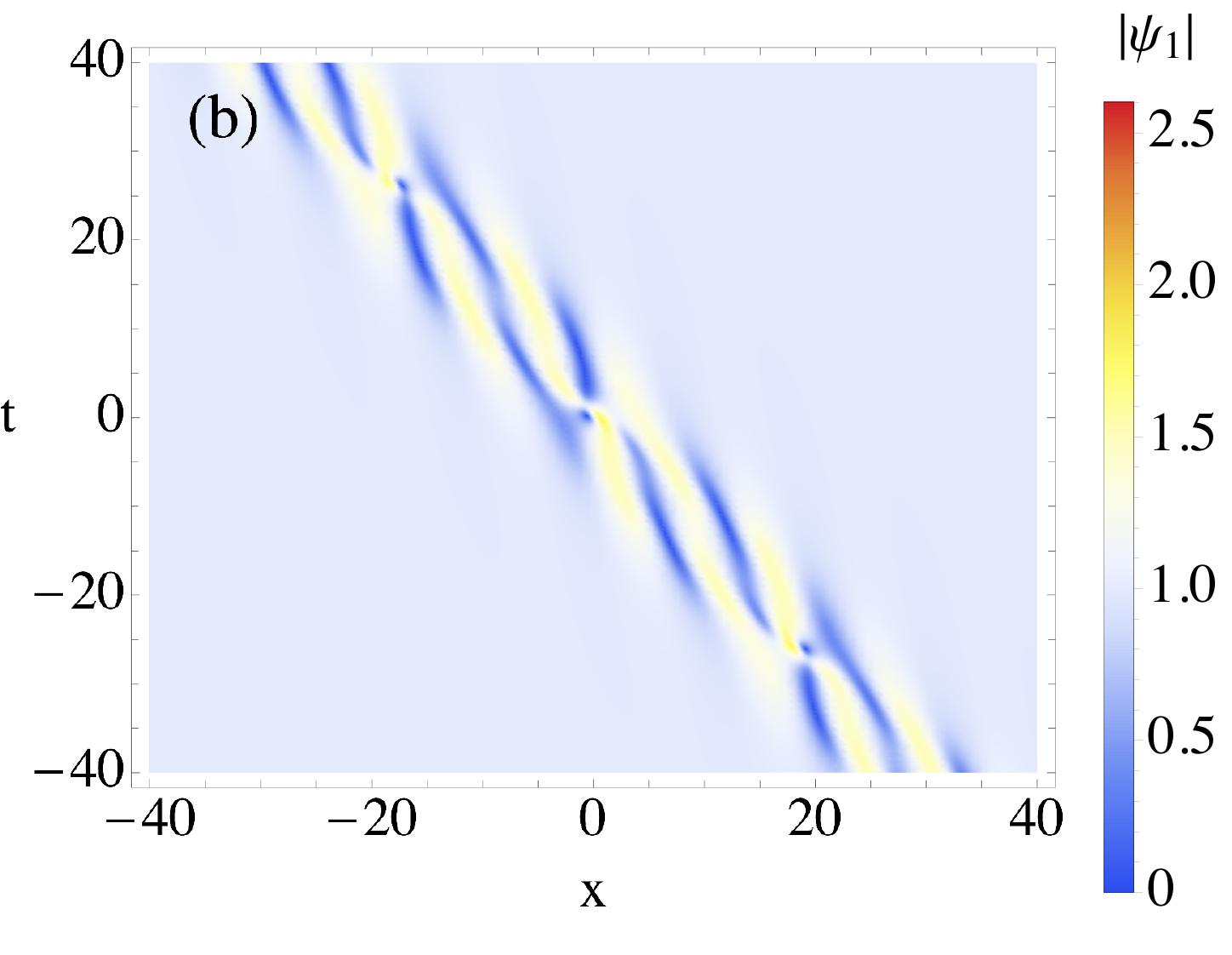}\,\,\,\,\,
    \includegraphics[width=0.32\linewidth]{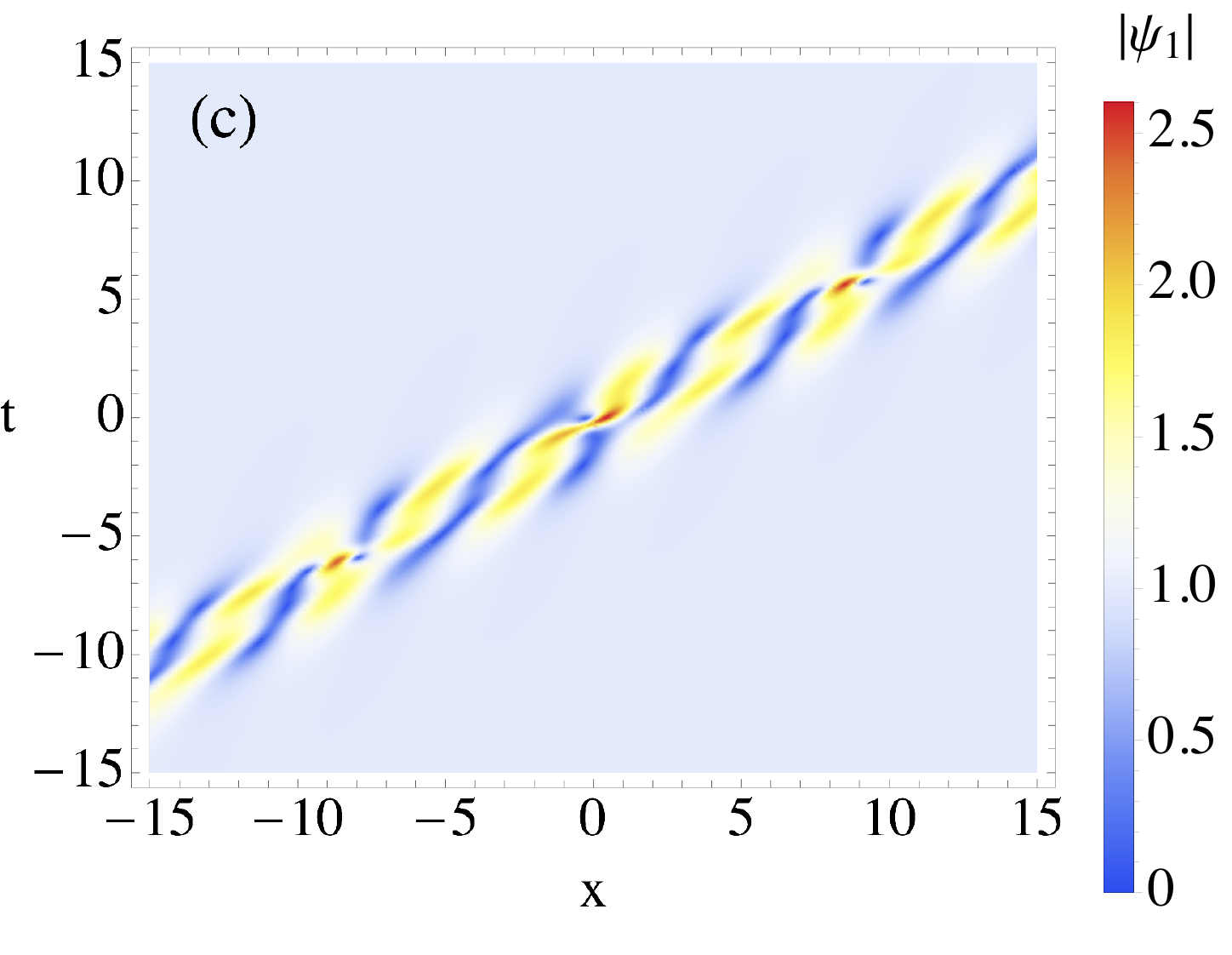}
\caption{Quasi-periodic vector breather molecules for the cases (a) $\mathrm{I}+\mathrm{I}$, $\xi_1=1/2$, $\xi_2=1/4$, $\alpha_1=\pi/5$, $\alpha_2=0.92$. (b) $\mathrm{II}+\mathrm{II}$, $\xi_1=1/2$, $\xi_2=1/4$, $\alpha_1=\pi/5$, $\alpha_2=0.89$, (c) $\mathrm{III}+\mathrm{III}$, $\xi_1=1/4$, $\xi_2=1/2$, $\alpha_1=\pi/5$, $\alpha_2=0.89$. All position and phase parameters of the breathers are zero; the values of $\alpha_2$ have been computed from the condition (\ref{bound_state}).}
\label{fig_13}
\end{figure}

For example, in the case $\mathrm{II}+\mathrm{II}$, and fixed parameters $\xi_1$, $\xi_2$, $\alpha_1$ the condition (\ref{bound_state}) results in $\alpha_2 = \arccos{\left( e^{-\xi_1+\xi_2}\cos{\alpha_2}\right)}$. Fig.~\ref{fig_13} shows typical two-breather molecules $\mathrm{I}+\mathrm{I}$, $\mathrm{II}+\mathrm{II}$ and $\mathrm{III}+\mathrm{III}$. In the general case, the breather molecule exhibits quasi-periodic oscillations because the breathers' individual oscillation frequencies are not commensurate, see in Fig.~\ref{fig_13}. As was shown in \cite{xu2019breather}, the commensurate condition for the breather oscillation frequencies in the scalar case leads to a high-order polynomial equation. In cases when the equation has solutions in the region of breather parameters validity, then the corresponding breather molecule is periodic, which was observed experimentally \cite{xu2019breather}. We leave the question of the construction of periodic vector breathers molecules to further studies.

Finally, we discuss one more important case -- the so-called superregular scenario of breather interactions \cite{zakharov2013nonlinear,gelash2014superregular,gelash2018formation}. The superregular interactions represent a near annihilation into a small-amplitude localized condensate perturbation of a pair of scalar NLSE breathers resulting from their collision. A reverse process -- the emergence of breathers -- is also possible, i.e., the formation of breathers due to the perturbation growth and evolution. The latter makes superregular breathers important exactly solvable scenarios of the nonlinear stage of modulation instability development, see also \cite{kibler2015superregular,wabnitz2017book,conforti2018auto}. Is it natural to ask whether finding nontrivial generalizations of the scalar superregular breathers in the vector case is possible? The recent study \cite{tian2021superregular} have not found nontrivial vector analogs of the superregular breathers emerging from small-amplitude condensate perturbations. Here we present our analysis of the question.

We use the mathematical interpretation of the breathers annihilation provided in \cite{gelash2014superregular}, according to which the folding of two breathers into one small localized condensate perturbation emerges due to the exact cancellation of the numerator of the two-breather solution in the case when,
\begin{equation}
\label{eugenvalue_cancellation}
    \xi_1 = \xi_2 = 0, \qquad \alpha_1 = -\alpha_2 = \alpha,
\end{equation}
so that the NLSE solution represent a pure unperturbed condensate. Then, in case when $\xi_1 = \xi_2 = \epsilon\ll 1$, the solution at the moment of collision has to be a small localized condensate perturbation because the breathers having opposite group velocities collide in the head-on manner so that no other continuous limit to the condensate solution at $\epsilon\to 0$ is possible, see details in \cite{gelash2014superregular}.

In the vector case the first numerator $\tilde{M}_{12}$ of the two-breather solution (\ref{two}) under the constrain (\ref{eugenvalue_cancellation}) simplifies as follows (for the second numerator $\tilde{M}_{13}$ the derivations are analogous),
\begin{equation}
\label{M_cancellation}
    \tilde{M}_{12} = i\tilde{m} (q_{12}q_{23} - q_{13}q_{22})(q^*_{11}q^*_{23} - q^*_{13}q^*_{21}),
\end{equation}
where $\tilde{m} = m_1 = m_2$. One can see that the numerator (\ref{M_cancellation}) is exactly cancelled at any $x$ and $t$, when $q_{12} = hq_{13}$, $q_{22} = hq_{23}$, where $h$ is an arbitrary constant. The latter happens only when $C_{0,1} = 0$, $C_{0,2} = 0$, so that $h=A_2/A_1$, i.e., when both breathers are of type $\mathrm{I}$ and the vector two-breather solution is the trivial generalization of the scalar one, see transformation (\ref{typeI_transformation}).

\begin{figure}[!t]
\centering
    \includegraphics[width=0.31\linewidth]{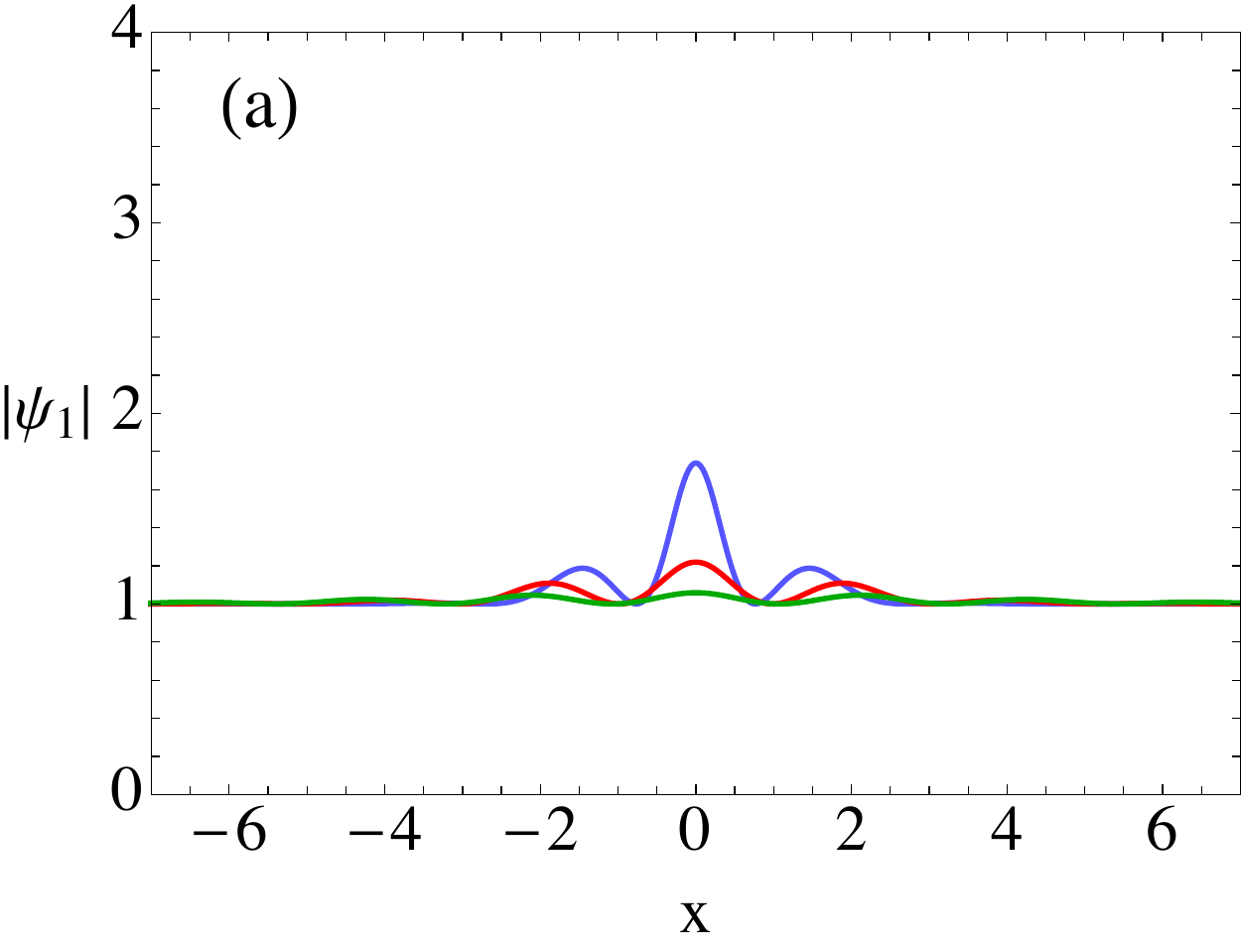}\,\,\,\,\,
    \includegraphics[width=0.31\linewidth]{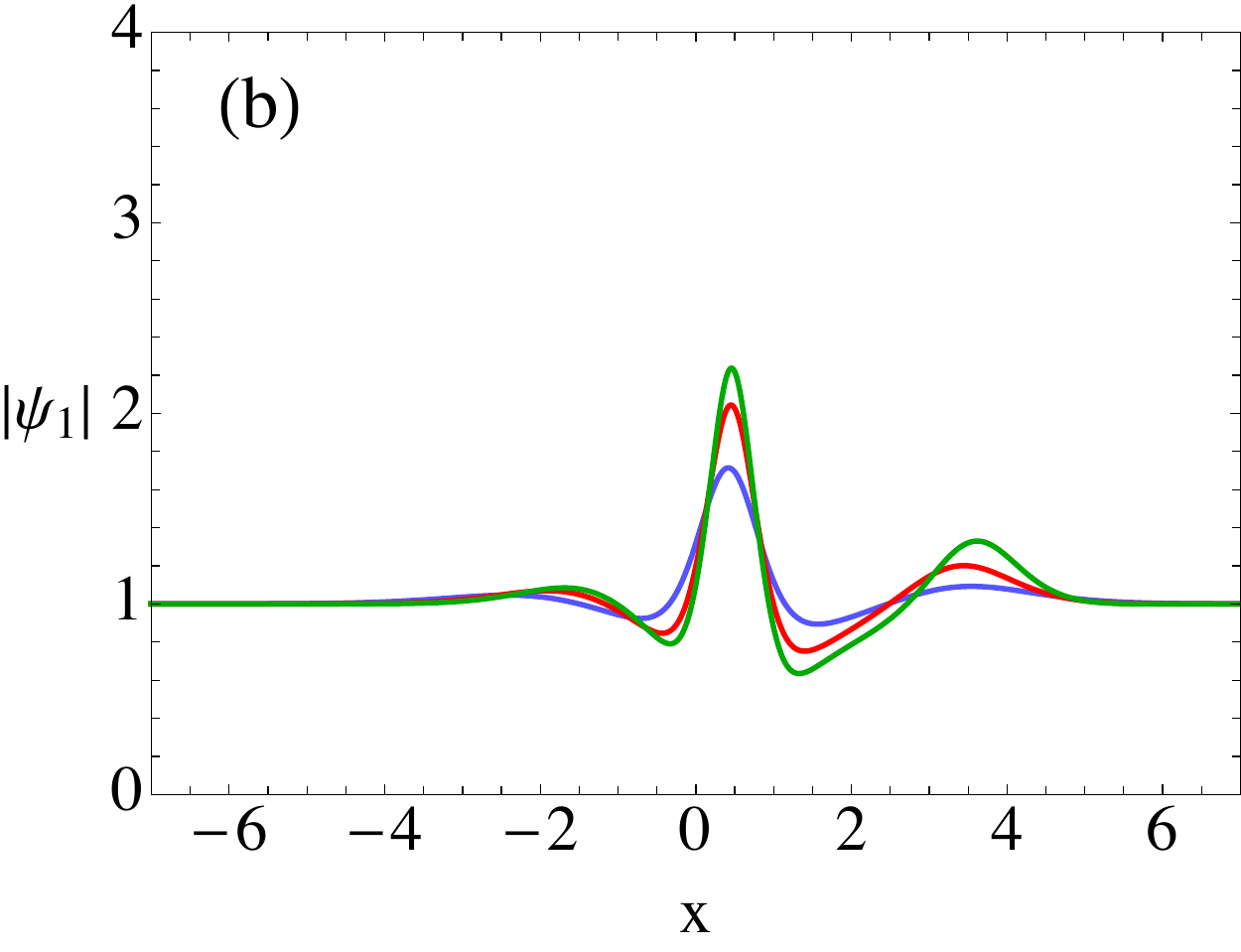}\,\,\,\,\,
    \includegraphics[width=0.31\linewidth]{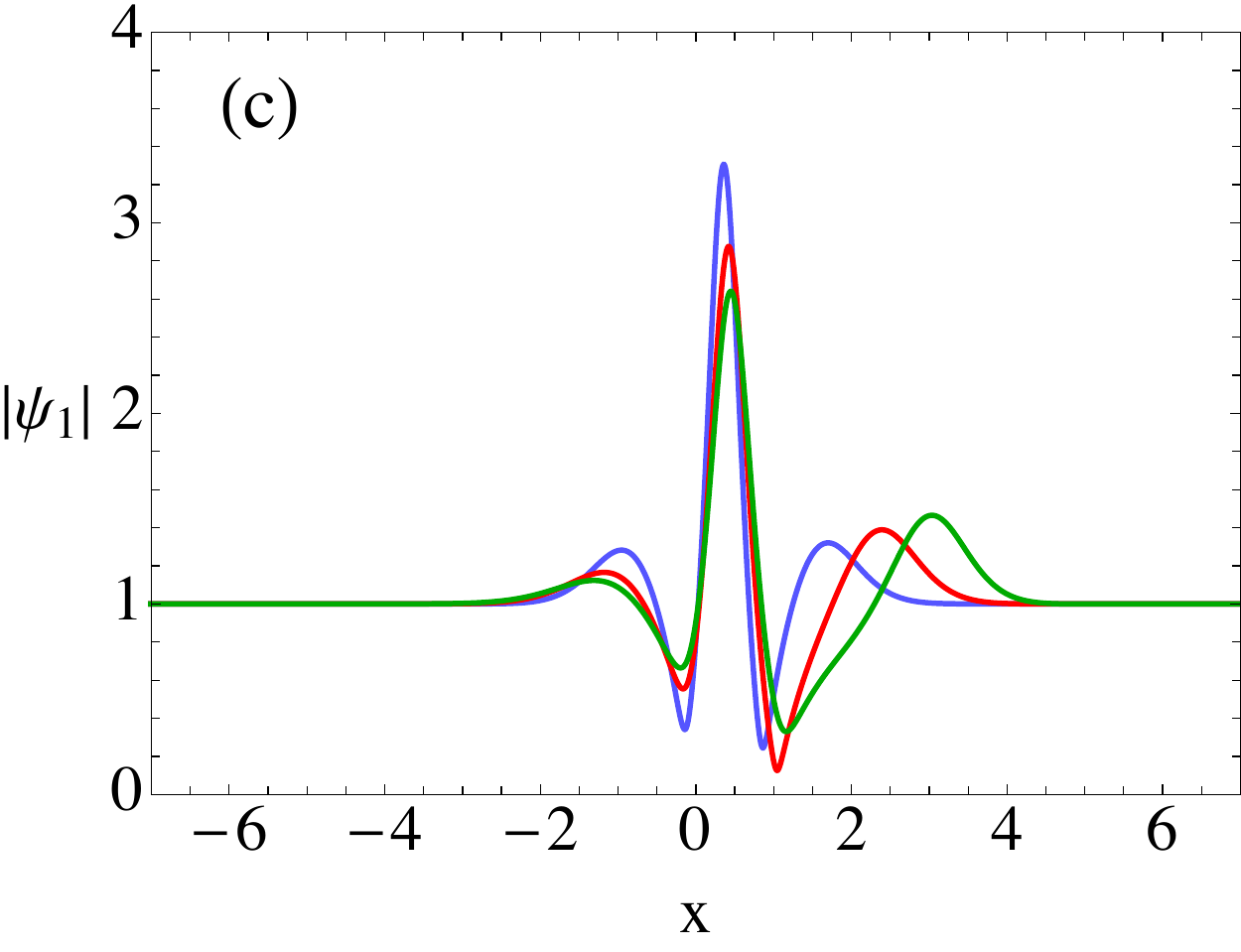}
\caption{Collision profiles of vector two-breather solutions of types (a) $\mathrm{I}+\mathrm{I}$, (b) $\mathrm{II}+\mathrm{II}$, (c) $\mathrm{III}+\mathrm{III}$ at $t=0$ and parameters $\xi_1 = \xi_2 = \epsilon$, $\alpha_1 = -\alpha_2 = \pi/3$, $\theta_{1,2}=\pi/2$,  $\delta_{1,2}=0$. The value of parameter $\epsilon$ is $0.4$ (blue curves), $0.2$ (red curves) and $0.1$ (green curves). Panel (a) shows a trivial vector analog of the superregular folding of a pair of scalar NLSE breathers into small-amplitude condensate perturbations. Panels (b) and (c) illustrate that there is no such folding for the breathers of type $\mathrm{II}$ and $\mathrm{III}$. Only the first wavefield component is shown; the behavior of $\psi_2$ is analogous.}
\label{fig_15}
\end{figure}

In the case when one or both breathers are of type $\mathrm{II}$ or $\mathrm{III}$, the numerator (\ref{M_cancellation}), together with denominator $M$, see (\ref{two}), can be canceled only at specific points in $x$, which location depend on the integration vectors. In this case, the vector two-breather solution transforms into a degenerate one (one needs to resolve the indeterminate form $0/0$); see examples of degenerate scalar breathers in \cite{kedziora2012second,gelash2014superregular}. Here we do not study the degenerate limit and instead focus on the behavior of vector two-breather solution at $\xi_1 = \xi_2 = \epsilon\ll 1$. Fig.~\ref{fig_15} shows two-breather solutions of different types at $t=0$, $\alpha=\pi/3$ and different values of $\epsilon$. In addition, we choose breather phases and positions as $\theta_{1,2}=\pi/2$ and $\delta_{1,2}=0$, which corresponds to the most efficient folding of the breathers in the scalar case, see details in \cite{gelash2014superregular}. One can see the trivial vector analog ($\mathrm{I}+\mathrm{I}$) of the superregular folding in Fig.~\ref{fig_15}(a), which shows that the amplitude of the condensate perturbation (produced by the breather collision) decreases when decreasing $\epsilon$. At the same time, in the cases of collisions $\mathrm{II}+\mathrm{II}$ and $\mathrm{III}+\mathrm{III}$ shown in Fig.~\ref{fig_15}(b,c), the amplitude of the wavefield remains large even at very small $\epsilon$. The latter means that instead of the superregular folding, the vector breathers at $\epsilon\to 0$ tend to a degenerate limit, as we discussed above. We conclude that the vector breathers of type $\mathrm{II}$ and $\mathrm{III}$ do not participate in the modulation instability development from small-amplitude perturbations, which is consistent with \cite{tian2021superregular} and also with the correspondence of these types of vector breathers to the stable branch of the dispersion law $\omega_{\mathrm{II}}$, see Sec. \ref{Sec:3}.

\section{Conclusions and discussions}\label{Sec:7}
In this work we have studied theoretically the vector breathers and their interactions in the framework of the two-component nonlinear Schrödinger equation -- the Manakov system. Our model imply the focusing-type nonlinearity in both system components, see Eq.~(\ref{VNLSE}), and the presence of a nonzero constant background; see also \cite{prinari2006inverse} for the defocusing case and gray vector solitons. As a starting point we take the vector variant of the dressing method \cite{raskovalov2022resonant}, the studies on the three fundamental breather types I, II, III \cite{kraus2015focusing}, and on the resonance vector breather interactions \cite{raskovalov2022resonant}. Then we reveal the connection between breather type and the branches of the dispersion law, analyse important particular cases of the breather solutions and, describe the asymptotic state of the breather interactions by computing spatial and phase shifts acquiring by breathers as a result of collisions. The three types of breathers generate a family of nine different shift expressions, which we summarize using Eq.~(\ref{2B_asymptotic_coeff_general}) and the corresponding to it Table \ref{table1}. We find that the spatial shifts of the vector breathers can change sign depending on the spectral parameters without changing the sign of the difference between the breather velocities. Finally the obtained shift expressions allowed us to interpret the resonance fusion and decay of breathers as a limiting case of infinite space shift in the case of merging breather eigenvalues. In the future, the shift expressions can be used to build a spectral theory of vector breather gases, similar to the recent scalar case studies, see \cite{el2020spectral}.

The breathers of types $\mathrm{II}$ and $\mathrm{III}$ exhibit fundamentally different wavefield dynamics when compared to the breathers of type $\mathrm{I}$. The latter, as a trivial vector generalization of the scalar NLSE breathers, see transformation (\ref{typeI_transformation}), describes particular scenarios of the modulation instability and formation of rogue waves. In contrast, the breathers of types $\mathrm{II}$ and $\mathrm{III}$, belonging to the stable branch of the dispersion law, do not participate in the development of modulation instability from small-amplitude perturbations. Indeed, for the Akhmediev-type eigenvalues, see Eq.~(\ref{Akhmediev_par_set}), as well as for superregular-type eigenvalues, see Sec. \ref{Sec:6}, the breathers of types $\mathrm{II}$ and $\mathrm{III}$ exhibit a localized condensate pulsations which are never small. On the other side, the breathers of types $\mathrm{II}$ and $\mathrm{III}$ represent an important class of localized pulsating exact solutions and, together with the type $\mathrm{I}$ breathers, are responsible for inelastic resonance interactions, see Sec. \ref{Sec:4}.

A fundamental question that needs further study is the eigenvalues portraits of localized small-amplitude arbitrary-shaped perturbations of the vector condensate. As was shown in \cite{conforti2018auto} for the scalar case, superregular eigenvalue pairs can be embedded into small amplitude and arbitrary-shaped condensate perturbations under certain conditions. Meanwhile, our present studies show that the $\mathrm{II}$ and $\mathrm{III}$ breathers cannot be folded in such a way, at least within the standard superregular scenario. All this leads us to a conjecture that only type $\mathrm{I}$ superregular breathers exist in the locally perturbed vector condensate. Our conjecture can be tested in the future using numerical computation of the eigenvalue spectrum of the auxiliary system (\ref{lax system 1}). Moreover, we think that the vector arbitrary-shaped perturbation evolution is driven by the interaction between type I breathers and the unstable continuous spectrum; see the works \cite{biondini2016universal,conforti2018auto,biondini2021long} explaining how it happens in the scalar case.

We believe that our study will also benefit to the rapidly developing area of statistical description of nonlinear waves in integrable systems -- the so-called integrable turbulence, see \cite{zakharov2009turbulence,pelinovsky2013two,randoux2014intermittency,agafontsev2015integrable,soto2016integrable,Gelash2018,gelash2019bound}. The first studies on the random polarized nonlinear waves have been recently obtained in \cite{manvcic2018statistics} and we think that our analysis of the vector breather interactions will provide new insights into this complex subject. Meanwhile, experimental observation of various aspects of the integrable scalar NLSE dynamics and statistics has been successfully performed in many different works, see e.g.  \cite{kibler2010peregrine,bailung2011observation,chabchoub2011rogue,kibler2012observation,frisquet2013collision,chabchoub2014hydrodynamics,chabchoub2019drifting,pierangeli2018observation,xu2019breather,Kraych2019Statistical,xu2020ghost}. In addition the development of vector modulation instability and vector dark rogue waves have been studied experimentally in a Manakov fiber system \cite{frisquet2015polarization,baronio2018observation}. At the same time the experimental observation of vector breathers represents a challenging task for further studies, see also \cite{baronio2018observation} where experimental conditions for experimental observation of vector breathers have been discussed.

\begin{acknowledgments}
The main part of the work was supported by the Russian Science Foundation (Grant No. 19-72-30028). The work of A.G. on Section \ref{Sec:6} and Appendix Section \ref{Sec:Appendix:2} was supported by RFBR Grant No. 19-31-60028. The work of A.R. on Appendix Sections \ref{Sec:Appendix:1} and \ref{Sec:Appendix:B1} was performed in the framework of the state assignment of the Russian Ministry of Science and Education ``Quantum'' No. AAAA-A18-118020190095-4. The authors thank participants of Prof. V.E. Zakharov’s seminar ``Nonlinear Waves'' and, especially, Prof. E.A. Kuznetsov for fruitful discussions.
\end{acknowledgments}

\section{Appendix}\label{Sec:Appendix}
In the Appendix section we provide details of theoretical derivations, additional illustrations and full expressions for the shift coefficients.
\subsection{Linearization of the Manakov system for small-amplitude perturbations}\label{Sec:Appendix:1}
In this section we derive the two branches of the dispersion law for the Manakov system, see Eq.~(\ref{dispersion_laws}) in the main part of the work. We consider small-amplitude perturbations of the condensate solution $\delta\psi_1$ and $\delta\psi_2$, so that the wavefield components are,
\begin{eqnarray}
\label{dpsi}
\psi_1 &=& A_1 + \delta\psi_1,
\\\nonumber
\psi_2 &=& A_2 + \delta\psi_2.
\end{eqnarray}

Then we substitute (\ref{dpsi}) into the Manakov system (\ref{VNLSE}) and linearize it by leaving only the first order terms proportional to $\delta\psi_1$ or $\delta\psi_2$. It is convenient to consider the following new variables,
\begin{eqnarray}
\label{f_g_variables}
\delta\psi_{j} + \delta\psi^*_{j} = f_{j},
\\\nonumber
\delta\psi_{j} - \delta\psi^*_{j} = g_{j},
\end{eqnarray}
where $j=1$ or $2$. Using (\ref{f_g_variables}) the linearized Manakov system can be obtained in the following form:
\begin{eqnarray}
\label{linearized_VNLSE}
i g_{jt}+\frac{1}{2}f_{jxx} + 2A_j (A_1 f_1 + A_2 f_2) = 0,
\\\nonumber
i f_{jt}+\frac{1}{2}g_{jxx} = 0.
\end{eqnarray}
We assume that the initial perturbations are simple linear harmonics, i.e. $\delta\psi_j \sim e^{i(kx -\omega t)}$, and, accordingly, choose,
\begin{eqnarray}
\label{f_and_g}
f_{j} =f^0_{j}\mathrm{Re}[e^{i(kx -\omega t)}] = f^0_{j}\cos(kx -\omega t),
\\\nonumber
g_{j} =ig^0_{j}\mathrm{Im}[e^{i(kx -\omega t)}] = i g^0_{j}\sin(kx -\omega t),
\end{eqnarray}
where $f^0_{j}$ and $g^0_{j}$ are arbitrary real-valued constants. Substituting expressions (\ref{f_and_g}) into Eq.~(\ref{linearized_VNLSE}) we obtain the two branches of the dispersion law $\omega_{\mathrm{I}}(k) = \pm k\sqrt{k^2/4 - A^2}$ and $\omega_{\mathrm{II}}(k) = \pm k^2/2$, see Eq.~(\ref{dispersion_laws}), as the consistency condition of  the linearized system (\ref{linearized_VNLSE}).

\subsection{Linearization of the breathers' tails}\label{Sec:Appendix:2}

Now we show how to derive the connection between the breather type and the branch of the dispersion law, which was discussed in Sec. \ref{Sec:3}. We consider the small-amplitude breathers' tails as condensate perturbations and study them in a linear approximation. Let us for definiteness assume that we are in the region of space far away from the breather center where $\mathrm{Im}[\zeta]x>0$. Then for the breather of type I we have the following small parameter,
\begin{equation}
    |e^{-\varphi}| \sim e^{-\mathrm{Im}[\zeta]x} = \varepsilon \ll 1,
\end{equation}
so that $|e^{\varphi}| \sim 1/\varepsilon \gg 1$. Using the smallest of $\varepsilon$ we present the first component (the computations are similar for the first and second components, so we present only the first ones) of the type I single-breather solution as,
\begin{equation}
\label{psiI_tail}
    \psi_{1} = A_{1} +\frac{2\,\mathrm{i}\,(\lambda_1 - \lambda_1^*)\, q_{1}^* q_{2}}{|\mathbf{q}|^2} \approx 
    A_{1} +  \frac{F+\delta f}{G+\delta g} \approx A_1 + \frac{F}{G} + \frac{\delta f}{G} - \frac{F \delta g}{G^2},
\end{equation}
where the leading order terms $F,\,G \sim 2/\varepsilon$, and the terms $\delta f,\,\delta g \sim 1$ can be written as follows,
\begin{eqnarray}
    && F = p_1 e^{\varphi + \varphi^*}, \qquad \delta f = p_2 e^{\varphi^* - \varphi} + p_3 e^{\varphi - \varphi^*}, \\\nonumber
    && G = s_1 e^{\varphi + \varphi^*}, \qquad \delta g = s_2 e^{\varphi^* - \varphi} + s_3 e^{\varphi - \varphi^*},
\end{eqnarray}
and $p_i$ and $s_i$ are coefficients of the order of unity, which particular composition is not important for this consideration. Note that to obtain the final result on the right hand side of (\ref{psiI_tail}) we neglected the terms $\delta f \delta g \sim \varepsilon^2$.

The leading term in (\ref{psiI_tail}) $A_1+F/G$ gives the asymptotic (\ref{asymptotics_I}), i.e. the background condensate with a changed phase. Meanwhile to obtain the connection with the dispersion relation we need the second order terms, i.e.,
\begin{equation}
\label{FG_second_order_terms}
    \frac{\delta f}{G} = \frac{p_2}{s_1}e^{-2\varphi} + \frac{p_3}{s_1}e^{-2\varphi^*}, \qquad
    -\frac{F}{G^2}\delta g = -\frac{p_1}{s_1}\left( \frac{s_2}{s_1}e^{-2\varphi} + \frac{s_3}{s_1}e^{-2\varphi^*}  \right).
\end{equation}

From formulas (\ref{FG_second_order_terms}) we find that the lenearization of the breather tail with respect to the background condensate gives a linear combinations of the first-order terms proportional to either $e^{2\varphi}$ or $e^{2\varphi^*}$. Now to retrieve the dispersion law we rewrite these exponents in the form $e^{ikx + i\omega t}$, see also Sec. \ref{Sec:3}. Considering, for example the first exponent $e^{2\varphi}$ , we obtain that $k = -2\zeta$ and $\omega=2\lambda\zeta$. Now using $\zeta(\lambda) = \sqrt{\lambda^2+A^2}$, see Eq.~(\ref{zeta_def}), we find $\omega(k) = \pm ik\sqrt{A^2-k^2/4}$. Thereby the breather tails in this case obey the first branch of the dispersion law $\omega_{\mathrm{I}}(k)$ with complex $k$ and $\omega$. The complexity of $k$ in (\ref{tails1}) means exponential decay of the breather tail. The same result can be obtained for the terms $\tilde{p} e^{2\varphi^*}$. One can repeat the presented procedure for type II and III breathers and in similar way find that they correspond to the second branch of the dispersion law $\omega_{\mathrm{II}}(k)$.

\subsection{Computation of the phase and position shifts}\label{Sec:Appendix:3}

Details of the computations of the phase and position shifts for the process $\mathrm{I}+\mathrm{I} \rightarrow \mathrm{I} + \mathrm{I}$ are similar to those for the scalar NLSE breathers presented in \cite{gelash2022breather}. The main idea boils down to asymptotic expansion of the two-breather solution (\ref{two}) at the moments of time far before/after the collision and regions of space where each of the breathers is located. Here we briefly consider the computational details for the full spatial shift in the process $\mathrm{II}+\mathrm{II} \rightarrow \mathrm{II} + \mathrm{II}$, while the other answers listed in Tables \ref{tablemain} and \ref{table1} can be obtained in a similar way. We choose the vectors $\mathbf{C}_1$ and $\mathbf{C}_2$ according to Eq.~(\ref{C_II_param}),
\begin{eqnarray}
\label{C_II_param}
    C_{n,0} = e^{-\mathrm{Im}[\lambda_n]\delta_n - i\theta_n /2}, \quad C_{n,1} = 0, \quad C_{n,2} = e^{-\mathrm{Im}[\zeta_n]\delta_n + i\theta_n/2}, \quad n=1,2.
\end{eqnarray}

Then, in addition we renormalize the vectors $\mathbf{q}_1$ and $\mathbf{q}_2$ using the property (\ref{Cn_transform}), so that they can be written in the following form,
\begin{eqnarray}
\label{qIInew}
q_{11} = \mathrm{i}\,r_1, \quad q_{12} = \frac{1}{A}\,(A_1 - A_2 \exp[u_{\mathrm{II},1}-\mathrm{i}\, v_{\mathrm{II,1}}]), \quad q_{13} = \frac{1}{A}\,(A_2 + A_1 \exp[u_{\mathrm{II},1}-\mathrm{i}\, v_{\mathrm{II,1}}]), \\
q_{21} = \mathrm{i}\,r_2, \quad q_{22} = \frac{1}{A}\,(A_1 - A_2 \exp[u_{\mathrm{II,2}}-\mathrm{i}\, v_{\mathrm{II},2}]), \quad q_{23} = \frac{1}{A}\,(A_2 + A_1 \exp[u_{\mathrm{II,2}}-\mathrm{i}\, v_{\mathrm{II},2}]),
\end{eqnarray}
where, in accordance with (\ref{uv_II}), $u_{\mathrm{II,n}}=(x- V_{\mathrm{II,n}}\,t-\delta_n)/(2\,l_{\mathrm{II,n}})$, $v_{\mathrm{II,n}} = p_{\mathrm{II},n} x- \omega_{\mathrm{II},n}\,t+\theta_n$, $n=1,2$. One can derive from (\ref{qIInew}) the following useful equality,
\begin{eqnarray}
|\mathbf{q}_{1,2}|^2=1+|r_{1,2}|^2+\exp({2\,u^{(1,2)}_{\mathrm{II}}}),
\\\nonumber
(\mathbf{q}_{1}, \mathbf{q}_{2}^*) = r_1 r_2^* + 1 + \exp[u^{(1)}_\mathrm{II} +u^{(2)}_\mathrm{II} -\mathrm{i}\,( v_{\mathrm{II}}^{(1)}-v_{\mathrm{II}}^{(2)})].
\end{eqnarray}

Let us for definiteness consider the breather with index $1$ and assume that $0 \leq \alpha_{1,2} \leq \pi$. We move to the reference frame associated with the first breather where $u_{\mathrm{II},1}=\mathrm{const}$. According to (\ref{2B_asymptotic_coeff_I+I}) we define for the first breather $s_1 = \mathrm{sign} (V_{\mathrm{II,2}}-V_{\mathrm{II,1}})$. Then in the limit $t\rightarrow\pm\infty$ we obtain
\begin{eqnarray}
u_{\mathrm{II},2} = (2\,l_{\mathrm{II},2})^{-1} \left[2\,l_{\mathrm{II},2} u_{\mathrm{II},1} +(V_{\mathrm{II},2}-V_{\mathrm{II},1})\,t +\delta_1-\delta_2 \right] \to \pm s_1\infty.
\end{eqnarray}
In the case $u_{\mathrm{II},2}\to\infty$ the denominator $M$ of the two-breather solution (\ref{two}) can be simplified as,
\begin{eqnarray}
\label{Mv1}
M = -[m\,|\mathbf{q}_1|^2 |\mathbf{q}_2|^2- n\, (\mathbf{q}_1, \mathbf{q}_2^*) (\mathbf{q}_2, \mathbf{q}_1^*)] =  - e^{2\, u_{\mathrm{II},2}}\, m \left[1+|r_1|^2+ e^{2\, u_{\mathrm{II},1}} \left(1- \frac{n}{m}\right) \right],
\end{eqnarray}
meanwhile in the case $u_{\mathrm{II},2}\to\infty$ as,
\begin{eqnarray}
\label{Mv2}
M = -e^{-2\, u_{\mathrm{II},2}} (1+|r_1|^2)^{-1} [m (1+|r_1|^2) (1+|r_2|^2) - n |1+r_1^* r_2|^2] \times\\\nonumber
\times \left[1+|r_1|^2+ e^{2\, u_{\mathrm{II},1}} \left(\frac{m (1+|r_1|^2) (1+|r_2|^2)}{m (1+|r_1|^2) (1+|r_2|^2) - n |1+r_1^* r_2|^2}\right) \right] .
\end{eqnarray}
Finally, comparing (\ref{Mv1}) and (\ref{Mv2}) with the denominator of the type $\mathrm{II}$ single breather solution (\ref{sol2}) we find that the shift of the center of the first breather after collision with the second breather is equal to
\[
\Delta \delta_{1}^{\mathrm{II},\mathrm{II}} = \frac{l_\mathrm{II}}{2} \left(\log \left[1 - \frac{n}{m} \right] +\log \left[1 - \frac{n |1+r_1 r_2^*|^2}{m (1+|r_1|^2)(1+|r_2|^2)} \right]\right) s_1,
\]
which is one of the answers presented in Table \ref{tablemain}.

\subsection{Additional illustrations}\label{Sec:Appendix:B1}

\begin{figure}[!h]
\centering
    \includegraphics[width=0.31\linewidth]{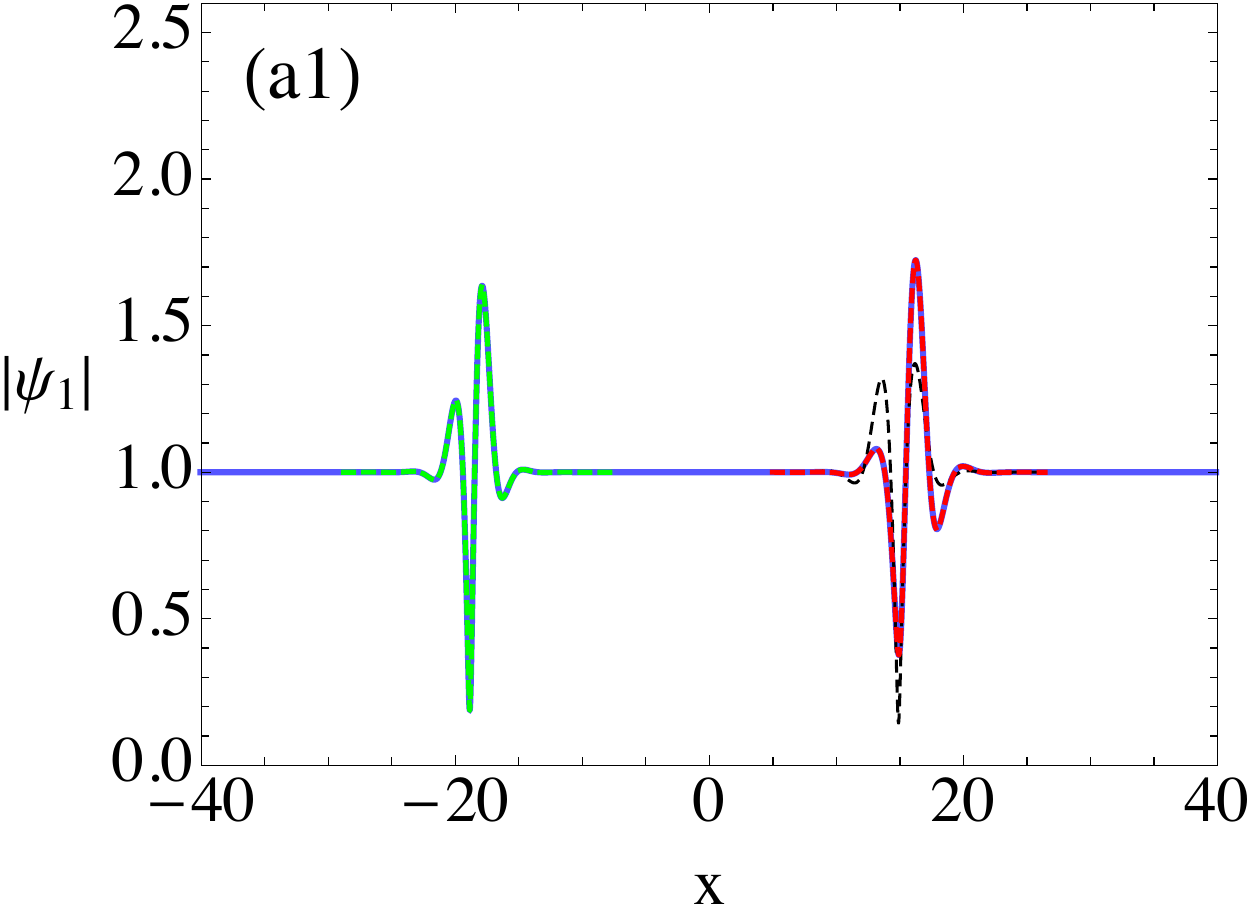}\,\,\,\,\,
    \includegraphics[width=0.31\linewidth]{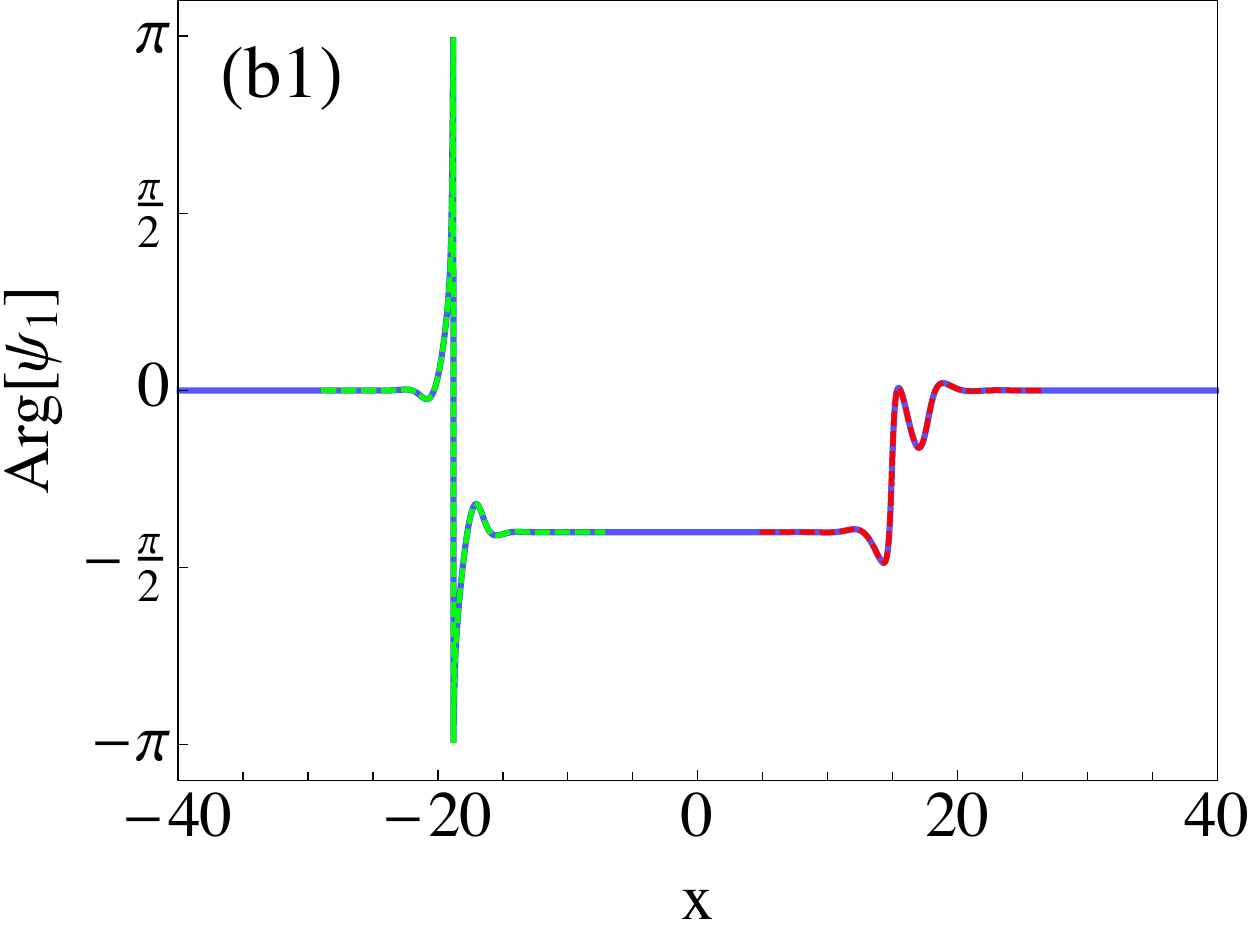}\,\,\,\,\,
    \includegraphics[width=0.32\linewidth]{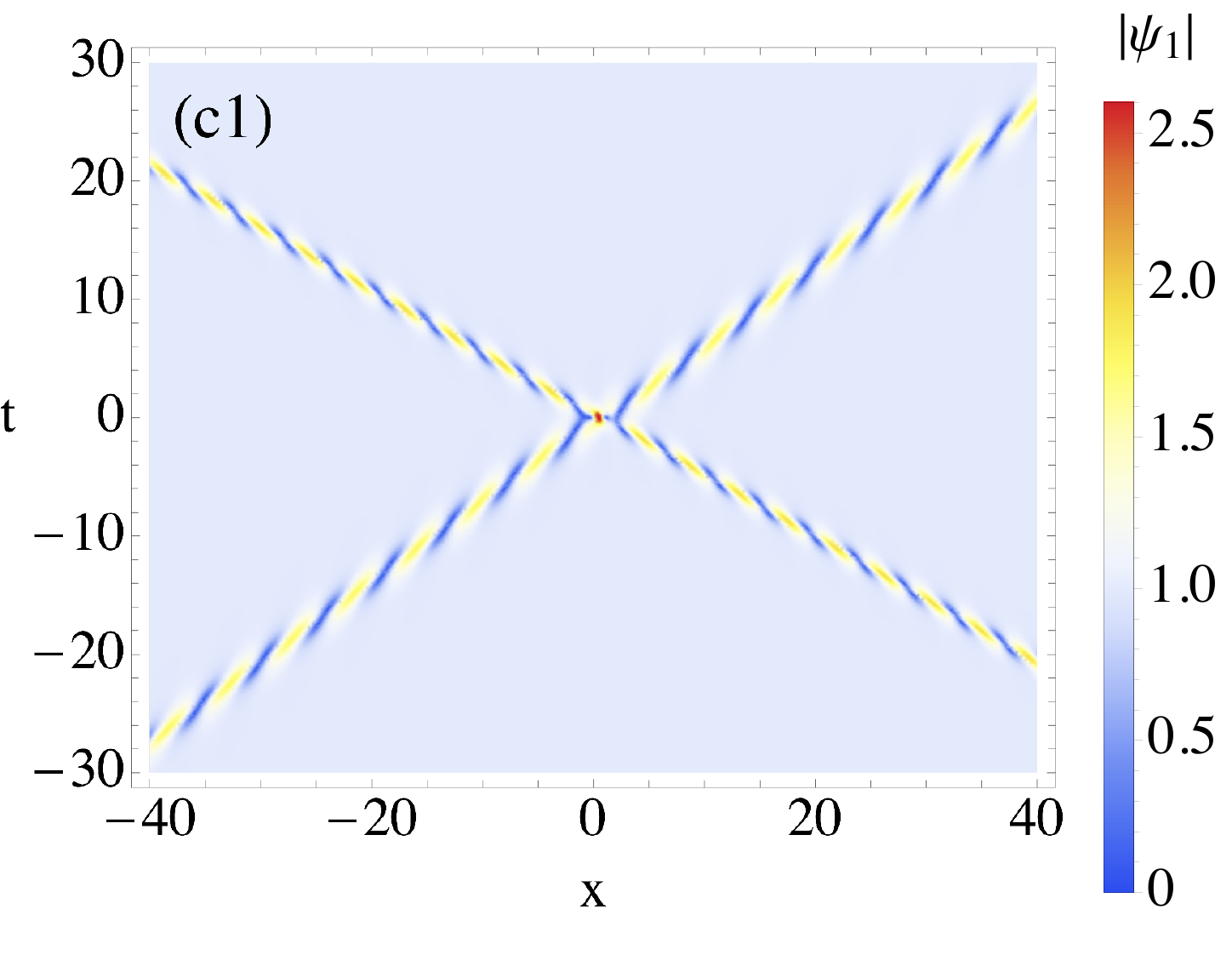}\\
    \includegraphics[width=0.31\linewidth]{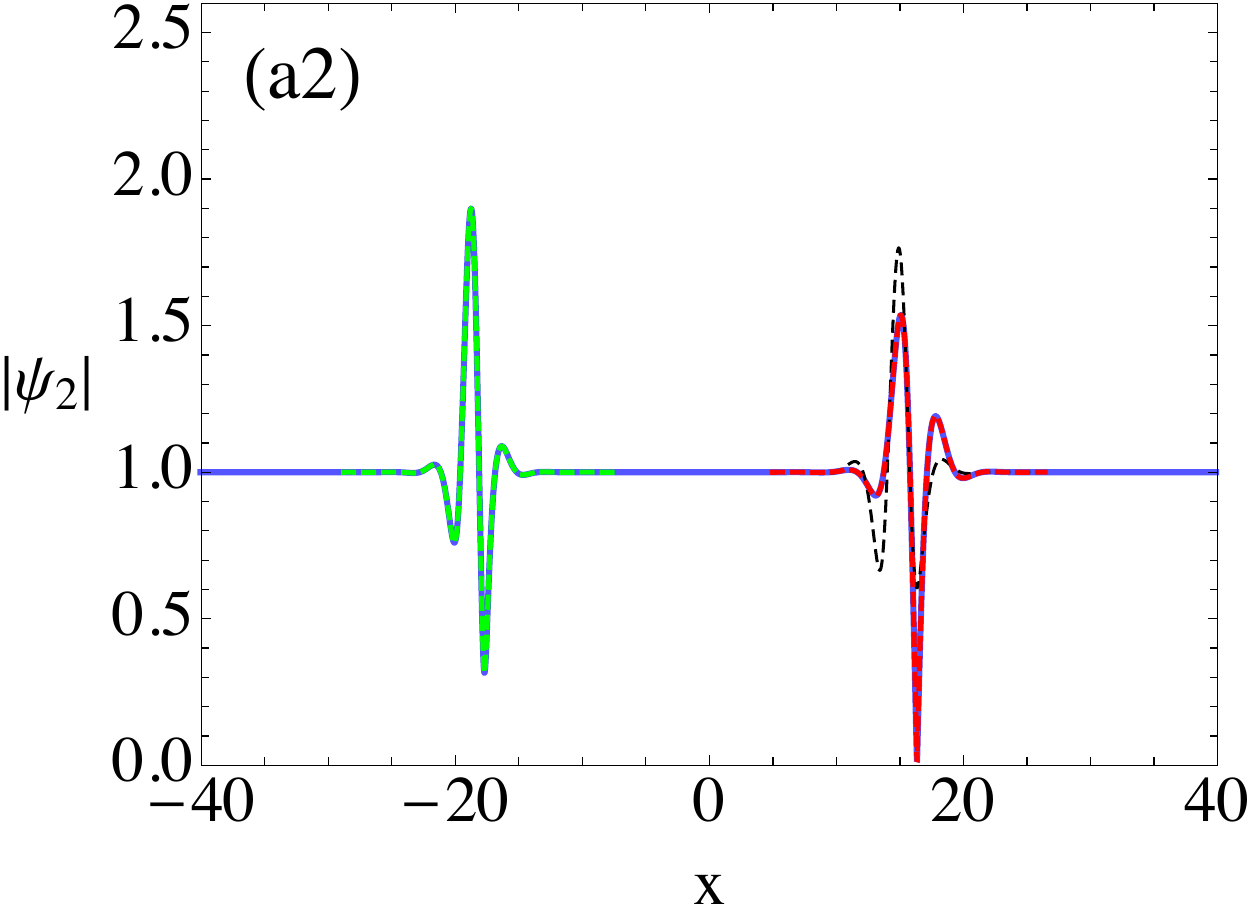}\,\,\,\,\,
    \includegraphics[width=0.31\linewidth]{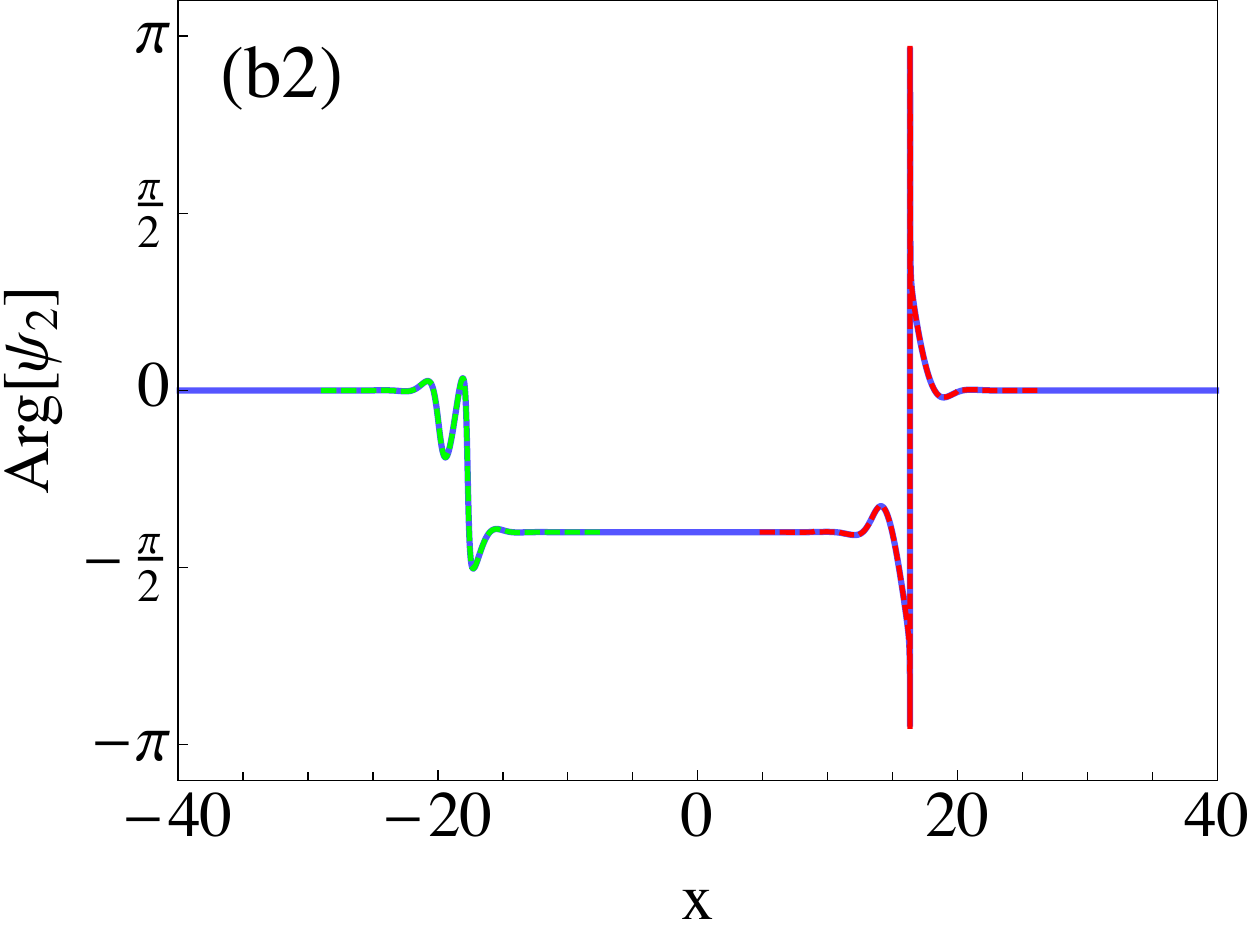}\,\,\,\,\,
    \includegraphics[width=0.32\linewidth]{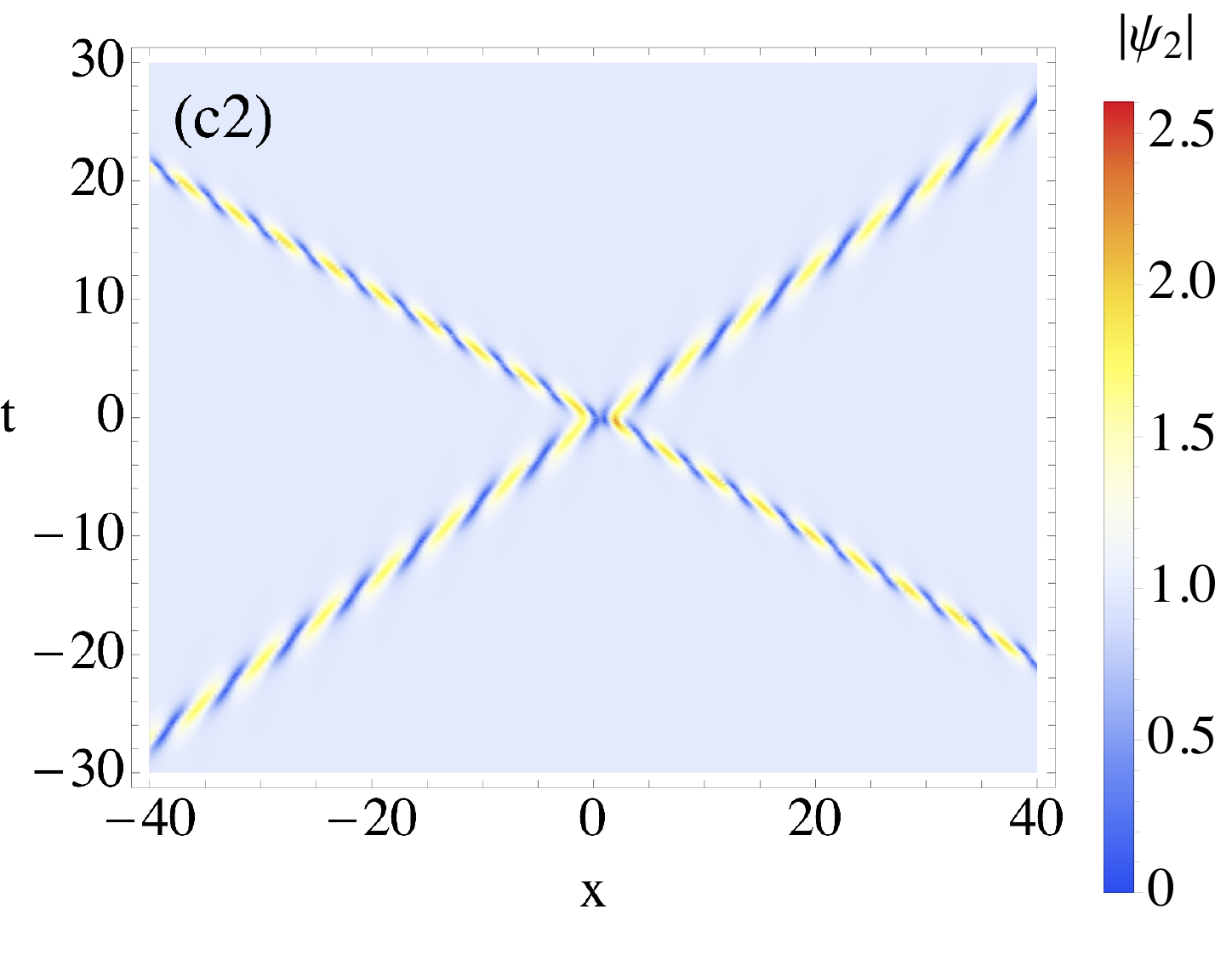}\\
\caption{
Elastic collision $\mathrm{III}+\mathrm{III} \rightarrow \mathrm{III} + \mathrm{III}$ of vector breathers with spectral parameters defined by Eq.~(\ref{parameters_2B}). (a,b) $|\psi_1|$ and $\mathrm{Arg}[\psi_{1}]$, where $\mathrm{Arg}$ means complex phase, after the breathers collision at $t=10.0$. (c) Spatio-temporal plot of the wave field evolution. The dotted green and red lines in (a,b) show a local approximation of the breathers after collision by single-breathers solutions from the asymptotic (\ref{2B_asymptotic_general}). Thin black dashed line in (a) shows how the first breather would have been if the it had been traveling alone.}
\label{fig_app1}
\end{figure}

\begin{figure}
\centering
    \includegraphics[width=0.31\linewidth]{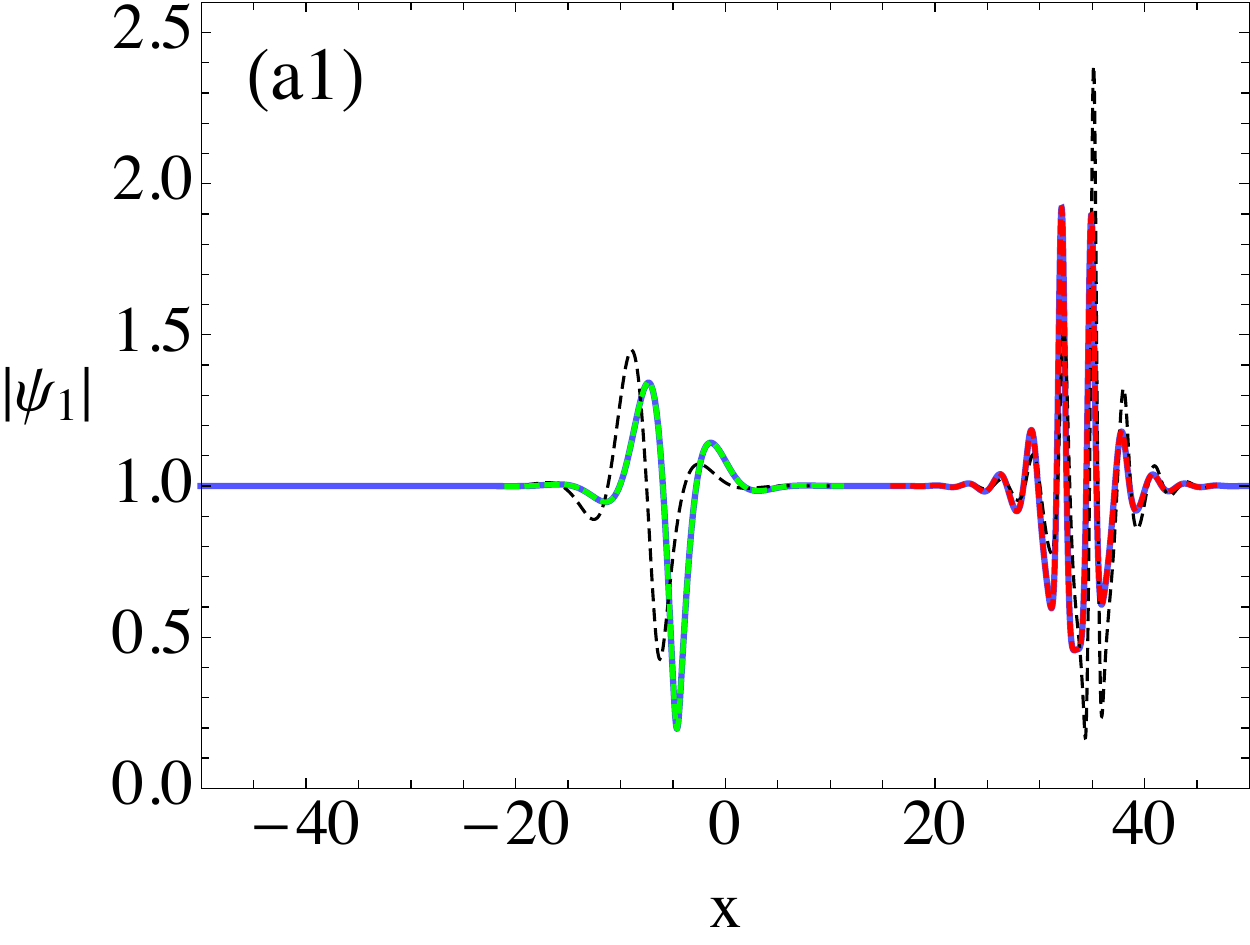}\,\,\,\,\,
    \includegraphics[width=0.31\linewidth]{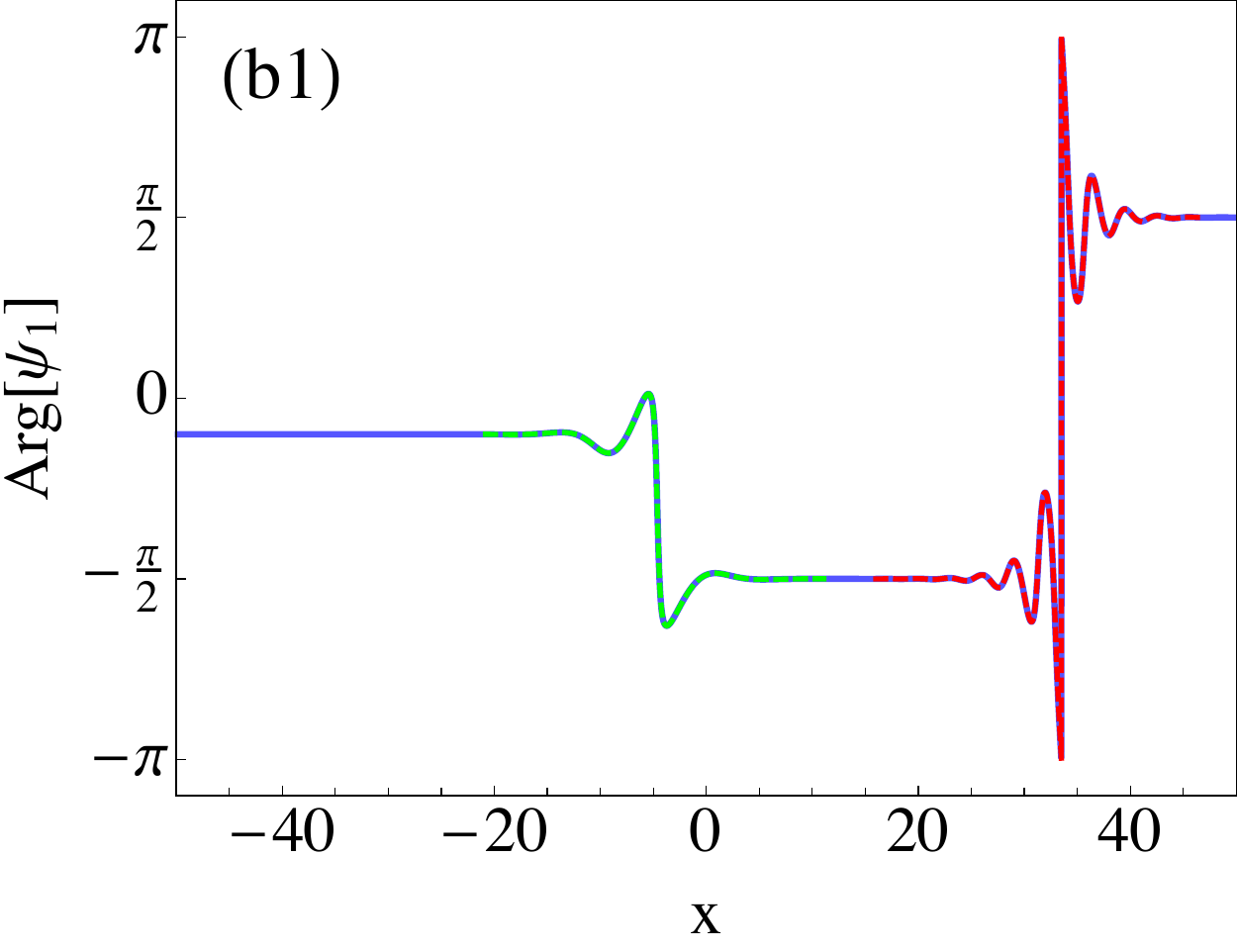}\,\,\,\,\,
    \includegraphics[width=0.32\linewidth]{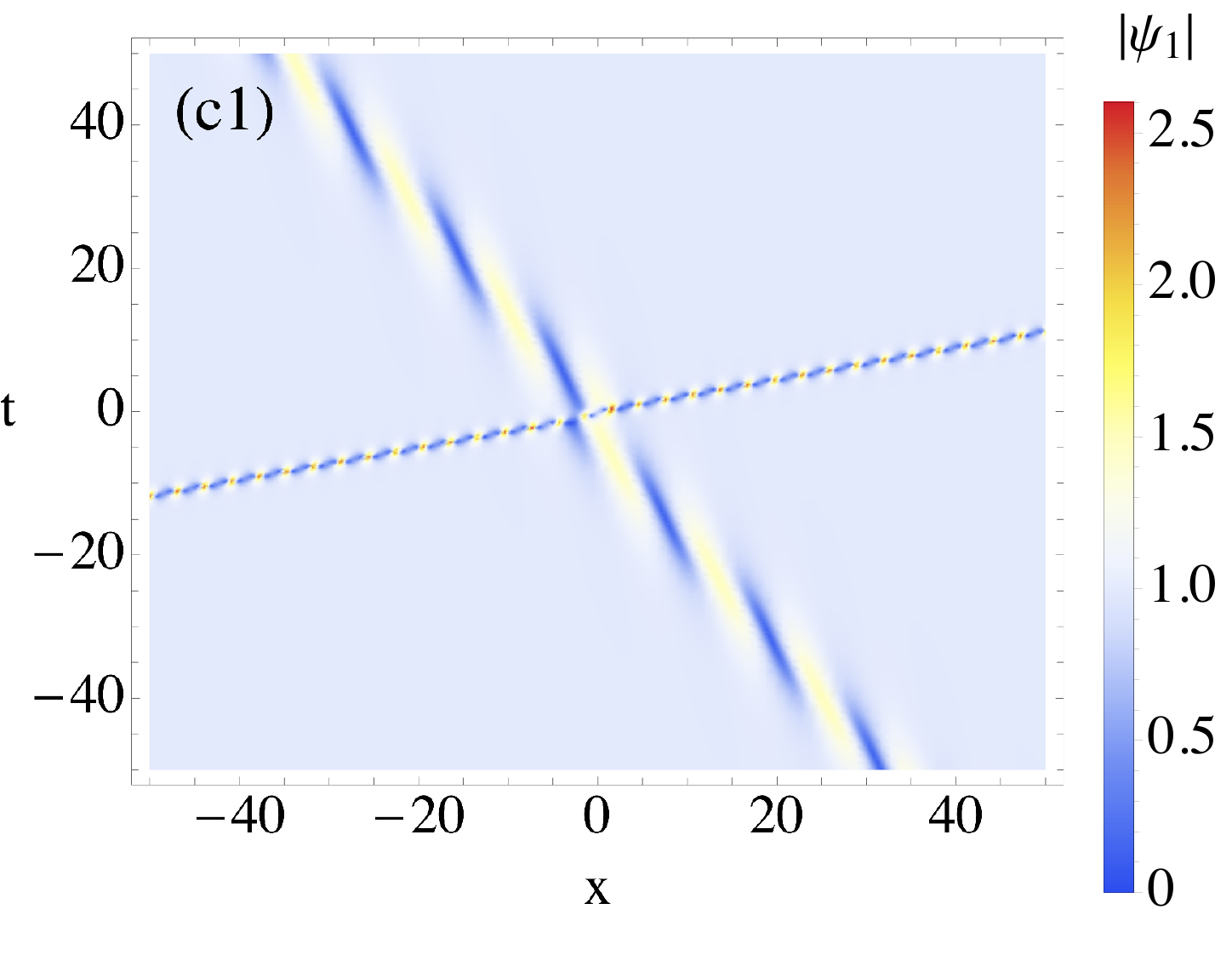}\\
    \includegraphics[width=0.31\linewidth]{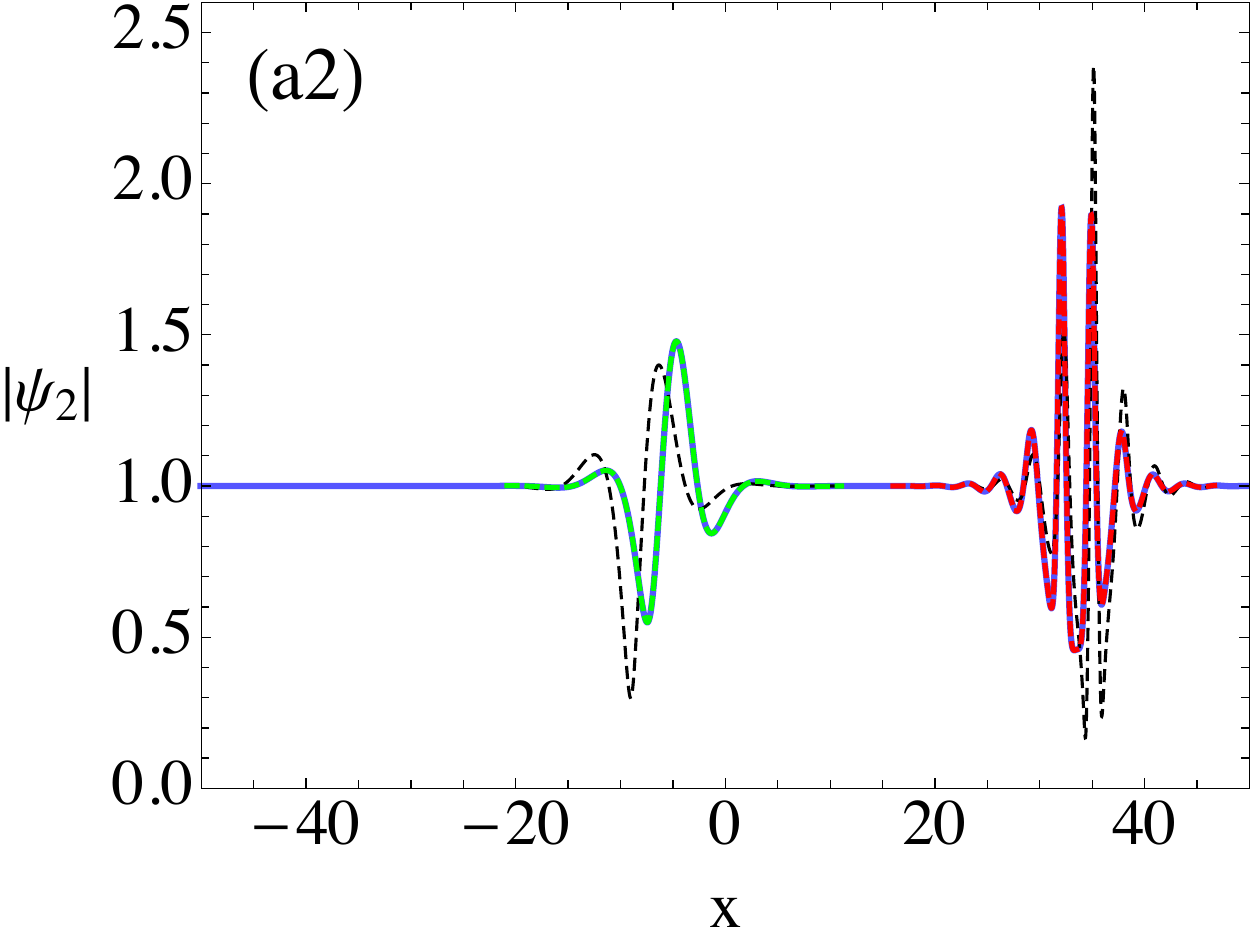}\,\,\,\,\,
    \includegraphics[width=0.31\linewidth]{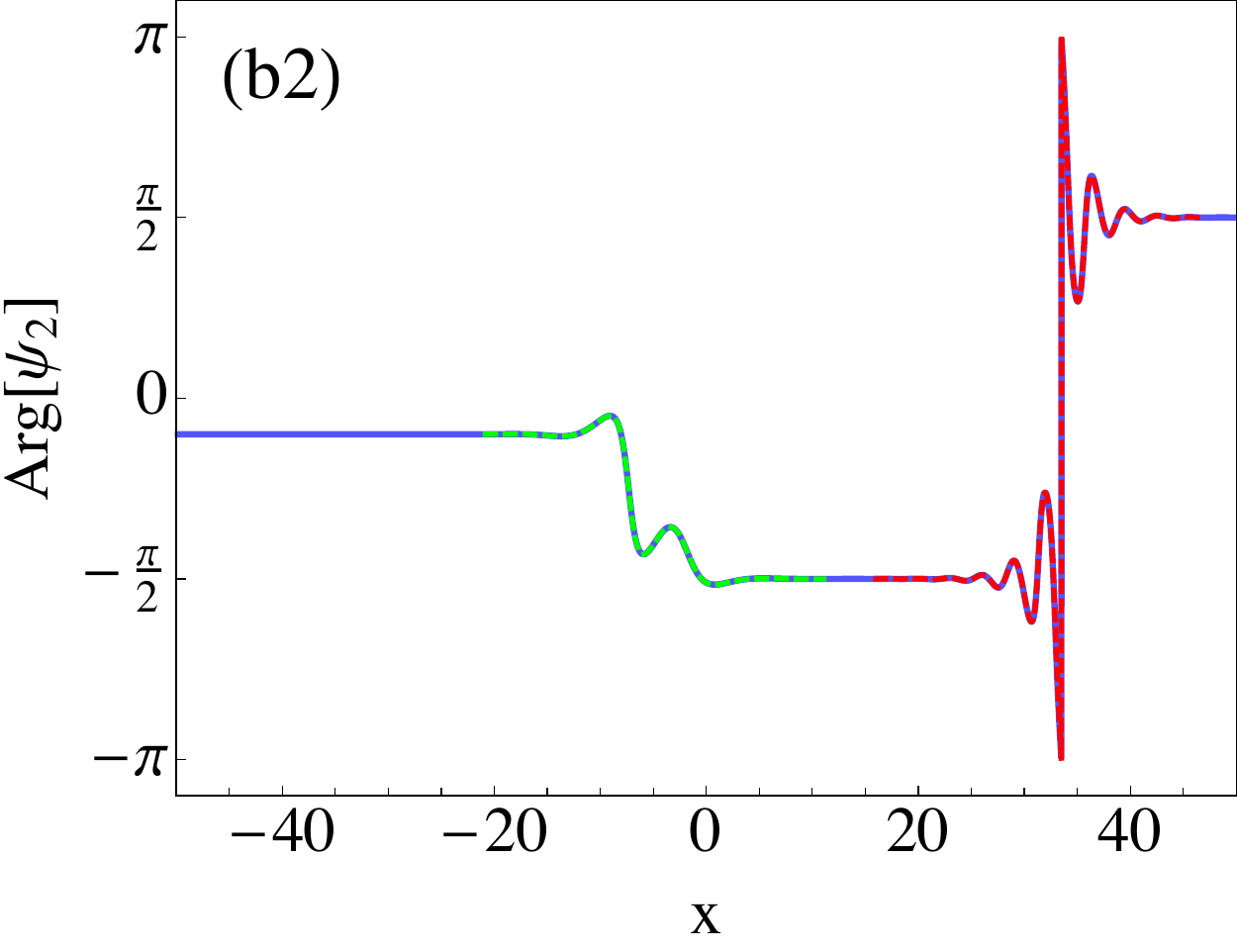}\,\,\,\,\,
    \includegraphics[width=0.32\linewidth]{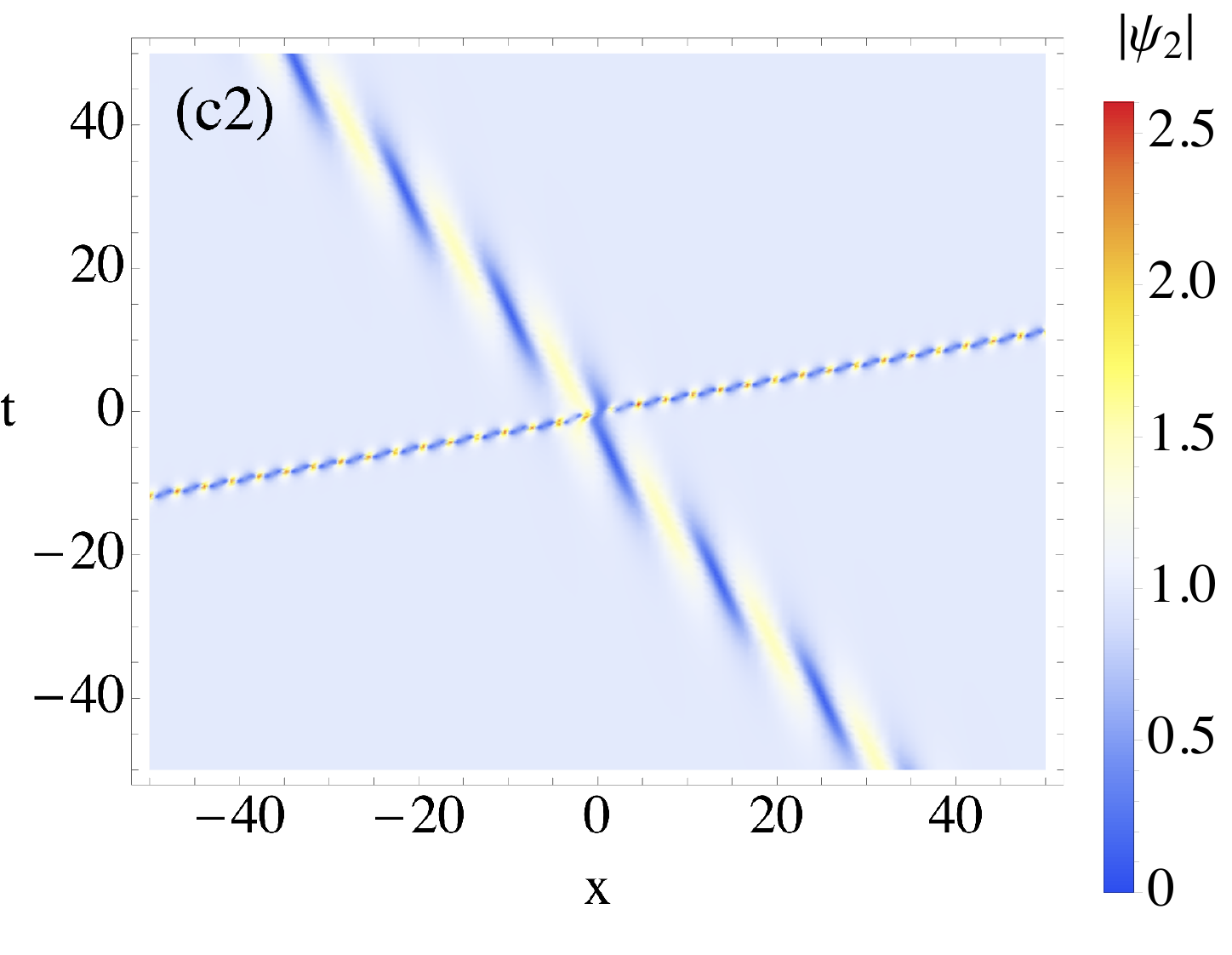}\\
\caption{
Elastic collision $\mathrm{I}+\mathrm{II} \rightarrow \mathrm{I} + \mathrm{II}$ of vector breathers with spectral parameters defined by Eq.~(\ref{parameters_2B_new}). (a,b) $|\psi_1|$ and $\mathrm{Arg}[\psi_{1}]$, where $\mathrm{Arg}$ means complex phase, after the breathers collision at $t=7.5$. (c) Spatio-temporal plot of the wave field evolution. The dotted green and red lines in (a,b) show a local approximation of the breathers after collision by single-breathers solutions from the asymptotic (\ref{2B_asymptotic_general}). Thin black dashed line in (a) shows how the first breather would have been if the it had been traveling alone.}
\label{fig_app2}
\end{figure}

\begin{figure}
\centering
    \includegraphics[width=0.31\linewidth]{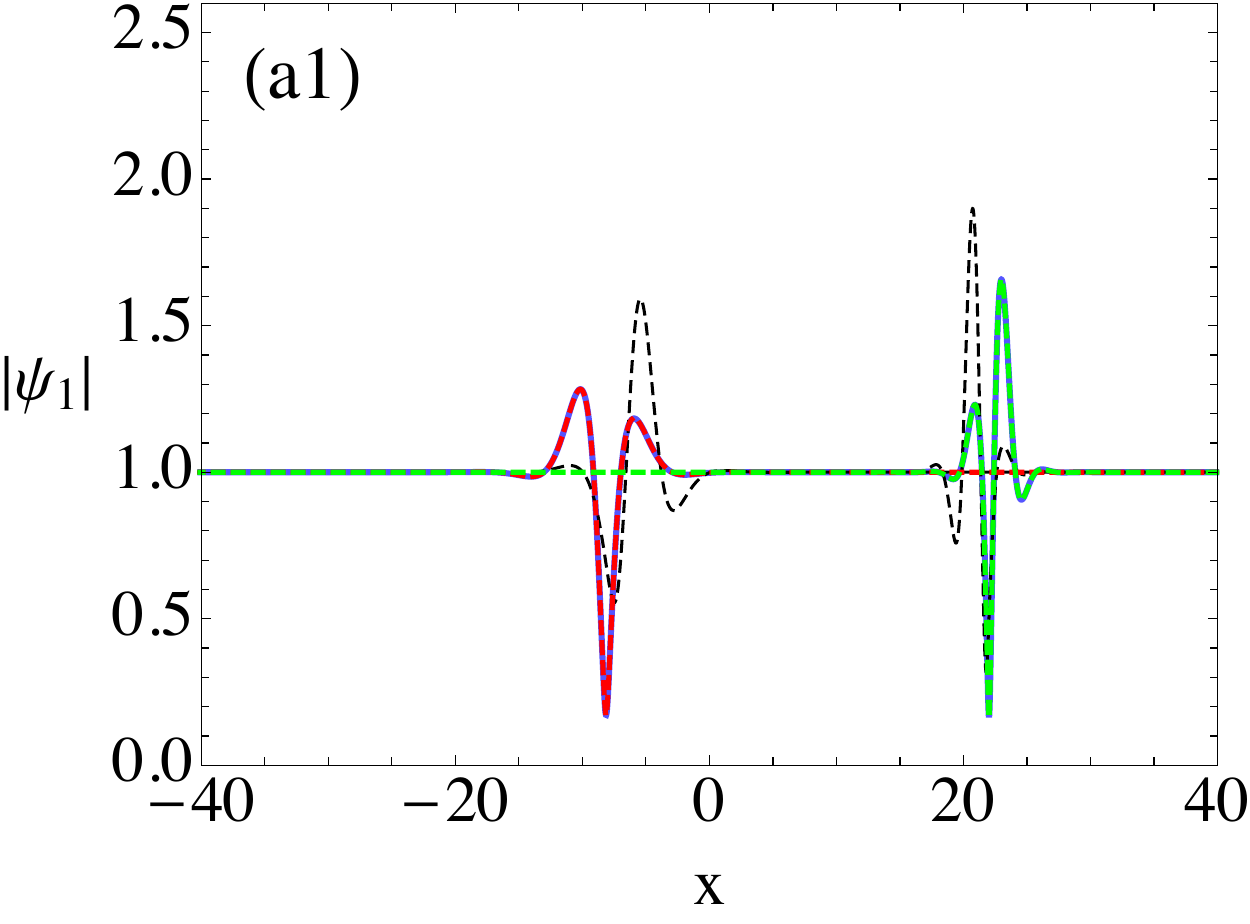}\,\,\,\,\,
    \includegraphics[width=0.31\linewidth]{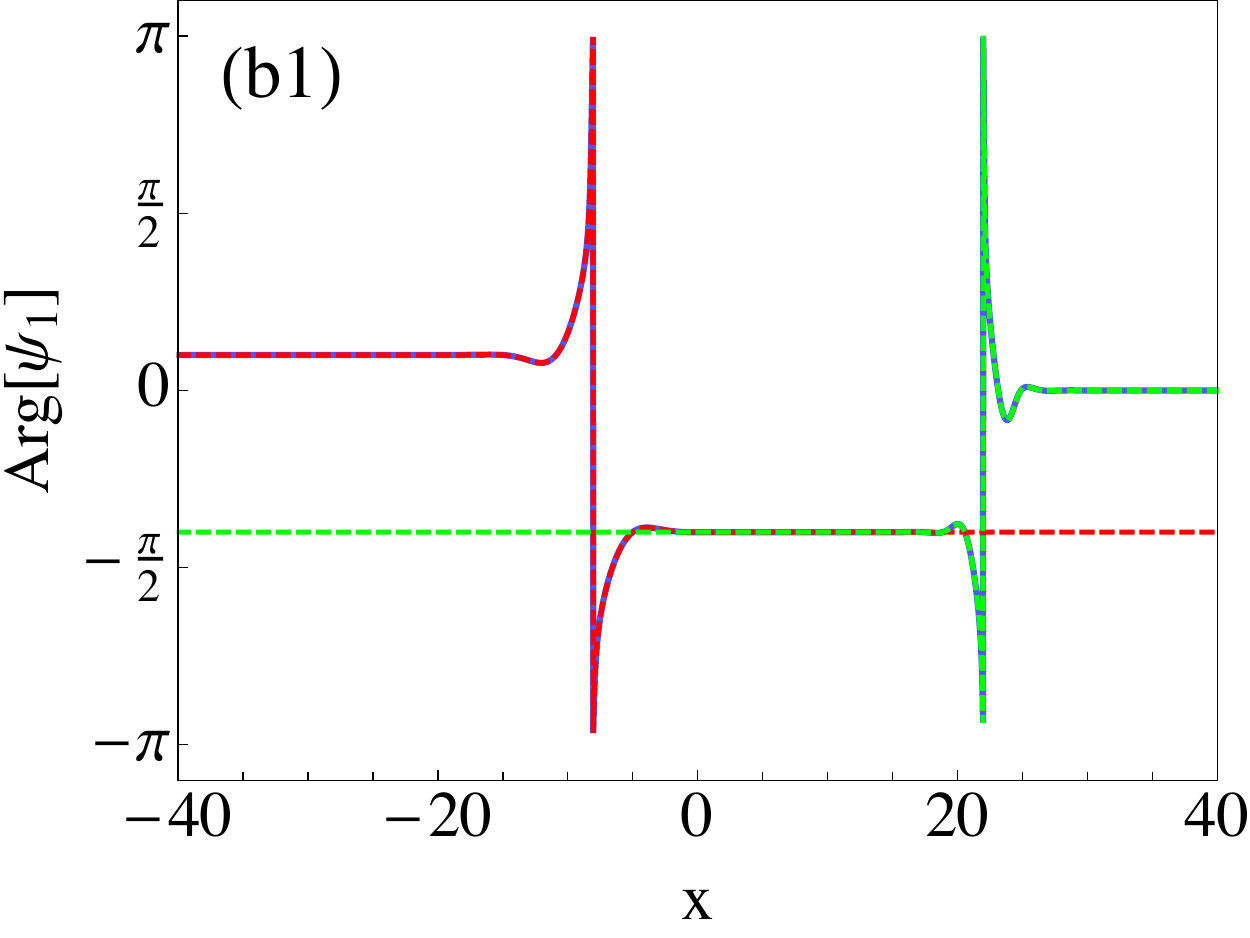}\,\,\,\,\,
    \includegraphics[width=0.32\linewidth]{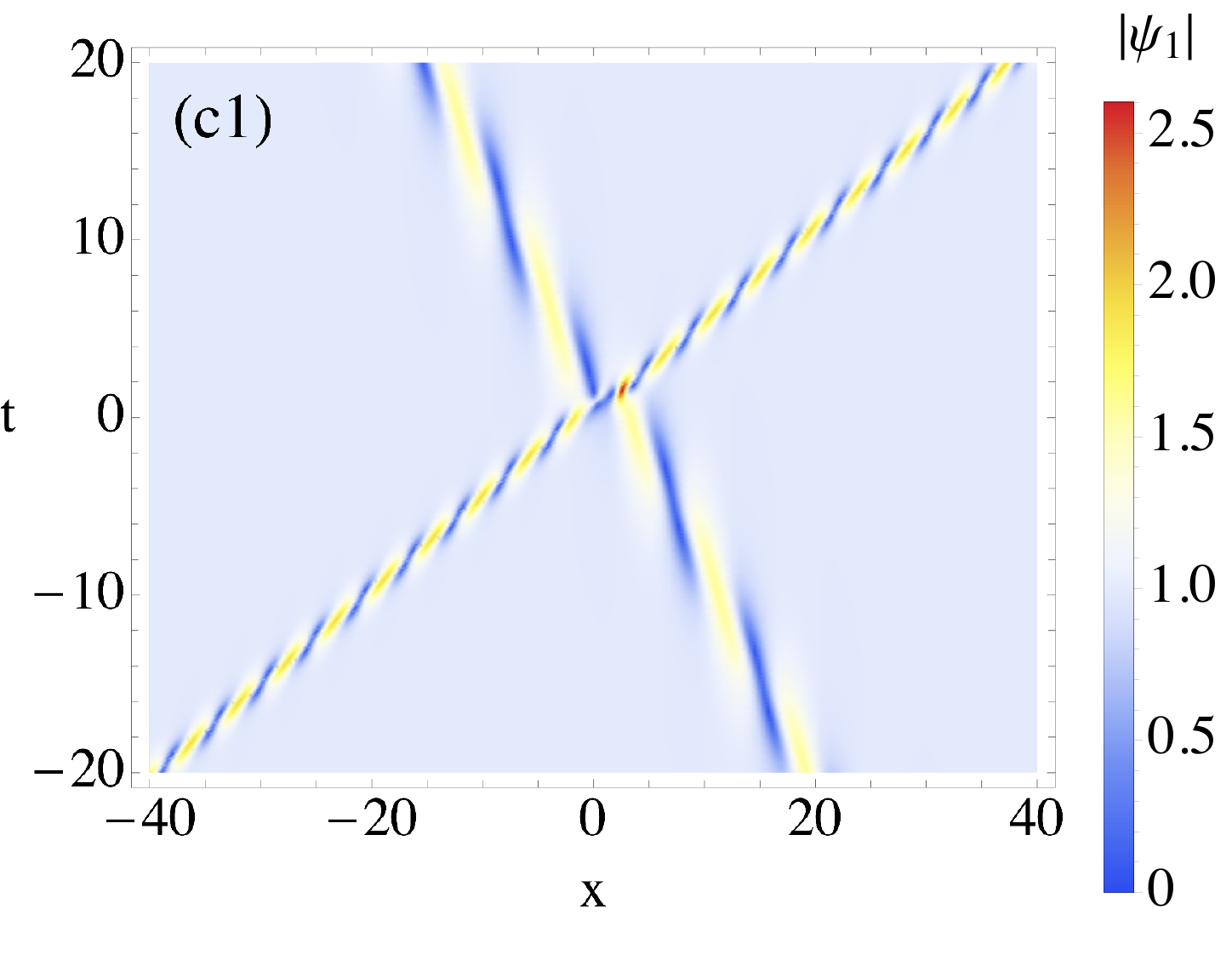}\\
    \includegraphics[width=0.31\linewidth]{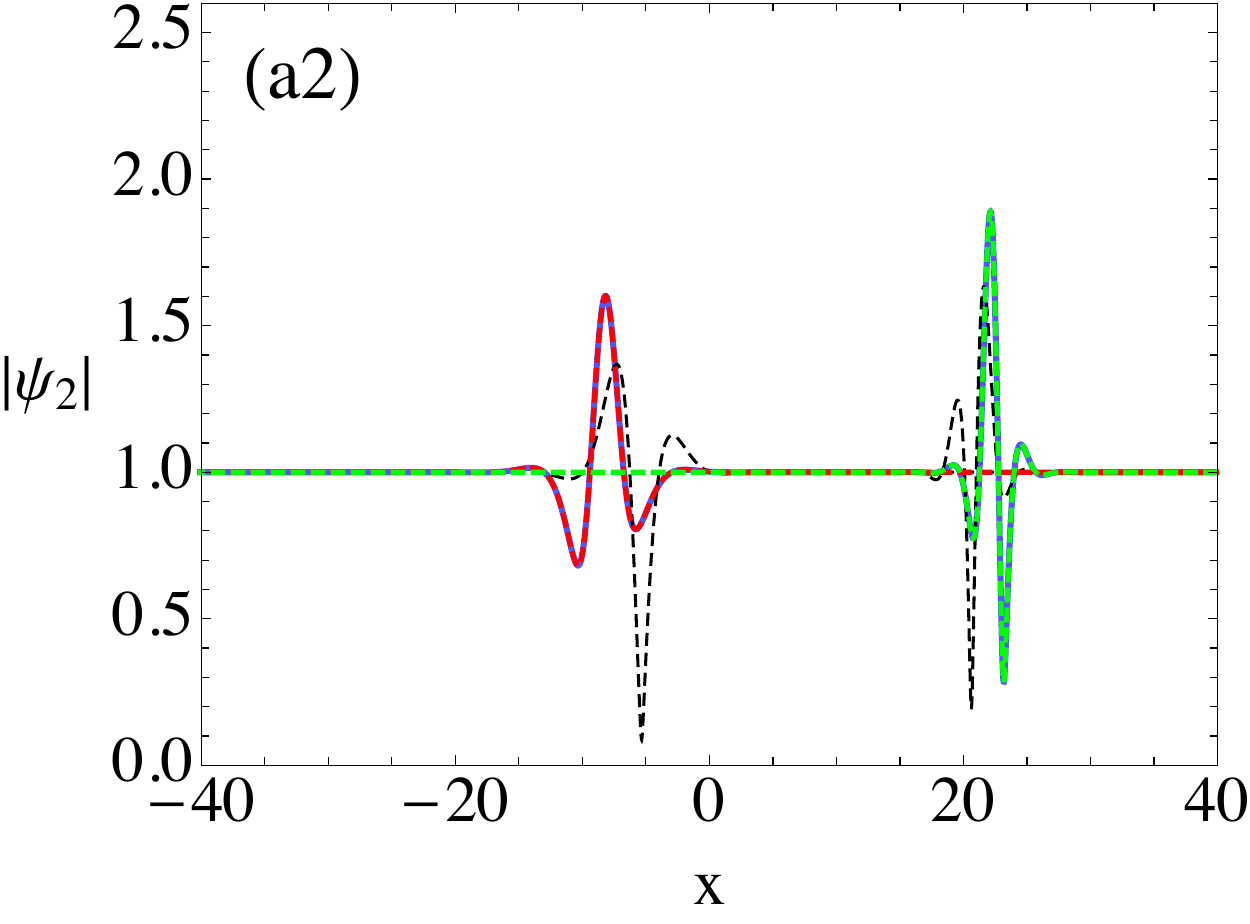}\,\,\,\,\,
    \includegraphics[width=0.31\linewidth]{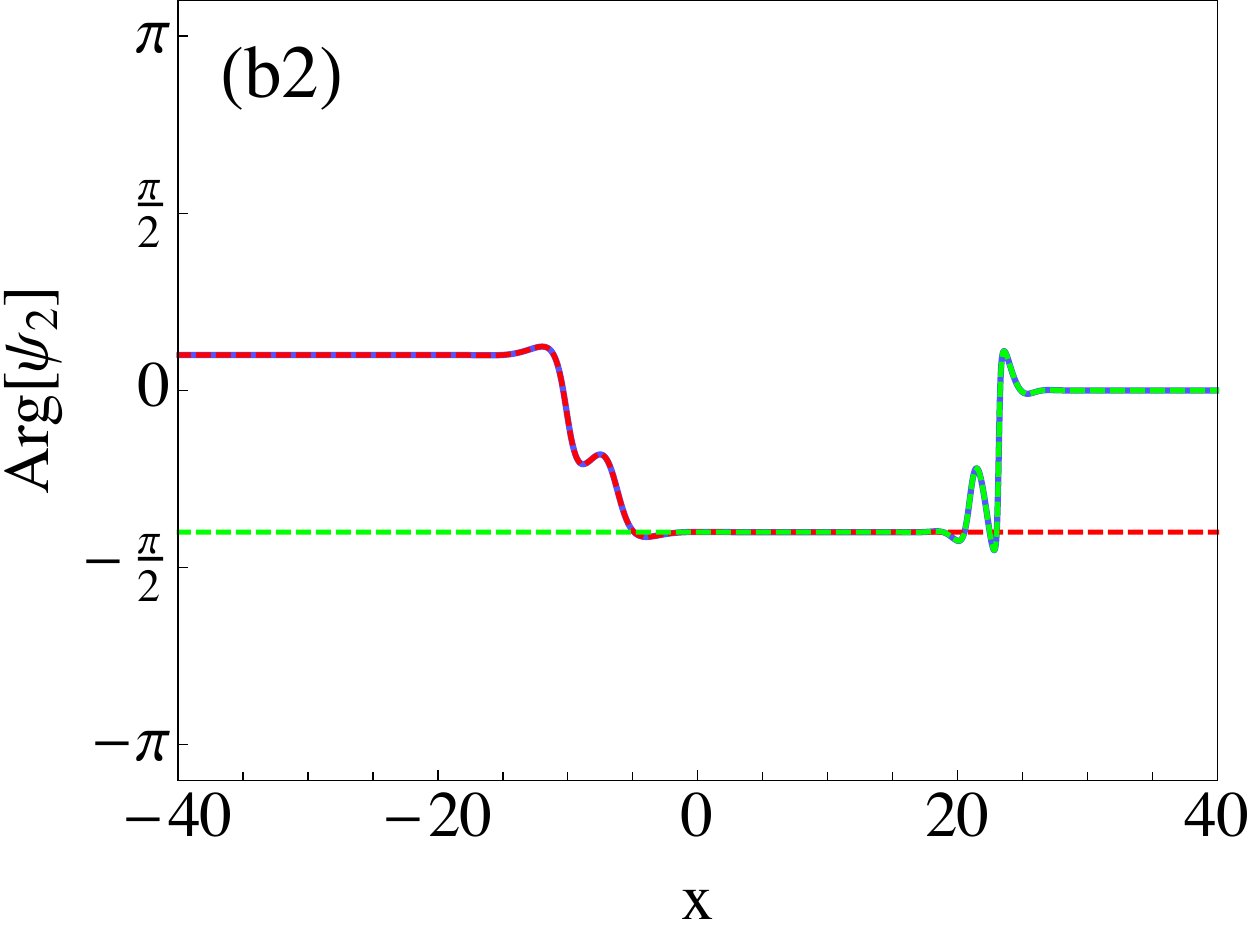}\,\,\,\,\,
    \includegraphics[width=0.32\linewidth]{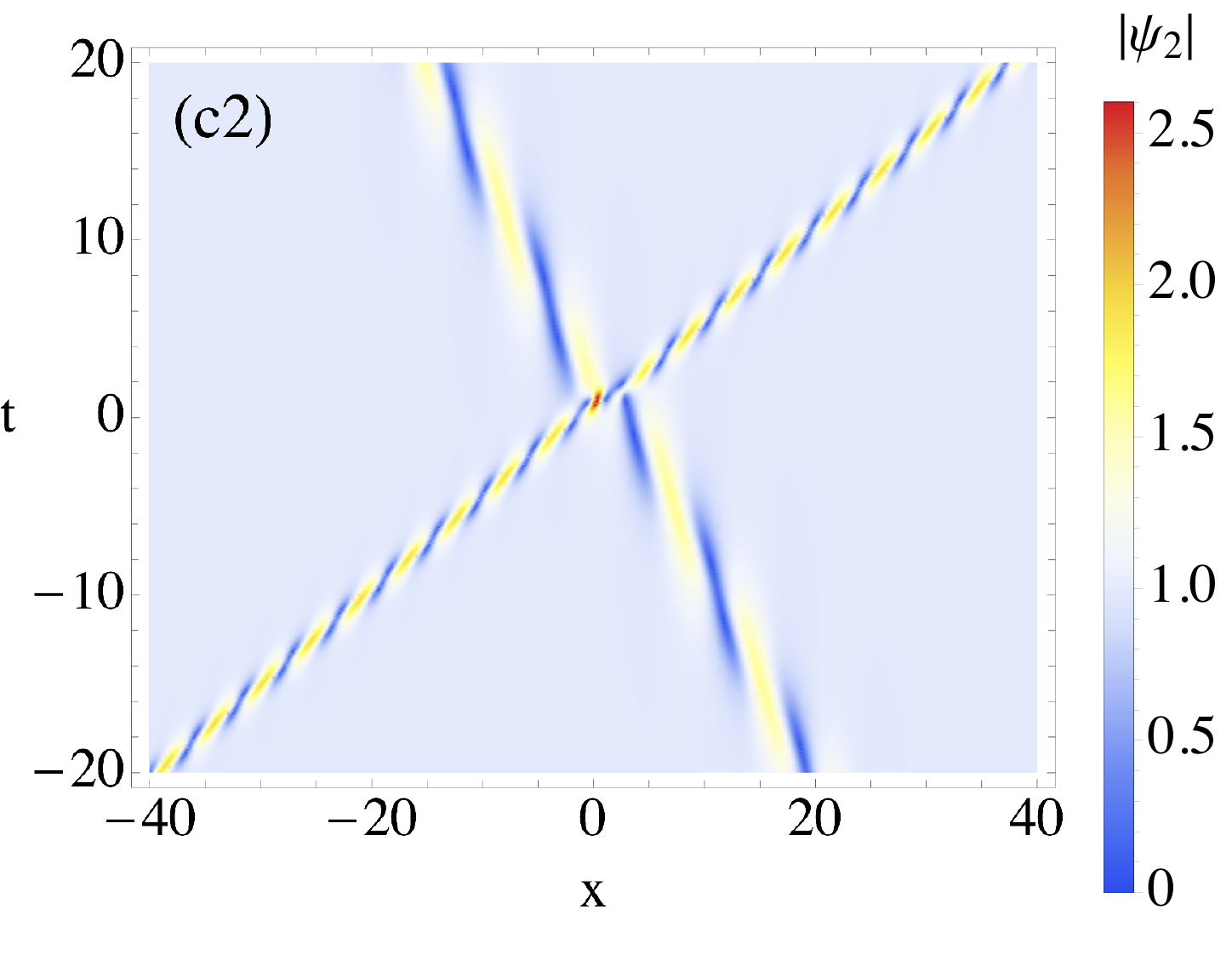}\\
\caption{
Elastic collision $\mathrm{II}+\mathrm{III} \rightarrow \mathrm{II} + \mathrm{III}$ of vector breathers with spectral parameters defined by Eq.~(\ref{parameters_2B_new}). (a,b) $|\psi_1|$ and $\mathrm{Arg}[\psi_{1}]$, where $\mathrm{Arg}$ means complex phase, after the breathers collision at $t=12.0$. (c) Spatio-temporal plot of the wave field evolution. The dotted green and red lines in (a,b) show a local approximation of the breathers after collision by single-breathers solutions from the asymptotic (\ref{2B_asymptotic_general}). Thin black dashed line in (a) shows how the first breather would have been if the it had been traveling alone.}
\label{fig_app3}
\end{figure}

In this section we provide additional to the shown in Sec. \ref{Sec:5} illustrations of the breather collisions and approximations of their asymptotic states using single breathers with shifted according to (\ref{2B_asymptotic_general}) parameters. Figs.~\ref{fig_app1}, \ref{fig_app2}, and \ref{fig_app3} show the processes $\mathrm{III}+\mathrm{III} \rightarrow \mathrm{III} + \mathrm{III}$, $\mathrm{I}+\mathrm{II} \rightarrow \mathrm{I} + \mathrm{II}$ and $\mathrm{II}+\mathrm{III} \rightarrow \mathrm{II} + \mathrm{III}$ respectively.

\newpage
\subsection{Tables with coefficients for the asymptotic state of elastic breather collision}\label{Sec:Appendix:B2}

In this section we complete the information on the asymptotic state of the two-breather collision, see Eq.~(\ref{2B_asymptotic_general}), providing the values of the coefficients $a_i^{\mathrm{B_i,\widetilde{B}_j}}$, $b_i^{\mathrm{B_i,\widetilde{B}_j}}$ etc. from Eq.~(\ref{2B_asymptotic_coeff_general}) and the wavefield asymptotics at $x\to\pm\infty$. Table \ref{table1} summarizes results of our derivations. Note, that the wavefield asymptotics of the two-breather solution represent a linear sum of the single breather wavefield asymptotics derived in the main part of the paper, see Eqs.~(\ref{asymptotics_I}), (\ref{asymptotics_II}), and (\ref{asymptotics_III}). We also remind, that the lower and the first upper indexes of the coefficients $a_i^{\mathrm{B_i,\widetilde{B}_j}}$, $b_i^{\mathrm{B_i,\widetilde{B}_j}}$, etc. indicate the breather for which the shift is presented, while the second upper index means the breather with which the studied one interacts. The lower index shows only what we call the studied breather, i.e., either the first or the second one. Meanwhile, the upper indexes also indicate the type of interacting breathers. For example, $b_i^{\mathrm{B_i,\widetilde{B}_j}}$ at $s_i=1$ ($a_i^{\mathrm{B_i,\widetilde{B}_j}}$ at $s_i=-1$) represents the correction $\delta_{0,i}^{+}$ for the position of the $i$-th breather of type $\mathrm{B}$ after the collision with the breather of type $\mathrm{\widetilde{B}}$.

We find that the coefficients $a_i^{\mathrm{B_i,\widetilde{B}_j}}$, $b_i^{\mathrm{B_i,\widetilde{B}_j}}$, etc. for the processes with different types of breathers often coincide with some of the coefficients computed for the process with the same types of breathers. For example, $b^{\mathrm{I}_i,\mathrm{II}_j}_i = a^{\mathrm{I}_i,\mathrm{I}_j}_i$, while $a^{\mathrm{II}_i,\mathrm{III}_j}_i = a^{\mathrm{II}_i,\mathrm{II}_j}_i$, see Table~\ref{table1}. In addition, in the expressions for $b^{\mathrm{II}_i,\mathrm{II}_j}_i$ and $b^{\mathrm{III}_i,\mathrm{III}_j}_i$ appears the same logarithm. Nevertheless we do not identify any symmetry properties of the coefficients and consider all of them as independent.

\newpage
\begin{table}[!h]
\centering 
\renewcommand{\arraystretch}{1.3}
\begin{tabular}{|c| c| c|}
\hline 
Process & $\begin{aligned} \text{Integration constants and} \\ \text{wavefield asymptotics} \end{aligned}$ & Space-phase shifts coefficients \\[0.5ex] 
\hline 
$\begin{aligned} \mathrm{I}_i+\mathrm{I}_j \rightarrow \\ \rightarrow \mathrm{I}_i + \mathrm{I}_j \end{aligned}$
&
$\begin{aligned}
C_{i,0} = 0,
\\
C_{i,1} = e^{\mathrm{Im}[\zeta_i]\delta_i + i\theta_i/2},
\\
C_{i,2} =  C_{i,1}^{-1}.
\\
\\
\psi_{1,2}^{\pm\infty} = A_{1,2} e^{\pm 2\,\mathrm{i}\,(\alpha_i+\alpha_j)}.
\end{aligned}$
&
$\begin{aligned}
a^{\mathrm{I}_i,\mathrm{I}_j}_i = l_\mathrm{I}(\lambda_i)\,\log \left|\frac{(r_i-r_j^*)(1+r_i r_j)}{(r_i-r_j)(1+r_i r_j^*)} \right|,
\\
b^{\mathrm{I}_i,\mathrm{I}_j}_i = -a^{\mathrm{I}_i,\mathrm{I}_j}_i,
\\
c^{\mathrm{I}_i,\mathrm{I}_j}_i = \mathrm{Arg} \left[\frac{(r_i^*-r_j^*)(1+r_i r_j)}{(r_i^*-r_j)(1+r_i r_j^*) \sin \alpha_j}\right]
\\
d^{\mathrm{I}_i,\mathrm{I}_j}_i = - c^{\mathrm{I}_i,\mathrm{I}_j}_i,
\qquad
e^{\mathrm{I}_i,\mathrm{I}_j}_i = \alpha_j, \qquad f^{\mathrm{I}_i,\mathrm{I}_j}_i = -\alpha_j.
\end{aligned}$
\\ 
\hline 
$\begin{aligned}\mathrm{II}_i+\mathrm{II}_j \rightarrow\\\rightarrow \mathrm{II}_i + \mathrm{II}_j\end{aligned}$ 
&
$\begin{aligned}
C_{i,0} = e^{-\mathrm{Im}[\lambda_i]\delta_i - i\theta_i /2},
\\
C_{i,1} = 0,
\\
C_{i,2} = e^{-\mathrm{Im}[\zeta_i]\delta_i + i\theta_i/2}.
\\
\\
\psi_{1,2}^{-\infty} = A_{1,2} e^{2\,\mathrm{i}\,(\alpha_i+\alpha_j)},
\\
\psi_{1,2}^{+\infty} = A_{1,2}.
\end{aligned}$
&
$\begin{aligned}
a^{\mathrm{II}_i,\mathrm{II}_j}_i = - l_\mathrm{II}(\lambda_i)\,\log 
\left[1 - \frac{n}{m}\right] ,
\\
b^{\mathrm{II}_i,\mathrm{II}_j}_i = l_\mathrm{II}(\lambda_i) \,\log 
\left[1 - \frac{n |1+r_i^* r_j |^2}{m (1+|r_i|^2) (1+|r_j|^2)}\right] ,
\\
c^{\mathrm{II}_i,\mathrm{II}_j}_i = -\mathrm{Arg}\left[1-\frac{n_i}{m_j}\right] ,
\qquad  d^{\mathrm{II}_i,\mathrm{II}_j}_i = \mathrm{Arg}\left[ 1-\frac{n_i (1+r_i r_j^*)}{m_j (1+|r_j|^2)} \right],
\\
e^{\mathrm{II}_i,\mathrm{II}_j}_i = 0, \qquad f^{\mathrm{II}_i,\mathrm{II}_j}_i = \alpha_j.
\end{aligned}$
\\    
\hline
$\begin{aligned}\mathrm{III}_i+\mathrm{III}_j \rightarrow\\\rightarrow \mathrm{III}_i + \mathrm{III}_j\end{aligned}$ 
&
$\begin{aligned}
C_{i,0} = e^{-\mathrm{Im}[\lambda_i]\delta_i - i\theta_i /2},
\\
\mathrm{i}r_iC_{i,1} = e^{\mathrm{Im}[\zeta_i]\delta_i + i\theta_i/2},
\\
C_{i,2} = 0.
\\
\\
\psi_{1,2}^{-\infty} = A_{1,2} e^{-2\,\mathrm{i}\,(\alpha_i+\alpha_j)},
\\
\psi_{1,2}^{+\infty} = A_{1,2}.
\end{aligned}$
&
$\begin{aligned}
a^{\mathrm{III}_i,\mathrm{III}_j}_i = -l_\mathrm{III}(\lambda_i)\,\log \left[1 - \frac{n}{m}\right],
\\
b^{\mathrm{III}_i,\mathrm{III}_j}_i = l_\mathrm{III}(\lambda_i) \,\log \left[1 - \frac{n |1+r_i^* r_j |^2}{m (1+|r_i|^2) (1+|r_j|^2)}\right],
\\
c^{\mathrm{III}_i,\mathrm{III}_j}_i = c^{\mathrm{II}_i,\mathrm{II}_j}_i,\qquad 
d^{\mathrm{III}_i,\mathrm{III}_j}_i = \mathrm{Arg}\left[ 1-\frac{n_i r_j (1+r_i r_j^*)}{m_j r_i (1+|r_j|^2)} \right] 
\\
e^{\mathrm{III}_i,\mathrm{III}_j}_i = 0,\qquad f^{\mathrm{III}_i,\mathrm{III}_j}_i = -\alpha_j.
\end{aligned}$
\\
\hline
$\begin{aligned}\mathrm{I}_i+\mathrm{II}_j \rightarrow\\\rightarrow \mathrm{I}_i + \mathrm{II}_j\end{aligned}$ 
&
$\begin{aligned}
C_{i,0} = 0,\quad
C_{i,1} = e^{\mathrm{Im}[\zeta_i]\delta_i + i\theta_i/2},
\\
C_{i,2} =  C_{i,1}^{-1}.
\\
\\
C_{j,0} = e^{-\mathrm{Im}[\lambda_j]\delta_j - i\theta_j /2},
\quad
C_{j,1} = 0,
\\
C_{j,2} = e^{-\mathrm{Im}[\zeta_j]\delta_j + i\theta_j/2}.
\\
\\
\psi_{1,2}^{-\infty} = A_{1,2} e^{2\,\mathrm{i}\,(\alpha_j-\alpha_i)},
\\
\psi_{1,2}^{+\infty} = A_{1,2} e^{2\,\mathrm{i}\,\alpha_i}.
\end{aligned}$
&
$\begin{aligned}
a^{\mathrm{I}_i,\mathrm{II}_j}_i = 0,
\qquad
b^{\mathrm{I}_i,\mathrm{II}_j}_i = a^{\mathrm{I}_i,\mathrm{I}_j}_i,
\qquad
c^{\mathrm{I}_i,\mathrm{II}_j}_i = 0,
\\
d^{\mathrm{I}_i,\mathrm{II}_j}_i = c^{\mathrm{I}_i,\mathrm{I}_j}_i,
\qquad
e^{\mathrm{I}_i,\mathrm{II}_j}_i = 0, \qquad f^{\mathrm{I}_i,\mathrm{II}_j}_i = \alpha_j.
\\
\\
a^{\mathrm{II}_j,\mathrm{I}_i}_j = b^{\mathrm{II}_j,\mathrm{II}_i}_j,
\quad
b^{\mathrm{II}_j,\mathrm{I}_i}_j = l_\mathrm{II}(\lambda_j) \,\log 
\left[1 - \frac{n |r_j - r^*_i |^2}{m (1+|r_j|^2) (1+|r_i|^2)}\right],
\\
c^{\mathrm{II}_j,\mathrm{I}_i}_j = d^{\mathrm{II}_j,\mathrm{II}_i}_j,
\quad
d^{\mathrm{II}_j,\mathrm{I}_i}_j = \mathrm{Arg}\left[ 1-\frac{n_j r_i (r^*_i - r_j)}{m_i (1+|r_i|^2)} \right],
\\
e^{\mathrm{II}_j,\mathrm{I}_i}_j = \alpha_i,
\quad f^{\mathrm{II}_j,\mathrm{I}_i}_j = -\alpha_i,
\end{aligned}$
\\ 
\hline
$\begin{aligned}\mathrm{I}_i+\mathrm{III}_j \rightarrow\\\rightarrow \mathrm{I}_i + \mathrm{III}_j\end{aligned}$ 
&
$\begin{aligned}
C_{i,0} = 0,\quad
C_{i,1} = e^{\mathrm{Im}[\zeta_i]\delta_i + i\theta_i/2},
\\
C_{i,2} =  C_{i,1}^{-1}.
\\
\\
C_{j,0} = e^{-\mathrm{Im}[\lambda_j]\delta_j - i\theta_j /2},
\\
\mathrm{i}r_jC_{j,1} = e^{\mathrm{Im}[\zeta_j]\delta_j + i\theta_j/2},
\quad
C_{j,2} = 0.
\\
\\
\psi_{1,2}^{-\infty} = A_{1,2} e^{-2\,\mathrm{i}\,(\alpha_i+\alpha_j)},
\\
\psi_{1,2}^{+\infty} = A_{1,2} e^{2\,\mathrm{i}\,\alpha_i}.
\end{aligned}$
&
$\begin{aligned}
a^{\mathrm{I}_i,\mathrm{III}_j}_i = 0,
\qquad
b^{\mathrm{I}_i,\mathrm{III}_j}_i = -a^{\mathrm{I}_i,\mathrm{I}_j}_i,
\qquad
c^{\mathrm{I}_i,\mathrm{III}_j}_i = 0,
\\
d^{\mathrm{I}_i,\mathrm{III}_j}_i = -c^{\mathrm{I}_i,\mathrm{I}_j}_i,
\qquad
e^{\mathrm{I}_i,\mathrm{III}_j}_i = 0, \qquad f^{\mathrm{I}_i,\mathrm{II}_j}_i = -\alpha_j.
\\
\\
a^{\mathrm{III}_j,\mathrm{I}_i}_j = l_\mathrm{III}(\lambda_j) \,\log 
\left[1 - \frac{n |r_j - r^*_i |^2}{m (1+|r_j|^2) (1+|r_i|^2)}\right],
\\
b^{\mathrm{III}_j,\mathrm{I}_i}_j = b^{\mathrm{III}_j,\mathrm{III}_i}_j,
\quad
c^{\mathrm{III}_j,\mathrm{I}_i}_j = \mathrm{Arg}\left[ 1-\frac{n_j (r_j - r^*_i)}{m_i r_j (1+|r_i|^2)} \right],
\\
d^{\mathrm{III}_j,\mathrm{I}_i}_j = d^{\mathrm{III}_j,\mathrm{III}_i}_j,
\quad
e^{\mathrm{II}_j,\mathrm{I}_i}_j = \alpha_i,
\quad
f^{\mathrm{II}_j,\mathrm{I}_i}_j = -\alpha_i.
\end{aligned}$
\\ 
\hline
$\begin{aligned}\mathrm{II}_i+\mathrm{III}_j \rightarrow\\\rightarrow \mathrm{II}_i + \mathrm{III}_j\end{aligned}$ 
&
$\begin{aligned}
C_{i,0} = e^{-\mathrm{Im}[\lambda_i]\delta_i - i\theta_i /2},
\quad
C_{i,1} = 0,
\\
C_{i,2} = e^{-\mathrm{Im}[\zeta_i]\delta_i + i\theta_i/2}.
\\
\\
C_{j,0} = e^{-\mathrm{Im}[\lambda_j]\delta_j - i\theta_j /2},
\\
\mathrm{i}r_jC_{j,1} = e^{\mathrm{Im}[\zeta_j]\delta_j + i\theta_j/2},
\quad
C_{j,2} = 0.
\\
\\
\psi_{1,2}^{-\infty} = A_{1,2} e^{2\,\mathrm{i}\,(\alpha_i-\alpha_j)},
\\
\psi_{1,2}^{+\infty} = A_{1,2}.
\end{aligned}$
&
$\begin{aligned}
\\
a^{\mathrm{II}_i,\mathrm{III}_j}_i = a^{\mathrm{II}_i,\mathrm{II}_j}_i,
\qquad
b^{\mathrm{II}_i,\mathrm{III}_j}_i = b^{\mathrm{II}_i,\mathrm{I}_j}_i,
\qquad
c^{\mathrm{II}_i,\mathrm{III}_j}_i = c^{\mathrm{II}_i,\mathrm{II}_j}_i,
\\
d^{\mathrm{II}_i,\mathrm{III}_j}_i = d^{\mathrm{II}_i,\mathrm{I}_j}_i,
\qquad
e^{\mathrm{II}_i,\mathrm{III}_j}_i = 0, 
\qquad
f^{\mathrm{II}_i,\mathrm{II}_j}_i = -\alpha_j.
\\
\\
a^{\mathrm{III}_j,\mathrm{II}_i}_j = a^{\mathrm{III}_j,\mathrm{III}_i}_j,
\qquad
b^{\mathrm{III}_j,\mathrm{II}_i}_j = a^{\mathrm{III}_j,\mathrm{I}_i}_j,
\qquad
c^{\mathrm{III}_j,\mathrm{II}_i}_j = c^{\mathrm{III}_j,\mathrm{III}_i}_j,
\\
d^{\mathrm{III}_j,\mathrm{II}_i}_j = c^{\mathrm{III}_j,\mathrm{I}_i}_j,
\qquad
e^{\mathrm{III}_j,\mathrm{II}_i}_j = 0, 
\qquad
f^{\mathrm{III}_j,\mathrm{I}_i}_j = \alpha_i.
\end{aligned}$
\\ 
\hline
\end{tabular}
\caption{Values of the integration constants, asymptotic phases, and space-phase shifts coefficients in collisions of vector breathers. Left column indicates the collision process type in the form $\mathrm{B_i}+\mathrm{\widetilde{B}_j} \rightarrow \mathrm{B_i} + \mathrm{\widetilde{B}_j}$, where $i=1\,\text{or}\,2$ while $j = 2\,\text{or}\,1$ respectively. The middle column shows the corresponding parametrization of the integration constants $\mathbf{C}_i$ and asymptotic values of the wavefield at $x\to\pm\infty$. Right column presents the values of the space-phase shifts coefficients defined in Eq.~(\ref{2B_asymptotic_coeff_general}).  When $\mathrm{B}$ and $\mathrm{\widetilde{B}}$ represent the same type we leave only $\Delta\delta^{\mathrm{B_i,B_j}}_i$ value in the table. The meaning of the indexes has been explained in the main text. For example, $b_i^{\mathrm{I_i,II_j}}$ at $s_i=1$ ($a_i^{\mathrm{I_i,II_j}}$ at $s_i=-1$) represents the correction to the position of the $i$-th breather of type $\mathrm{I}$ at large time after the collision with the breather of type $\mathrm{II}$.}
\label{table1}
\end{table}

\end{document}